\documentclass[twocolumn]{aastex63}
\begin{document}

\newcommand{\logg} {\log \textsl{\textrm{g}}}
\newcommand{\qh} {q({\rm H})}
\newcommand{\qhe} {q({\rm He})}
\newcommand{\Te} {T_{\rm eff}}
\newcommand{\mv} {$M_V$}
\newcommand{\msun} {$M_\odot$}
\newcommand{\mbol} {M_{\rm bol}}
\newcommand\gta{\lower 0.5ex\hbox{$\buildrel > \over \sim\ $}} 
\newcommand\lta{\lower 0.5ex\hbox{$\buildrel < \over \sim\ $}} 
\newcommand{\nh} {N({\rm H})/N({\rm He})}
\newcommand{\nhe} {N({\rm He})/N({\rm H})}
\newcommand{\ha} {$\rm{H}{\alpha}$}
\newcommand{\hb} {$\rm{H}{\beta}$}
\newcommand{\hgamma} {H$\gamma$}
\newcommand{\hdelta} {H$\delta$}
\newcommand{\hepsilon} {H$\epsilon$}
\newcommand{\nuv} {${\rm{NUV}}-V$}
\newcommand{\rsun} {$R_{\odot}$}
\newcommand{\lsun} {$L_{\odot}$}
\newcommand{\loghe} {$\log$ He/H}
\newcommand{\logh} {\log {\rm H/He}}
\newcommand{\heii} {He {\sc ii} $\lambda$4686}
\newcommand{\hei} {He {\sc i} $\lambda$4471}
\newcommand{\heif} {He {\sc i} $\lambda$5877}
\newcommand{\civ} {C {\sc iv} $\lambda$4658}

\title{On the Nature of Ultracool White Dwarfs: Not so Cool Afterall}

\author[0000-0003-2368-345X]{P. Bergeron}
\affiliation{D{\'e}partement de Physique, Universit{\'e} de Montr{\'e}al, C.P. 6128, Succ. Centre-Ville, 
Montr{\'e}al, Quebec H3C 3J7, Canada}

\author[0000-0001-6098-2235]{Mukremin Kilic}
\affiliation{Homer L. Dodge Department  of Physics and Astronomy, University of Oklahoma,
440 W. Brooks St., Norman, OK, 73019 USA}

\author[0000-0002-9632-1436]{Simon Blouin}\affil{Department of Physics and
Astronomy, University of Victoria, Victoria, BC V8W 2Y2, Canada}

\author[0000-0002-2384-1326]{A. B\'edard}\affiliation{D{\'e}partement de 
Physique, Universit{\'e} de Montr{\'e}al, C.P. 6128, Succ. Centre-Ville, 
Montr{\'e}al, Quebec H3C 3J7, Canada}

\author[0000-0002-3681-2989]{S. K. Leggett}\affil{Gemini Observatory/NSF's NOIRLab,
670 N. A'ohoku Place, Hilo, HI 96720, USA}

\author[0000-0002-4462-2341]{Warren R.\ Brown} \affiliation{Smithsonian
Astrophysical Observatory, 60 Garden Street, Cambridge, MA 02138 USA}

\shortauthors{Bergeron et al.}
\shorttitle{The Nature of Ultracool WDs}

\begin{abstract}

A recent analysis of the 100 pc white dwarf sample in the SDSS
footprint demonstrated for the first time the existence of a well
defined ultracool --- or IR-faint --- white dwarf sequence in the
Hertzsprung-Russell diagram. Here we take advantage of this discovery
to enlarge the IR-faint white dwarf sample threefold. We expand our
selection to the entire Pan-STARRS survey footprint as well as the
Montreal White Dwarf Database 100 pc sample, and identify 37
candidates with strong flux deficits in the optical.  We present
follow-up Gemini optical spectroscopy of 30 of these systems, and
confirm all of them as IR-faint white dwarfs. We identify an
additional set of 33 objects as candidates based on their colors and
magnitudes. We present a detailed model atmosphere analysis of all 70
newly identified IR-faint white dwarfs together with 35 previously
known objects reported in the literature. We discuss the physics of
model atmospheres and show that the key physical ingredient missing in
our previous generation of model atmospheres was the high-density
correction to the He$^-$ free-free absorption coefficient. With new
model atmospheres calculated for the purpose of this analysis, we now
obtain significantly higher effective temperatures and larger stellar
masses for these IR-faint white dwarfs than the $\Te$ and $M$ values
reported in previous analyses, thus solving a two decade old problem.
In particular, we identify in our sample a group of ultramassive white
dwarfs in the Debye cooling phase with stellar parameters never
measured before.

\end{abstract}


\section{INTRODUCTION}

The Hertzsprung-Russel (H-R) diagram for white dwarfs obtained by the
Gaia mission \citep{gaia18} revealed a wealth of information and
helped to identify particularly interesting structures. Most
noteworthy was the so-called Q-branch, a nearly horizontal sequence
composed mostly of massive DA and DQ stars, which has successfully
been interpreted as evidence for crystallization \citep{tremblay19},
when the release of latent heat and chemical fractionation decrease
the cooling rate of a white dwarf, leading to a pile up of objects in
a color-magnitude diagram.

Also of interest was the B-sequence in the H-R
diagram, a bifurcation between the non-DA and DA white dwarfs in the
range $0.0<(G_{\rm BP}-G_{\rm RP})<0.8$ that could not be reproduced
by synthetic colors from model atmospheres with pure helium
compositions and normal mass. To match the observed sequence, one needed to
invoke either problems with the physics of pure helium model
atmospheres, or a higher than average mass for the non-DA sequence.
\citet{bergeron19} indeed showed that cool ($\Te\lesssim 10,000$~K)
non-DA white dwarfs had significantly larger than average masses
($M\sim 0.7-0.8$ \msun) when analyzed with the photometric technique
using pure helium atmospheres. However, they
also convincingly demonstrated that normal masses could be inferred
(see their Figure 11) when using helium-rich models containing a small
trace of hydrogen (H/He $=10^{-5}$; see also Figure 4 of
\citealt{bedard22} for a prediction in the color-magnitude diagram).

\begin{table*}
\scriptsize
\centering
\caption{Previously Known IR-faint White Dwarfs\label{tabold}}
\begin{tabular}{clccc}
\hline
Name & Source ID & Other Name &  Spectral Type & Reference    \\
\hline
J0041$-$2221 & Gaia DR2 2349916559152267008 &  LHS 1126                     & DQ & \citet{bergeron94}\\
J0146+1404  & Gaia DR2 2587993017344962688 &  WD0143+138               & DC & \citet{harris08} \\
J0224$-$2854 & Gaia DR2 5068532996689788544 &  WD0222$-$291             & DC & \citet{oppenheimer01}  \\
J0309+0025 & Gaia DR2 3266873724451739776 &   SDSS J030924.87+002525.1 &  DC & \citet{kilic06a} \\
J0346+2455 & Gaia DR2 66837563803594880   &  WD0343+247               & DC &  \citet{hambly99} \\
J0804+2239 & Gaia DR2 680099824985004288  &   WD0801+228               & DZ & \citet{blouin18b}  \\
J0840+0515 & Gaia DR2 582509857257561472  &  SDSS J084001.43+051529.2 &  DC & \citet{leggett11}  \\
J0853$-$2446 & Gaia DR2 5652718097353105664 &   LHS 2068                      & DC & \citet{ruiz01}  \\
J0854+3503 & Gaia DR2 716743042845256576  &   WD0851+352               & DC & \citet{gates04}  \\
J0909+4700 & Gaia DR2 1011466005095102464 &  WD0905+472              & DC & \citet{kilic10b}  \\
J0925+0018 & Gaia DR2 3840846114438361984 &   LHS 2139                     & D: & \citet{leggett18}  \\
J0928+6049 & Gaia DR2 1039078998380506880 &  SDSS J092803.85+604903.3 & DC & \citet{kilic20}  \\
J0947+4459 & Gaia DR2 820969357814798080  &   WD0944+452               & DC & \citet{gates04} \\
J1001+3903 & Gaia DR2 803211596486728064  &   WD0958+393               & DC & \citet{gates04}   \\
J1147+2220 & Gaia DR2 3979751266665795456 &   SDSS J114713.33+222049.0 & DC & \citet{kilic20}  \\
J1203+0426 & Gaia DR2 3894780007343533184 &   WD1200+047               & DC & \citet{kilic06a}\\
J1220+0914 & Gaia EDR3 3905335598144227200&  WD1218+095               & DC & \citet{gates04} \\
J1238+3502 & Gaia DR2 1518373537314807552 &  WD1235+353               & DC & \citet{harris08} \\ 
J1251+4403 & Gaia DR2 1528861748669458432 &  WD1248+443               & DC & \citet{harris08}  \\
J1320+0836& Gaia DR2 3731477241851793152 &   SDSS J132016.50+083644.3 &  DC & \citet{leggett11}  \\
J1337+0001 & Gaia DR2 3662779171232754688 &   WD1335+002               & DC & \citet{harris01}  \\
J1403+4533 & Gaia DR2 1505825635741455872 &  WD1401+457               & DC & \citet{gates04} \\
J1403$-$1514 & Gaia DR2 6300991145225638272 &   WD1401$-$149             &  DC & \citet{leggett18} \\
J1404+1330 & Gaia DR2 1229916112012470528 &  SDSS J140451.90+133055.1 & DC & \citet{leggett11}  \\
J1437+4151 & Gaia DR2 1492944375984949504 &  WD1435+420               & DC & \citet{kilic10b} \\
J1556$-$0806 & Gaia DR2 4348098485293072128 &   SSS J1556$-$0806         & DC & \citet{rowell08}  \\
J1632+2426 & Gaia DR2 1300727345195414272 &  WD1630+245               & DC & \citet{harris08}  \\
J1653+6253 & Gaia DR2 1631578537252535040 &  LHS 3250                     & DC & \citet{harris99}  \\
J1722+5752 & Gaia DR2 1433166540130924544 &  WD1722+579               & DC & \citet{kilic06a}  \\
J1727+0808 & Gaia DR2 4490300553197280256 &  SDSS J172748.71+080819.6 &  DC & \citet{kilic20} \\
J1824+1213 & Gaia DR2 4484289866726156160 &                     & DZ & \citet{hollands21}\\
J2138$-$0056 & Gaia DR2 2686607906002083328 &  SDSS J213805.13$-$005615.8 & DC & \citet{kilic20}  \\
J2239+0018 & Gaia DR2 2654379433485461632 &  SDSS J223954.12+001847.3 & DC & \citet{harris08} \\ 
J2242+0048 & Gaia DR2 2654423998066862464 &  WD2239+005               & DC & \citet{kilic06a} \\
J2317+1830 & Gaia DR2 2818957013992481280 &  SDSS J231726.72+183049.7 & DZ & \citet{hollands21}\\
\hline
\end{tabular}
\end{table*}

Another previously undetected feature in the color-magnitude diagram
was reported by \citet{kilic20} who identified an almost horizontal
branch with $M_g\sim 15.5$ in the $M_g$ vs $(g-z)$ Pan-STARRS
color-magnitude diagram (see their Figure 21) composed of white dwarfs
with optical and near-infrared flux deficits. This infrared flux
deficiency is most likely caused by collision-induced absorption by
molecular hydrogen due to collisions with helium (H$_2$-He CIA).

At the low temperatures and high densities of cool white dwarf
atmospheres, CIA becomes the dominant source of opacity
\citep[e.g.,][]{borysow01}. In pure hydrogen atmosphere white dwarfs,
CIA appears at effective temperatures below 4000 K. Hence, white
dwarfs that show significant near-infrared flux deficits have
traditionally been classified as ultracool white dwarfs. However, cool
helium-rich white dwarfs have lower opacities and higher atmospheric
pressures, and the infrared flux deficiency due to H$_2$-He CIA starts
to dominate at higher effective temperatures.

\citet{bergeron02} presented a model atmosphere analysis of the
prototype of this class, LHS 3250, and demonstrated that the pure
hydrogen atmosphere models fail to reproduce the observed energy
distribution. Even though none of their models provided perfect fits,
they concluded that LHS 3250 is better explained with a helium-rich
composition \citep[see also][]{gianninas15}. Fortunately, a few of the
white dwarfs with near-infrared flux deficits are DZ white
dwarfs. \citet{blouin18b} simultaneously fit the metal absorption
lines and the CIA in one of these DZ white dwarfs, WD 0801+228
\citep[see also][]{blouin19b}, and found an excellent fit to both the
spectroscopic and photometric data with a model that has $T_{\rm eff}
= 4970$ K and $\logh = -1.6$, confirming a mixed atmosphere
composition and a relatively warm temperature. Hence, it is more
appropriate to classify these objects as ``IR-faint''. And as we shall
demonstrate below, the term ``ultracool'' should definitely be
abandoned to describe these white dwarfs showing strong CIA
absorption.

The fact that the observed IR-faint white dwarf sequence is so tight
indicates that these stars likely have similar hydrogen abundances,
and that this must be the cooling sequence of white dwarfs with mixed
H/He atmospheres. \citet{bergeron19} and \citet{bedard22} showed that
many of the warmer DC stars (above 6000 K) have trace amounts of
hydrogen in their atmospheres. Hence, an IR-faint white dwarf
sequence due to CIA seems unavoidable, but was detected only recently.

We now have a fantastic opportunity to explore this sequence thanks to
large scale photometric and astrometric surveys. Here we take
advantage of the Gaia DR2 and EDR3 astrometry to expand the IR-faint
white dwarf sample to the entire Pan-STARRS footprint and the Montreal
White Dwarf Database (MWDD) 100 pc sample \citep{dufour17}. We
discuss our sample selection in Section \ref{sec:sample}, and present
follow-up optical spectroscopy of 30 candidates in Section
\ref{sec:data}. We describe the theoretical framework used in our
analysis in Section \ref{sec:models}, including new model atmospheres
and evolutionary models. We then provide model atmosphere fits for all
confirmed and candidate IR-faint white dwarfs in Section
\ref{sec:results}. We discuss the overall sample properties in Section
\ref{sec:discussion}, and conclude in Section \ref{sec:conclusion}.

\section{SAMPLE SELECTION}\label{sec:sample}

Since the discovery of the first IR-faint white dwarfs, LHS 1126 and
LHS 3250 \citep{bergeron94,harris99}, we have been able to find only
35 such stars in the past two decades.  Table \ref{tabold} presents
this list. Many of these white dwarfs were found serendipitously by
the Sloan Digital Sky Survey (SDSS) spectroscopy
\citep{gates04,harris08}, which targeted them due to their unusually
blue colors. Most show flux deficits in the redder SDSS bands, but a
significant fraction show significant flux deficits only in the
near-infrared range beyond 1~$\mu$m \citep[see for
  example][]{kilic10b,gianninas15}.  All 35 objects in Table
\ref{tabold} show significant flux deficits compared to the
best-fitting pure hydrogen and pure helium atmosphere models. However,
this list excludes white dwarfs like SDSS J004506.40-060825.7
\citep{oppenheimer01}, SDSS J110217.48+411315.4 \citep{hall08}, and
SDSS J145239.00+452238.3 \citep{harris08}, where the spectral energy
distributions do not show significant flux deficits compared to the
pure hydrogen or pure helium models, according to our own analysis of
these objects.

\begin{figure}
\hspace{-0.2in}
\includegraphics[width=3.5in, clip=true, trim=0.3in 1.2in 1.1in 1.7in]{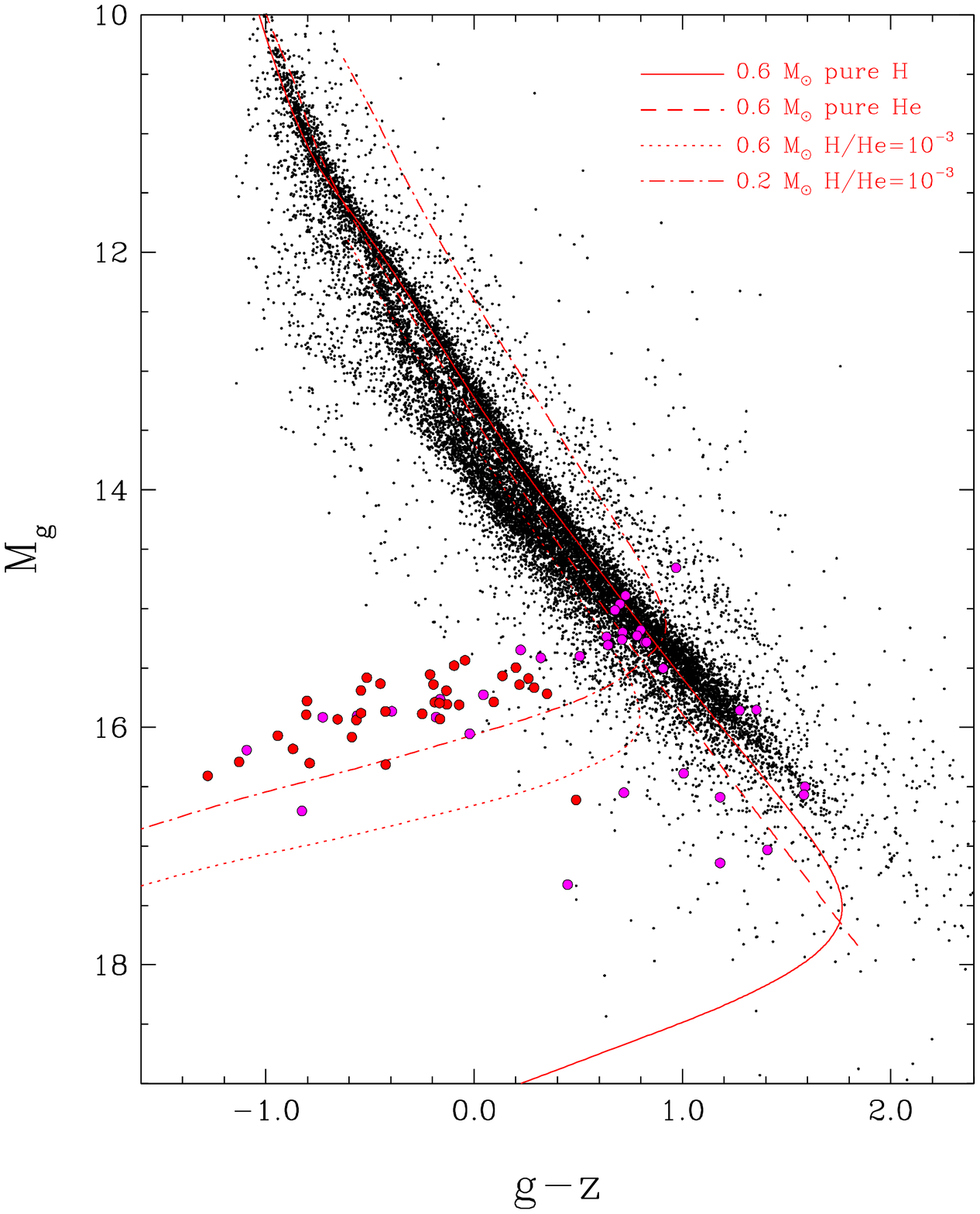}
\caption{Color-magnitude diagram of the Pan-STARRS white dwarfs within
  120 pc and with $5\sigma$ parallax measurements from Gaia
  DR2. Previously known IR-faint white dwarfs are marked by magenta
  circles. Newly identified IR-faint white dwarfs that were targeted
  at Gemini are marked by red circles. Also shown are same cooling
  sequences as those displayed in Figure 21 of \citet{kilic20} for
  various masses and atmospheric compositions, as indicated in the
  figure.
\label{colormag}}
\end{figure}

The previous surveys for IR-faint white dwarfs have been mostly
limited to the SDSS footprint \citep{gates04,harris08,kilic20}.  To
take advantage of the significantly larger sky coverage of the
Pan-STARRS $3\pi$ survey, we selected all white dwarf candidates with
$5\sigma$ significant parallax measurements from Gaia DR2 that are
also in the Pan-STARRS footprint. We used the same color-magnitude and
quality cuts as in \citet{kilic20} to select a relatively clean white
dwarf sample.

Figure \ref{colormag} shows a color-magnitude diagram for the
Pan-STARRS + Gaia DR2 sample with $\varpi/\sigma_{\varpi}\geq5$ and
within 120 pc.  Previously known IR-faint white dwarfs are marked by
magenta dots. We also show for reference the same cooling sequences
for pure hydrogen, pure helium, and mixed H/He atmosphere white dwarfs
as in Figure 21 of \citet{kilic20}. Because IR-faint white dwarfs
have $M_g$ ranging from 15 to 17 mag, and Gaia's limiting magnitude is
$\approx21$ mag, expanding our search beyond the local $\sim$100 pc
volume is currently not feasible.

We selected candidates in the H-R diagram along the nearly horizontal
IR-faint white dwarf sequence identified by \citet{kilic20}, $\pm 1$
mag in $M_g$, and fit their Pan-STARRS photometry and Gaia DR2
parallaxes with pure hydrogen and pure helium models to identify
objects with significant flux deficits compared to these models. Some
of these candidates appear to be consistent with being massive pure
hydrogen atmosphere white dwarfs. However, there are many that clearly
bear the signature of IR-faint white dwarfs; they are significantly
fainter than expected from pure hydrogen or pure helium atmosphere
models in the redder Pan-STARRS bands.

In addition to the Pan-STARRS footprint, we also searched for white
dwarfs with strong flux deficits in the MWDD 100 pc sample using Gaia
colors. The MWDD selection is based on Gaia DR2 \citep{gaia18}, and
includes all candidates with $10\sigma$ significant parallax
($\varpi$), $G_{\rm BP}$ and $G_{\rm RP}$ photometry, and $\varpi +
\sigma_{\varpi}>10$ mas.  Non-Gaussian outliers in color and absolute
magnitude are removed using the recommendations from
\citet{lindegren18}, and a cut in Gaia color and absolute magnitude is
used to select the white dwarf candidates. We identify several
additional IR-faint white dwarf candidates outside of the Pan-STARRS
footprint. However, Gaia photometry is not sufficient to reliably
confirm them as IR-faint white dwarfs. Two of these objects have
$grizy$ photometry available from the Dark Energy Survey
\citep{dark16}, and clearly show significant flux deficits in the
$izy$ bands.

In total, we identify 37 IR-faint white dwarf candidates with strong
flux deficits in the optical. We mark these with red circles in Figure
\ref{colormag} and present their Gaia source IDs, astrometry, and
Pan-STARRS or Dark Energy Survey $grizy$ photometry in Table
\ref{tabnew}.  Four of these objects are missing from Figure
\ref{colormag} because they are either outside of the Pan-STARRS
footprint (the two DES objects) or they lack $z$-band photometry. Only
one of these newly identified candidates, SDSS J003908.33+303538.9,
has spectroscopy available in the SDSS, confirming it to be a DC white
dwarf. A comparison with Figure 21 of \citet{kilic20} shows the
substantial increase in the number of IR-faint white dwarf candidates
based on our selection.

\begin{table*}
\tiny
\centering
\caption{Newly Identified IR-faint White Dwarfs with Strong Flux Deficits in the Optical Bands\label{tabnew}}
\begin{tabular}{clcccccccccc}
\hline
Name & Source ID &  $\varpi$ & $\mu$ & Pan-STARRS or & $g$ & $r$ & $i$ & $z$ & $y$ & Sp.Type & Spectral\\
 & & (mas) & (mas yr$^{-1}$) & DES Name & (mag) & (mag) & (mag) & (mag) & (mag) &  &Source\\
\hline
J0039+3035 & Gaia EDR3 2858553485723741312 & 16.36 & 93.9 & PSOJ009.7847+30.5941 & 20.34 & 20.39 & 20.92 & 21.62 & \nodata & DC & SDSS \\
J0231+3254 & Gaia EDR3 134315687614338432  &  8.67 & 198.8 & PSOJ037.7706+32.9068 & 20.81 & 20.23 & 20.23 & 20.61 & 20.08 & DC & Gemini \\
J0235$-$3032 &Gaia EDR3 5064259336725948672 & 30.64 & 319.4 & PSOJ038.9116$-$30.5410 &18.86 & 18.89 & 19.59 & 19.99 & \nodata & \nodata & \nodata \\
J0320+2948 & Gaia EDR3 122341593670893440  & 10.54 & 108.1 & PSOJ050.1452+29.8081 & 20.52 & 20.22 & 20.46 & 20.97 & \nodata & DC & Gemini\\
J0414+0309 & Gaia EDR3 3259405227298444544 & 14.91 &  89.0 & PSOJ063.7109+03.1651 & 19.82 & 19.29 & 19.43 & 19.96 & 20.55 & DC & Gemini\\
J0416$-$1826 & Gaia EDR3 5093305994390659072 & 10.52 & 215.1 & PSOJ064.1636$-$18.4451 & 20.46 & 19.96 & 20.17 & 20.55 & 20.49 & DC & Gemini\\
J0437$-$5946 & Gaia EDR3 4678027595110360832 & 14.47 & 254.6 & DESJ043729.88$-$594601.8 & 20.14	& 19.87 & 20.35 & 20.68 & 20.85 & \nodata & \nodata\\
J0440$-$0414 & Gaia EDR3 3201530847924700544 & 10.39 &  80.9 & PSOJ070.2465$-$04.2480 & 20.48 & 19.96 & 19.98 & 20.35 & 20.33 & DC & Gemini\\
J0445$-$4906 & Gaia EDR3 4784964034443286656 & 16.11 &  69.5 & DESJ044520.28$-$490603.3 & 20.14 & 20.04 & 20.58 & 21.01 & 21.29 & \nodata & \nodata \\
J0448+3206 & Gaia EDR3 161053615673941248  & 12.65 & 193.2 & PSOJ072.1315+32.1142 & 20.18 & 19.86 & 20.27 & 20.72 & \nodata & DC & Gemini\\
J0551$-$2652 & Gaia EDR3 2910863064947630848 & 10.60 &  93.7 & PSOJ087.7684$-$26.8807 & 20.61 & 20.21 & 20.33 & 20.87 & \nodata & DC & Gemini\\
J0559+0731 & Gaia EDR3 3323118004821936768 & 11.97 & 234.5 & PSOJ089.8401+07.5291 & 20.39 & 20.11 & 20.65 & 21.19 & \nodata & DC & Gemini\\
J0756$-$2001 & Gaia EDR3 5714040261719110784 & 23.76 & 128.3 & PSOJ119.1135$-$20.0198 & 19.01 & 18.72 & 19.34 & 19.82 & 20.00 & DC & Gemini\\
J0814+3300 & Gaia EDR3 902414964384353408  & 13.85 & 158.0 & PSOJ123.6780+33.0062 & 19.96 & 19.41 & 19.38 & 19.67 & 19.76 & DC & Gemini\\
J0910$-$0222 & Gaia EDR3 5763109404082525696 & 16.28 & 142.8 & PSOJ137.5026$-$02.3666 & 20.25 & 20.46 & 20.95 & 20.68 & \nodata & DC & Gemini\\
J1105$-$2114 & Gaia EDR3 3551152013432136832 & 12.92 & 207.9 & PSOJ166.2734$-$21.2458 & 20.32 & 20.04 & 20.54 & 20.87 & \nodata & DC & Gemini\\
J1121+1417 & Gaia EDR3 3966679722679277824 &15.59 & 317.5 & PSOJ170.2507+14.2906 & 19.62 & 19.18 & 19.51 & 20.13 & 20.26 & \nodata & \nodata\\
J1125+0941 & Gaia EDR3 3914946356266960896 & 12.38 & 209.9 & PSOJ171.4320+09.6949 & 20.35 & 19.87 &19.93 & 20.42 & 19.94 & \nodata & \nodata\\
J1136$-$1057 & Gaia EDR3 3586879608689430400 & 17.57 & 219.4 & PSOJ174.1987$-$10.9569 & 20.08 & 20.36 & 20.91 & 20.87 & \nodata & DC & Gemini\\
J1142$-$1315 & Gaia EDR3 3585053427252374272 & 16.87 &  86.8 & PSOJ175.6462$-$13.2641 & 19.93 & 19.94 & 20.58 & 20.88 & \nodata & DC & Gemini\\
J1304+0126 & Gaia EDR3 3691095100341397632 & 13.52 &  22.3 & PSOJ196.1226+01.4366 & 19.99 & 19.42 & 19.38 & 19.77 & 19.98 & DC & Gemini\\
J1336+0748 & Gaia EDR3 3718763318316606208 & 11.94 & 266.8 & PSOJ204.1774+07.8072 & 20.42 & 20.01 & 20.36 & 20.55 & \nodata & DC & Gemini\\
J1355-2600 & Gaia EDR3 6178573689547383168 & 17.10 & 266.9 & PSOJ208.9363$-$26.0075 & 20.20 & 20.25 & \nodata & \nodata & \nodata & DC & Gemini\\
J1503$-$3005 & Gaia EDR3 6211904903507006336 & 15.41 & 209.6 & PSOJ225.7513$-$30.0969 & 19.99 & 19.68 & 20.17 & 20.65 & 20.59 & DC & Gemini\\
J1531+4421 & Gaia EDR3 1394479501945576064 & 11.10 & 137.4 & PSOJ232.9708+44.3618 & 20.36 & 19.82 & 19.81 & 20.10 & 19.89 & DC & Gemini\\
J1542+2750 & Gaia EDR3 1224133608563103872 &  9.56 &  82.1 & PSOJ235.5150+27.8394 & 20.74 & 20.28 & 20.51 & 20.93 & \nodata & DC & Gemini\\
J1602+0856 & Gaia EDR3 4454676827432347904 & 11.59 & 77.3 & PSOJ240.6872+08.9411 & 20.47 & 19.92 & 20.05 & 20.37 & 20.33 & \nodata & \nodata\\
J1610+0619 & Gaia EDR3 4449818459207085696 & 11.57 &  93.2 & PSOJ242.7270+06.3175 & 20.55 & 20.24 & 20.57 & 20.98 & \nodata & DC & Gemini\\
J1612+5128 & Gaia EDR3 1424656526287583744 & 11.52 & 260.5 & PSOJ243.1532+51.4695 & 20.63 & 20.64 & 21.07 & \nodata & \nodata & DC & Gemini\\
J1633+3829 & Gaia EDR3 1331687458035163776 &  8.49 &  64.1 & PSOJ248.3530+38.4903 & 21.07 & 20.54 & 20.52 & 20.72 & \nodata & DC & Gemini\\
J1830+2529 & Gaia EDR3 4537112917888780416 & 11.28 & 109.0 & PSOJ277.5951+25.4861 & 20.67 & 20.25 & 20.54 & 20.83 & \nodata & DC & Gemini\\
J1922+0233 & Gaia EDR3 4288942973032203904 & 25.38 &  75.5 & PSOJ290.5261+02.5536 & 19.59 & 19.06 & 18.94 & 19.10 & 19.47 & DZ & Gemini\\
J2056+7218 & Gaia EDR3 2275065002988477440 & 14.22 & 123.1 & PSOJ314.0267+72.3156 & 20.32 & 20.18 & 20.67 & 20.90 & \nodata & DC & Gemini\\
J2148$-$2821 & Gaia EDR3 6809702159983236992 & 11.29 & 178.7 & PSOJ327.0429$-$28.3556 & 20.62 & 20.17 & 20.52 & 20.87 & \nodata & DC & Gemini\\
J2305+3922 & Gaia EDR3 1929838143078434432 & 27.86 & 487.7 & PSOJ346.4583+39.3742 & 18.21 & 17.77 & 17.88 & 18.25 & 18.45 & DC & Gemini\\
J2340+6117 & Gaia EDR3 2012467915077073024 & 14.04 &  96.3 & PSOJ355.0451+61.2995 & 19.74 & 19.31 & 19.47 & 19.84 & 19.94 & DC & Gemini\\
J2340+6902 & Gaia EDR3 2214375294029777408 & 13.47 &  77.2 & PSOJ355.1328+69.0430 & 20.15 & 19.73 & 20.06 & 20.31 & \nodata & DC & Gemini\\
\hline
\end{tabular}
\end{table*}

\section{OBSERVATIONAL DATA}\label{sec:data}

We tested the efficiency of our sample selection using the Fast
Turnaround observing mode on Gemini.  We obtained optical spectroscopy
for twelve targets using the 8-m Gemini North telescope equipped with
the Gemini Multi-Object Spectrograph (GMOS) as part of the program
GN-2020B-FT-104.  We used the B600 grating and a 1$\arcsec$ slit, and
binned the CCD by $4\times4$. We initially centered the grating at
5300 \AA\ and observed J0414+0309 and J1922+0233 in this setup, which
provided spectra over the wavelength range 3730-6920 \AA, with a
resolution of 2.05 \AA\ per pixel.

IR-faint white dwarfs have spectral energy distributions that usually
peak beyond 5000 \AA.  Our initial observations showed that the
signal-to-noise ratio of the spectra in the blue, below 5000 \AA, was
poor and the blue portion of the spectra did not provide any
meaningful constraints on the spectral types of the first two objects
observed. To sample the peak of the spectral energy distribution, we
changed our observing setup and shifted the central wavelength to 6350
\AA\ for the rest of the targets. This setup provided spectra over the
wavelength range 4770-7980 \AA\ (J1922+0233, mentioned above, was
reobserved with this alternative setup). We reduced the Gemini data
using the {\sc gmos} package under {\sc IRAF}.

Our Fast Turnaround program was a success, confirming all of the
observed targets as IR-faint white dwarfs.  We observed 18 additional
objects as part of the queue programs GN-2021A-Q-203 and
GS-2021A-Q-300 using the same setup as above, with a central
wavelength of 6350 \AA. Three more targets were included in the queue,
but did not get observed. In total, we obtained spectra for 30
candidates at Gemini.

\begin{figure}
\hspace{-0.3in}
\includegraphics[width=3.6in, clip=true, trim=0.3in 0.8in 0.6in 0.3in]{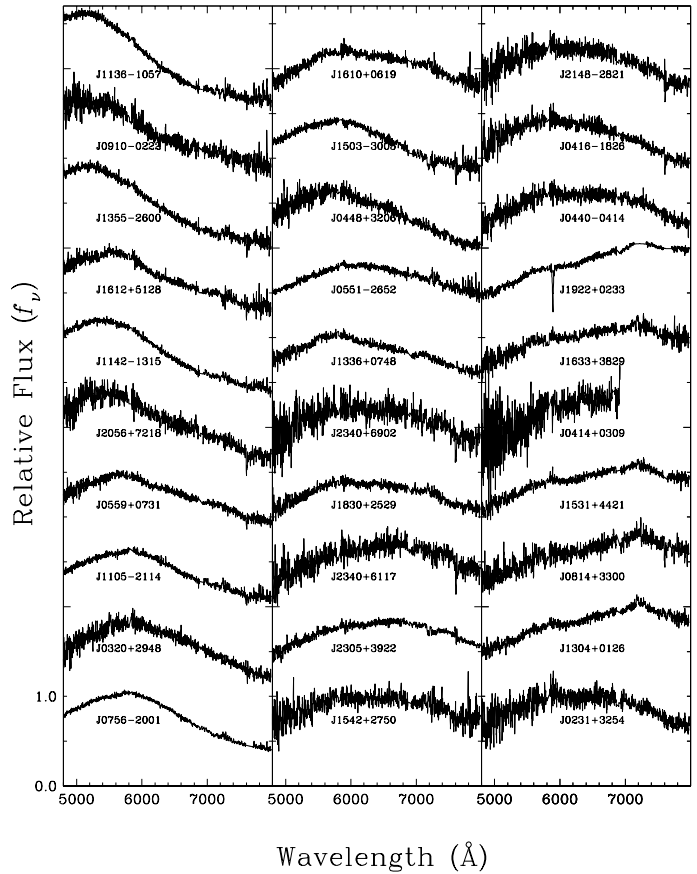}
\caption{Gemini GMOS spectroscopy of 30 newly identified IR-faint white dwarfs.
Objects are plotted based on their $g-r$ colors, increasing from top left to the bottom
right. This sample essentially doubles the number of spectroscopically confirmed IR-faint
white dwarfs known. All but one of these objects, J1922+0233, are confirmed to
be DC white dwarfs with featureless spectra.
\label{allspec}}
\end{figure}

Figure \ref{allspec} shows the spectra for all 30 candidates observed
at Gemini\footnote{All spectra are available on the MWDD Web site
  (http://montrealwhitedwarfdatabase.org/).}. The spectra are organized
in increasing $g-r$ color from the top left to the bottom right. All
but one of these objects are confirmed to be DC white dwarfs with
featureless spectra. The exception is J1922+0233, which is a DZ white
dwarf with a strong Na D absorption feature \citep[see
  also][]{tremblay20}.

The Gemini spectra confirm the unusual spectral energy distributions
of all 30 targets; they peak between 5000 and 7000 \AA, and show
significant absorption in the red. For comparison, the spectral energy
distribution of the prototype IR-faint white dwarf LHS 3250 peaks at
around 6000 \AA. About half of the targets in Figure \ref{allspec}
have spectral energy distributions that are even more extreme than LHS
3250.

\section{THEORETICAL FRAMEWORK}\label{sec:models}

\subsection{Photometric Technique}

We use the photometric technique as detailed in \citet{bergeron19},
and follow the same approach and use the SDSS $u$, Pan-STARRS $grizy$
photometry, and Gaia EDR3 parallaxes in our analysis. If available, we
also supplement these data with near-infrared photometry from the
UKIRT Infrared Deep Sky Survey \citep[UKIDSS,][]{ukidss} and the VISTA
Hemisphere Survey \citep[VHS,][]{mcmahon21}. Because CIA dominates in
the near-infrared, the majority of our targets are too faint to be
detected in the UKIDSS and VHS. Only three targets have near-infrared
$J$-band photometry available, but no $H$- or $K$-band data.

We convert the observed magnitudes into average fluxes using the
appropriate zero points, and compare with the average synthetic fluxes
calculated from model atmospheres with the appropriate chemical
composition. Since our sample is restricted to $\sim$100 pc, we do not
correct for reddening. We fit for the effective temperature and the
solid angle, $\pi (R/D)^2$, where $R$ is the radius of the star and
$D$ is its distance. Given the distance measurements from Gaia, we
constrain the radius of the star directly, and its mass based on
evolutionary models.  The details of our fitting method are further
discussed in \citet{gianninas15}, \citet{bergeron19}, and
\citet{kilic20}.

\subsection{Evolutionary Models}

We use detailed mass-radius relations for CO-core ($X_{\rm C}=X_{\rm
  O}=0.5$) white dwarfs with masses in the range $M=0.2-1.3\ M_\odot$
\citep{bedard20}. For the purpose of this analysis, we also calculated
a few He-core models with STELUM \citep{bedard22}. These He-core
sequences were started from artificial static white dwarf models, but
the initial structures are quickly ``forgotten'' and thus have no
impact on the mass-radius relation at the low effective temperatures
of interest here. However, this means that our He-core sequences do
not provide reliable cooling times, which can only be obtained through
a detailed modeling of pre-white dwarf evolutionary phases
\citep{althaus13,istrate16}.

\begin{figure}
\hspace{-0.3in}
\includegraphics[width=3.7in, clip=true, trim=0.5in 3in 0.1in 3in]{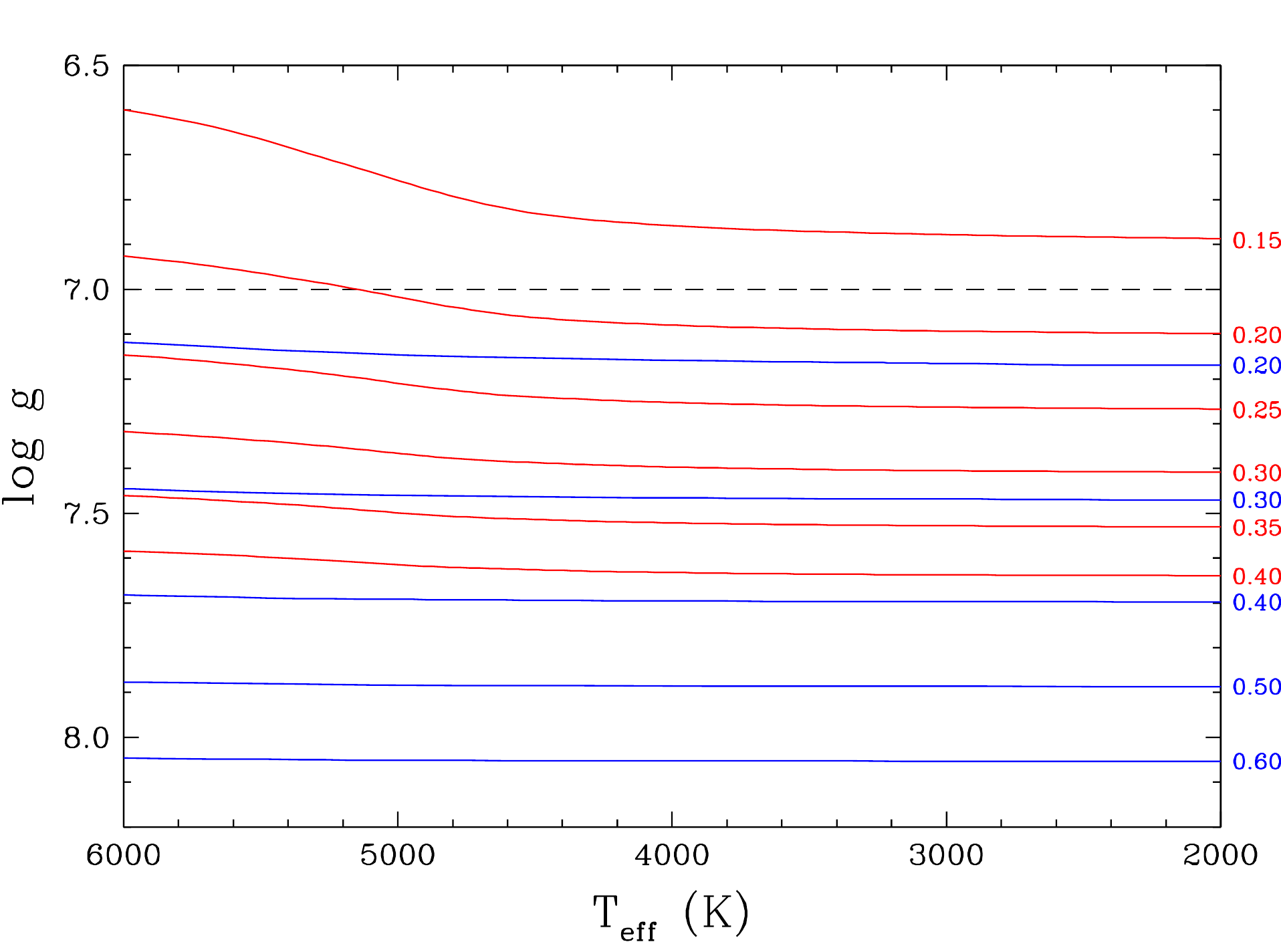}
\caption{Cooling sequences for CO- (blue) and He-core (red) white dwarfs based on the STELUM code
\citep{bedard22}. The masses (in $M_{\odot}$) are given on the right. The dashed line shows the lowest
surface gravity ($\logg = 7$) model available in our model atmosphere grid.
\label{grid}}
\end{figure}

Figure \ref{grid} shows the evolutionary models for He-core (red) and
CO-core (blue) white dwarfs.  As expected, a He-core white dwarf is
larger and more luminous (which corresponds to a lower surface
gravity) than a CO-core white dwarf at a given mass. The dashed line
shows the lowest surface gravity ($\logg = 7$) model available in our
model atmosphere grid. 

\subsection{Model Atmospheres}\label{modatm}

The first grid of model atmospheres we used is that described in the
analysis of IR-faint white dwarfs by \citet[][also used in
  \citealt{kilic20}]{gianninas15}, based on the original calculations
of \citet{bsw95}, but with improvements to the H$_2$-He CIA
calculations from \citet{jorgensen00}, including the density
correction from \cite{hare58}\footnote{In what follows this correction
  is always included when using the profiles from
  \citet{jorgensen00}.}. These models assume an ideal gas equation of
state. Here we have significantly extended this model grid in $\logh =
-5.0$, $-4.0$ (0.5) $-1.0$ (1.0) $+2.0$, where the numbers in parentheses
indicate the step size, and $\logg = 7.0$ (0.5) 9.0.

\begin{figure}
\hspace{-0.2in}
\includegraphics[width=3.5in, clip=true, trim=0.3in 1.2in 1.1in 2in]{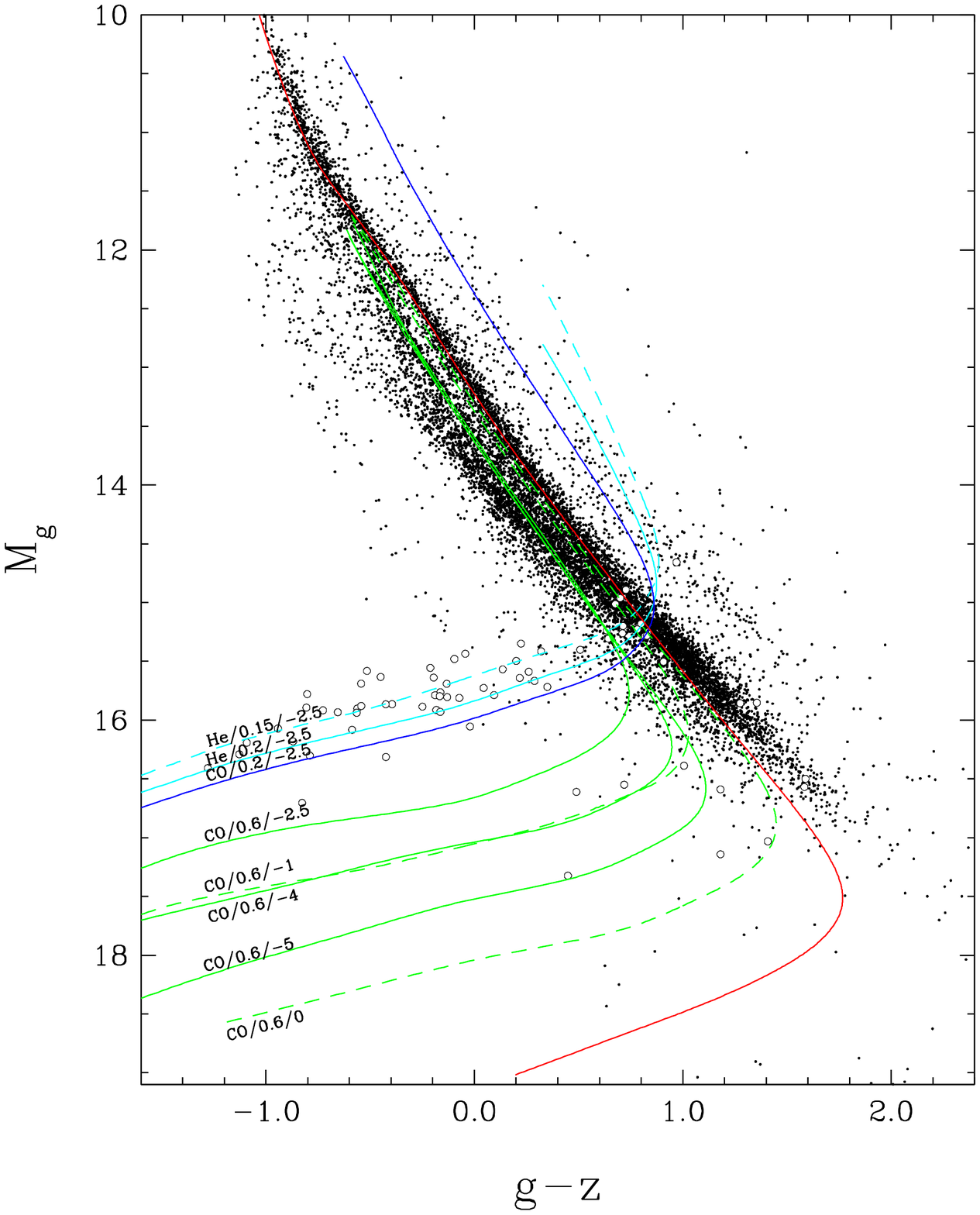}
\caption{Color-magnitude diagram for the same white dwarf sample as in
  Figure \ref{colormag} along with the evolutionary sequences for pure
  hydrogen atmosphere CO-core white dwarfs with 0.6 \msun\ shown in
  red, while green, blue, and cyan lines show the evolutionary
  sequences for CO- and He-core white dwarfs with a variety of masses
  and $\logh$ ratios using our old grid of model atmospheres (see
  text). Each sequence is labeled based on its core composition,
  mass, and atmospheric composition.  He/0.15/$-2.5$ means a He-core
  white dwarf with $M = 0.15$ \msun\ and $\logh = -2.5$.
\label{theory}}
\end{figure}

As was the case for LHS 3250 analyzed in detail by \citet{bergeron02},
or similar IR-faint white dwarfs reported by \citet{kilic20}, we
measured extremely low temperatures ($T_{\rm eff}\,\lta4000$ K) for
most objects in our sample, but more importantly, the masses we
inferred were excessively small ($M\sim 0.15$ to
$0.35~M_{\odot}$). The reason for such low masses can be understood by
looking at Figure \ref{theory} where we show the same color-magnitude
diagram as in Figure \ref{colormag}, but with evolutionary sequences
for various masses, core compositions, and $\logh$ ratios. The maximum
CIA absorption in these models occurs around $\logh\sim-2.5$, where
the models reach their maximum luminosity in the optical.
\citet{blouin18b} discussed the reason for this maximum: a large H/He
ratio means that there is more H$_2$ to cause CIA, but a very small
H/He ratio means the density is higher in a helium-dominated
atmosphere, which also leads to stronger CIA. At $\logh\sim-2.5$,
however, the predicted colors at 0.6 \msun\ fall short of matching the
observed sequence of IR-faint white dwarfs in Figure \ref{theory} by a
full magnitude.  Even at a mass of 0.2 \msun, the predicted CO-core
sequence is still not luminous enough. Only with He-core models with a
mass as low as 0.15 \msun\ are we able to match the observed IR-faint
sequence.

Another feature of interest in this figure is the almost perfect
overlap of the 0.6 \msun, CO-core models with $\logh = -1$ and
$-4$. Such a degeneracy was also reported by \citet{blouin18b} who
found two solutions at $\logh = -1.6$ and $-3.5$ for the DZ white
dwarf WD 0801+228. Even though the latter solution produced a slightly
better fit to the photometry, it was incompatible with the observed
metal lines in this star and therefore ruled out. We come back to this
degeneracy issue below.

In addition to the extreme stellar parameters we measure using these
models, the peak of the energy distribution is always predicted too
narrow, even though our solutions provide a reasonable match to the
overall distribution, as shown for instance in Figure 20 of
\citealt{kilic20}. As concluded by \citet{bergeron02}, although there
is little doubt that these IR-faint white dwarfs have helium-rich
compositions, the problem with these extreme stellar parameters
probably lie in the physics included in our model atmospheres, which
is either inadequate or incomplete, the most obvious being the
non-ideal effects of the equation-of-state at the high atmospheric
pressures that characterize these helium-rich atmospheres.

With this idea in mind, \citet{blouin18a} presented a new generation
of cool white dwarf atmosphere models that include the improved
H$_2$-He CIA opacity profiles from \citet{abel12} with a high-density
correction at $\lambda \gtrsim 1.5\,\mu$m based on the ab initio
molecular dynamics simulations from \citet{blouin17}, as well as other
physical improvements. As we shall see below, the most relevant of
these physical improvements, in the context of our study, is the
correction to the He$^-$ free-free absorption coefficient by
\citet{iglesias02}. The differences between the old and new models are
further discussed in detail by \citet{blouin18a}. Unfortunately, their
model grid is incomplete at very low effective temperatures ($T_{\rm
  eff}\,\lta3000$ K) due to numerical issues (convergence of the
temperature structure), and could not be used to analyze our sample of
(mostly cool) IR-faint white dwarfs.

In order to provide a qualitative and quantitative comparison between
both model grids, we first compared the best-fit parameters in the
temperature range where they overlap near $T_{\rm eff}\sim4800$~K. We
found small differences in the best-fitting parameters from the two
model grids, but the fits are qualitatively the same.  We then
performed an additional comparison by calculating a small set of
models similar to those of \citet{blouin18a} but this time at much
lower temperatures around $T_{\rm eff}=3000$~K and log H/He = $-3$,
and compared those with our old models. In this case, the models
showed significant differences, with the peak of the energy
distribution in the new models considerably shifted to the red with
respect to our old models.

In fact, it was impossible to adjust the parameters of the old
models to match even approximately the new models. Given that the old
models provide fairly decent fits to the observed spectral energy
distributions, at least qualitatively, we are thus forced to conclude
that the new models would fail to match the photometric observations
of the coolest IR-faint white dwarfs in our sample.

\begin{figure}
\includegraphics[width=3.3in, clip=true, trim=0.3in 2.5in 0.3in 2.7in]{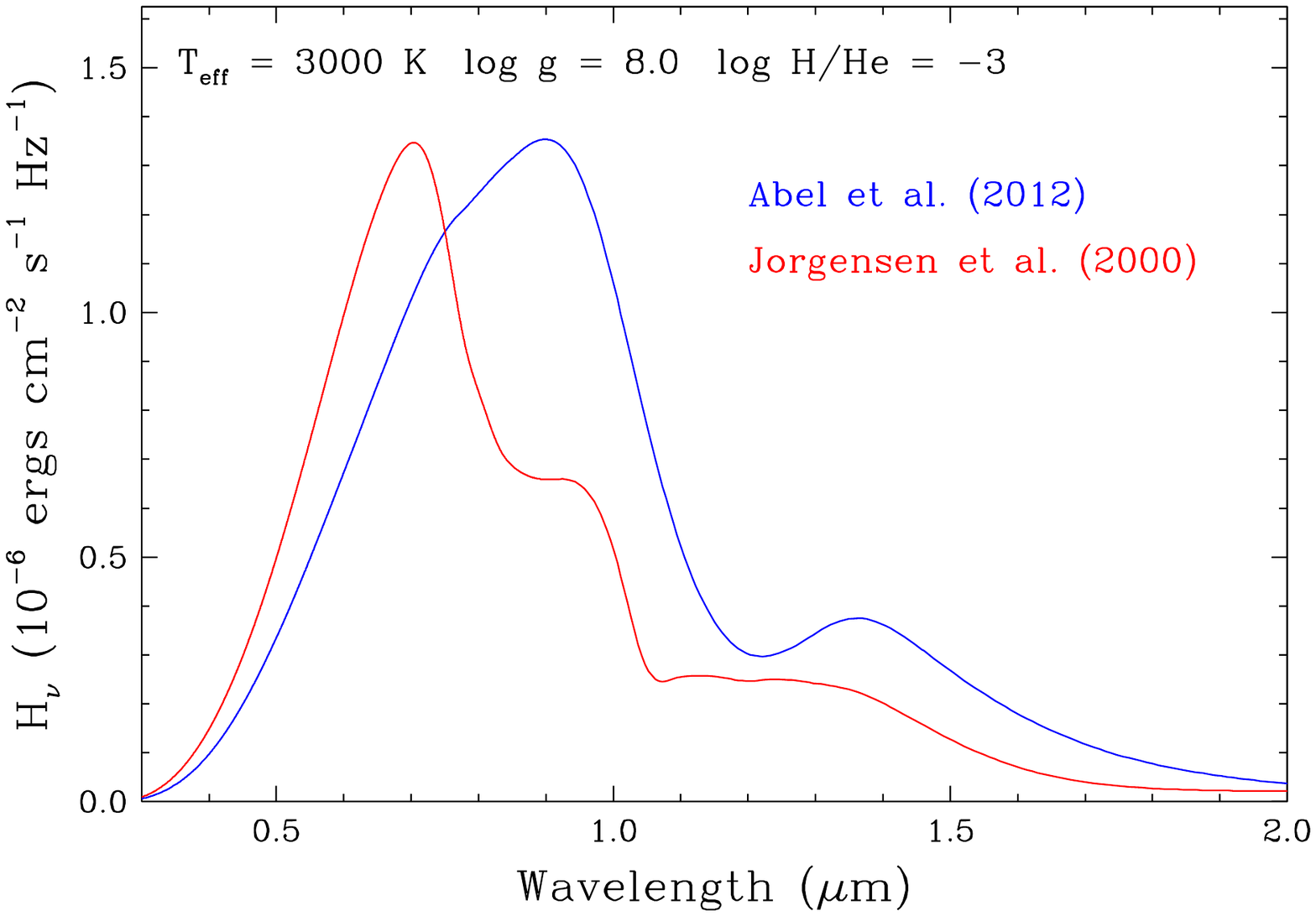}
\caption{
Energy distribution of a model from \citet{blouin18a}, which
  relies on the improved H$_2$-He CIA opacity profiles from
  \citet[][in blue]{abel12}, compared to a similar model but
  calculated with the earlier CIA profiles from \citet[][in
    red]{jorgensen00}. Note that in this last case, we also include the density
  correction from \cite{hare58}.
\label{figAbel}}
\end{figure}

We traced back the problem to the use of the improved H$_2$-He CIA
opacity profiles from \citet{abel12} used in the new models, while our
old models rely on the earlier calculations from \citet{jorgensen00}.
A comparison between two models calculated with these two sets of CIA
opacities, everything else being equal, is displayed in Figure
\ref{figAbel}. The energy distribution with the CIA profiles from
Jorgensen et al.~peaks at much shorter wavelengths, in better
agreement with the observations, no matter whether we try to vary the
effective temperature of the models relying on the calculations from
Abel et al. It is worth mentioning that a large difference at short
wavelengths ($\lambda<2\,\mu$m) between these two sets of opacity
calculations was noted before by \citet[][see their Figure 6]{abel12}.
It is difficult to interpret our result any further given that the
calculations from Abel et al.~represent a significant improvement over
Jorgensen et al., and a deeper investigation of this discrepancy is
clearly outside the scope of this paper.

\begin{figure}
\hspace{-0.4in}
\includegraphics[width=4.1in, clip=true, trim=0.in 6.0in 0.in 0.2in]{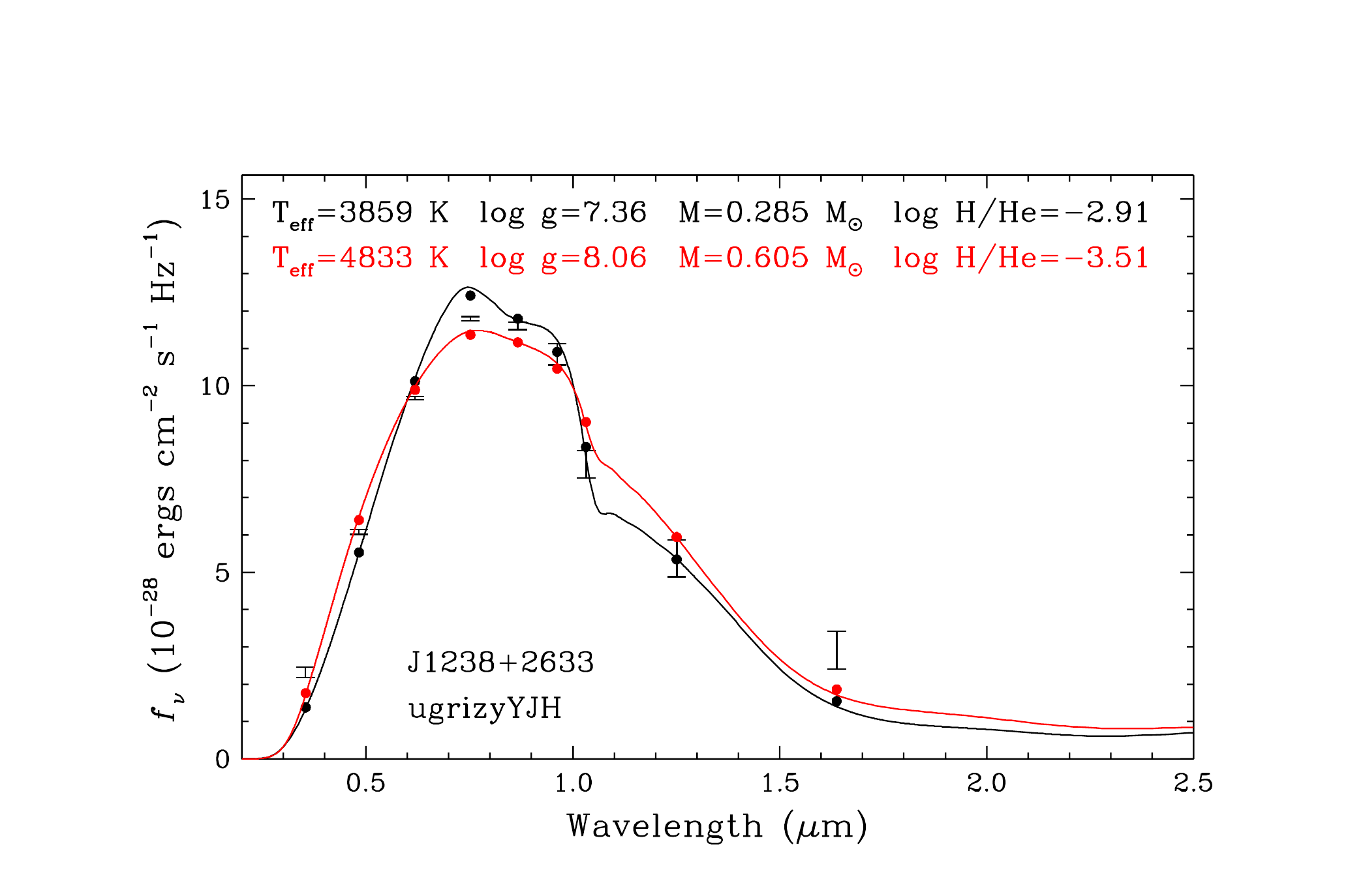}
\caption{Fits to the spectral energy distribution of the IR-faint
  white dwarf J1238+2633 (shown by error bars) using two different
  sets of model atmospheres; solid lines show the monochromatic fluxes
  for the best-fit model for each star, and dots show the synthetic
  photometry of those models in each filter. The corresponding stellar
  parameters are given at the top of the figure. The solution shown in
  black has been obtained with our original model grid (see text),
  while that shown in red comes from models similar to those of
  \citet{blouin18a} where the H$_2$-He CIA opacity profiles from
  \citet{abel12} have been replaced with those of \citet{jorgensen00}.
\label{figJ1238}}
\end{figure}

Despite this situation, we decided to continue our model comparison but
this time by using the CIA profiles from \citet{jorgensen00}
throughout. We calculated a small grid of models using the same
atmosphere code as in \citet{blouin18a}, except for the adopted
CIA opacity, and fitted J1238+2633, a fairly warm IR-faint white dwarf
in our extended sample (see Section \ref{mild}). The comparison of our
fits using both model grids is displayed in Figure
\ref{figJ1238}. While both sets of models fail to provide perfect fits
to the observed energy distribution, the most striking feature is that
the derived stellar parameters are drastically different. In
particular, our old models yield an effective temperature that is
almost 1000 K cooler, and a stellar mass that is $\sim$0.3 $M_{\odot}$
smaller than the solution obtained with this test model grid, entirely consistent
with the odd results we obtain from the photometric analysis of
our IR-faint white dwarf sample.

\begin{figure}
\hspace{0.in}
\includegraphics[width=3.3in, clip=true, trim=0.3in 2.4in 0.3in 3.0in]{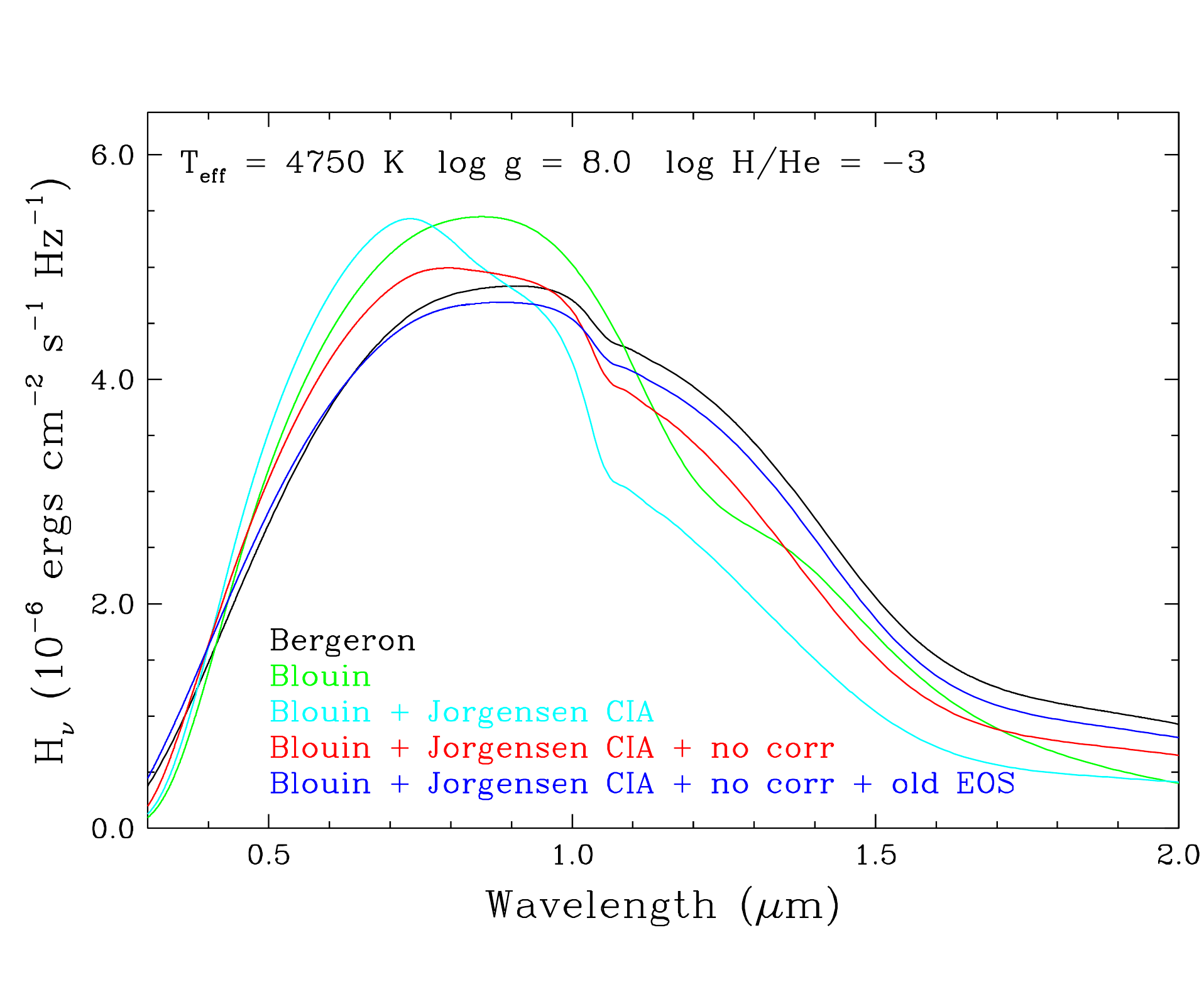}
\caption{Spectral energy distributions from model atmospheres
  calculated with different assumptions related to the input physics. The
  black (Bergeron) and green (Blouin) lines correspond respectively to
  our old models (see text) and those of \citet{blouin18a}, while the
  remaining three energy distributions are similar to Blouin's models
  but with the H$_2$-He CIA opacity profiles from \citet{jorgensen00}
  instead of \citet{abel12}. In addition, we have removed the correction to
  the He$^-$ free-free absorption coefficient from \citet{iglesias02}
  in the model shown in red (no corr), and used the ideal gas equation-of-state 
  in the model shown in blue (old EOS).
\label{figBlouin}}
\end{figure}

In order to understand which of the physical effects is responsible
for the observed discrepancy between our solutions, we compare in
Figure \ref{figBlouin} various spectral energy distributions with
different assumptions in the input physics. The black (labeled
Bergeron) and green (labeled Blouin) lines correspond respectively to
our old models and those of \citet{blouin18a}. As discussed above,
even though these two energy distributions appear qualitatively
similar, the differences are much more important at lower
temperatures. In the model shown in cyan, we used the same input
physics as in Blouin et al., but we rely on the H$_2$-He CIA opacity
profiles from \citet{jorgensen00} instead of \citet{abel12}; the
effect here is significant (see also Figure \ref{figAbel}). Then, in
the model shown in red, we removed the correction to the He$^-$
free-free absorption coefficient from \citet{iglesias02}; as can be
seen here, this correction has a major effect on the predicted energy
distribution.  And finally, in the model shown in blue, we reverted
back to the ideal gas equation of state; in the particular physical
regime explored here, the nonideal effects due to the equation of
state are small but non-negligible compared to the changes in H$_2$-He
CIA opacity and He$^-$ correction. We can also see that this last
model is practically identical to our old model (shown in
black). These comparisons indicate that for all models calculated with
the H$_2$-He CIA opacity profiles from \citet{jorgensen00}, the most
important effect is the inclusion of the correction to the He$^-$
free-free absorption coefficient.

\begin{figure}
\hspace{0.in}
\includegraphics[width=3.3in, clip=true, trim=0.3in 2.4in 0.3in 2.7in]{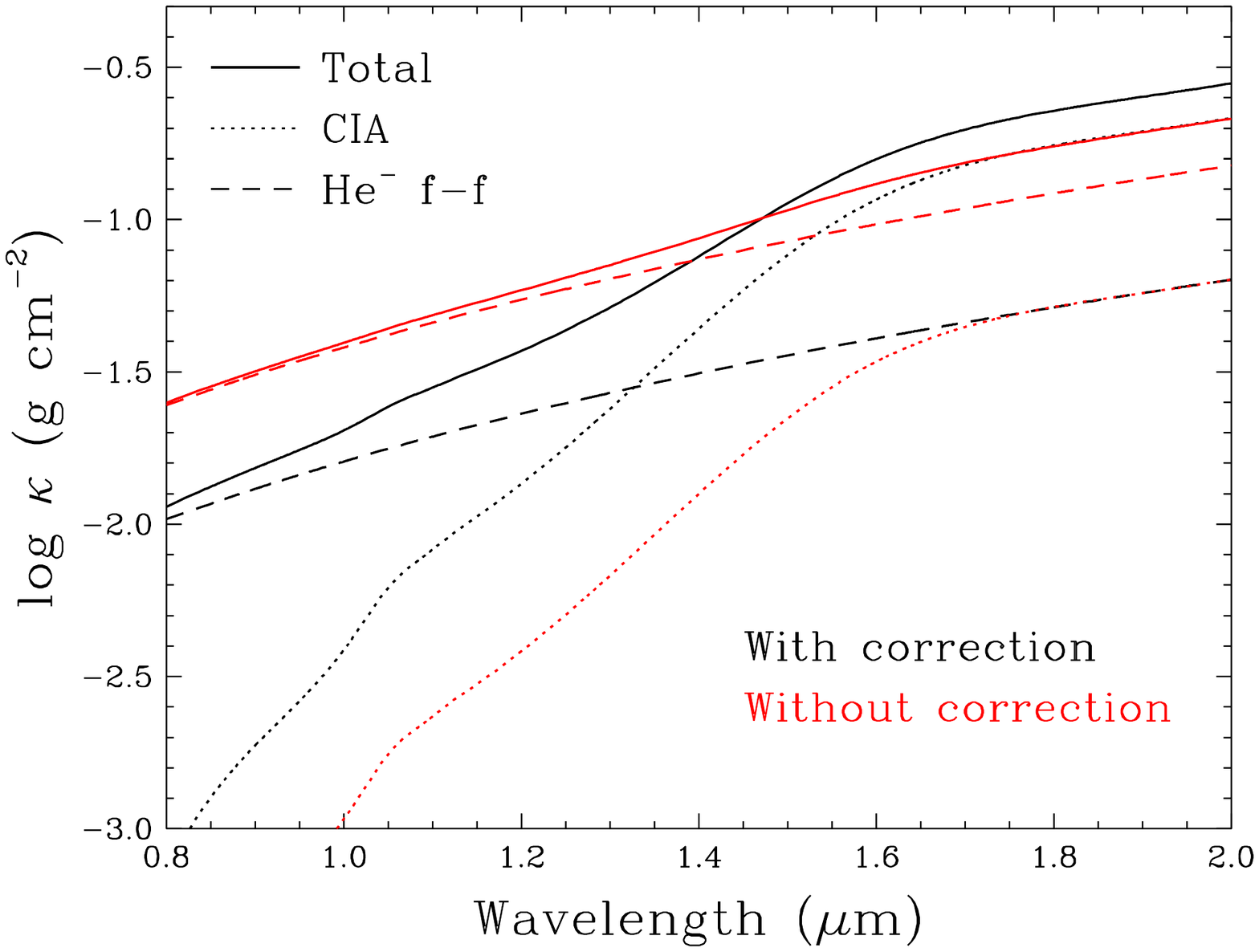}
\caption{Comparison of the H$_2$-He CIA, He$^-$ free-free, and total
  opacity as a function of wavelength at the photosphere of models
  at $T_{\rm eff}=4750$~K, $\logg=8$, and log H/He = $-3$, with and
  without the high-density correction to the He$^-$ free-free absorption
  coefficient from \citet{iglesias02} taken into account.
\label{plot_Kilic}}
\end{figure}

This particular behavior can be explained in terms of a competition
between the H$_2$-He CIA opacity and the He$^-$ free-free opacity, as
illustrated in Figure \ref{plot_Kilic}, where we show the contributions
of each opacity source at the photosphere of our test model, with and
without the correction to the He$^-$ free-free opacity taken into
account. The correction has the effect of reducing significantly the
relative contribution of the He$^-$ free-free opacity with respect to
that from the H$_2$-He CIA opacity, as can also be appreciated by
comparing the cyan and red models in Figure \ref{figBlouin}. In
addition, the overall total opacity is reduced, resulting in larger
densities that affect the overall atmospheric structure. Given these
results, we have thus decided to recalculate our original model grid
described at the beginning of this subsection but by properly
including this high-density correction factor to the He$^-$ free-free
absorption coefficient.

\begin{figure}
\hspace{-0.4in}
\includegraphics[width=4.1in, clip=true, trim=0.0in 6.0in 0.in 1.0in]{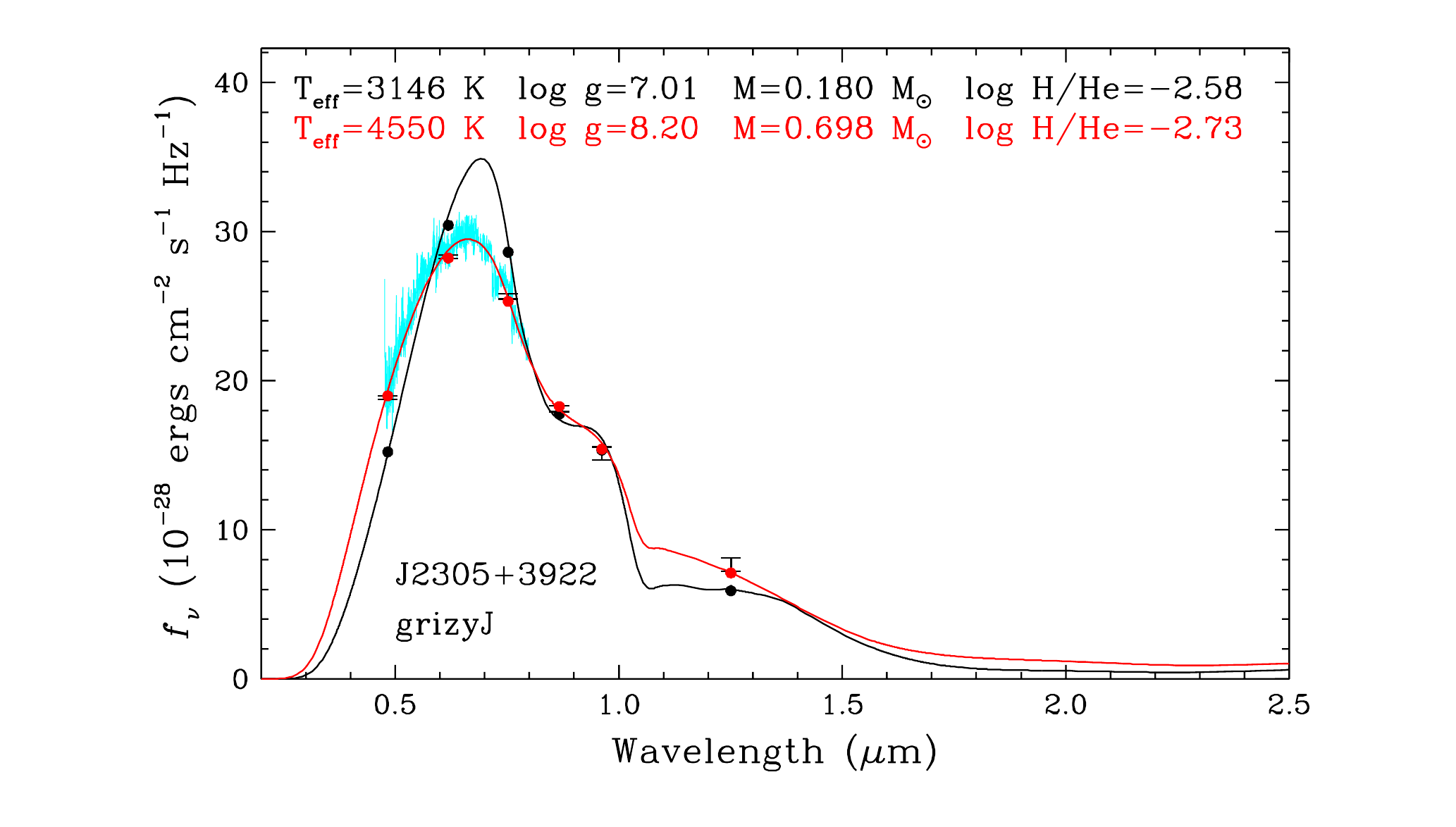}
\caption{Fits to the spectral energy distribution of the IR-faint
  white dwarf J2305+3922. This object was also spectroscopically confirmed
  at Gemini, and the spectrum is shown in cyan. Here we show two
  solutions, one with our original model grid (black), and the other
  solution using our revised model grid where the correction to the He$^-$
  free-free absorption coefficient from \citet{iglesias02} has been
  taken into account (red).
\label{fig_improve}}
\end{figure}

We compare in Figure \ref{fig_improve} the fits to J2305+3922, a
typical IR-faint white dwarf in our sample, obtained with our old
model grid to those achieved with our revised models (a full
comparison for our complete sample will be discussed in Section
\ref{global}). While the photometric fit with the old models is a poor
match to the measured photometry --- similar in quality to the fits
displayed in Figure 20 of \citet{kilic20} ---, the fit to the
photometry (and spectroscopy as well) using our revised model grid is
significantly superior. More importantly, the stellar parameters
(given at the top of the figure) have changed drastically, going from
an ultracool ($T_{\rm eff}=3146$~K) and extremely low-mass (0.18
$M_{\odot}$) IR-faint white dwarf, to a much hotter ($4550$~K) and
massive (0.70 $M_{\odot}$) white dwarf; the H/He abundance ratio
remains essentially unchanged, however. The increase by more than 1400
K in temperature makes the object much more luminous, thus requiring a
smaller stellar radius and larger mass. These revised models will have
important consequences on the results of our analysis.

Note also that despite the fact that we now obtain much higher
effective temperatures for our objects, well within the range of
$T_{\rm eff}$ values calculated by \citet{blouin18a}, those models
would still fail to match the photometry of the objects in our sample
with strong IR-flux deficiencies because of the problem related to
the use of the H$_2$-He CIA opacity profiles from \citet{abel12}, as
discussed above.

We end this section by stressing that even though our revised models
represent a significant improvement over our previous model grid, both
qualitatively and quantitatively, they represent by no means the best
models that could be achieved. First, we have neglected the nonideal
effects in the equation of state. Second, instead of using the
improved H$_2$-He CIA opacity calculations of \citet{abel12}, we
relied on the more approximate calculations from \citet{jorgensen00},
which for some reason provide a much better description of the
observed energy distribution of the coolest IR-faint objects in our
sample. Clearly, this deserves further investigation.

\section{RESULTS}\label{sec:results}

\begin{figure*}
\includegraphics[width=2.4in]{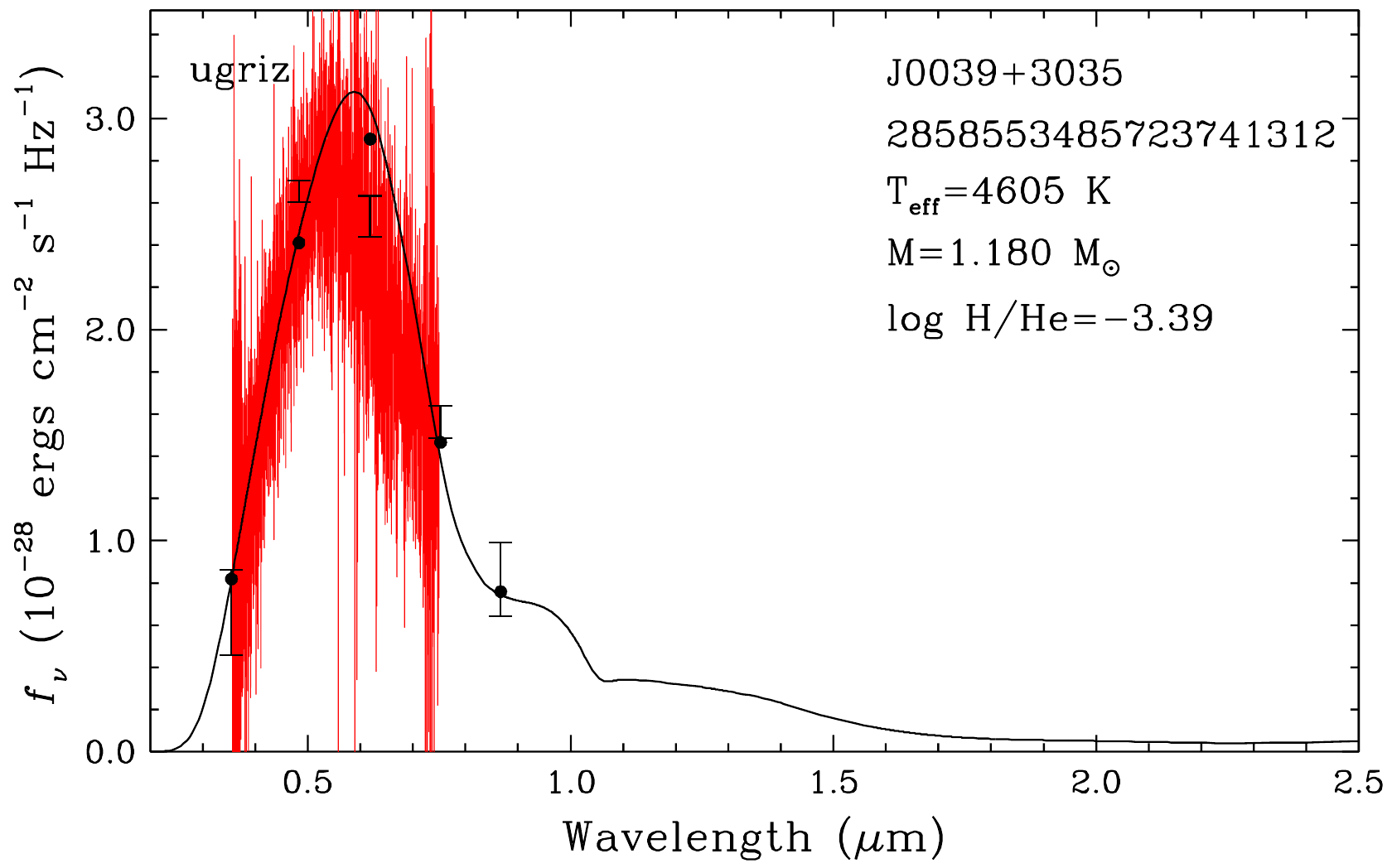}
\includegraphics[width=2.4in]{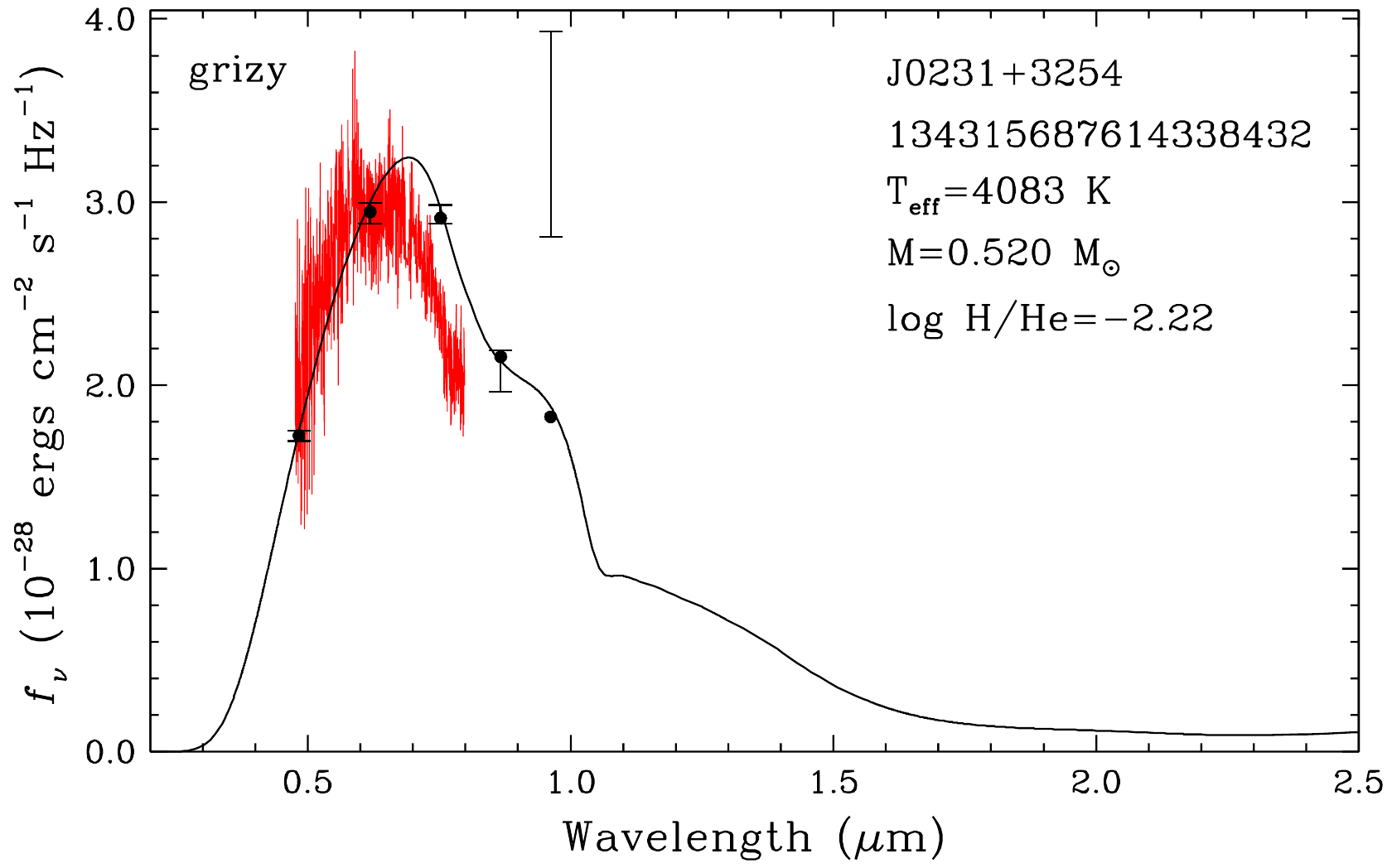}
\includegraphics[width=2.4in]{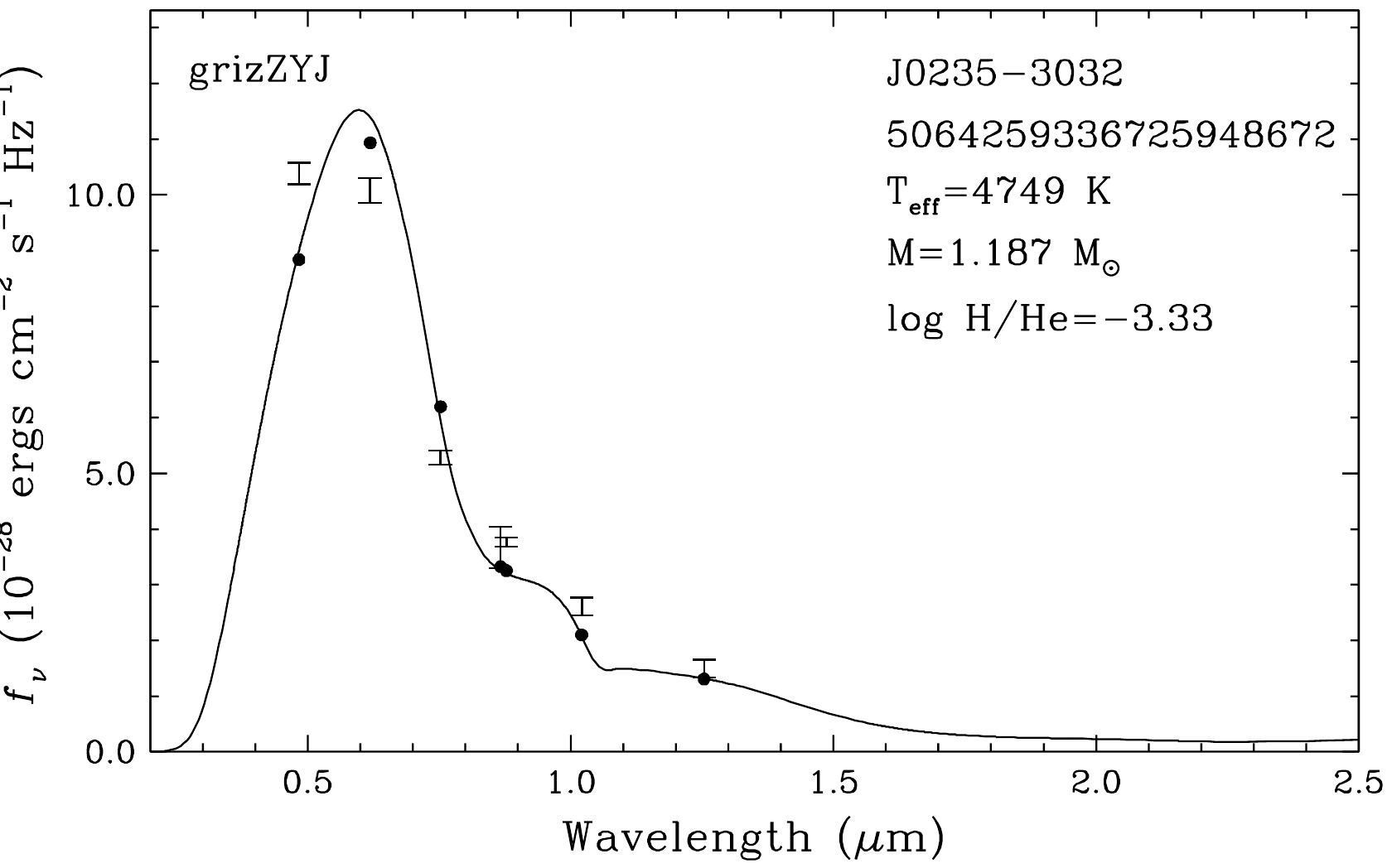}
\includegraphics[width=2.4in]{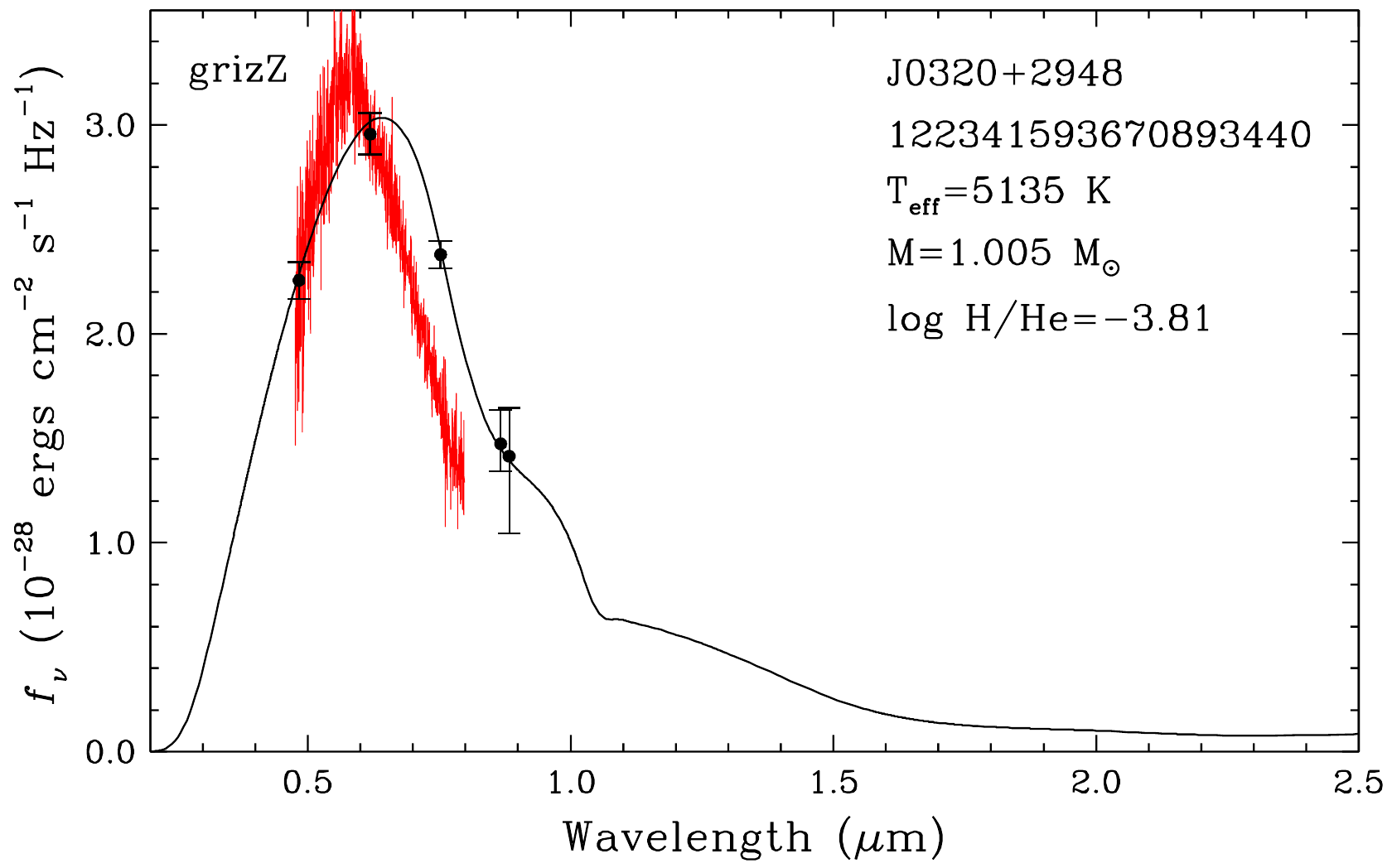}
\includegraphics[width=2.4in]{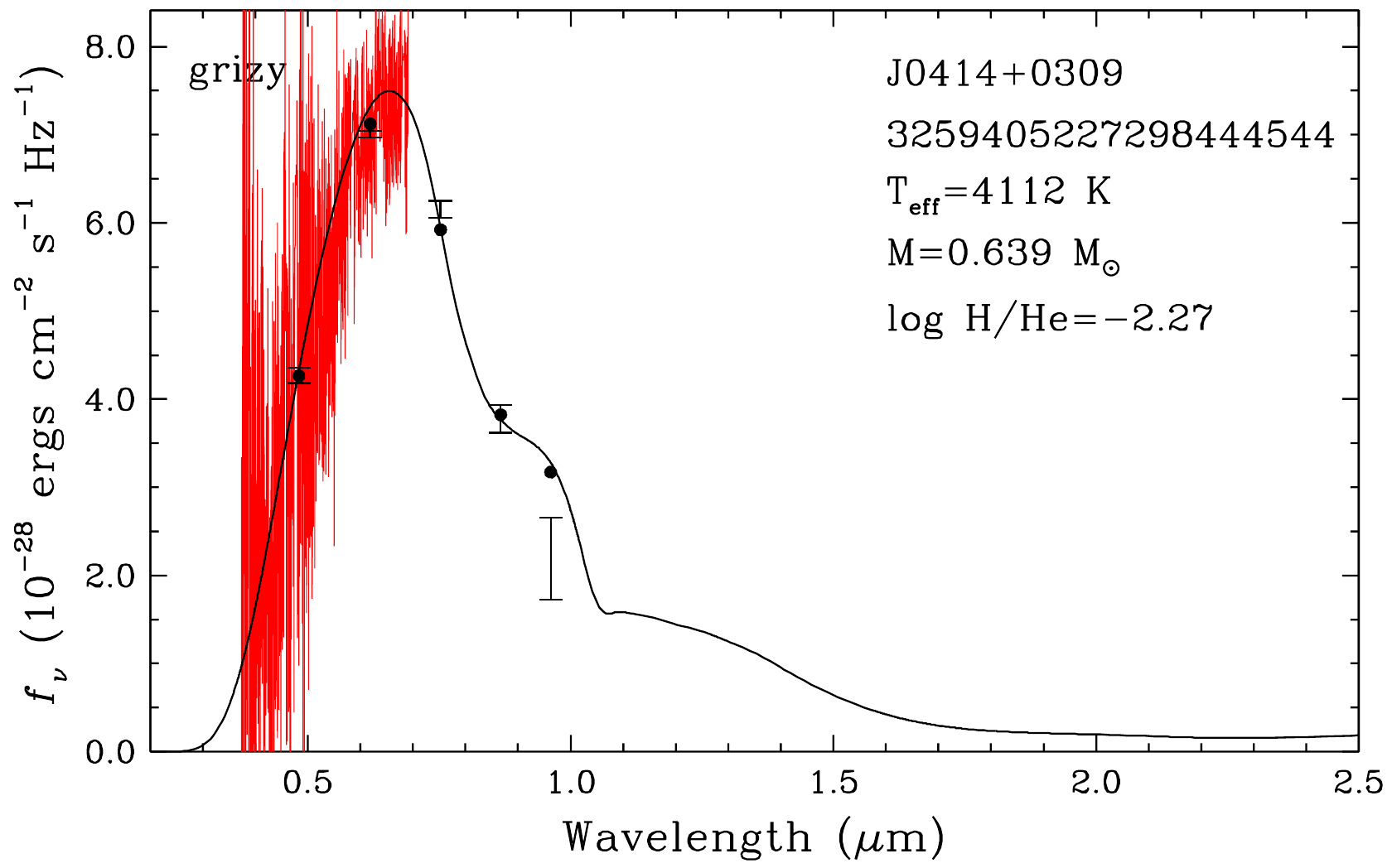}
\includegraphics[width=2.4in]{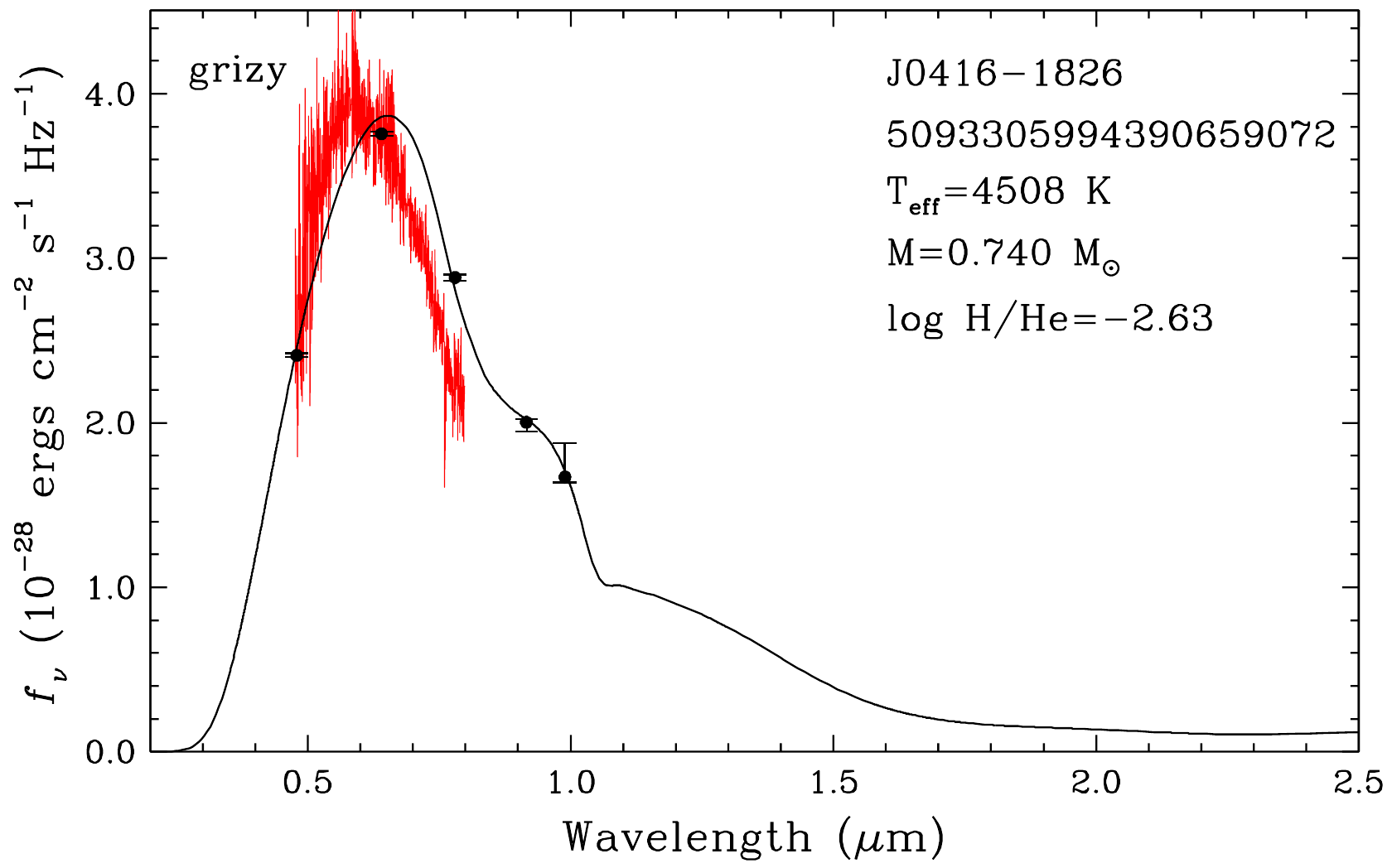}
\includegraphics[width=2.4in]{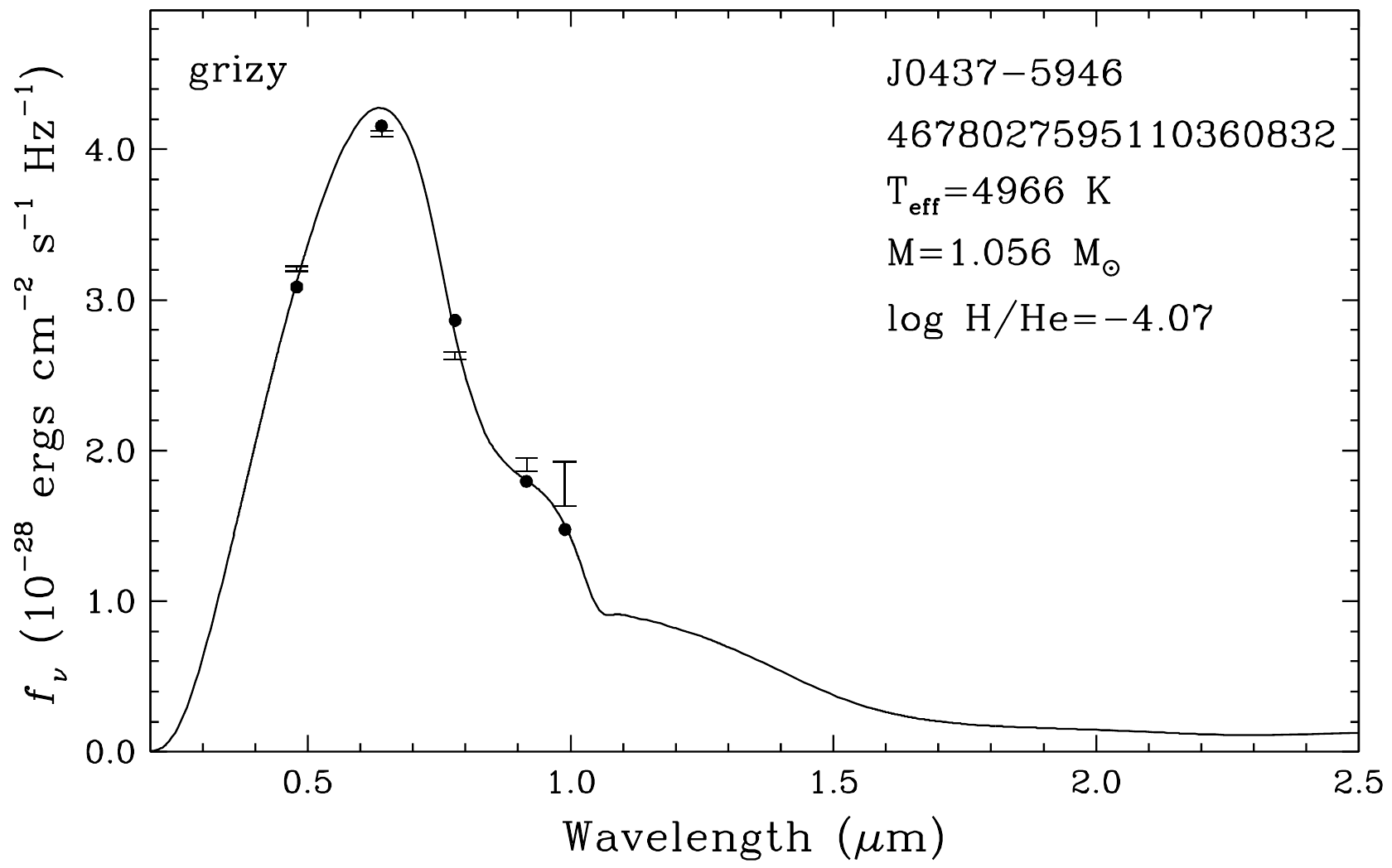}
\includegraphics[width=2.4in]{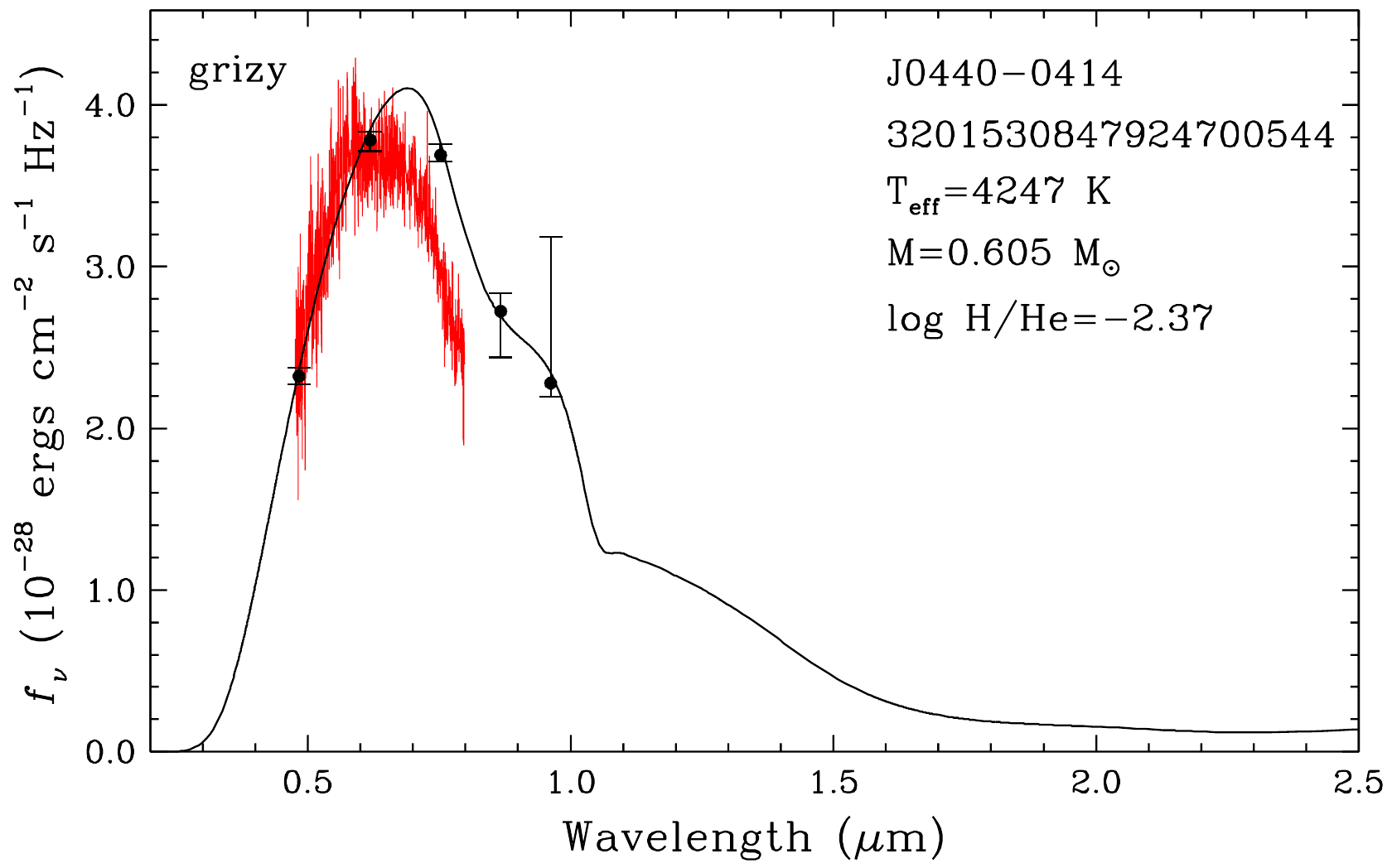}
\includegraphics[width=2.4in]{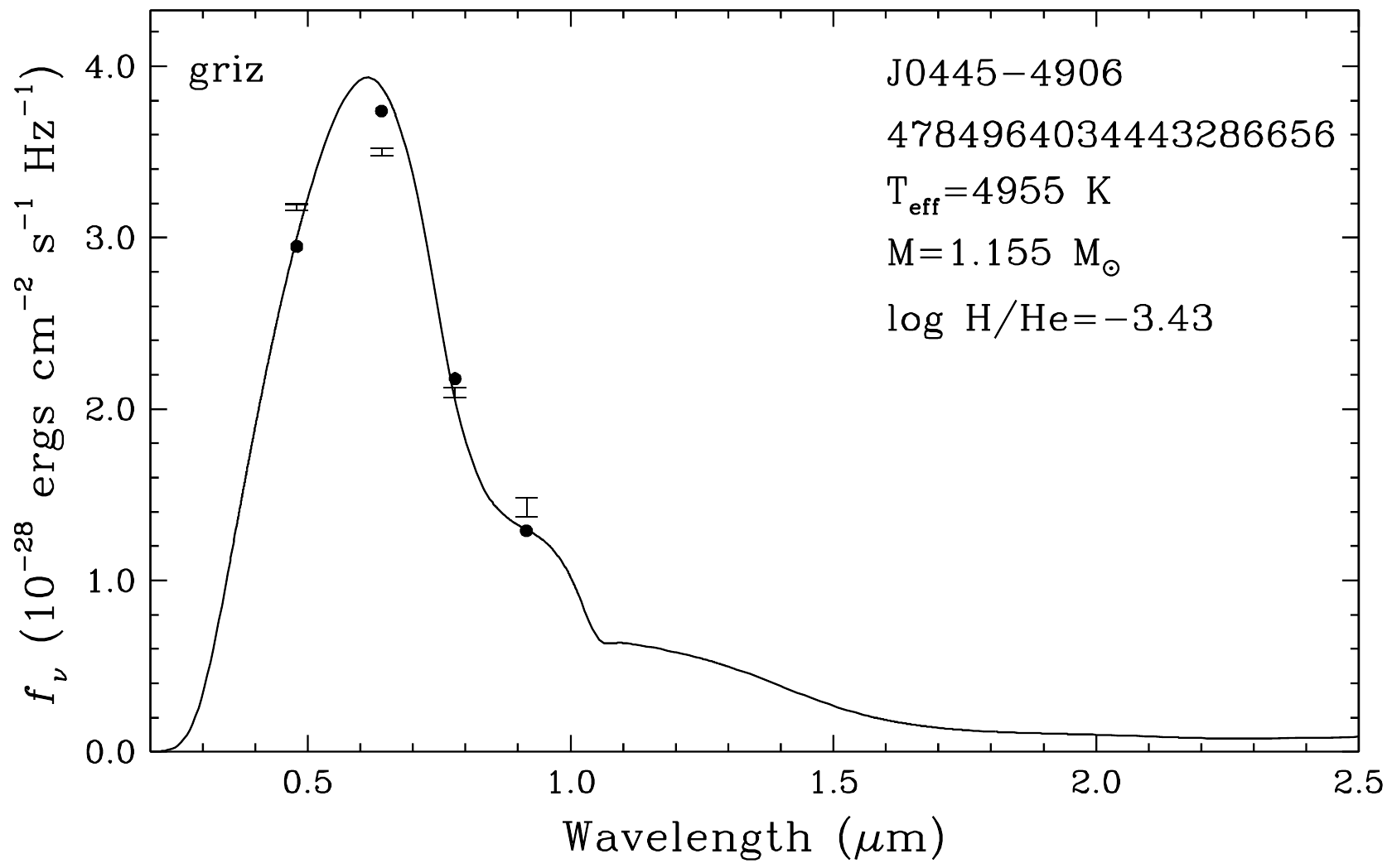}
\includegraphics[width=2.4in]{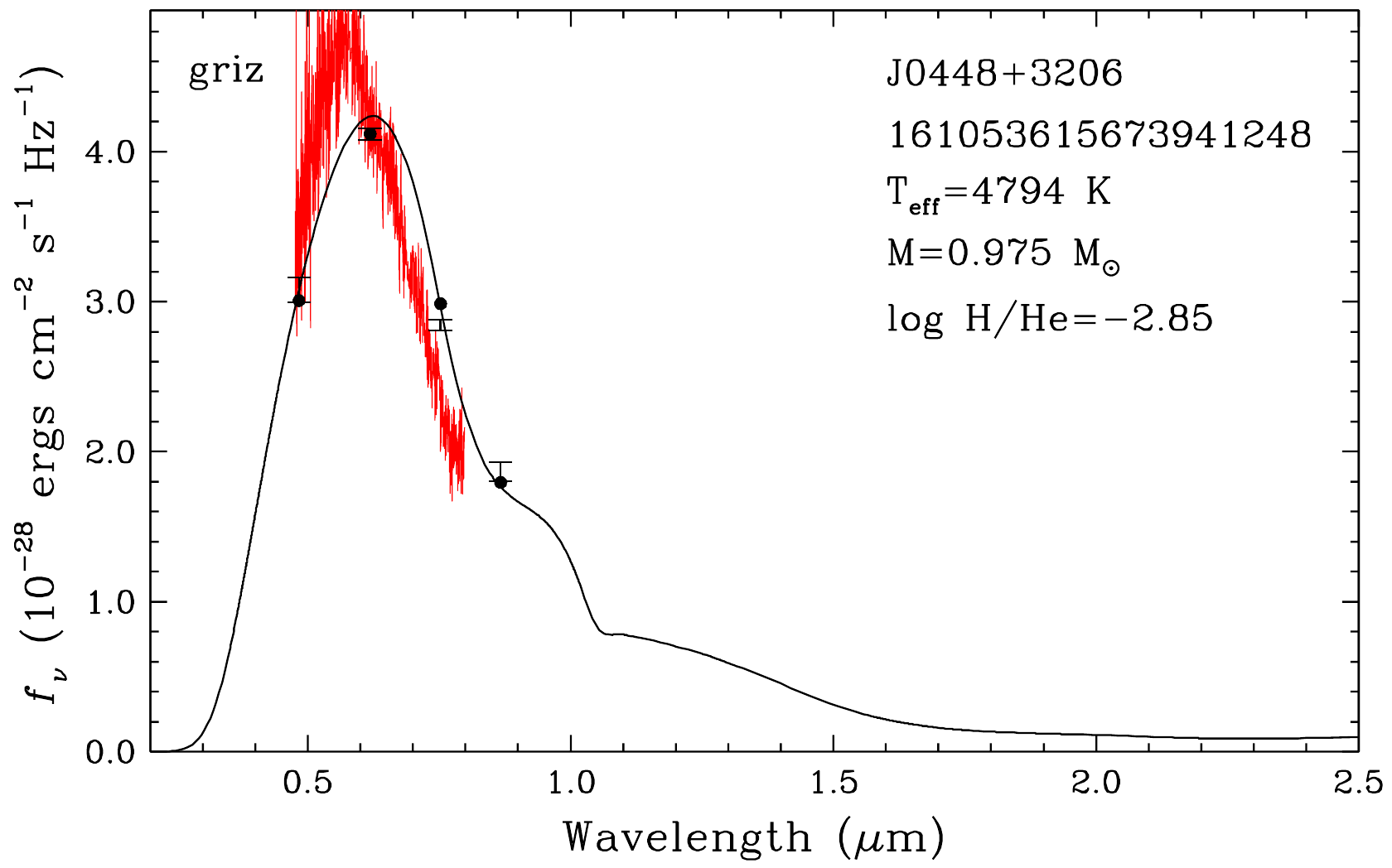}
\includegraphics[width=2.4in]{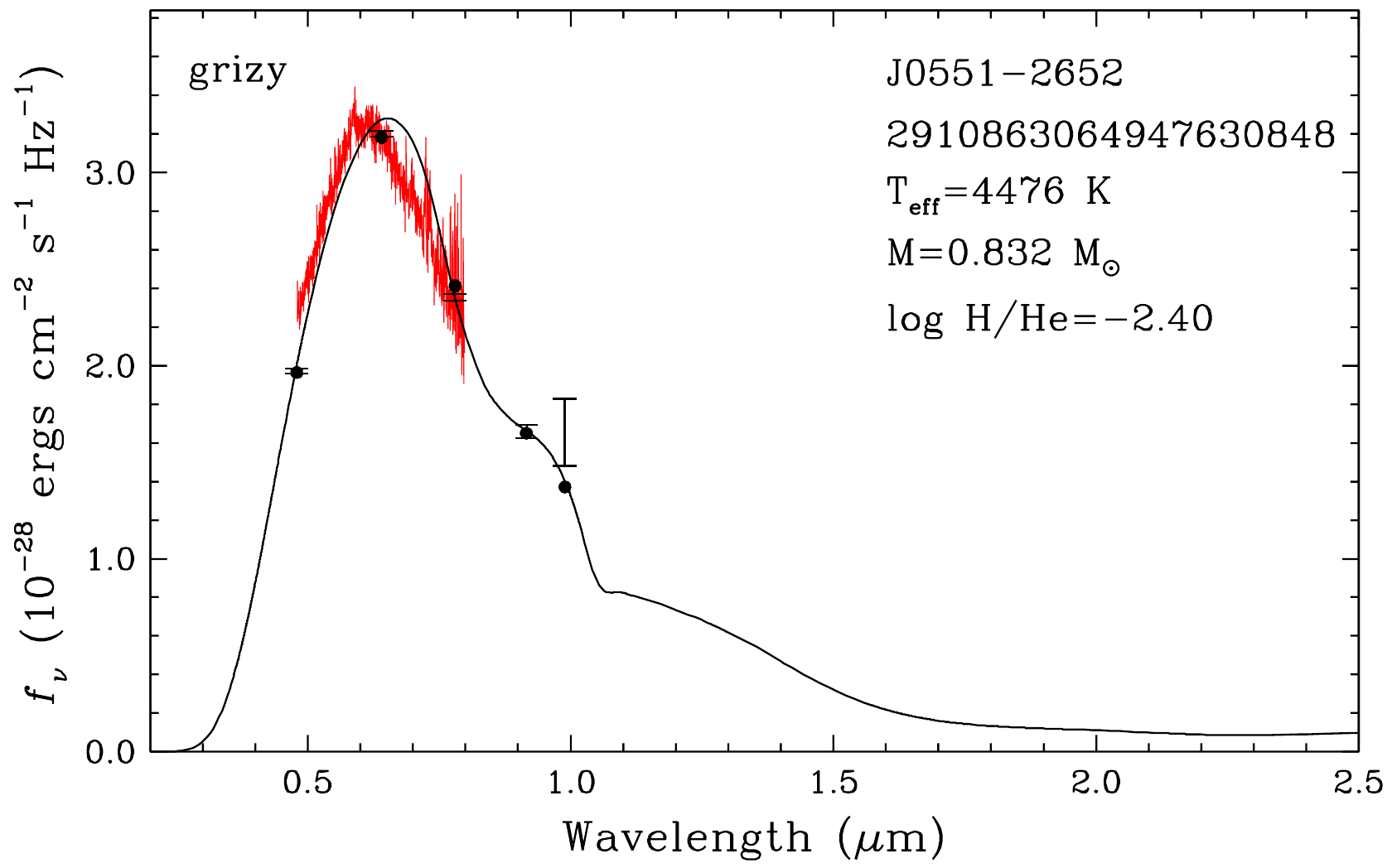}
\includegraphics[width=2.4in]{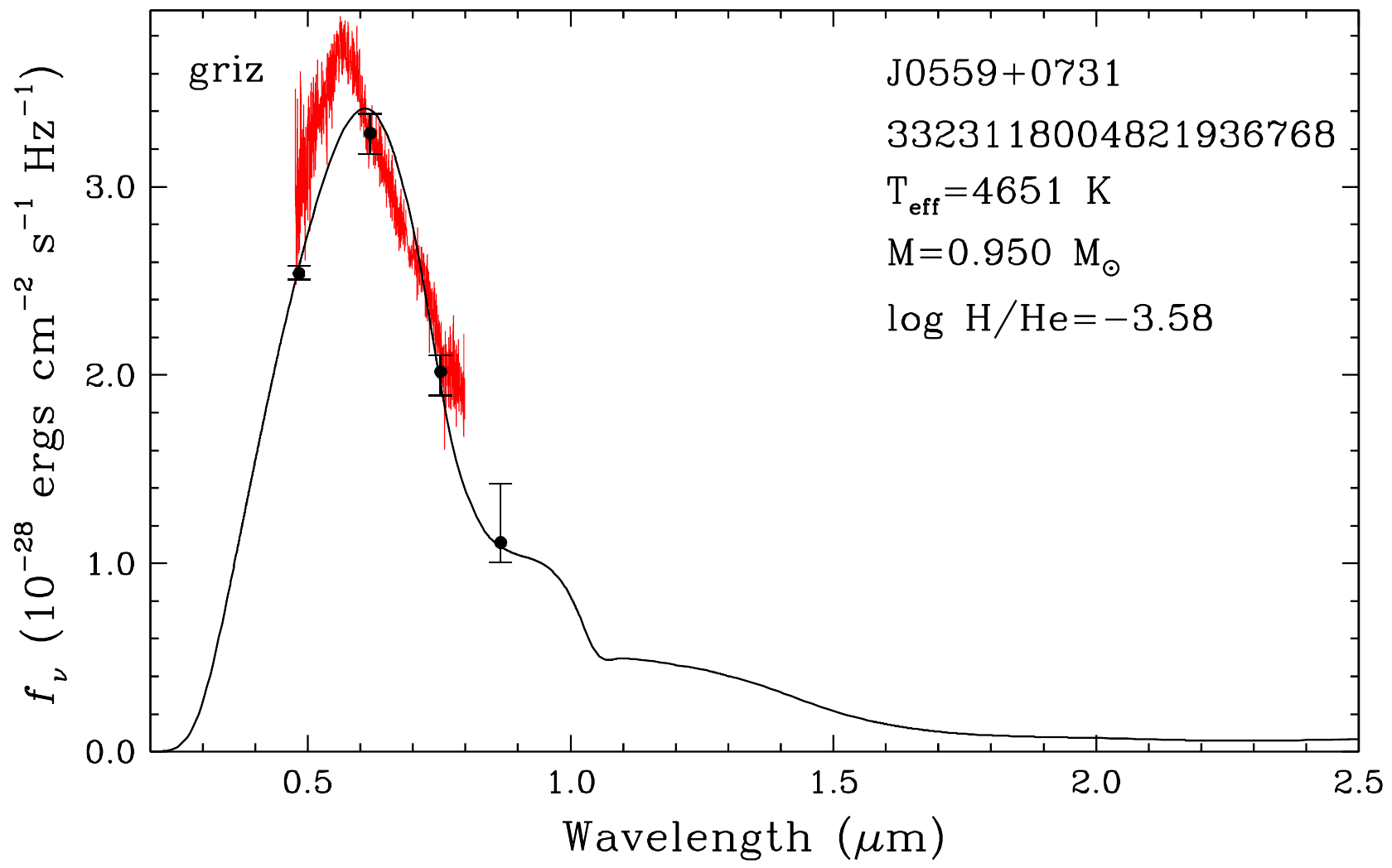}
\includegraphics[width=2.4in]{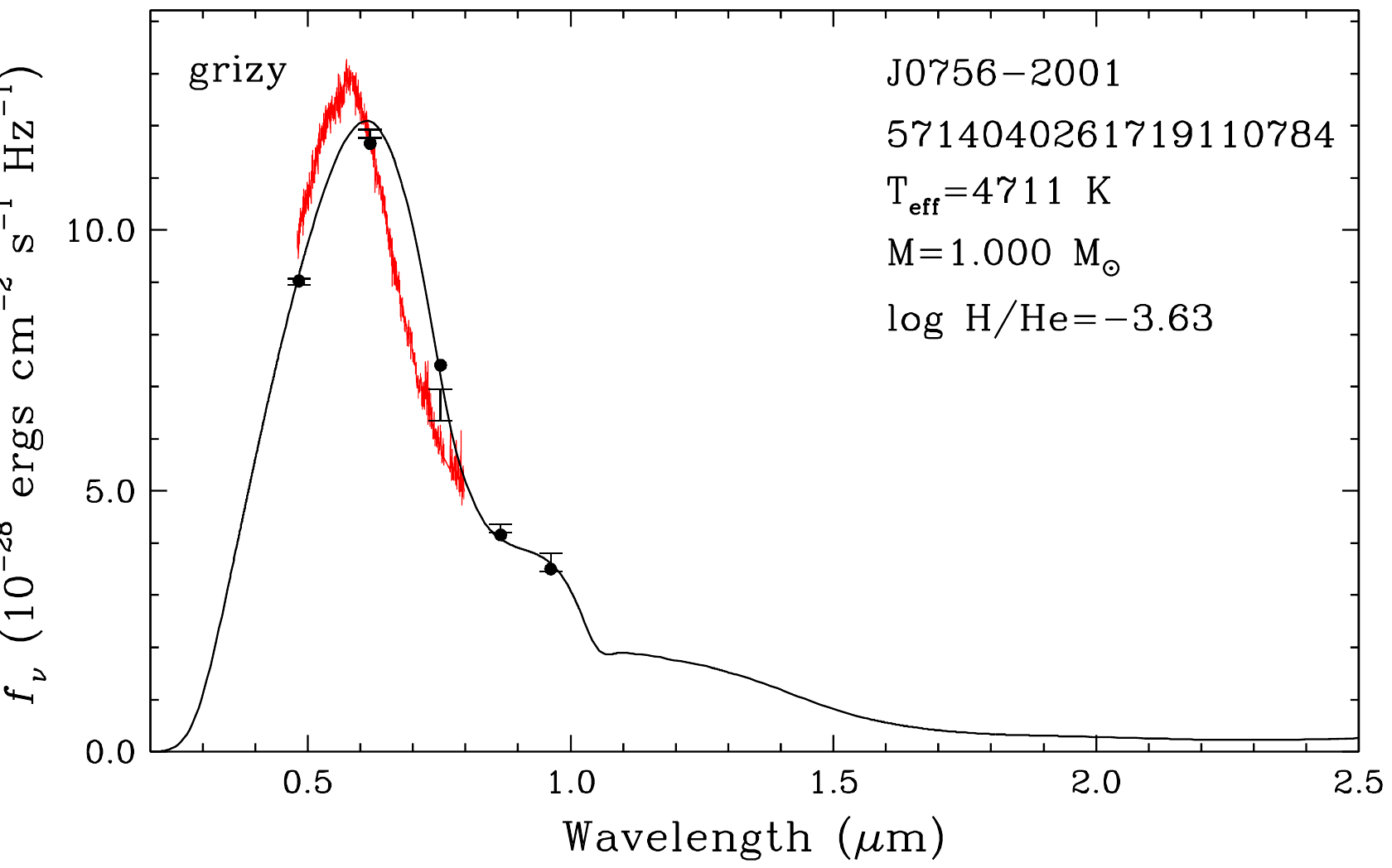}
\includegraphics[width=2.4in]{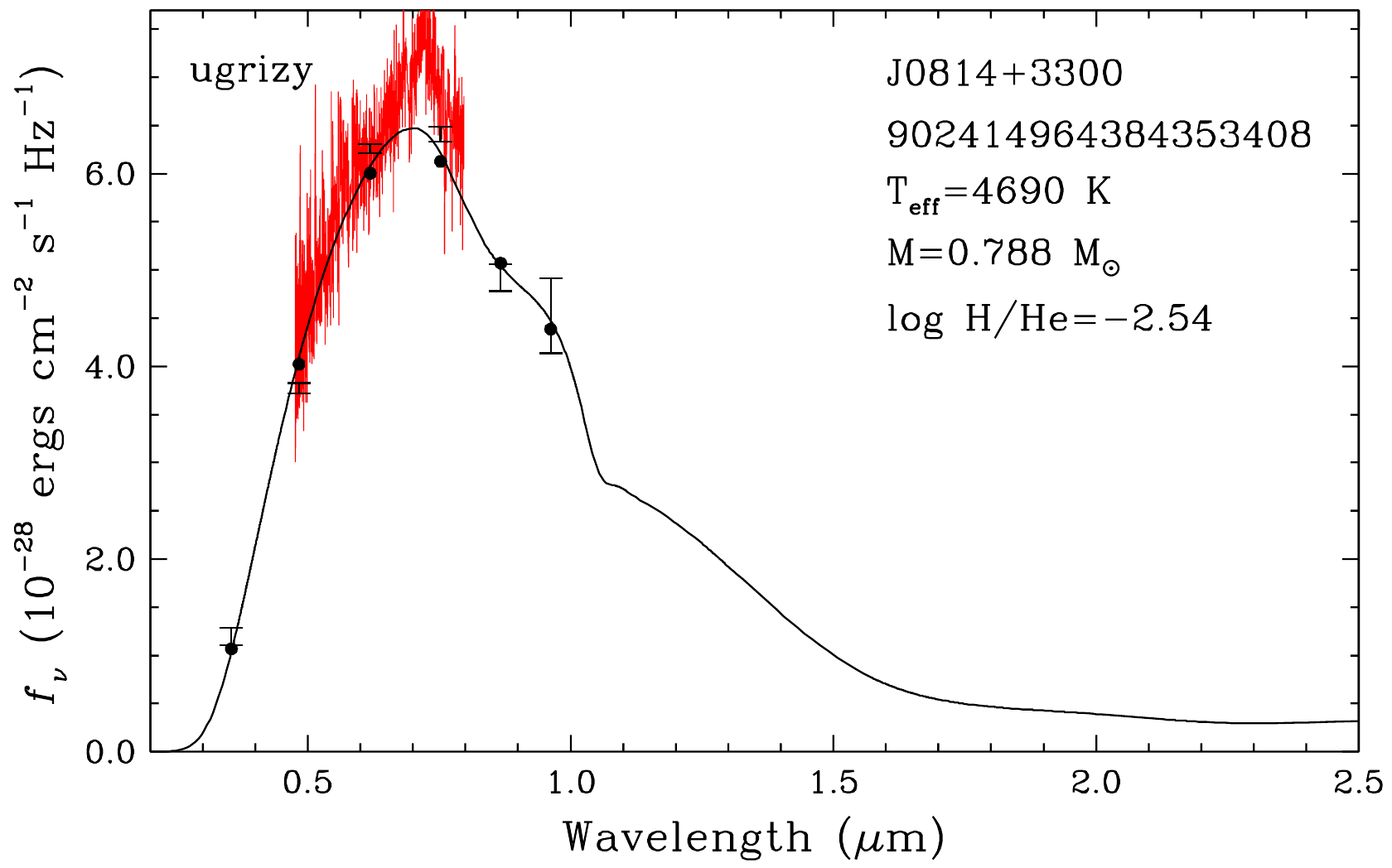}
\includegraphics[width=2.4in]{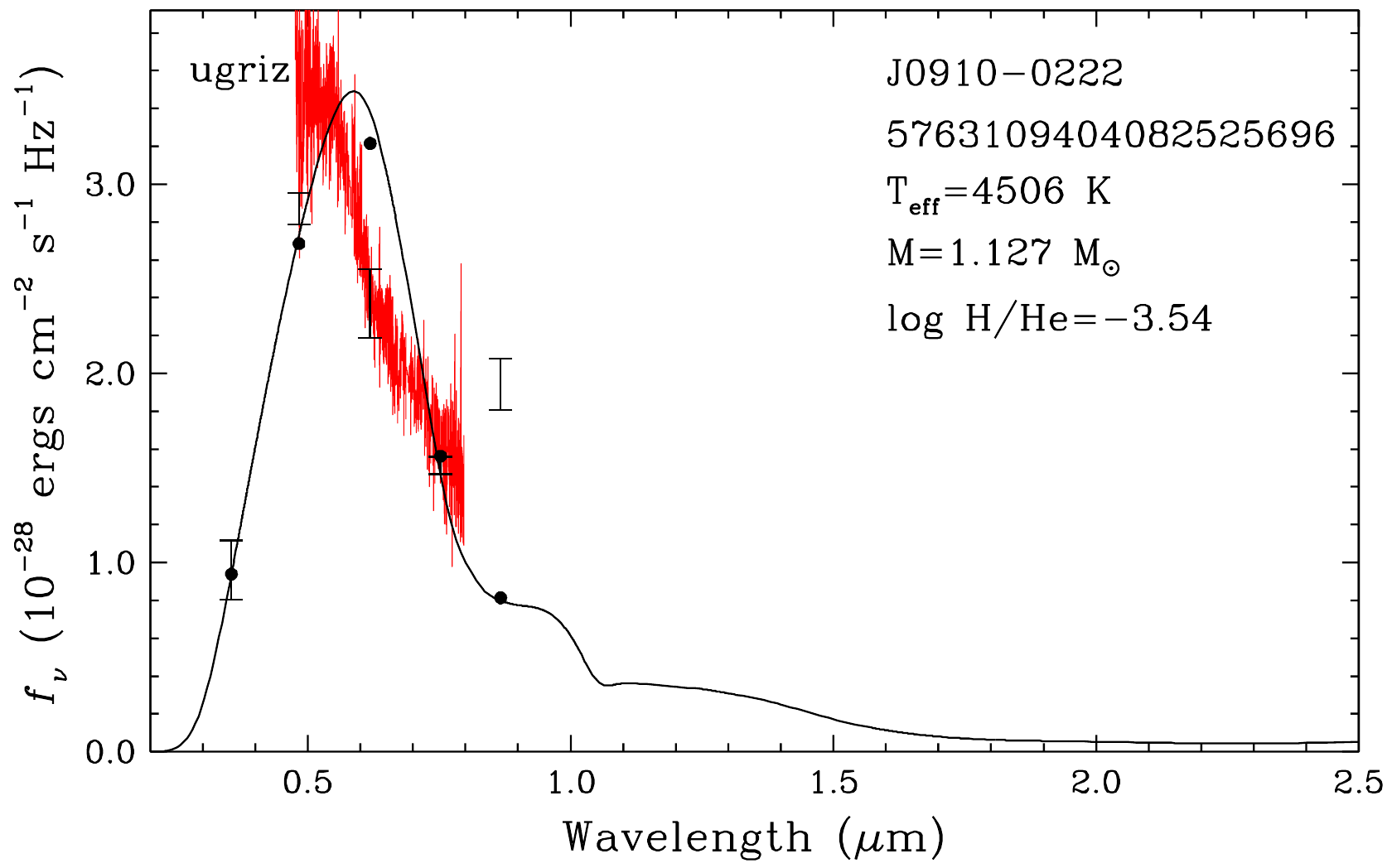}
\caption{Fits to the spectral energy distributions of 37 new IR-faint
  white dwarfs with strong flux deficits in the optical.  Black lines
  show the monochromatic fluxes for the best-fit model for each star,
  and dots show the synthetic photometry of those models in each
  filter. Thirty of these objects were spectroscopically confirmed at
  Gemini, and one (J0039+3035) has a spectrum in the SDSS. These
  spectra are shown in red. The remaining six targets currently lack
  optical spectroscopy.  All 37 targets have spectral energy
  distributions that peak in the optical and display significant
  absorption in the near-infrared bands, and are best explained by
  mixed H/He atmospheres.
\label{fitgem}}
\end{figure*}

\begin{figure*}
\addtocounter{figure}{-1}
\includegraphics[width=2.4in]{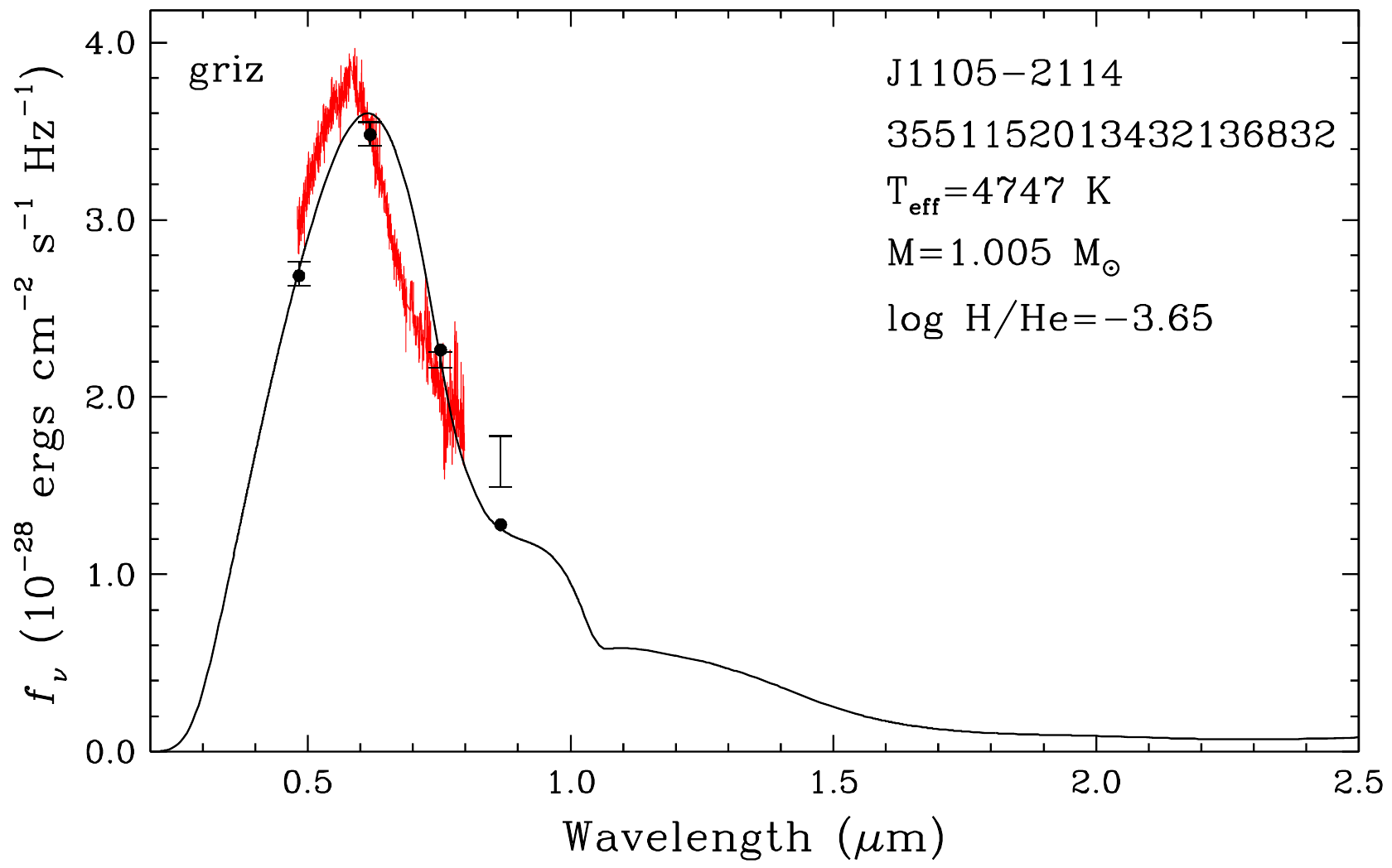}
\includegraphics[width=2.4in]{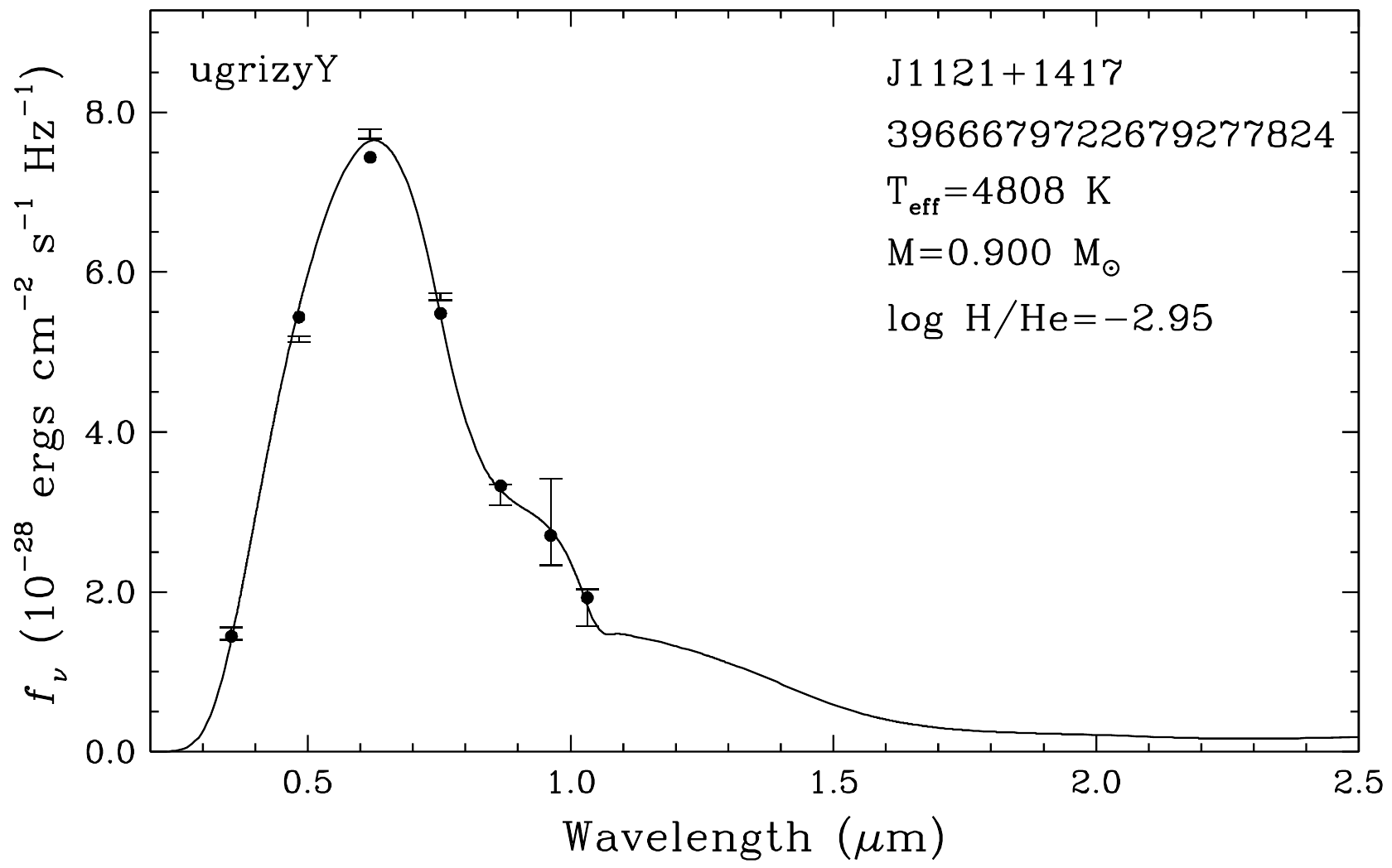}
\includegraphics[width=2.4in]{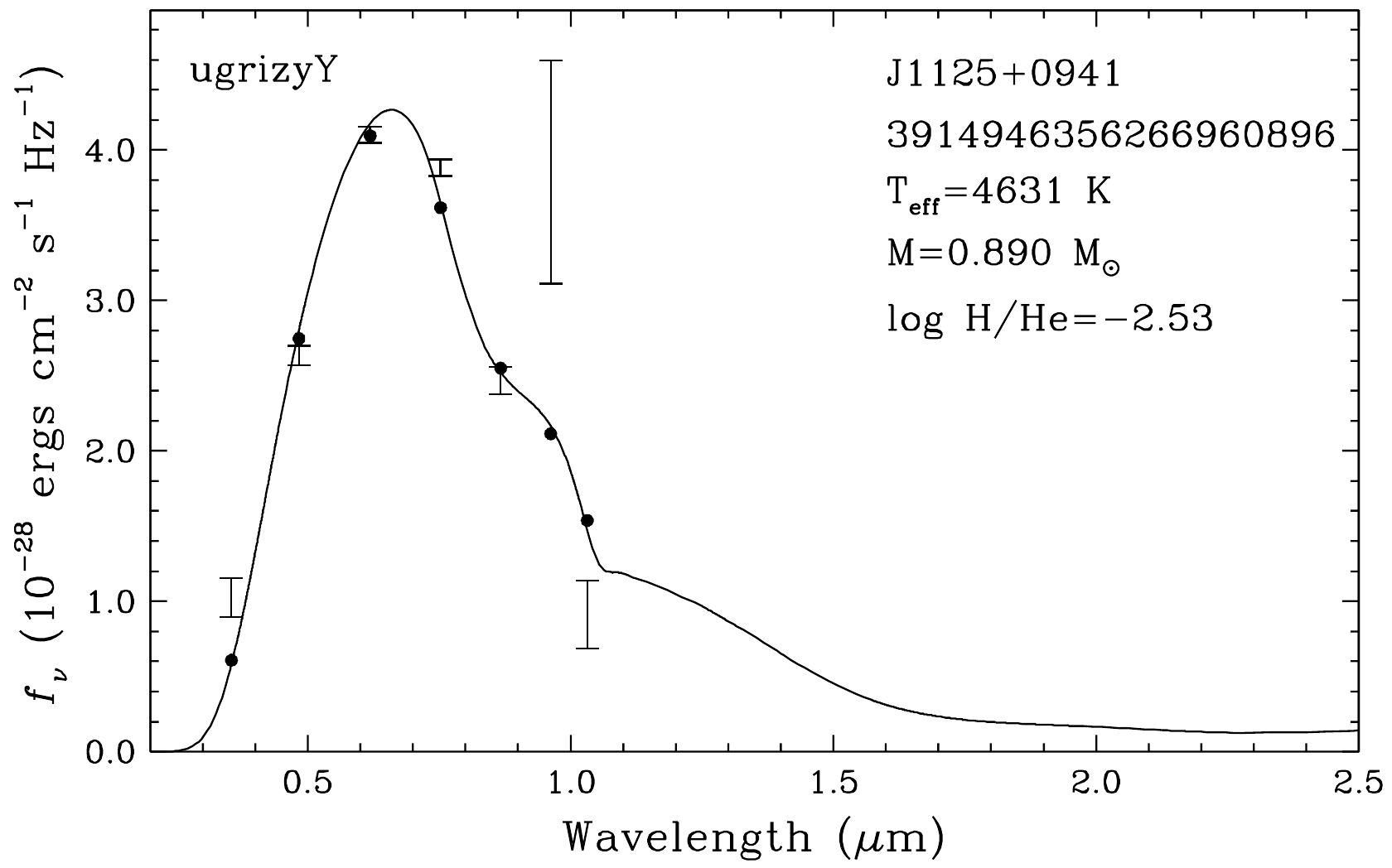}
\includegraphics[width=2.4in]{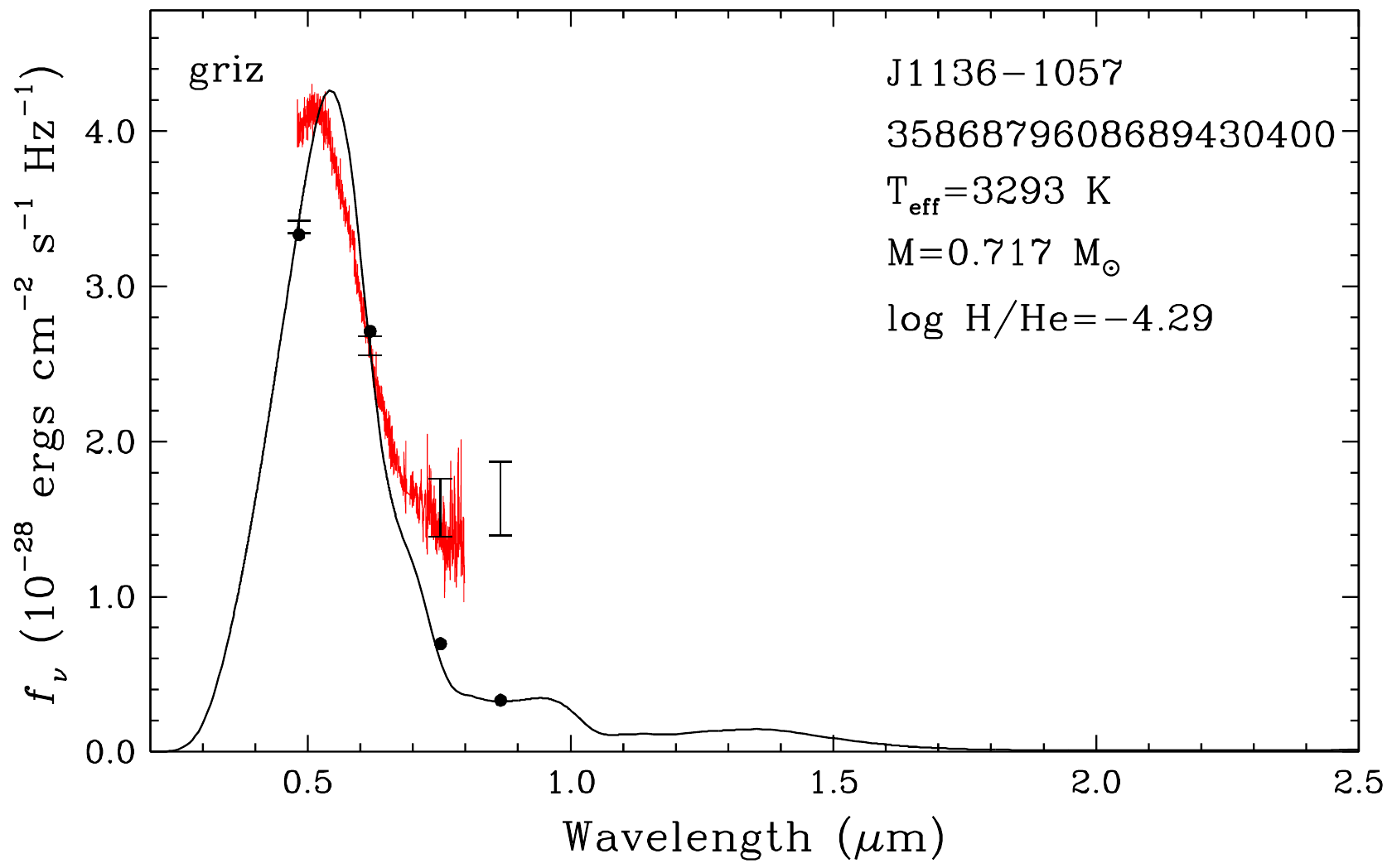}
\includegraphics[width=2.4in]{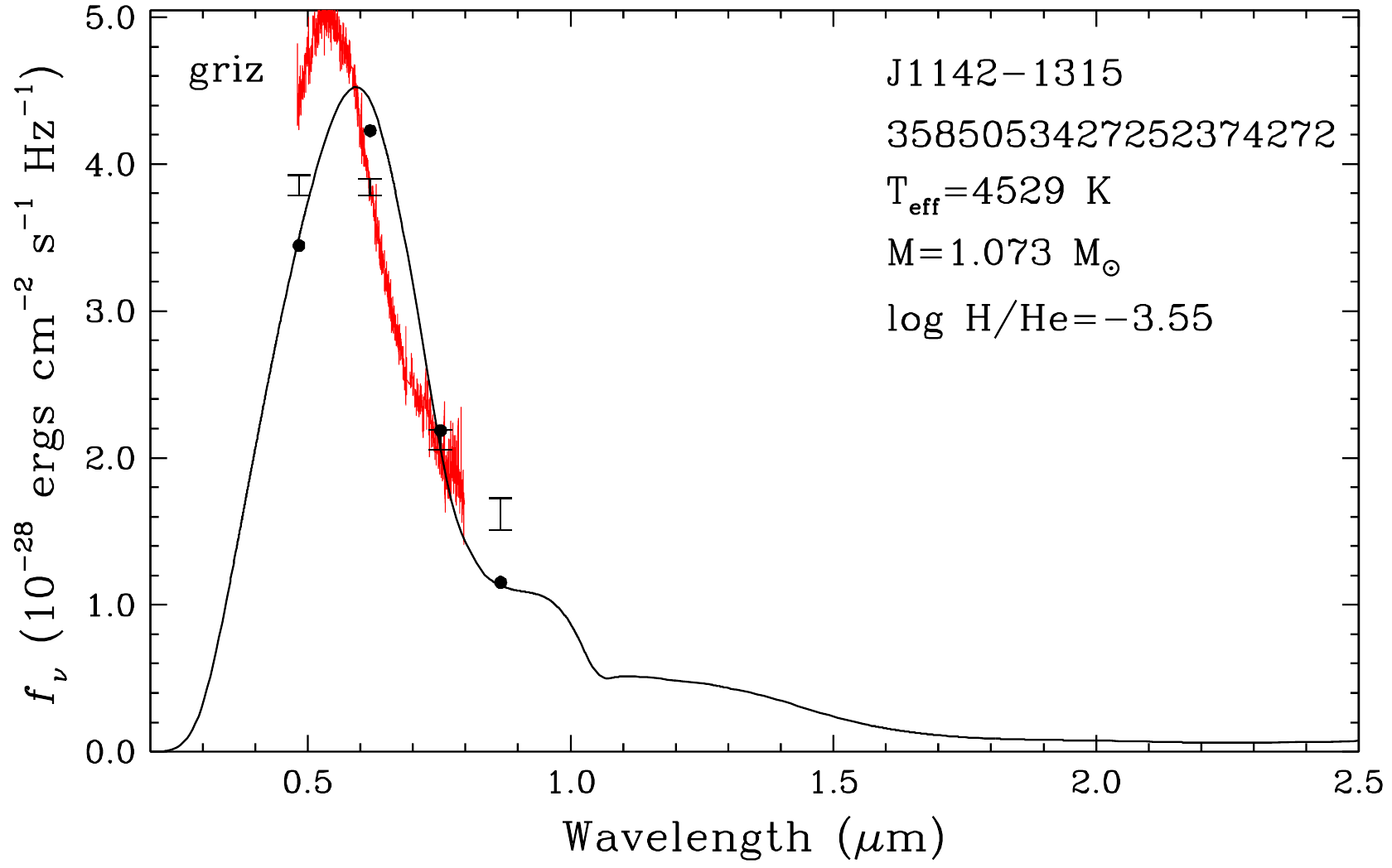}
\includegraphics[width=2.4in]{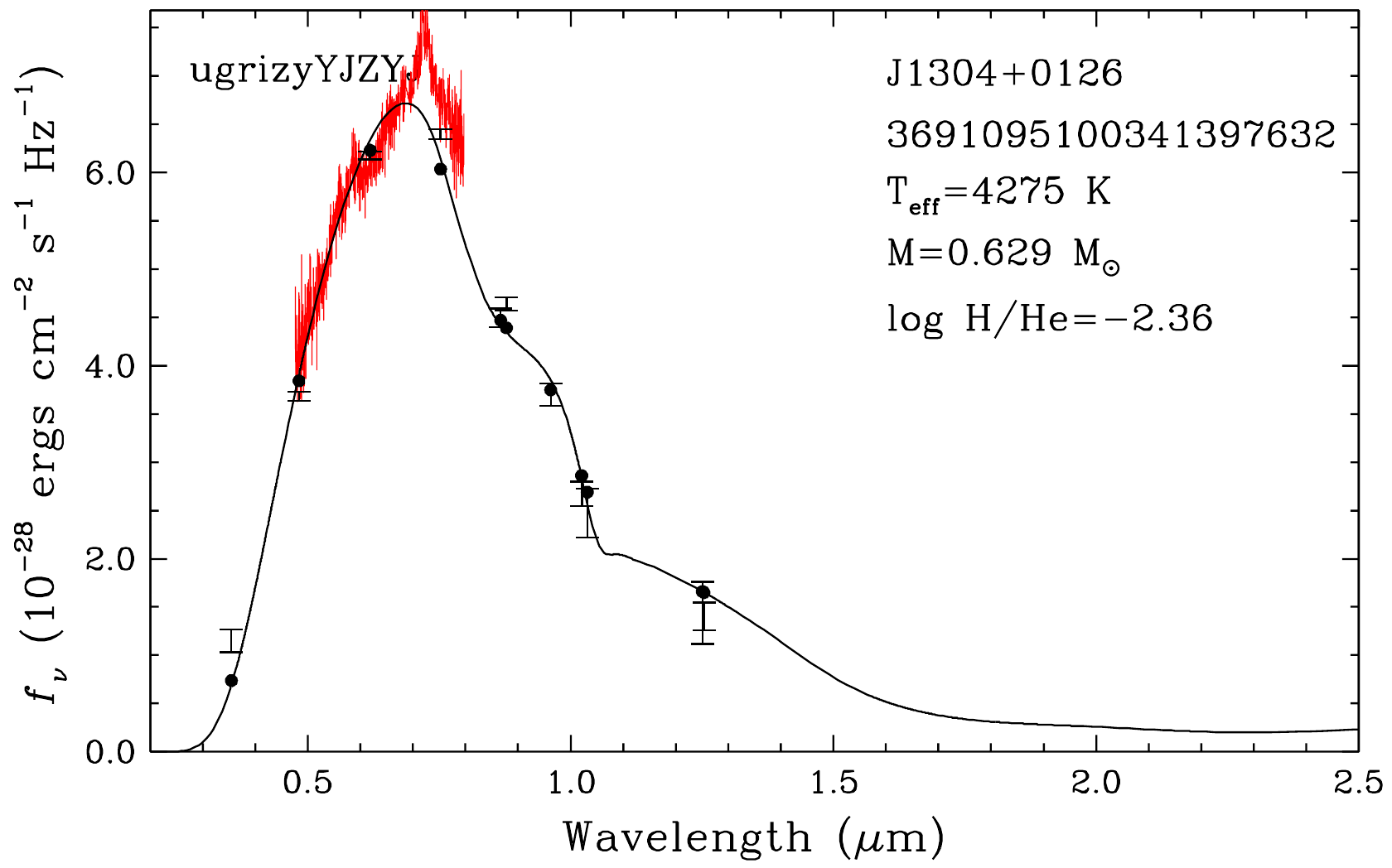}
\includegraphics[width=2.4in]{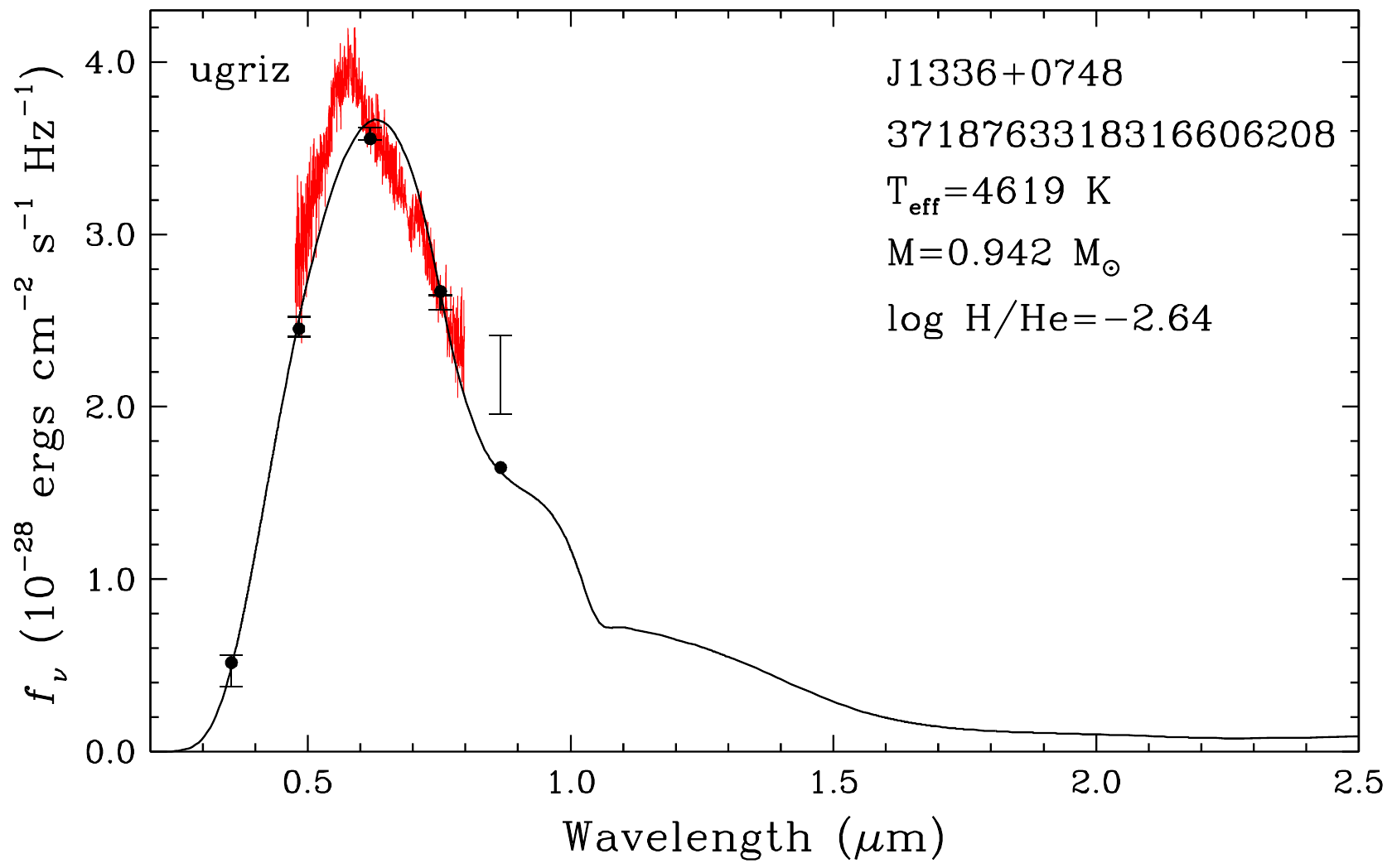}
\includegraphics[width=2.4in]{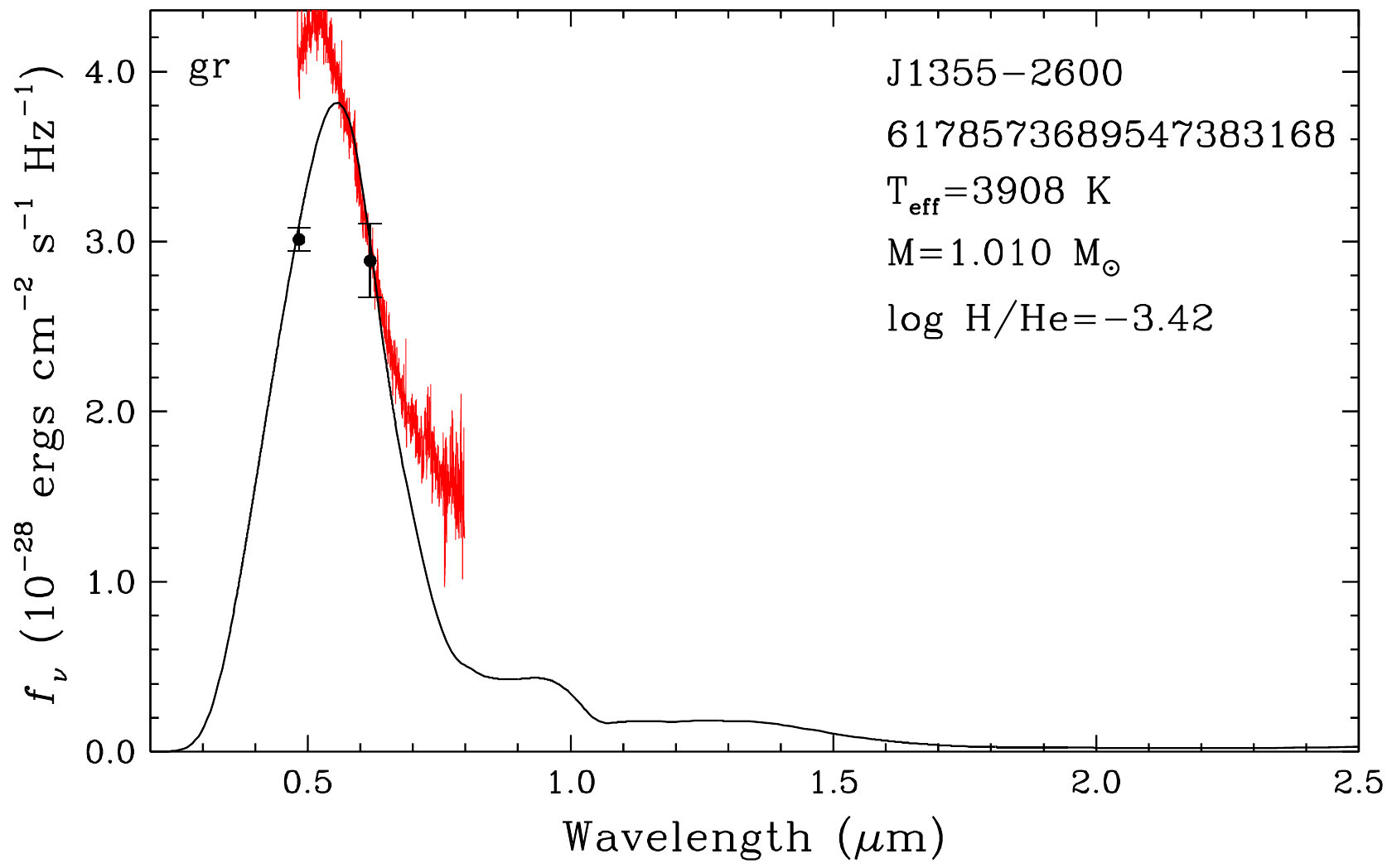}
\includegraphics[width=2.4in]{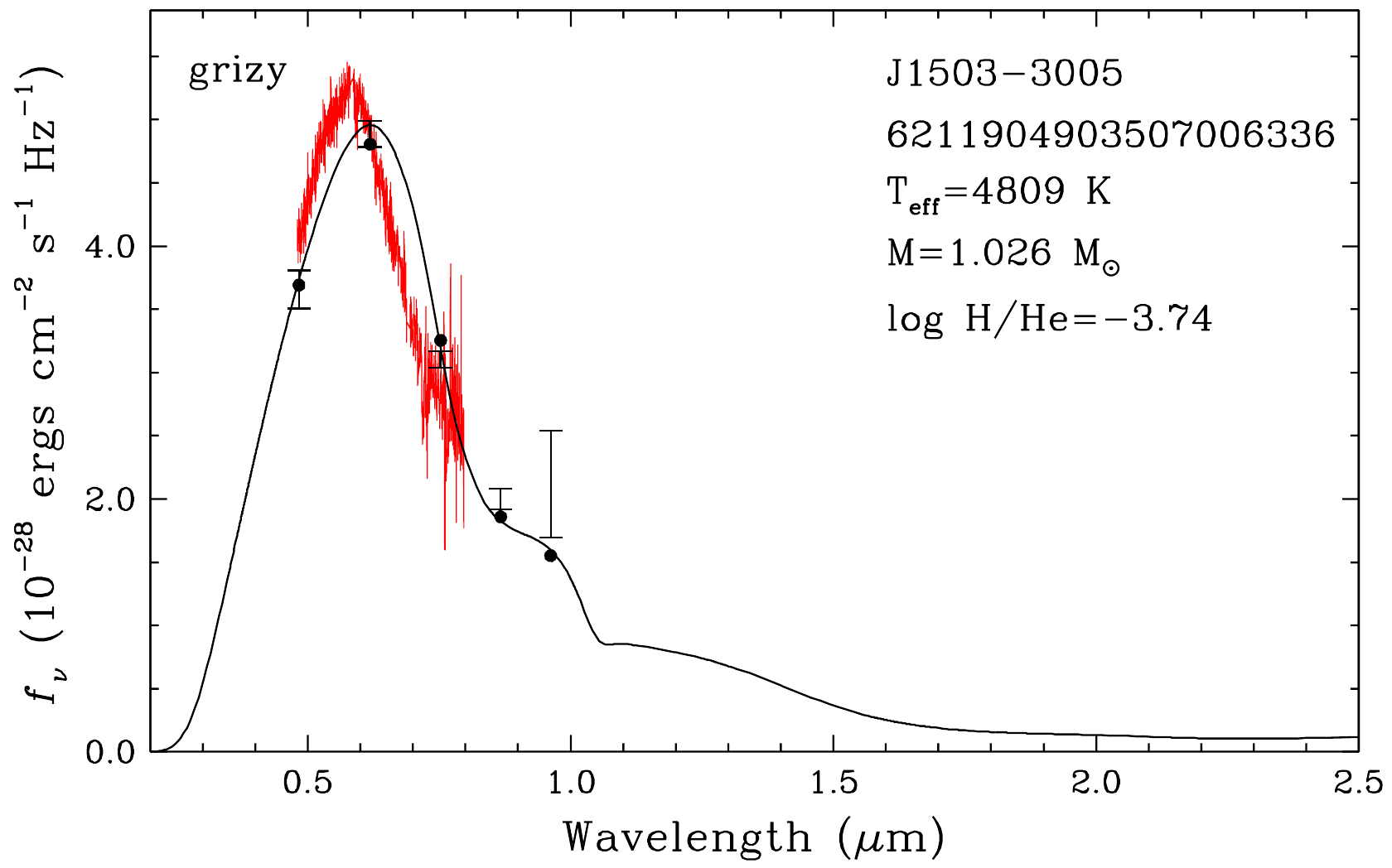}
\includegraphics[width=2.4in]{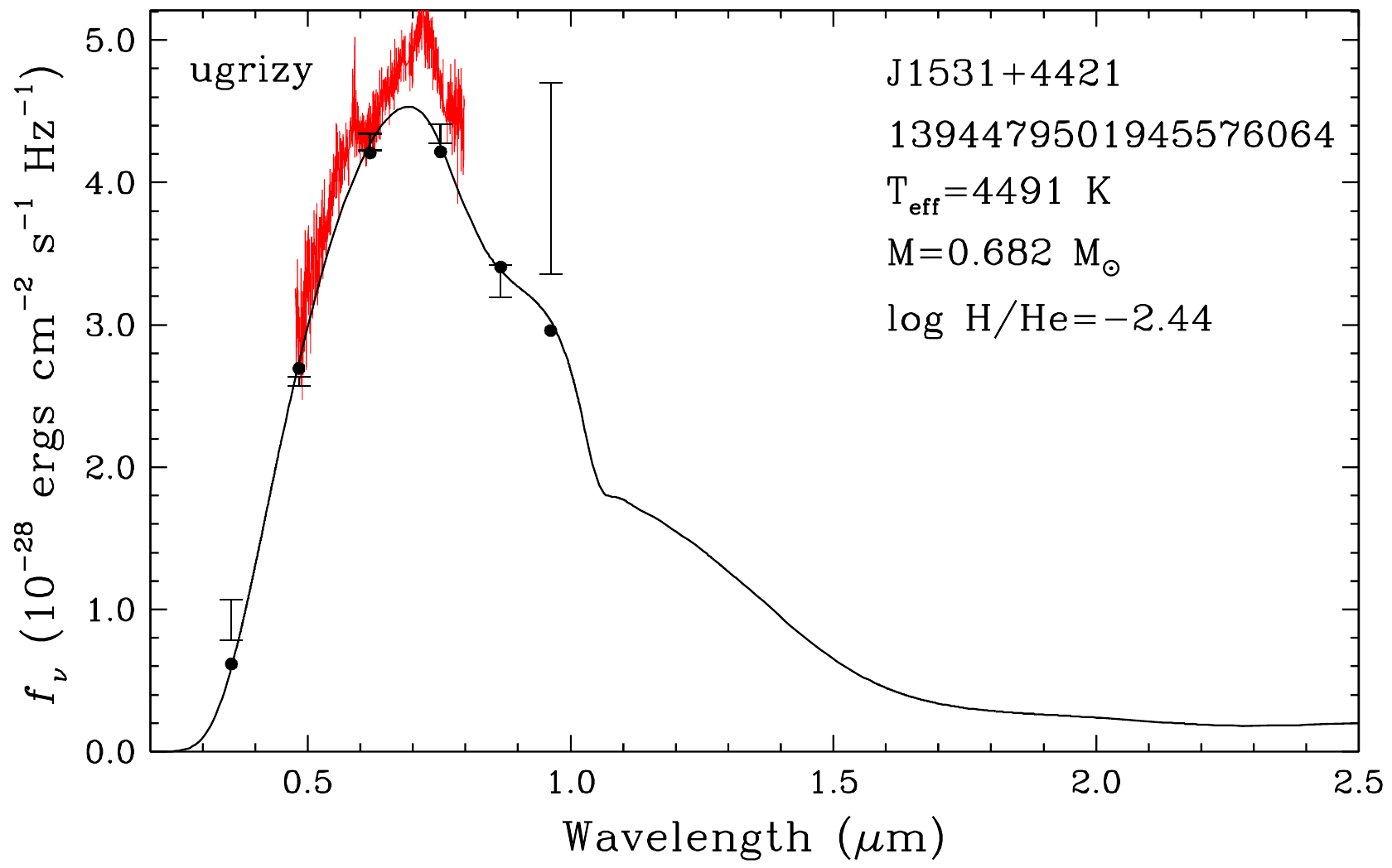}
\includegraphics[width=2.4in]{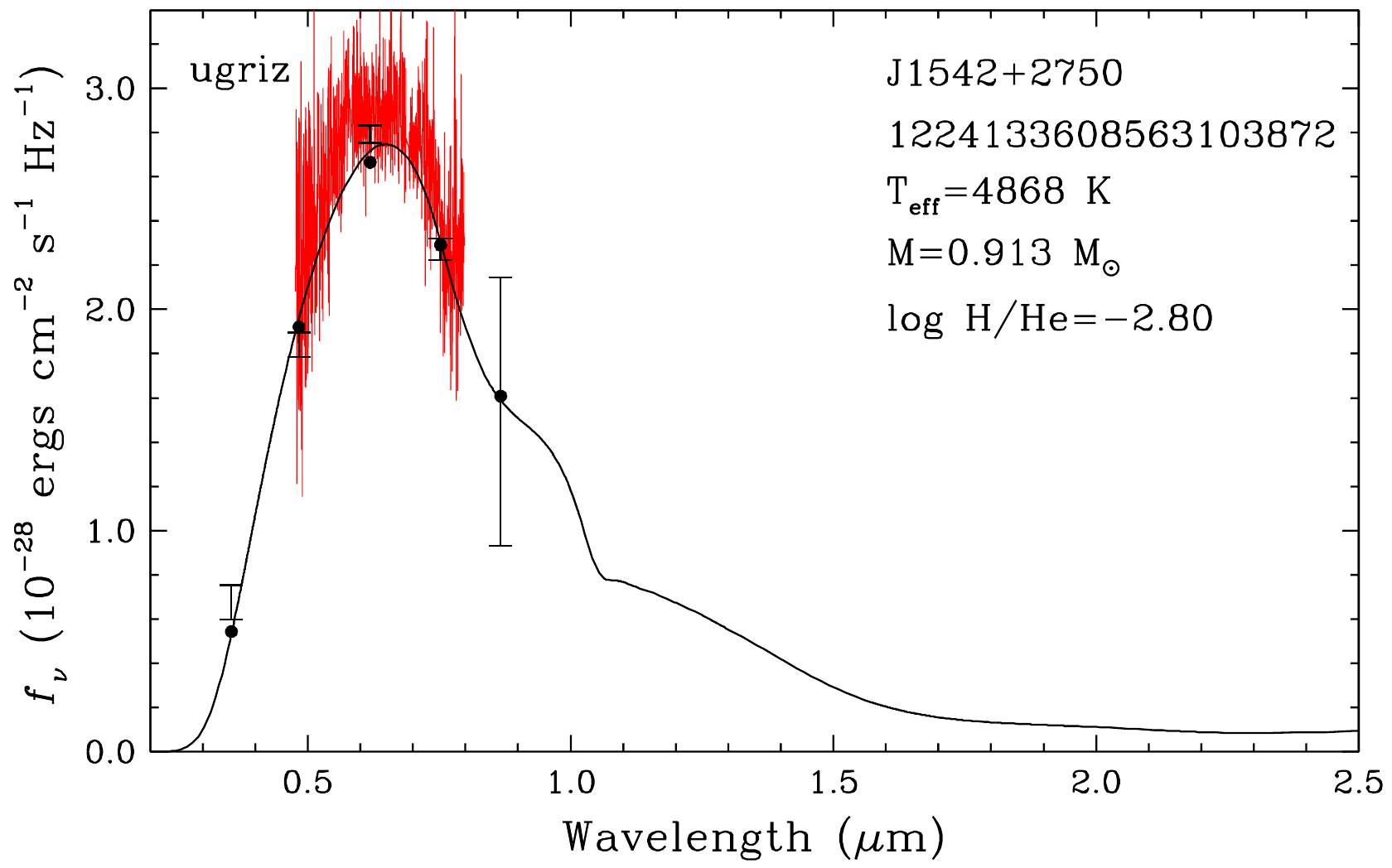}
\includegraphics[width=2.4in]{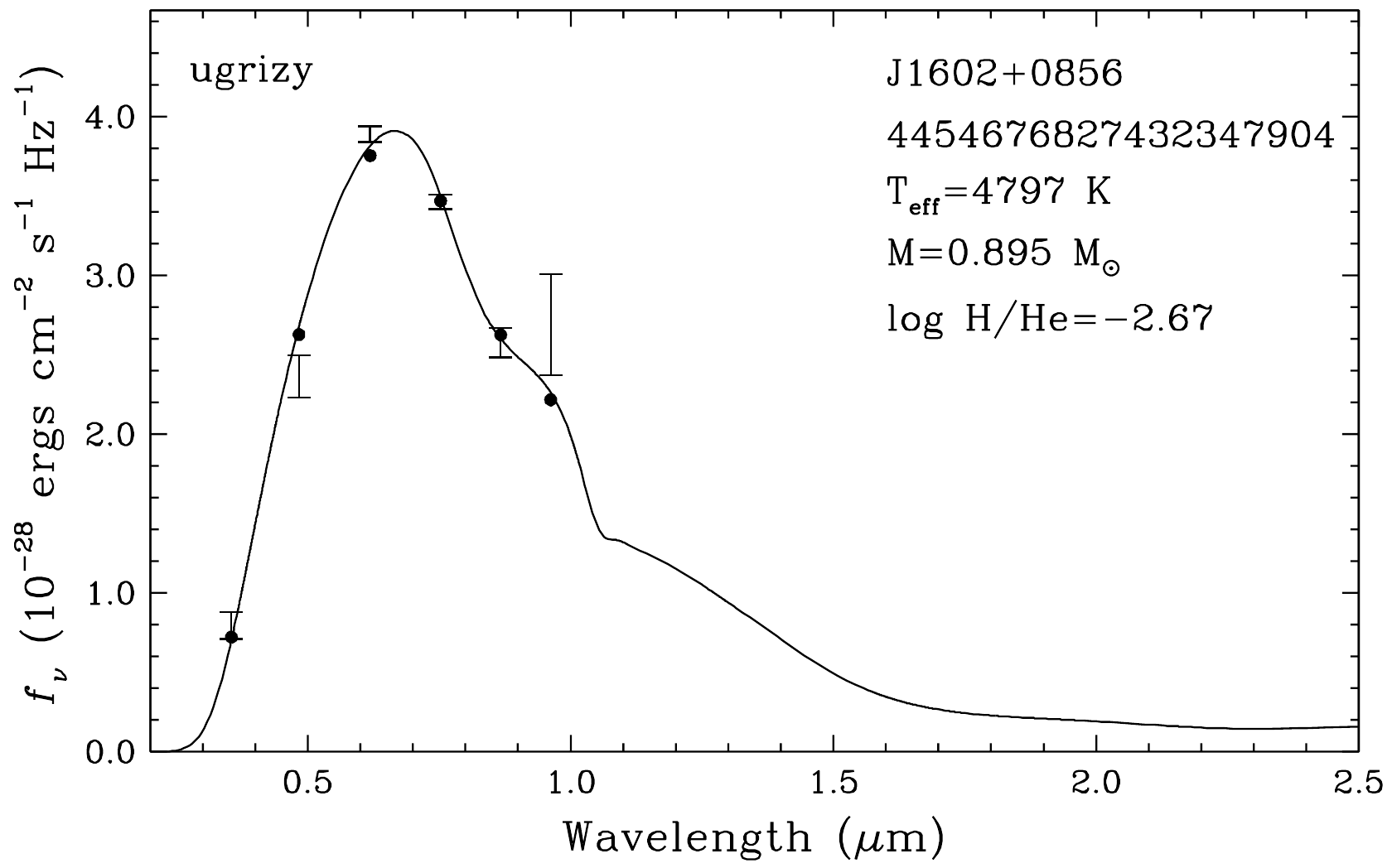}
\includegraphics[width=2.4in]{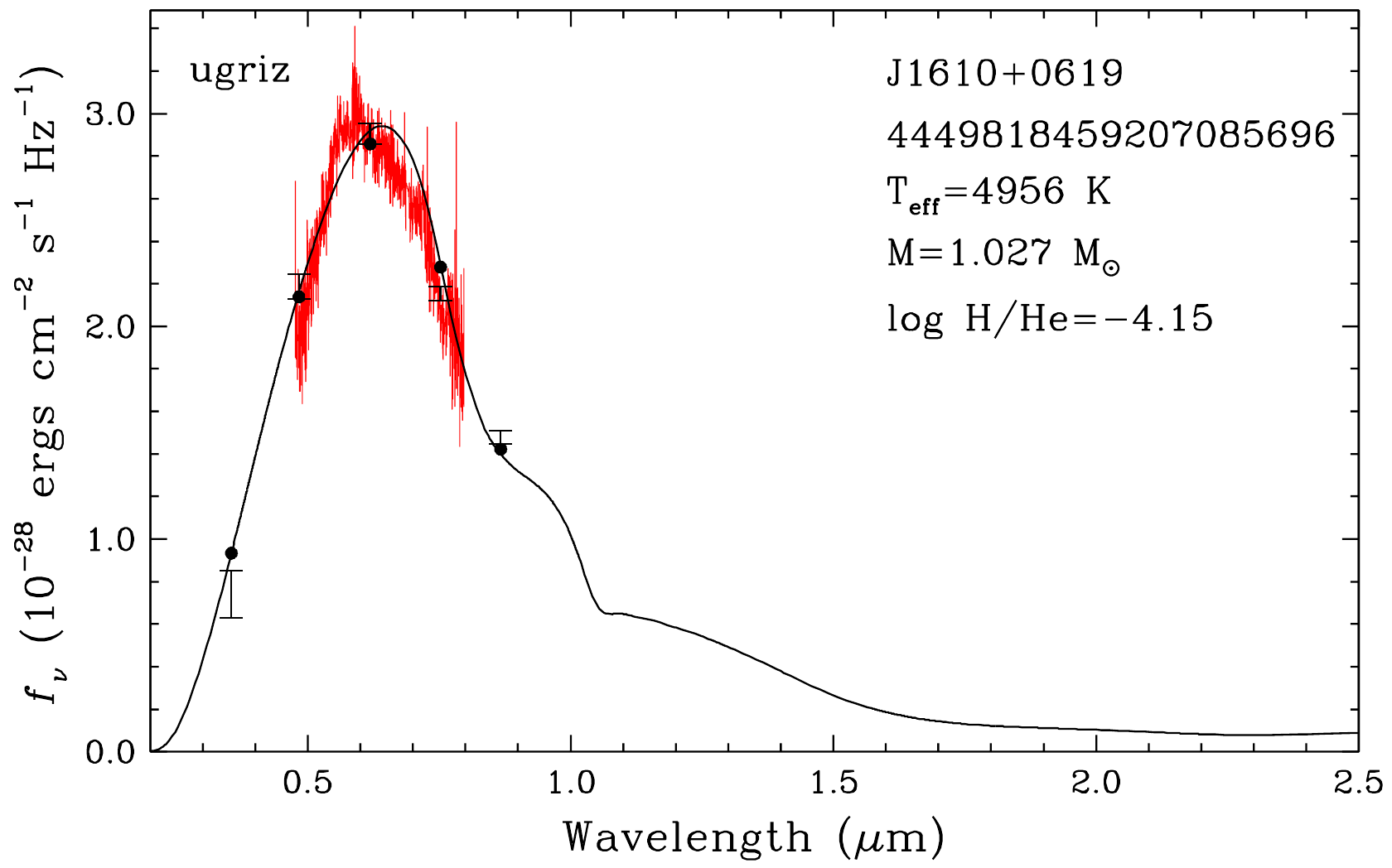}
\includegraphics[width=2.4in]{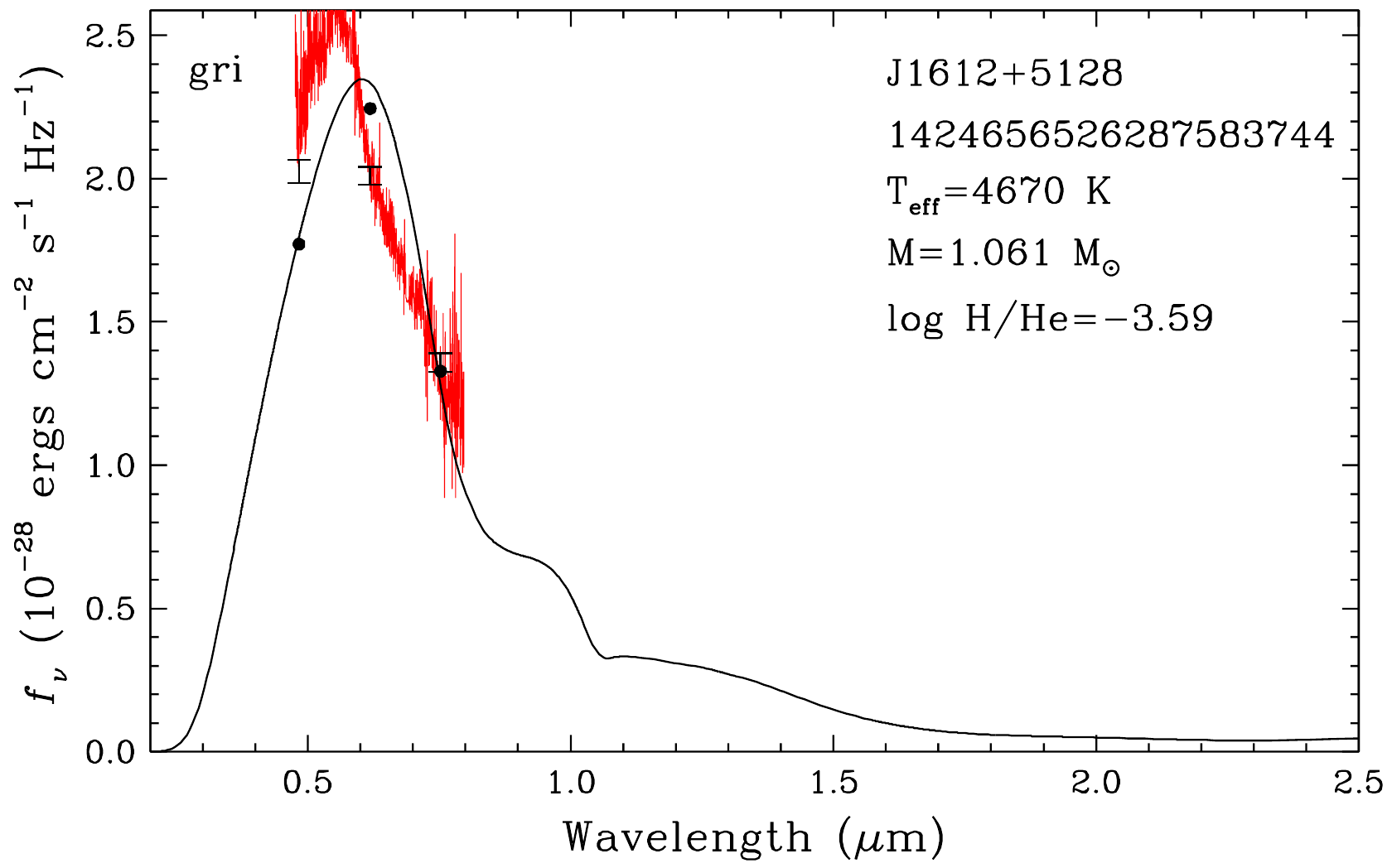}
\includegraphics[width=2.4in]{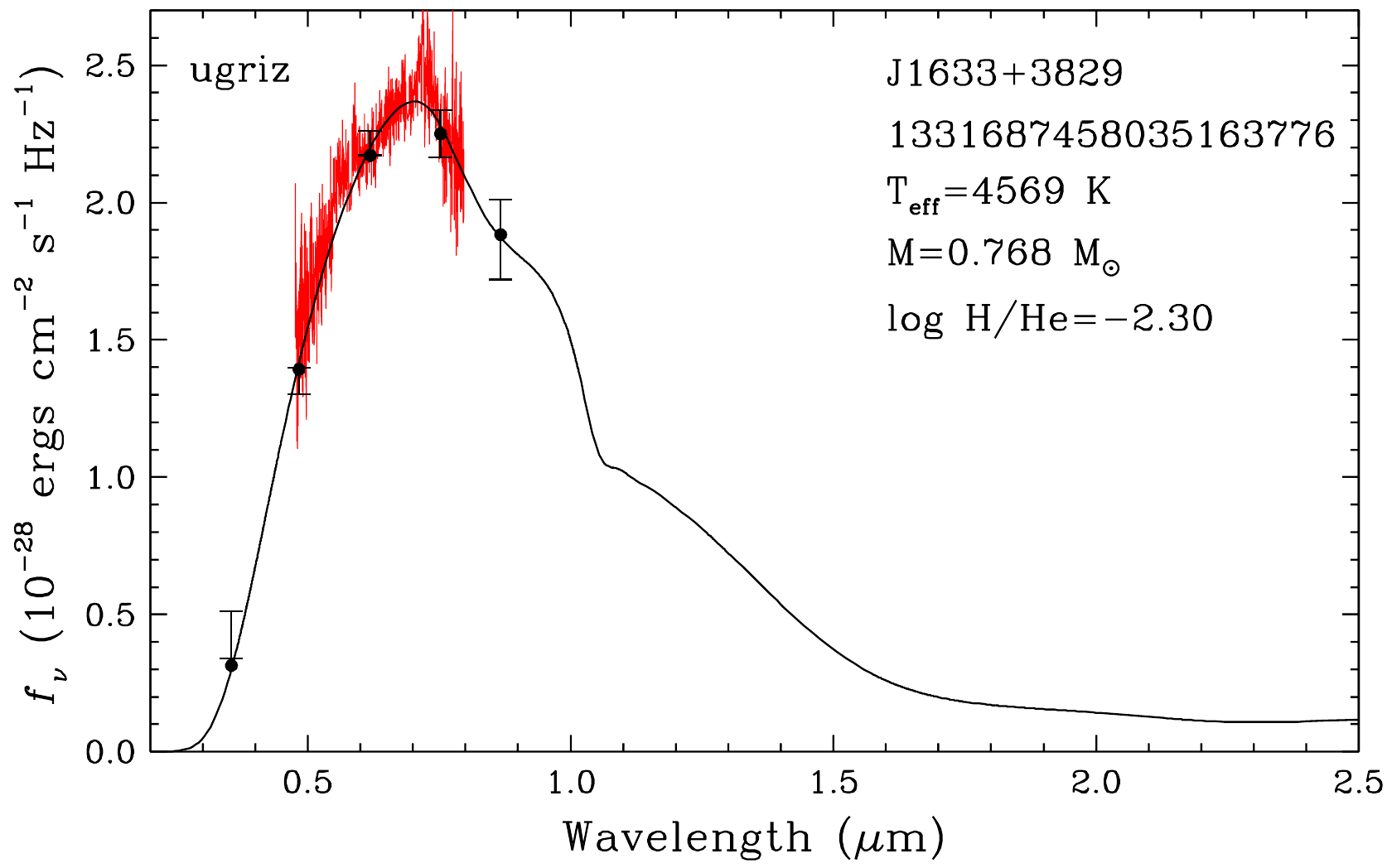}
\includegraphics[width=2.4in]{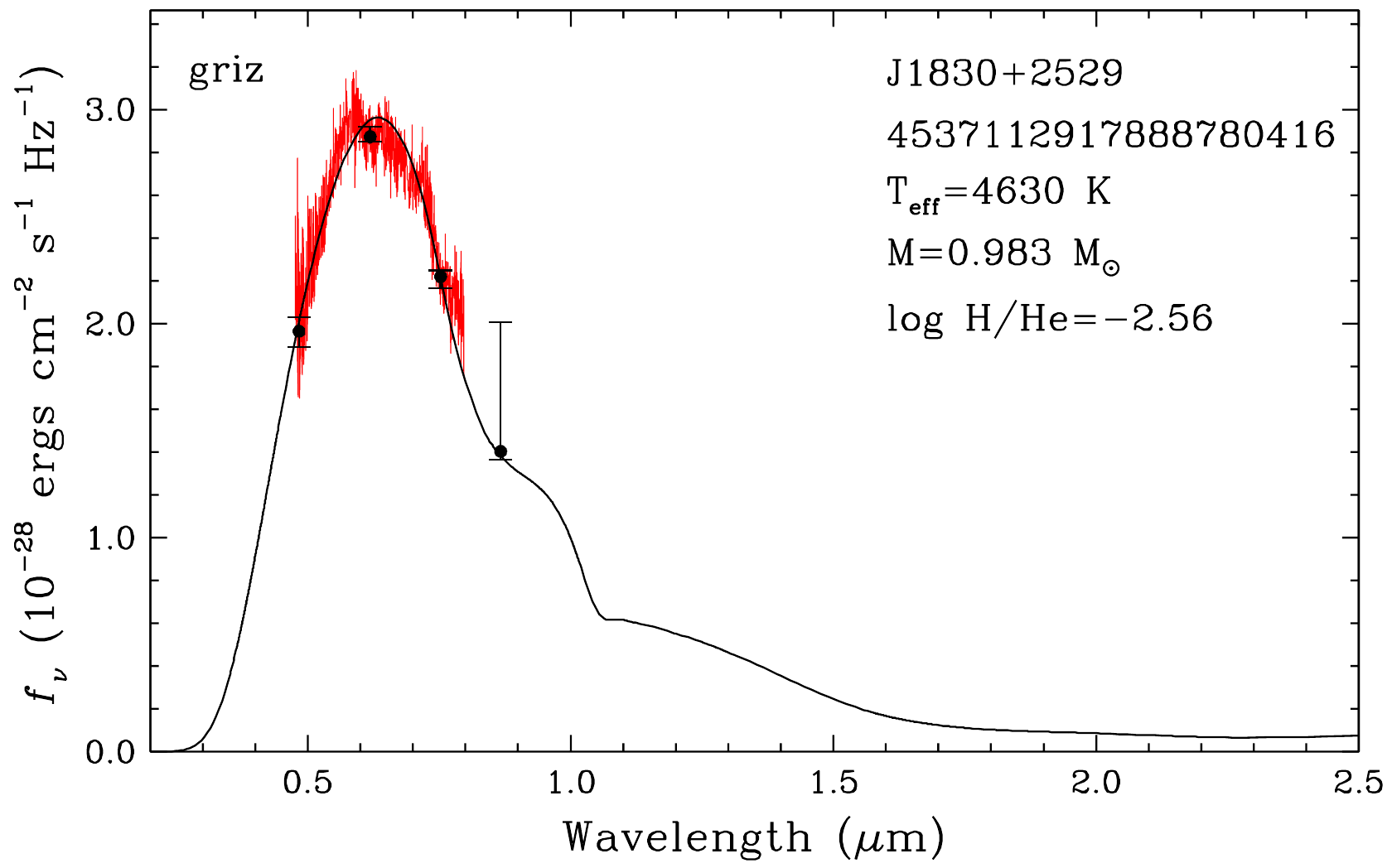}
\includegraphics[width=2.4in]{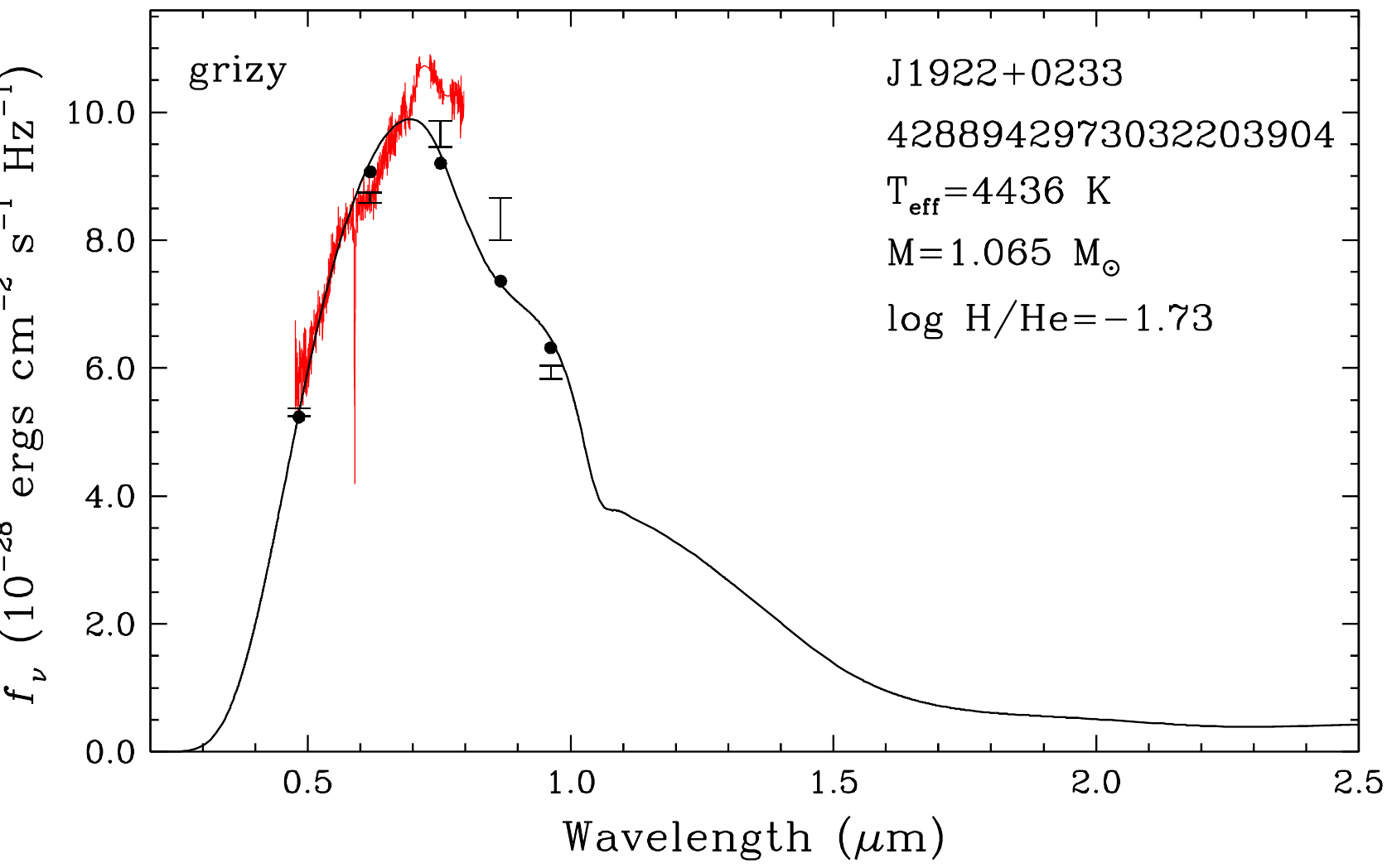}
\includegraphics[width=2.4in]{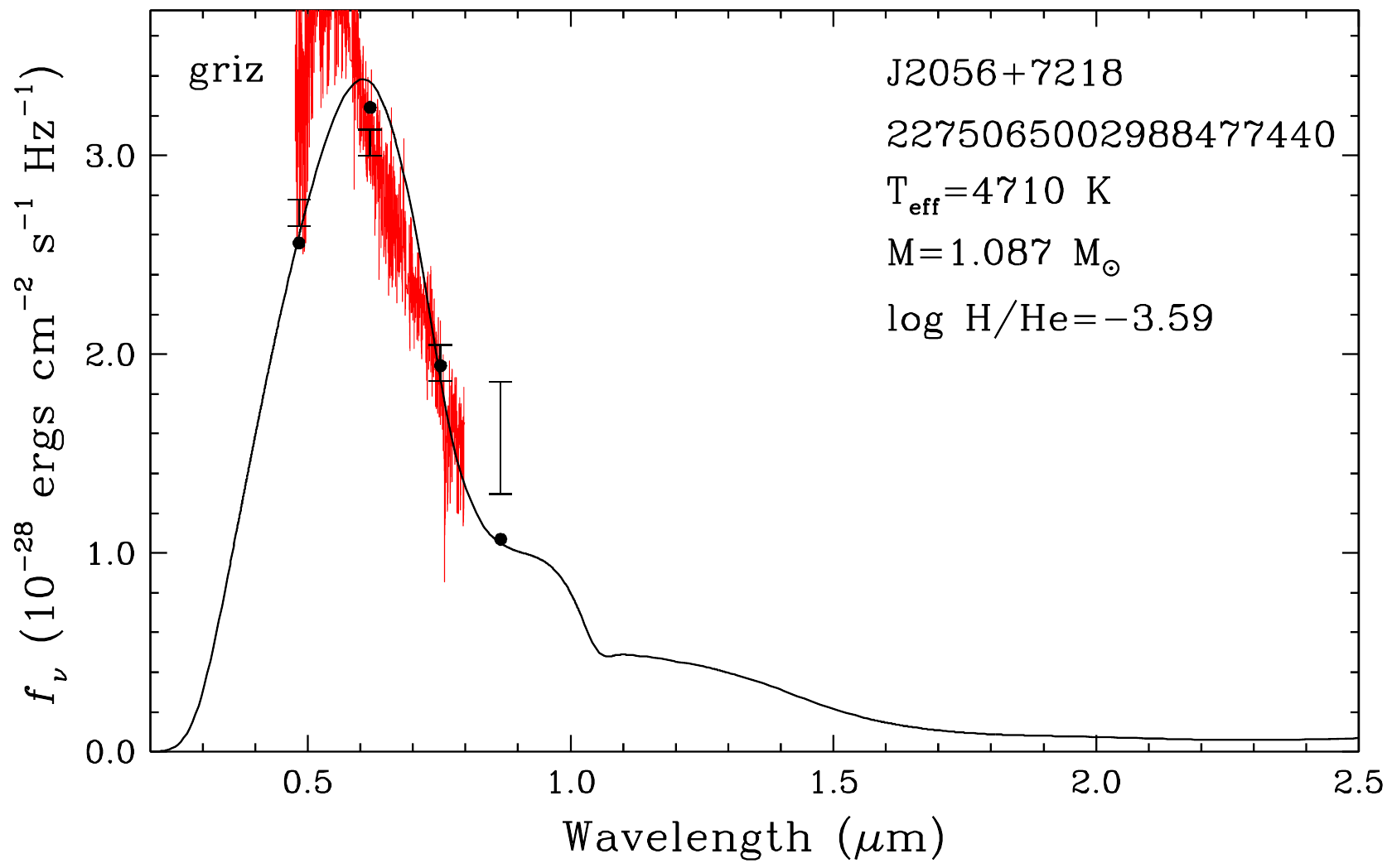}
\caption{Continued.}
\end{figure*}

\begin{figure*}
\addtocounter{figure}{-1}
\includegraphics[width=3.5in]{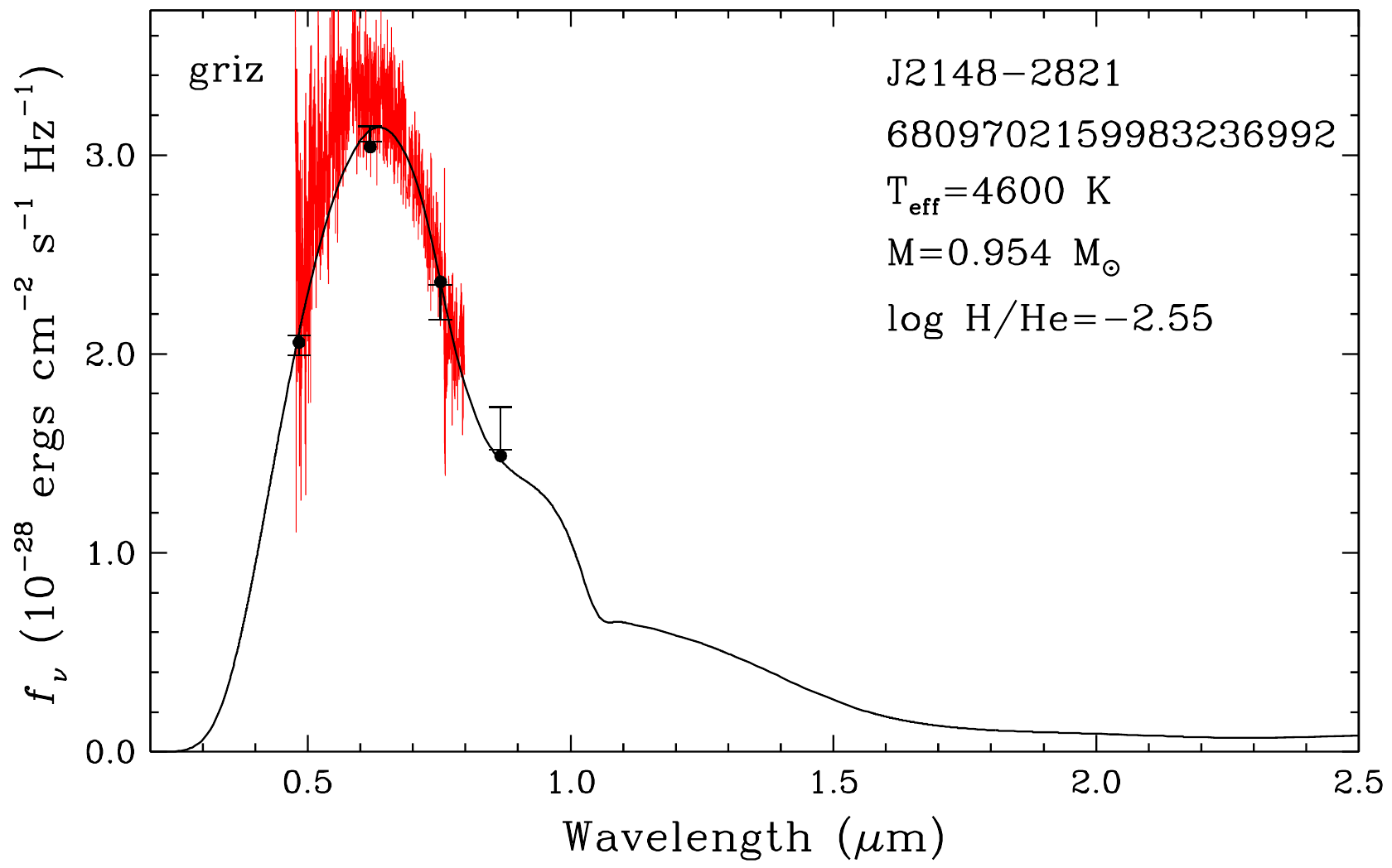}
\includegraphics[width=3.5in]{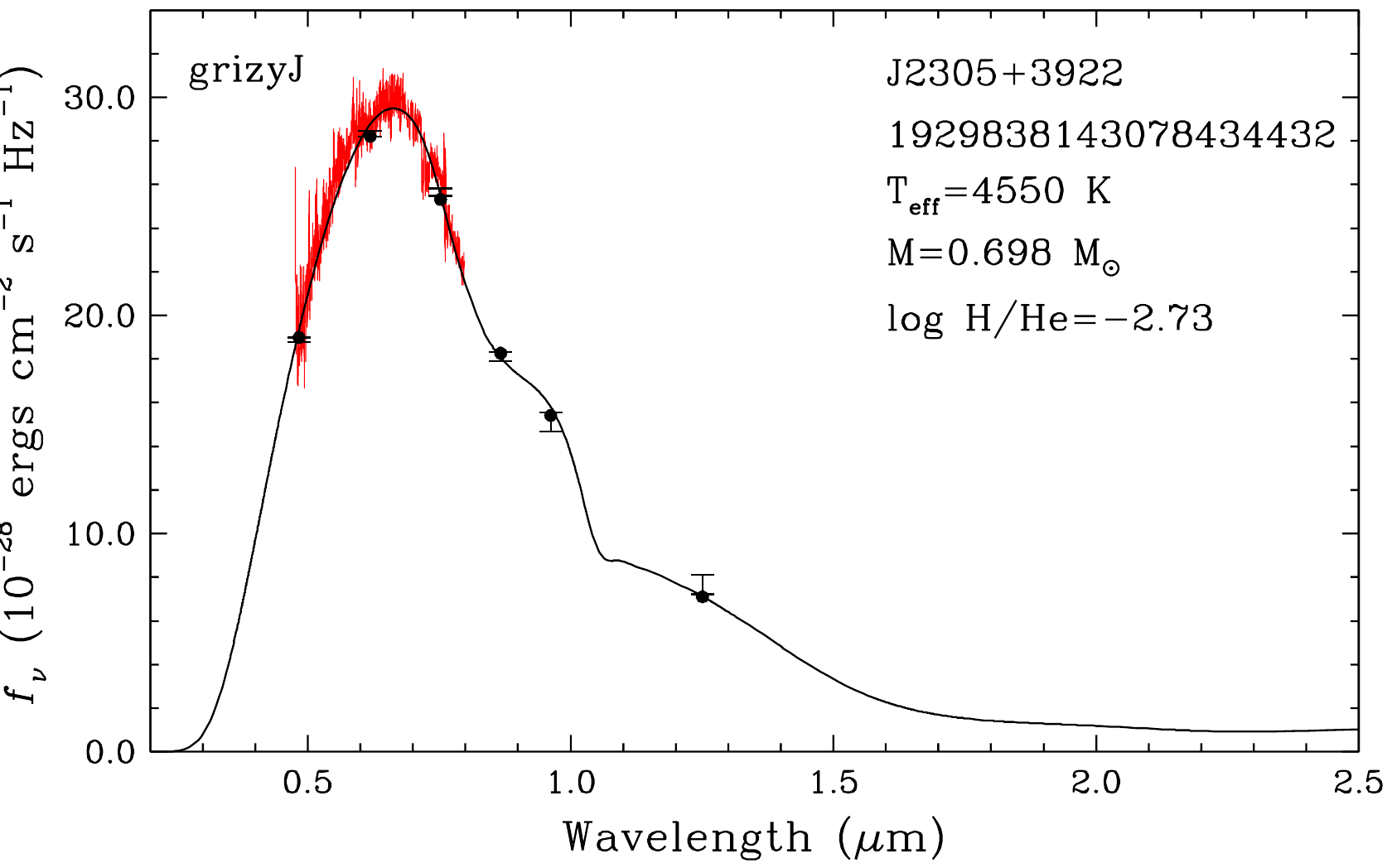}
\includegraphics[width=3.5in]{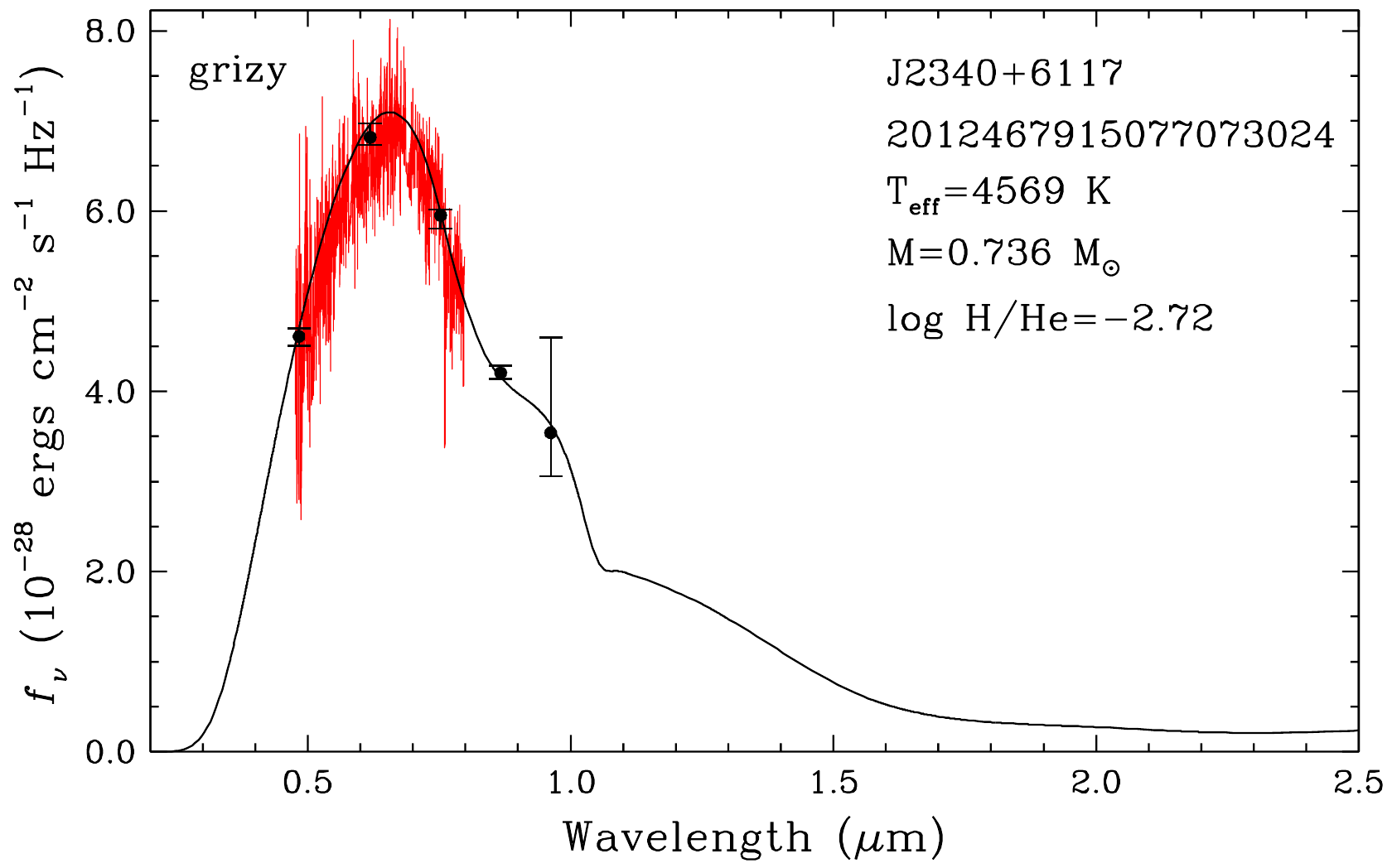}
\includegraphics[width=3.5in]{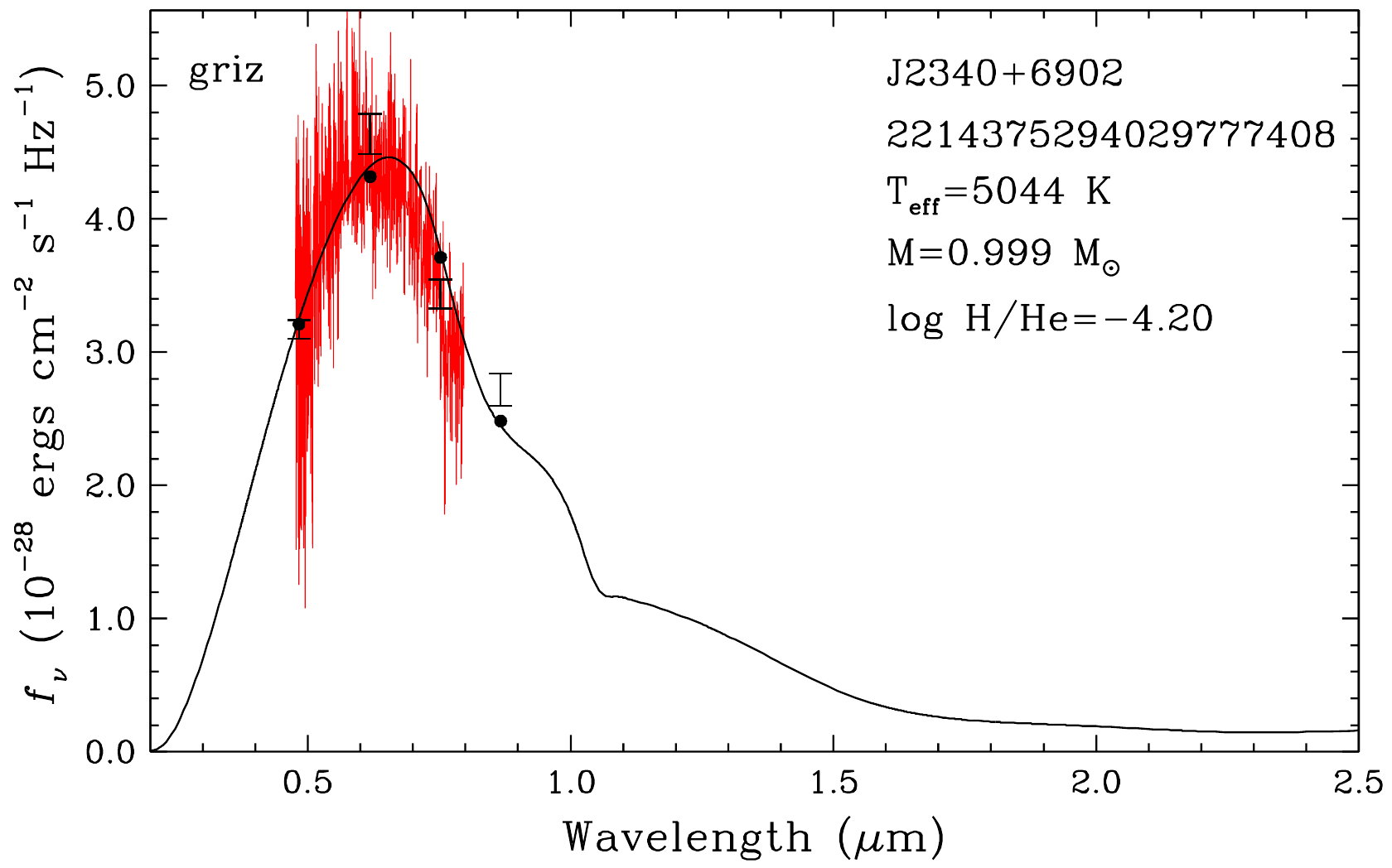}
\caption{Continued.}
\end{figure*}

\subsection{IR-Faint White Dwarfs with Strong Flux Deficits}\label{strong}

Figure \ref{fitgem} shows the spectral energy distributions of the 37
newly identified IR-faint white dwarfs with strong flux deficits in
the optical. Error bars show the observed photometry and the red lines
show the spectra, normalized at $\sim$6000 \AA. Optical spectra are
available for 31 of these objects, and in most cases the spectra
follow the photometric spectral energy distributions, revealing
relatively narrow peaks with significant absorption in the red.  Each
object is labeled based on its Gaia Source ID, object name based on
Gaia DR2 coordinates, and the photometry used in the fitting: $ugrizy$
means SDSS $u$ + Pan-STARRS $grizy$.  Solid black lines and dots show
the monochromatic fluxes and the synthetic photometry for the best-fit
model, respectively. The parameters of this model are included in each
panel. For reasons discussed in Section \ref{cmd}, with this
particular model grid, we always found a unique and robust solution to
all the IR-faint white dwarfs analyzed here and below. In particular,
we did not find any degeneracy in our solutions similar to those
reported in previous investigations (see Section \ref{modatm}).

In contrast with the fits obtained with our old model grid (see, e.g.,
\citealt{bergeron02}, \citealt{gianninas15}, \citealt{kilic20}), those
displayed in Figure \ref{fitgem} show a remarkably good agreement with
the optical and infrared photometry, as well as with the optical
spectroscopy. In those cases where the predicted monochromatic fluxes
do not match the observed spectrum, there is also a discrepancy
between the spectrum and the measured photometry (see, e.g.,
J0231$+$3254, J0440$-$0414, etc.), suggesting that some of the spectra
suffer from flux calibration issues. But in most cases, the agreement
is excellent.  While the solutions for these IR-faint white dwarfs
obtained with our old model grid (not shown here) span a range of
$\Te=1950$ to 3370 K and $M\sim 0.15$ to 0.35 \msun, the best fits
with our revised models yield much higher temperatures well above 4000
K for most objects (see also Section \ref{global}), and significantly
larger masses, with more than half of the objects in our sample in the
range $M\sim 0.8-1.2$ \msun.  We discuss further the global properties
of our sample in Section \ref{sec:discussion}.

Since the objects displayed in Figure \ref{fitgem} represent extreme
cases of white dwarfs with the strongest observed infrared flux
deficiencies, we are forced to conclude that IR-faint white dwarfs are
not ultracool afterall. Also, they do not require unusually large
radii and low masses either, as was the case with the prototype LHS
3250 analyzed by \citet{bergeron02}. Obviously, the key physical
ingredient missing in our previous generation of model atmospheres was
the correction to the He$^-$ free-free absorption coefficient
described in \citet{iglesias02}.

\subsection{An IR-Faint DZ White Dwarf}

J1922+0233 is the first IR-faint white dwarf discovered with an
absorption feature and strong flux deficits in the optical bands
\citep[see Figure \ref{fitgem}, and also][]{mccleery20,tremblay20}. It
shows a strong Na doublet feature, which can be used as an independent
diagnostic of the H/He ratio. However, our attempts to fit both the
sodium doublet and the CIA feature in J1922+0233 failed.

\begin{figure}
\includegraphics[width=3.3in, clip=true, trim=0.3in 2.2in 0.4in 2.7in]{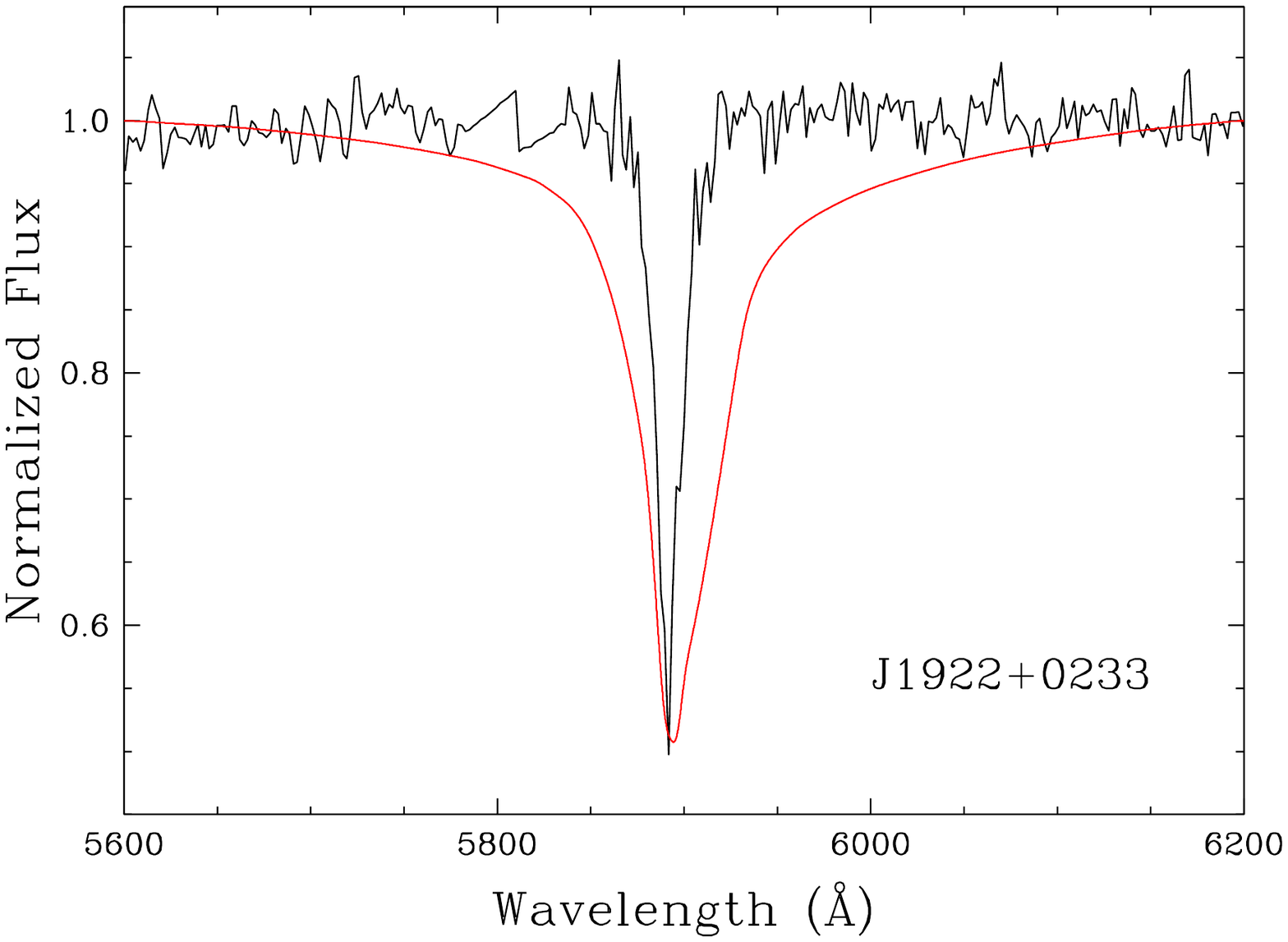}
\caption{Na D region of J1922+0233. The atmospheric parameters are set
  to the values given in Figure \ref{fitgem}, and the Na abundance of
  the synthetic spectrum is adjusted ($\log\ {\rm Na/He}=-9.7$) to
  match the depth of the observed Na D doublet. We find that the
  inferred He-rich composition leads to too much broadening of the Na D
  doublet.
\label{sim1922}}
\end{figure}

Figure \ref{sim1922} shows a comparison between the observed sodium
feature and a synthetic spectrum calculated using the unified
Na profiles described in \citet{blouin19a} with the stellar parameters
given in Figure \ref{fitgem} ($\Te=4436$~K, $M=1.065$ \msun,
$\log\ {\rm H/He}=-1.73$), and by adjusting the Na abundance (only in
the synthetic spectrum calculation) to match the depth of the observed
Na D doublet. As can be seen in the figure, the inferred abundance of
$\log\ {\rm Na/He}=-9.7$ predicts way too much broadening of the Na D
doublet as a result of the high photospheric pressure encountered in
He-rich atmospheres.  The relatively narrow sodium feature in
J1922+0233 can only be explained by lowering the density
significantly, at least in the line-forming region. For instance, it
is possible to achieve a good fit to the sodium feature by assuming a
pure hydrogen atmosphere, where the photospheric density is two orders
of magnitude lower than in our best-fit model, but in this case the
predicted energy distribution is totally inconsistent with the
observations, as hydrogen-dominated atmospheres do not produce strong
CIA features that are required to fit the photometry.
 
We know of other cool DZs with mixed H/He atmospheres and Na D
features. For example, \citet{blouin19a} was able to fit successfully
both the CIA and the Na D features in WD J2356$-$209 with a model that
has $T_{\rm eff} = 4040 \pm 110$ K and $\logh = -1.5 \pm 0.2$. Hence,
it is surprising in the case of J1922+0233 to see simultaneously very
strong CIA (which demands a low hydrogen abundance) and a narrow Na D
feature (which demands a high hydrogen abundance). It is difficult to
reconcile those two observations.  It is also interesting that only
one out of the 37 objects presented here is a DZ, whereas $\sim$30\%
of helium atmosphere white dwarfs in the 4000-5000 K range are
DZs. Hence, J1922+0233 seems to be rather different than the general
population of IR-faint white dwarfs with strong flux deficits in the
optical. J1922+0233 may be an exception, rather than the rule, among
the IR-faint white dwarf sequence.

\subsection{Known IR-Faint White Dwarfs}\label{known}

Given our revised model atmospheres, we felt it was worth reanalyzing
the previously known IR-faint white dwarfs published in the literature
and summarized in Table \ref{tabold}. Our best fits to these 35
IR-faint white dwarfs are presented in Figure \ref{pre} (in
Appendix). Again, the quality of these fits is excellent in most
cases. For J1238+3502, we found that we could achieve a much better
fit to this object (also displayed in Figure \ref{pre}) by dropping
the $J$ magnitude reported by \citet{kilic10b}. However, since there
is no reason to believe this measurement is erroneous, we provide here
both solutions for this star.

Some of the previously known IR-faint white dwarfs in Figure \ref{pre}
show significantly less absorption in the infrared than those
displayed in Figure \ref{fitgem} as a result of their larger hydrogen
abundances, and thus smaller H$_2$-He CIA opacity (see, e.g.,
J0041$-$2221, J0309+0025, J0854+3503, etc.). Otherwise, they are found
in the same temperature and mass range as those reported in our
spectroscopic sample. Again, we defer the discussion of the global
properties of this sample to Section \ref{sec:discussion}.  Also worth
mentioning is the case of the prototype LHS 3250 (J1653+6253), for
which we obtain $\Te=4993$~K, $M=1.049$ \msun, and $\log\ {\rm
  H/He}=-2.74$ (see the fit in Figure \ref{pre}), while
\citet{bergeron02} reported significantly different values of
$\Te=3480$~K, $M=0.23$ \msun, and $\log\ {\rm H/He}=-4.7$, and a
considerably worse fit (see their Figure 7). Our improved models have
successfully solved this two decade old problem.

\subsection{Additional IR-Faint White Dwarfs}\label{mild}

Our follow-up spectroscopy presented in Section \ref{strong}
specifically targeted objects with strong
flux deficits in the Pan-STARRS bands (see Figure \ref{colormag}).
However, this sample represents only the tip of the iceberg, and there
are likely many other IR-faint white dwarfs hiding in the 100 pc
sample. In order to identify those missing white dwarfs with weaker
flux deficits in the optical, we first investigated the spectral
energy distributions of all targets below the observed white dwarf
sequence, and then expanded our search to all white dwarfs in the MWDD
100 pc sample.

We fitted the spectral energy distribution of each target with pure
hydrogen and pure helium atmosphere models, and compared the resulting
$\chi^2$ to the best-fitting mixed H/He atmosphere model fits.  We
identified 210 candidates as potential IR-faint white dwarfs. We then
cross-correlated this list with the UKIDSS and VHS, and found
near-infrared photometry for 71 targets, and updated our model fits
for those stars with both optical and near-infrared data. Four of
these objects have follow-up spectroscopy available in the literature,
and all four are confirmed to be DC white dwarfs by
\citet{tremblay20}.

\begin{figure}
\hspace{-0.4in}
\includegraphics[width=4.1in, clip=true, trim=0.0in 6.0in 0.in 0.9in]{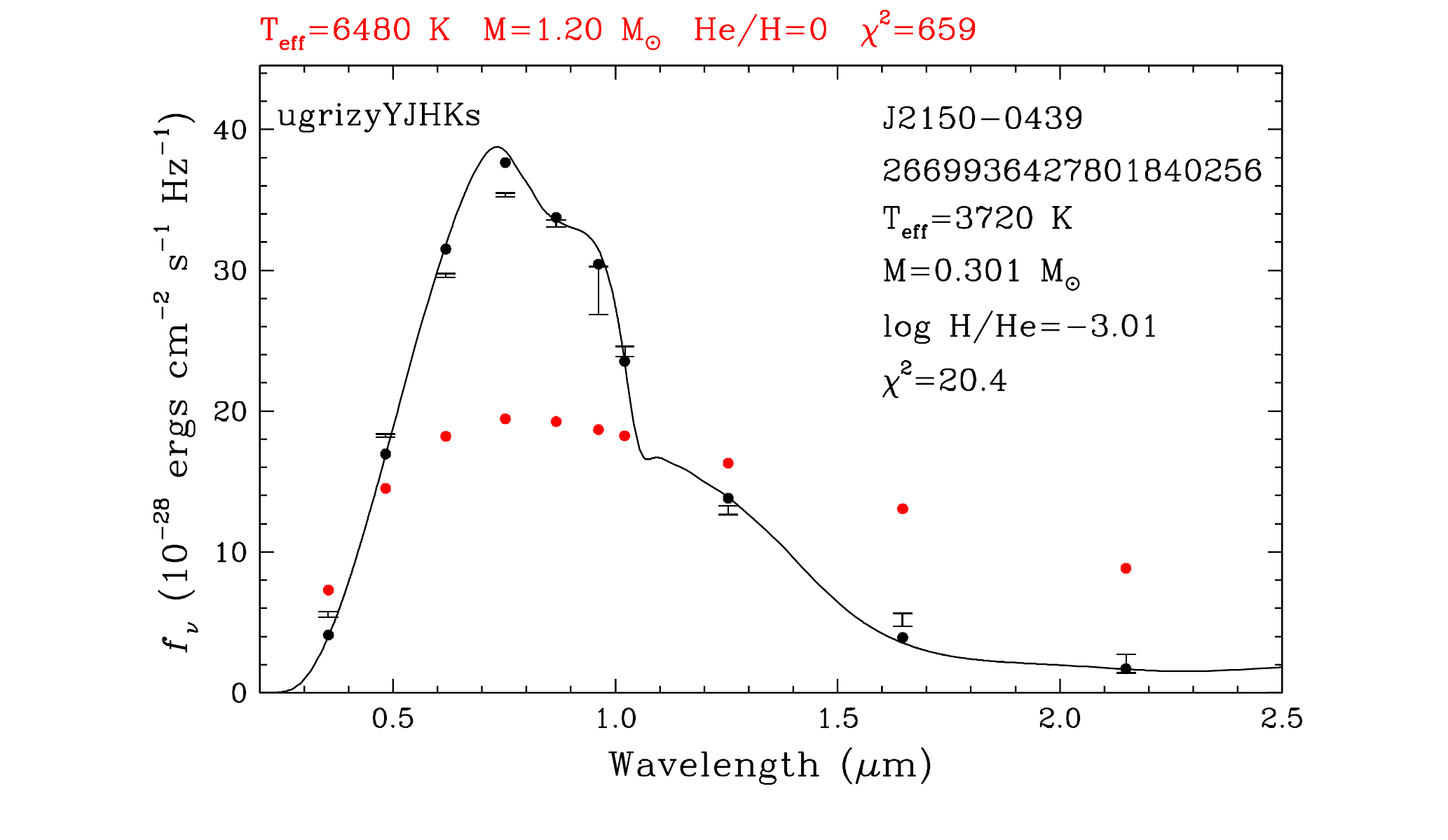}
\caption{Fits to the spectral energy distribution of a newly
  identified IR-faint white dwarf with mild flux deficits in the
  Pan-STARRS bands; note that we rely here on our old model
  atmospheres. J2150$-$0439 is a spectroscopically confirmed DC white
  dwarf within 40 pc \citep{tremblay20}. The symbols are the same as
  in Figure \ref{fitgem}. The red dots show the best-fitting pure
  hydrogen atmosphere model with parameters also highlighted in red.
  Near-infrared photometry reveals strong flux deficits compared to
  the pure H atmosphere models, confirming the IR-faint nature of this
  object.
\label{add}}
\end{figure}

Figure \ref{add} shows our model fits to one of these DC white dwarfs,
J2150$-$0439. Note that because our search for these additional
candidates was performed prior to upgrading our model atmospheres, the
fits displayed here are based on our old model atmosphere grid.
Error bars show the observed photometry, and the red dots show the
best-fitting pure hydrogen atmosphere model. The parameters for this
model along with the reduced $\chi^2$ of the fit are presented in
red. IR-faint white dwarfs occupy the same region of the
color-magnitude diagram as massive white dwarfs; the pure hydrogen
atmosphere solution for J2150$-$0439 requires a mass of
$1.2~M_{\odot}$.  However, this solution is clearly bad, with a
reduced $\chi^2=659$.

The solid black line and dots show the monochromatic fluxes and the
synthetic photometry for the best-fitting mixed H/He atmosphere model,
respectively. The best-fitting mixed atmosphere model parameters are
included in the figure for comparison.  Even though the reduced
$\chi^2$ value for our helium-dominated model fit is not great (but
recall that we are using our old model grid here), it provides a much
better fit than the pure hydrogen model. J2150$-$0439 shows relatively
mild absorption in the optical compared to the IR-faint white dwarfs
discussed in the previous section.  However, near-infrared photometry
clearly shows significant flux deficits compared to the pure hydrogen
atmosphere model; it is undoubtedly IR-faint.

We identified 33 additional IR-faint candidates using this procedure,
including four confirmed DC white dwarfs, where the available
photometry clearly favors an IR-faint classification. These IR-faint
white dwarfs are listed in Table \ref{tabcan} and our best fits using
our revised model atmospheres are displayed in Figure \ref{fitcan} (in
Appendix). One remarkable feature of this particular subsample is that
all objects have more normal masses below 0.8 \msun, the reason of
which is discussed in the next section. Also, we uncovered a few
IR-faint white dwarfs with extremely low masses below $\sim$0.3
\msun\ (J0357$-$2606, J0406$-$0333, J1448+2935, and J2035+4054).

\begin{table*}
\scriptsize
\caption{Additional IR-faint White Dwarf Candidates\label{tabcan}}
\begin{tabular}{clccccccc}
\hline
Name & Source ID & Sp.Type &  $\varpi$ & $g$ & $r$ & $i$ & $z$ & $y$ \\
 & & & (mas) & (mag) & (mag) & (mag) & (mag) & (mag) \\
\hline
J0027+0554 & Gaia DR2 2747384888699406080 &  DC      & 27.06 & 18.57 & 17.96 & 17.72 & 17.75 & 17.87 \\
J0035+2009 & Gaia DR2 2796224099985984768 &  \nodata & 21.92 & 18.60 & 18.08 & 17.90 & 17.94 & 18.08 \\
J0146+2122 & Gaia DR2   98035961426035712 &  \nodata & 19.01 & 19.17 & 18.64 & 18.49 & 18.64 & 18.87 \\
J0259$-$0455 & Gaia DR2 5184384997855024384 &  \nodata & 10.47 & 20.04 & 19.66 & 19.94 & 20.18 & 20.11 \\
J0311+5257 & Gaia DR2  446199342020477184 &  \nodata & 12.21 & 19.91 & 19.33 & 19.17 & 19.35 & 19.75 \\
J0357$-$2606 & Gaia DR2 5082529165532368000 &  \nodata & 15.86 & 18.79 & 18.25 & 18.21 & 18.44 & 18.70 \\
J0400+2138 & Gaia DR2   53022195903806976 &  \nodata & 19.54 & 19.06 & 18.50 & 18.41 & 18.69 & 19.01 \\
J0406$-$0333 & Gaia DR2 3251942081669060992 &  \nodata & 18.37 & 19.94 & 18.95 & 18.64 & 18.79 & 19.14 \\
J0439$-$2016 & Gaia DR2 2977377569898585344 &  \nodata & 19.72 & 18.82 & 18.30 & 18.15 & 18.27 & 18.48 \\
J0758$-$1711 & Gaia DR2 5718015859550552704 &  \nodata & 12.31 & 20.02 & 19.47 & 19.30 & 19.37 & 19.53 \\
J1004$-$0506 & Gaia DR2 3822028007288795264 &  DC      & 28.30 & 19.12 & 18.16 & 17.78 & 17.70 & 17.73 \\
J1015$-$2009 & Gaia DR2 5668952347880121472 &  \nodata & 13.53 & 19.86 & 19.31 & 19.18 & 19.38 & 19.61 \\
J1102+3523 & Gaia DR2  761660188783588864 &  \nodata &  9.65 & 20.62 & 20.06 & 20.02 & 20.21 & 20.64 \\
J1238+2633 & Gaia DR2 3962123479637203200 &  \nodata & 14.86 & 19.44 & 18.94 & 18.72 & 18.74 & 18.81 \\
J1326$-$1558 & Gaia DR2 3604874422147728256 &  \nodata & 18.04 & 19.06 & 18.56 & 18.49 & 18.75 & 19.01 \\
J1448+2935 & Gaia DR2 1281810110201241344 &  \nodata & 12.36 & 19.40 & 18.74 & 18.55 & 18.56 & 18.75 \\
J1525+6247 & Gaia DR2 1640531649982657280 &  \nodata & 10.63 & 20.14 & 19.69 & 19.70 & 20.00 & \nodata \\
J1546+2054 & Gaia DR2 1216212932955510528 &  \nodata & 14.21 & 19.57 & 19.09 & 18.95 & 19.09 & 19.29 \\
J1559+7314 & Gaia DR2 1702458378242137088 &  \nodata & 18.70 & 19.02 & 18.50 & 18.41 & 18.63 & 18.87 \\
J1639+0106 & Gaia DR2 4384015024746850432 &  \nodata & 11.10 & 20.17 & 19.67 & 19.64 & 19.86 & 20.22 \\
J1753+0758 & Gaia DR2 4475975364094688512 &  \nodata & 23.46 & 18.68 & 18.14 & 17.99 & 18.18 & 18.54 \\
J1922$-$0402 & Gaia DR2 4211947536675142656 &  \nodata & 12.97 & 20.14 & 19.64 & 19.49 & 19.71 & 19.77 \\
J1944$-$0425 & Gaia DR2 4209580601680083456 &  \nodata &  9.98 & 20.08 & 19.62 & 19.49 & 19.59 & 19.70 \\
J1951+4026 & Gaia DR2 2073772770741915264 &  DC      & 25.03 & 18.76 & 18.04 & 17.83 & 17.87 & 18.01 \\
J2012$-$2720 & Gaia DR2 6846795314323118464 &  \nodata & 15.88 & 19.37 & 18.86 & 18.65 & 18.72 & 18.83 \\
J2035+4054 & Gaia DR2 2064838031864808832 &  \nodata & 20.03 & 20.05 & 18.92 & 18.67 & 18.91 & 19.33 \\
J2148+2601 & Gaia DR2 1799385928868168960 &  \nodata &  9.77 & 20.15 & 19.67 & 19.60 & 19.73 & 19.92 \\
J2150$-$0439 & Gaia DR2 2669936427801840256 &  DC      & 28.17 & 18.25 & 17.72 & 17.53 & 17.59 & 17.76 \\
J2217+4241 & Gaia DR2 1958785020063453312 &  \nodata & 12.72 & 20.18 & 19.57 & 19.41 & 19.53 & 19.67 \\
J2237+2220 & Gaia DR2 1874330118886582016 &  \nodata & 10.92 & 19.94 & 19.49 & 19.39 & 19.53 & 19.77 \\
J2332+0959 & Gaia DR2 2761731595589115520 &  \nodata & 13.40 & 19.82 & 19.26 & 19.05 & 19.12 & 19.31 \\
J2337$-$2158 & Gaia DR2 2387846647997936640 &  \nodata & 12.71 & 19.62 & 19.13 & 18.93 & 18.93 & 19.02 \\
J2355+4419 & Gaia DR2 1922994640971085184 &  \nodata & 13.09 & 19.93 & 19.40 & 19.25 & 19.49 & 19.80 \\
\hline
\end{tabular}
\end{table*}

Even though there are other white dwarfs with mild flux
deficits in the Pan-STARRS bands, many would require infrared
photometry and optical spectroscopy to confirm their nature. In order
to avoid mis-classifying objects based on noisy $z$- or $y$-band data,
we adopted a conservative approach and only included objects where
available optical and near-infrared photometry clearly shows a strong
flux depression in the red filters. For example, 11 of the 30
candidates without follow-up spectroscopy have near-infrared
photometry available, and they are unambiguously IR-faint. Similarly,
14 of these objects have SDSS $u$-band photometry available, which
anchors the spectral energy distribution in the blue and helps
identify significant absorption in the red. Follow-up spectroscopy of
all of these objects would be helpful in confirming their nature.

\section{DISCUSSION}\label{sec:discussion}

\subsection{Color Diagrams}\label{cmd}

We used Gaia astrometry and Pan-STARRS photometry to identify 37
IR-faint white dwarfs with strong flux deficits in the optical, and 33
additional white dwarfs with milder deficits. We have follow-up
spectroscopy available for 30 of them from Gemini and five from the
literature. The optical spectra are featureless for all but one of
these white dwarfs. The spectra follow the photometric spectral energy
distributions with significant absorption in the red, confirming the
nature of these targets as IR-faint white dwarfs. Hence, these two
samples increase the number of IR-faint white dwarfs from 35 to 105,
thus tripling the sample size. 

\begin{figure}
\includegraphics[width=3.2in, clip=true, trim=0.3in 1.2in 1.1in 2in]{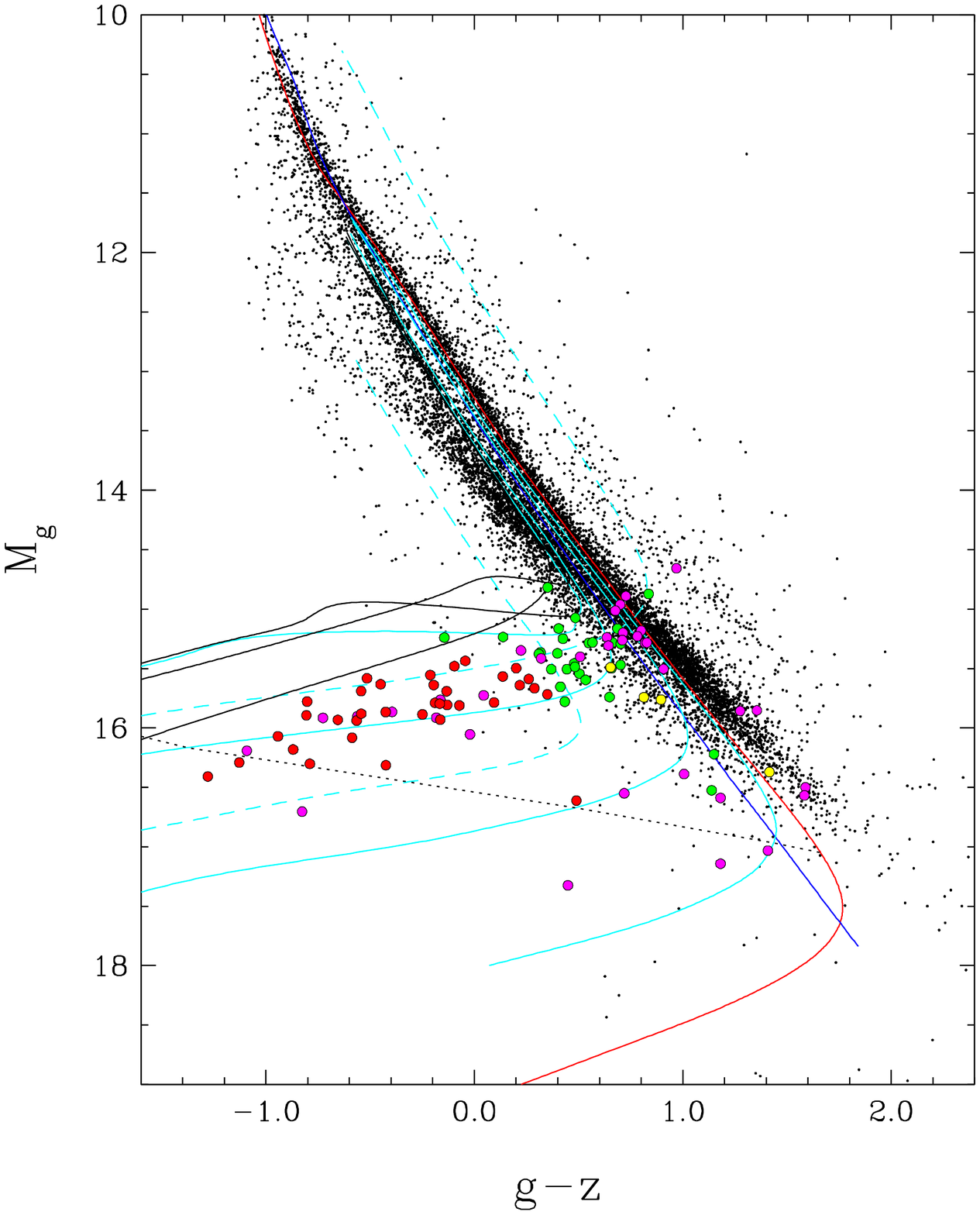}
\caption{Color-magnitude diagram of the MWDD 100 pc sample (black
  dots) along with the previously known IR-faint white dwarfs (magenta
  dots, Section \ref{known}) and the newly identified IR-faint white
  dwarfs with strong (red dots, Section \ref{strong}) and mild (green
  dots, Section \ref{mild}) CIA absorption.  Yellow symbols mark the
  four spectroscopically confirmed DC white dwarfs with mild
  absorption. The red and blue curves show the cooling sequences for
  0.6 \msun\ CO-core white dwarfs with pure H and pure He atmospheres,
  respectively. The solid cyan curves have the same mass but with
  $\logh = 0$, $-1$, $-2$, and $-3$, from bottom to top, while the
  solid black curves are with $\logh = -3.5$, $-4$, and $-5$, from top
  to bottom (starting from the left of the diagram); $\logh = -5$ is
  the smallest hydrogen abundance in our model grid. The dashed cyan
  lines show the mixed H/He sequences with $\logh = -2$ but at 0.2
  \msun\ and 1.0 \msun\ (top and bottom curves, respectively). Our new
  model atmospheres are used throughout. The black dotted line
  indicates Gaia's limiting magnitude at $G\sim21$ (i.e., $M_G=16$ for
  $D=100$~pc) in our 0.6 \msun\ models.\label{colorgyz}}
\end{figure}

\begin{figure*}
\hspace{-0.3in}
\includegraphics[width=2.5in, clip=true, trim=0.3in 2in 0.5in 1.4in]{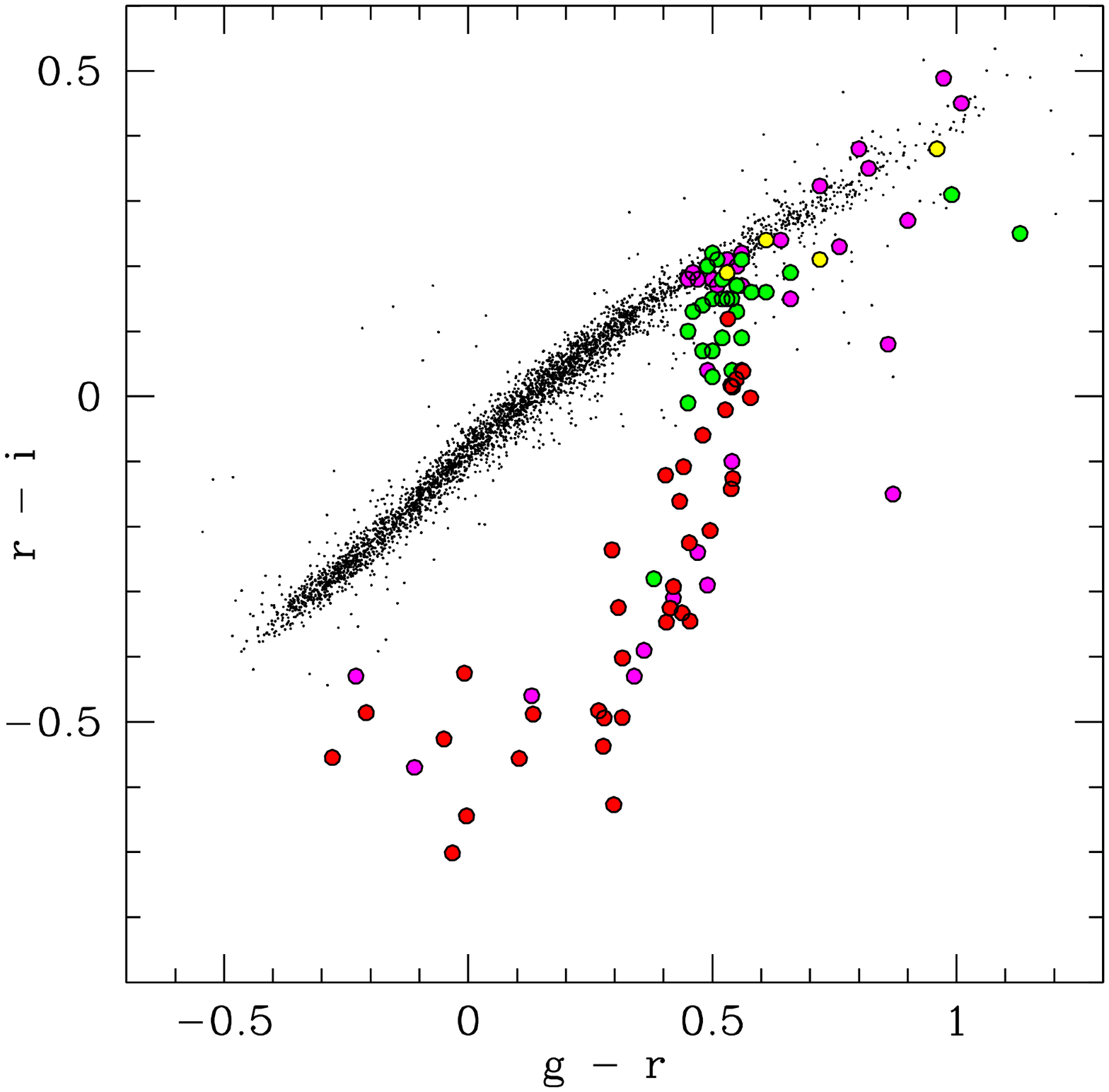}
\includegraphics[width=2.5in, clip=true, trim=0.3in 2in 0.5in 1.4in]{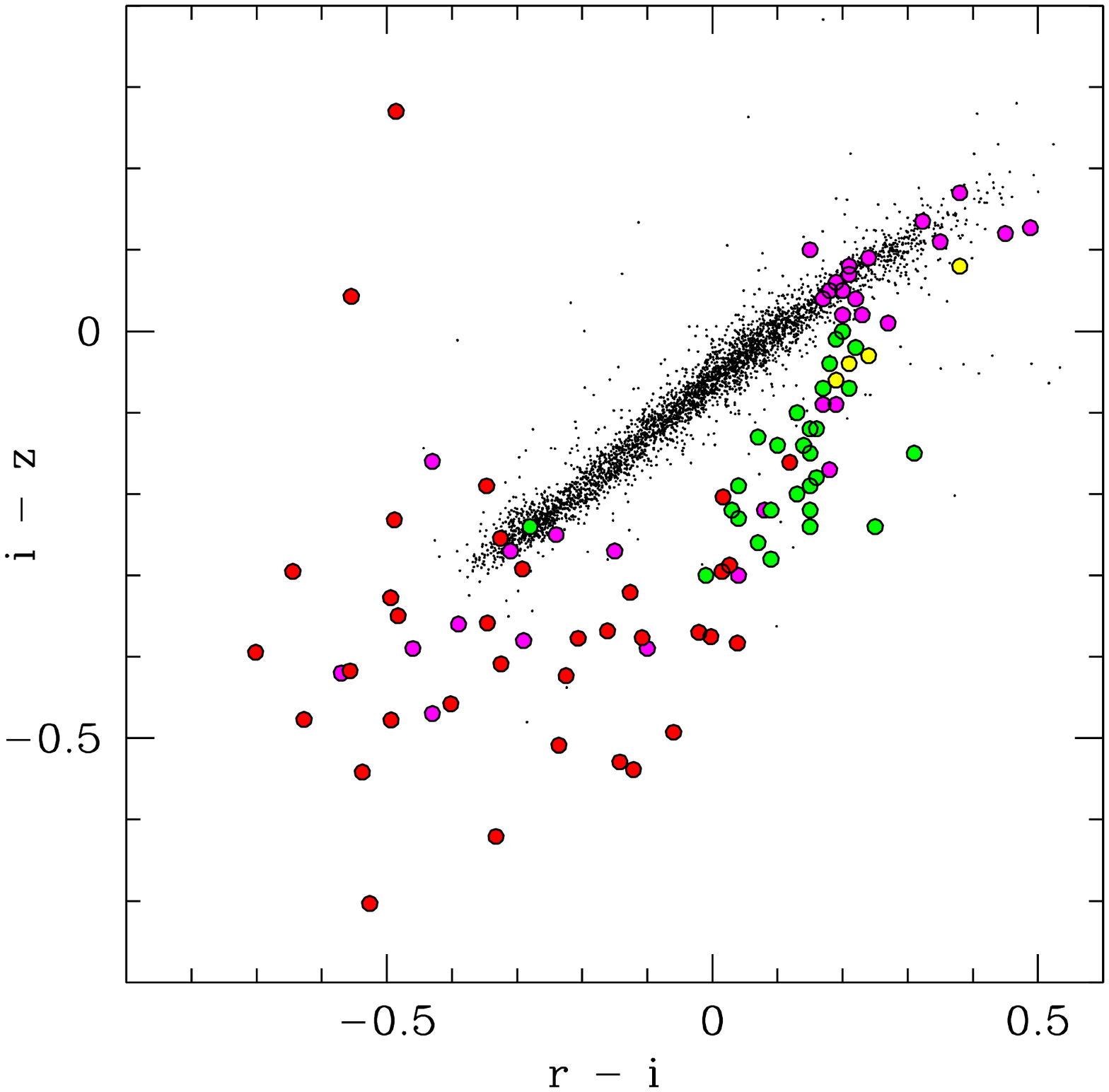}
\includegraphics[width=2.5in, clip=true, trim=0.3in 2in 0.5in 1.4in]{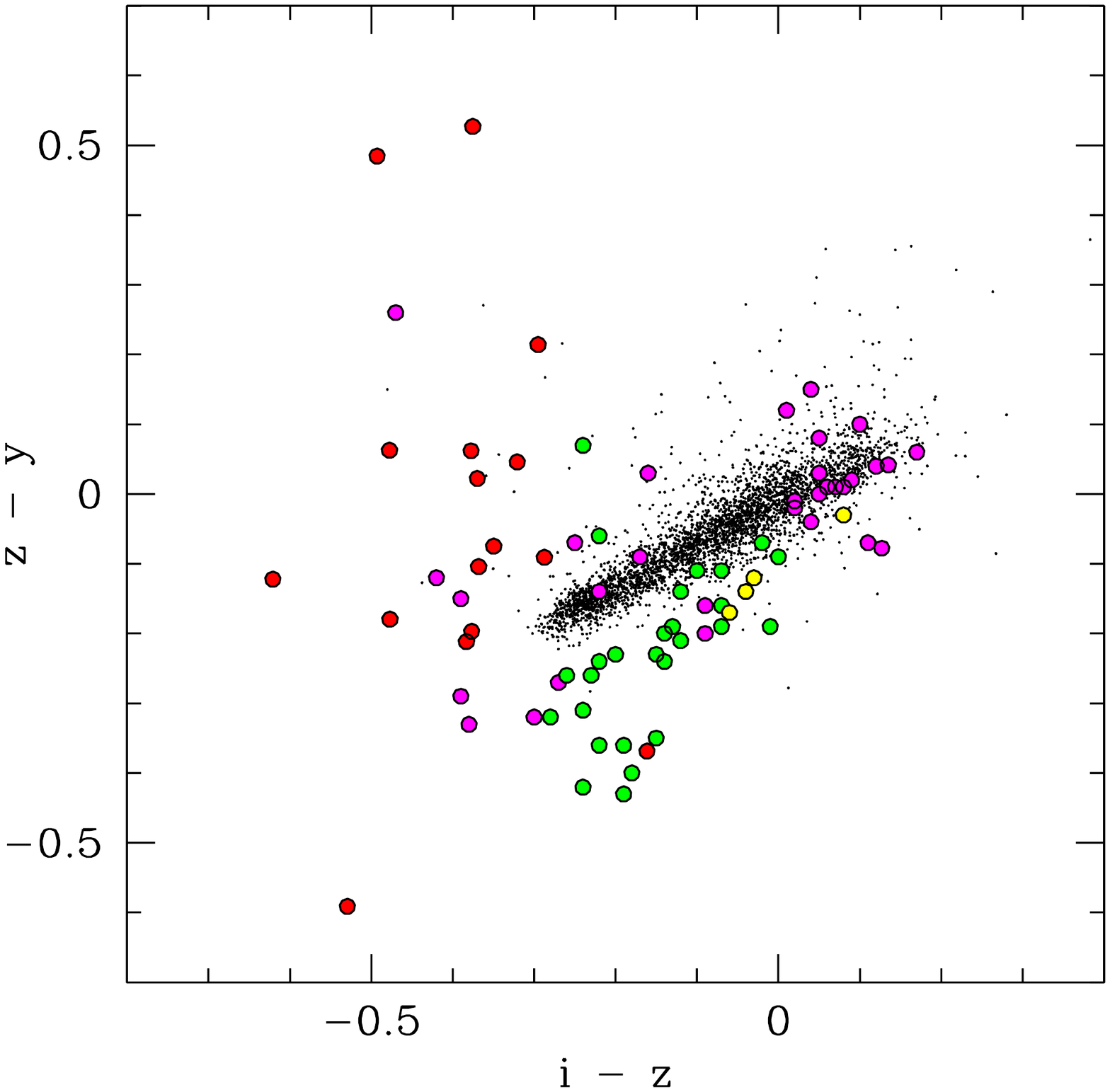}
\caption{Pan-STARRS color-color diagrams for the spectroscopically
  confirmed white dwarfs in the Montreal White Dwarf Database 100 pc
  sample (black dots) along with the IR-faint white dwarfs discussed
  here (colored symbols). The symbols are the same as in Figure
  \ref{colorgyz}.
\label{colors}}
\end{figure*}

Figure \ref{colorgyz} shows the color-magnitude diagram of the MWDD
100 pc sample (black dots) along with evolutionary sequences described
further below.  The previously known IR-faint white dwarfs and the newly
identified IR-faint white dwarfs with strong and mild absorption are
marked by magenta, red, and green dots, respectively. The four
spectroscopically confirmed DC white dwarfs with mild absorption (see
Figure \ref{add}) are marked by yellow dots.  We also show color-color
diagrams of the spectroscopically confirmed white dwarfs in the 100 pc
sample in various Pan-STARRS filters in Figure \ref{colors}.

These two figures clearly show that the 33 additional IR-faint white
dwarf candidates with mild optical flux deficits (green and yellow
dots) are an extension of the IR-faint white dwarf sequence. The
previously known 35 IR-faint white dwarfs (magenta), newly identified
37 with strong flux deficits (red), and the 33 objects with mild CIA
in the optical bands (green and yellow) clearly form a sequence in the
color-magnitude diagrams and also in the $g-r$ versus $r-i$
color-color diagram. Even though many of the mild CIA cases are
difficult to identify based on the $gri$ photometry alone, these
objects are clearly outliers in their $i-z$ and $z-y$ colors. The
IR-faint white dwarf sequence around $M_g \sim 16$ is rather tight,
both in the color-magnitude diagrams and the $g-r$ versus $r-i$
color-color diagram, suggesting that these objects probably have
similar masses and compositions. We also find a dozen objects or so
with luminosities below this tight sequence; we come back to these
objects later. Also indicated as a black dotted line in Figure
\ref{colorgyz} is Gaia's limiting magnitude at $G\sim21$ in our 0.6
\msun\ models. This limiting magnitude may partially explain why we
observe a lack of continuous IR-faint sequences in the color-magnitude
diagram.

Even with our detailed analysis of the 100 pc sample, we have only
scratched the surface in terms of the IR-faint white dwarfs.
Near-infrared observations of 112 cool DC white dwarfs by
\citet{kilic10b} found eight previously unknown IR-faint white dwarfs
that show significant absorption in the $H$- and $K$-bands or only in
the $K$-band. As the name indicates, given that these objects are
relatively faint in the infrared, they are usually missing in the
wide-field surveys like the UKIDSS and VHS. There are likely many more
IR-faint white dwarfs within 100 pc that can only be identified
through follow-up near-infrared observations.

Also shown in Figure \ref{colorgyz} are evolutionary sequences,
calculated with our new model atmospheres, for $0.6~M_{\odot}$ CO-core
white dwarfs with pure H, pure He, and mixed H/He compositions (red,
blue, and cyan/black lines). In the case of mixed H/He atmospheres, we
show the effects on the predicted colors of varying the atmospheric
composition ($\logh$ between $-5$ and 0) as well as the stellar mass
($M=0.2$ \msun\ and 1.0 \msun\ at $\logh = -2$). Our new models reach
a maximum optical luminosity ($M_g$) around $\logh = -4.0$ to $-3.5$
at 0.6 \msun\ depending on the temperature. Even though our old models
displayed in Figure \ref{theory} reach their maximum luminosity at
roughly the same H/He value, the predicted luminosities are well below
the observed sequence of IR-faint white dwarfs, even with a mass as
low as 0.2 \msun\ with a CO-core.  For this reason, our fitting code
with the old models always pick the solution where the CIA is
strongest in the model grid (see, e.g., Figure 20 of
\citealt{kilic20}), precisely because these models underestimate the
CIA opacities compared to the observations, as discussed above.  In
contrast, some of our new models are now more luminous than the
observed sequence of IR-faint white dwarfs, even at 0.6 \msun. This
should lead to a larger spread in the best-fitted parameters, in
particular the stellar mass and the hydrogen-to-helium abundance
ratio.  Also, the degeneracy observed in Figure \ref{theory} between
the 0.6 \msun, CO-core models with $\logh = -1$ and $-4$ is no longer
observed in Figure \ref{colorgyz} with our new model grid. This is the
reason why we always find a unique solution to all the IR-faint white
dwarfs in our sample. Finally, even though the results displayed in
Figure \ref{colorgyz} illustrate how a change in H/He can be
compensated to some extent by a change in mass, this occurs only in
this color-magnitude diagram. Indeed, while the luminosity in the
optical is controlled by the radius, and thus the mass of the star,
the CIA opacity, which dominates in the infrared, is controlled by the
H/He ratio. One parameter cannot be simply compensated by the other
when the overall energy distribution is considered.

\subsection{Global Properties of the Sample}\label{global}

Table \ref{tabfit} presents the best-fit parameters for all 105
IR-faint white dwarfs in our sample, including the previously known
and newly identified systems.  Despite the fact that our revised
models show significant improvements over our previous model grid,
those models remain approximate, as mentioned above, and the numbers
given in Table \ref{tabfit} should be taken with caution. As discussed
in Section \ref{known}, we provide in Table \ref{tabfit} two solutions
for J1238+3502, including one with the $J$ magnitude omitted from the fit (see
Figure \ref{pre}). In both cases, the solutions appear way too cool
for the inferred mass, which leads to cooling ages that are largely
extrapolated. 

\begin{figure}
\hspace{-0.1in}
\includegraphics[width=3.9in, clip=true, trim=1.0in 0.5in 0.5in
  0.7in]{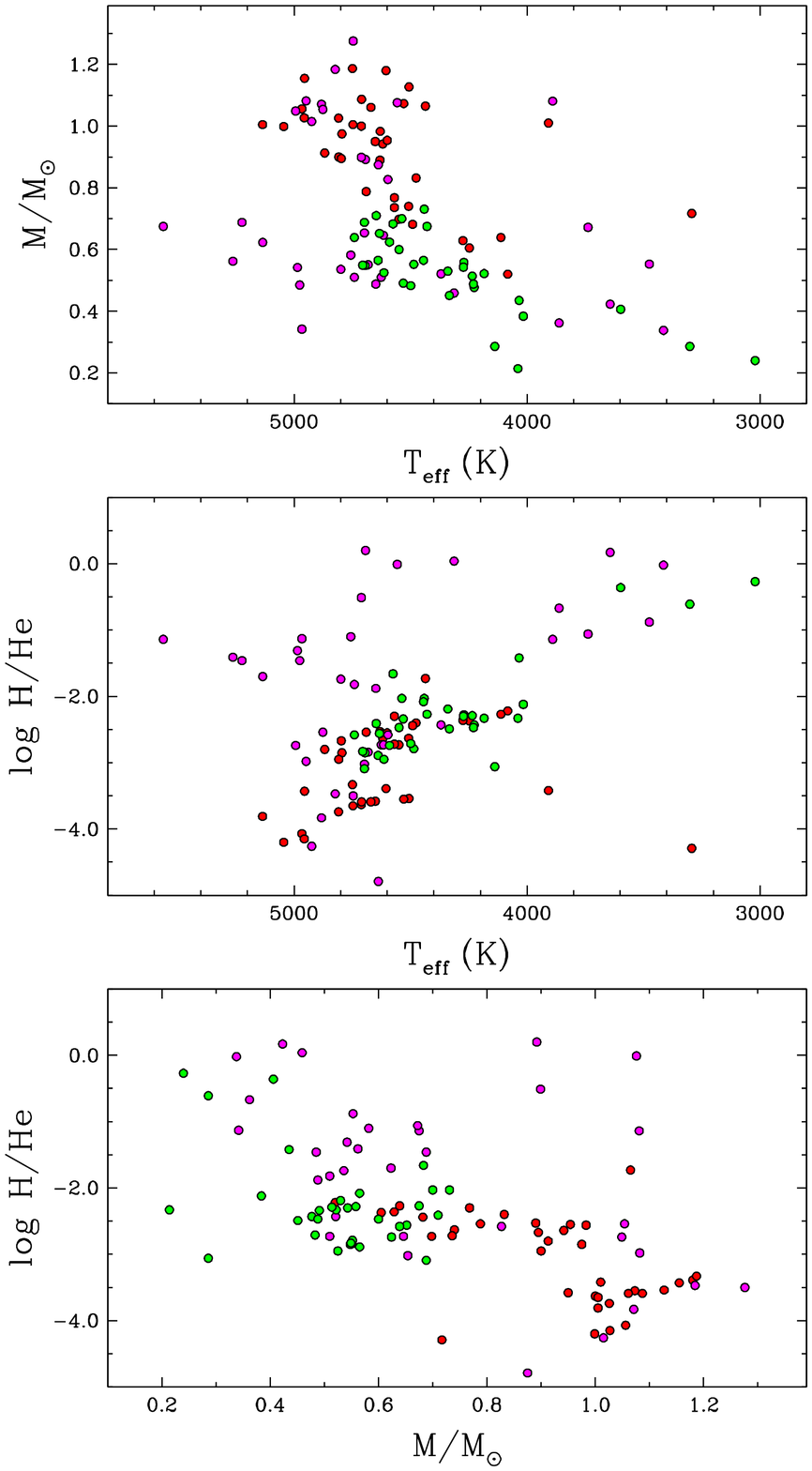}
\caption{Masses, effective temperatures, and H/He ratios for our IR-faint white dwarf
  sample. The symbols are the same as in Figure \ref{colorgyz}.
\label{correl}}
\end{figure}

In Figure \ref{correl}, we compare the physical parameters, $\Te$,
$M$, and $\logh$ one against each other, using the same color symbols
as in Figure \ref{colorgyz}. The IR-faint white dwarfs in our sample
have a range in $\Te$ between roughly 3000 K and 5600 K, although most
objects are clustered around $4600\pm200$ K, reinforcing our
conclusion that most IR-faint white dwarfs are not ultracool
afterall. The masses span an interval between 0.2 and 1.3 \msun, as
usually found for other mass distributions, although here we have a
strong excess of massive ($M> 0.8$ \msun) white dwarfs. Note that all
these massive objects consist of previously known IR-faint white
dwarfs (magenta) and newly IR-faint white dwarfs with strong CIA
absorption (red) identified in our spectroscopic sample.  In contrast,
those with mild IR absorption (green) have normal masses around
$\sim0.6$ \msun, indicating that more normal mass white dwarfs with
their lower photospheric densities produce, in general, less CIA
absorption, and they are thus more difficult to identify. This
interpretation is consistent with their location in Figure
\ref{colorgyz}, where they lie much closer to the main white dwarf
sequence. Incidentally, these white dwarfs with mild CIA absorption
tend to have similar hydrogen abundances around $\logh\sim-2.5$. The
fact that we find normal masses for these IR-faint white dwarfs gives
us confidence in our model atmospheres, at least at these photospheric
densities.

\startlongtable
\begin{deluxetable*}{lccccc}
\tablecolumns{6}
\tablewidth{0pt}
\tablecaption{Best-fit parameters for IR-faint White Dwarfs\label{tabfit}}
\tablehead{\\Name & $T_{\rm eff}$ & Mass & $\log{\rm H/He}$ & $\log{L/L_{\odot}}$ & Cooling Age\\
& (K) & ($M_{\odot}$) & & & (Gyr)}
\startdata
J0027+0554  & $ 4575 \pm  22 $ & $ 0.683^{+0.009}_{-0.009} $ & $ -1.66 $ & $ -4.311 $ & $ 7.444 $ \\
J0035+2009  & $ 4695 \pm  29 $ & $ 0.548^{+0.009}_{-0.009} $ & $ -2.85 $ & $ -4.148 $ & $ 6.028 $ \\
J0039+3035  & $ 4605 \pm  77 $ & $ 1.180^{+0.019}_{-0.022} $ & $ -3.39 $ & $ -4.869 $ & $ 5.123 $ \\
J0041$-$2221  & $ 5262 \pm  23 $ & $ 0.562^{+0.000}_{-0.000} $ & $ -1.41 $ & $ -3.961 $ & $ 4.672 $ \\
J0146+1404  & $ 4369 \pm  74 $ & $ 0.521^{+0.035}_{-0.036} $ & $ -2.43 $ & $ -4.250 $ & $ 6.329 $ \\
J0146+2122  & $ 4647 \pm  51 $ & $ 0.710^{+0.025}_{-0.025} $ & $ -2.41 $ & $ -4.308 $ & $ 7.464 $ \\
J0224$-$2854  & $ 4882 \pm 162 $ & $ 1.071^{+0.002}_{-0.003} $ & $ -3.83 $ & $ -4.598 $ & $ 6.241 $ \\
J0231+3254  & $ 4083 \pm 122 $ & $ 0.520^{+0.032}_{-0.076} $ & $ -2.22 $ & $ -4.334 $ & $ 14.91 $ \\
J0235$-$3032  & $ 4749 \pm 130 $ & $ 1.187^{+0.003}_{-0.003} $ & $ -3.33 $ & $ -4.831 $ & $ 4.973 $ \\
J0259$-$0455  & $ 4698 \pm 166 $ & $ 0.688^{+0.054}_{-0.057} $ & $ -3.09 $ & $ -4.270 $ & $ 7.277 $ \\
J0309+0025  & $ 5561 \pm  31 $ & $ 0.675^{+0.006}_{-0.006} $ & $ -1.14 $ & $ -3.963 $ & $ 5.056 $ \\
J0311+5257  & $ 4016 \pm  72 $ & $ 0.384^{+0.026}_{-0.026} $ & $ -2.12 $ & $ -4.205 $ & $ 9.648 $ \\
J0320+2948  & $ 5135 \pm 574 $ & $ 1.005^{+0.052}_{-0.062} $ & $ -3.81 $ & $ -4.427 $ & $ 6.535 $ \\
J0346+2455  & $ 3643 \pm  55 $ & $ 0.423^{+0.007}_{-0.007} $ & $ +0.17 $ & $ -4.417 $ & $ 13.33 $ \\
J0357$-$2606  & $ 4138 \pm  33 $ & $ 0.286^{+0.006}_{-0.006} $ & $ -3.06 $ & $ -4.039 $ & $ 6.058 $ \\
J0400+2138  & $ 4184 \pm  37 $ & $ 0.522^{+0.012}_{-0.013} $ & $ -2.33 $ & $ -4.326 $ & $ 6.752 $ \\
J0406$-$0333  & $ 3301 \pm  80 $ & $ 0.286^{+0.016}_{-0.015} $ & $ -0.61 $ & $ -4.442 $ & $ 9.653 $ \\
J0414+0309  & $ 4112 \pm  79 $ & $ 0.639^{+0.026}_{-0.027} $ & $ -2.27 $ & $ -4.459 $ & $ 7.903 $ \\
J0416$-$1826  & $ 4508 \pm  24 $ & $ 0.740^{+0.041}_{-0.043} $ & $ -2.63 $ & $ -4.388 $ & $ 7.763 $ \\
J0437$-$5946  & $ 4966 \pm  82 $ & $ 1.056^{+0.014}_{-0.015} $ & $ -4.07 $ & $ -4.548 $ & $ 6.313 $ \\
J0439$-$2016  & $ 4333 \pm  30 $ & $ 0.451^{+0.007}_{-0.008} $ & $ -2.49 $ & $ -4.200 $ & $ 5.402 $ \\
J0440$-$0414  & $ 4247 \pm 112 $ & $ 0.605^{+0.043}_{-0.045} $ & $ -2.37 $ & $ -4.372 $ & $ 7.441 $ \\
J0445$-$4906  & $ 4955 \pm 303 $ & $ 1.155^{+0.012}_{-0.014} $ & $ -3.43 $ & $ -4.708 $ & $ 5.326 $ \\
J0448+3206  & $ 4794 \pm 159 $ & $ 0.975^{+0.039}_{-0.043} $ & $ -2.85 $ & $ -4.513 $ & $ 7.004 $ \\
J0551$-$2652  & $ 4476 \pm  39 $ & $ 0.832^{+0.039}_{-0.042} $ & $ -2.40 $ & $ -4.485 $ & $ 7.855 $ \\
J0559+0731  & $ 4651\pm 2832 $ & $ 0.950^{+0.048}_{-0.054} $ & $ -3.58 $ & $ -4.537 $ & $ 7.266 $ \\
J0756$-$2001  & $ 4711 \pm 278 $ & $ 1.000^{+0.006}_{-0.007} $ & $ -3.63 $ & $ -4.570 $ & $ 6.933 $ \\
J0758$-$1711  & $ 4340 \pm  54 $ & $ 0.530^{+0.034}_{-0.034} $ & $ -2.19 $ & $ -4.270 $ & $ 6.516 $ \\
J0804+2239  & $ 4985 \pm  27 $ & $ 0.542^{+0.005}_{-0.004} $ & $ -1.31 $ & $ -4.039 $ & $ 5.327 $ \\
J0814+3300  & $ 4690 \pm  50 $ & $ 0.788^{+0.023}_{-0.025} $ & $ -2.54 $ & $ -4.362 $ & $ 7.574 $ \\
J0840+0515  & $ 4756 \pm  60 $ & $ 0.582^{+0.072}_{-0.076} $ & $ -1.10 $ & $ -4.156 $ & $ 6.288 $ \\
J0853$-$2446  & $ 3738 \pm  37 $ & $ 0.672^{+0.003}_{-0.003} $ & $ -1.06 $ & $ -4.653 $ & $ 8.641 $ \\
J0854+3503  & $ 4693 \pm  73 $ & $ 0.892^{+0.016}_{-0.016} $ & $ +0.20 $ & $ -4.461 $ & $ 7.457 $ \\
J0909+4700  & $ 4649 \pm  37 $ & $ 0.488^{+0.013}_{-0.012} $ & $ -1.88 $ & $ -4.112 $ & $ 5.228 $ \\
J0910$-$0222  & $ 4506 \pm  70 $ & $ 1.127^{+0.030}_{-0.035} $ & $ -3.54 $ & $ -4.818 $ & $ 5.854 $ \\
J0925+0018  & $ 4313 \pm  60 $ & $ 0.459^{+0.017}_{-0.019} $ & $ +0.04 $ & $ -4.158 $ & $ 10.46 $ \\
J0928+6049  & $ 4638 \pm 211 $ & $ 0.875^{+0.040}_{-0.042} $ & $ -4.79 $ & $ -4.465 $ & $ 7.569 $ \\
J0947+4459  & $ 4597 \pm  44 $ & $ 0.827^{+0.010}_{-0.011} $ & $ -2.58 $ & $ -4.434 $ & $ 7.706 $ \\
J1001+3903  & $ 4924 \pm 109 $ & $ 1.015^{+0.020}_{-0.021} $ & $ -4.26 $ & $ -4.512 $ & $ 6.654 $ \\
J1004$-$0506  & $ 3598 \pm  17 $ & $ 0.406^{+0.007}_{-0.007} $ & $ -0.36 $ & $ -4.422 $ & $ 12.94 $ \\
J1015$-$2009  & $ 4235 \pm  82 $ & $ 0.514^{+0.029}_{-0.030} $ & $ -2.29 $ & $ -4.298 $ & $ 6.542 $ \\
J1102+3523  & $ 4429 \pm  90 $ & $ 0.675^{+0.080}_{-0.089} $ & $ -2.27 $ & $ -4.361 $ & $ 7.625 $ \\
J1105$-$2114  & $ 4747 \pm 899 $ & $ 1.005^{+0.035}_{-0.040} $ & $ -3.65 $ & $ -4.563 $ & $ 6.87 $ \\
J1121+1417  & $ 4808 \pm  52 $ & $ 0.900^{+0.018}_{-0.017} $ & $ -2.95 $ & $ -4.429 $ & $ 7.305 $ \\
J1125+0941  & $ 4631 \pm  79 $ & $ 0.890^{+0.040}_{-0.045} $ & $ -2.53 $ & $ -4.482 $ & $ 7.538 $ \\
J1136$-$1057  & $ 3293\pm 2675 $ & $ 0.717^{+0.069}_{-0.077} $ & $ -4.29 $ & $ -4.912 $ & $ 9.31 $ \\
J1142$-$1315  & $ 4529 \pm 862 $ & $ 1.073^{+0.017}_{-0.018} $ & $ -3.55 $ & $ -4.731 $ & $ 6.43 $ \\
J1147+2220  & $ 4698 \pm  53 $ & $ 0.654^{+0.013}_{-0.013} $ & $ -3.02 $ & $ -4.240 $ & $ 7.054 $ \\
J1203+0426  & $ 4966 \pm  21 $ & $ 0.342^{+0.004}_{-0.005} $ & $ -1.13 $ & $ -3.769 $ & $ 4.021 $ \\
J1220+0914  & $ 3890 \pm  60 $ & $ 1.081^{+0.008}_{-0.008} $ & $ -1.14 $ & $ -5.006 $ & $ 6.668 $ \\
J1238+2633  & $ 4705 \pm  38 $ & $ 0.549^{+0.021}_{-0.021} $ & $ -2.83 $ & $ -4.146 $ & $ 6.021 $ \\
J1238+3502  & $ 3189 \pm 157 $ & $ 0.430^{+0.045}_{-0.051} $ & $ +0.00 $ & $ -4.657 $ & $ 16.95 $ \\
J1238+3502  & $ 3475 \pm 151 $ & $ 0.553^{+0.000}_{-0.014} $ & $ -0.88 $ & $ -4.702 $ & $ 23.96 $ \\
J1251+4403  & $ 4746 \pm  42 $ & $ 1.276^{+0.008}_{-0.009} $ & $ -3.50 $ & $ -5.035 $ & $ 3.61 $ \\
J1304+0126  & $ 4275 \pm  49 $ & $ 0.629^{+0.028}_{-0.028} $ & $ -2.36 $ & $ -4.383 $ & $ 7.587 $ \\
J1320+0836  & $ 4975 \pm 118 $ & $ 0.485^{+0.040}_{-0.066} $ & $ -1.46 $ & $ -3.937 $ & $ 7.08 $ \\
J1326$-$1558  & $ 4230 \pm  43 $ & $ 0.488^{+0.013}_{-0.014} $ & $ -2.47 $ & $ -4.276 $ & $ 6.222 $ \\
J1336+0748  & $ 4619 \pm  83 $ & $ 0.942^{+0.037}_{-0.041} $ & $ -2.64 $ & $ -4.540 $ & $ 7.338 $ \\
J1337+0001  & $ 4948 \pm  26 $ & $ 1.082^{+0.011}_{-0.011} $ & $ -2.98 $ & $ -4.590 $ & $ 6.094 $ \\
J1355$-$2600  & $ 3908 \pm \dots $ & $ 1.010^{+0.038}_{-0.043} $ & $ -3.42 $ & $ -4.907 $ & $ 7.403 $ \\
J1403+4533  & $ 4823 \pm  17 $ & $ 1.184^{+0.003}_{-0.004} $ & $ -3.47 $ & $ -4.797 $ & $ 4.992 $ \\
J1403$-$1514  & $ 4741 \pm  41 $ & $ 0.510^{+0.006}_{-0.007} $ & $ -1.82 $ & $ -4.097 $ & $ 5.392 $ \\
J1404+1330  & $ 4626 \pm  44 $ & $ 0.510^{+0.020}_{-0.021} $ & $ -2.73 $ & $ -4.140 $ & $ 5.628 $ \\
J1437+4151  & $ 5134 \pm  44 $ & $ 0.623^{+0.007}_{-0.007} $ & $ -1.70 $ & $ -4.058 $ & $ 5.843 $ \\
J1448+2935  & $ 4039 \pm  44 $ & $ 0.214^{+0.011}_{-0.011} $ & $ -2.33 $ & $ -3.982 $ & $ 4.57 $ \\
J1503$-$3005  & $ 4809 \pm 591 $ & $ 1.026^{+0.022}_{-0.024} $ & $ -3.74 $ & $ -4.567 $ & $ 6.664 $ \\
J1525+6247  & $ 4486 \pm 207 $ & $ 0.552^{+0.032}_{-0.033} $ & $ -2.79 $ & $ -4.231 $ & $ 6.484 $ \\
J1531+4421  & $ 4491 \pm  92 $ & $ 0.682^{+0.028}_{-0.028} $ & $ -2.44 $ & $ -4.344 $ & $ 7.576 $ \\
J1542+2750  & $ 4868 \pm 148 $ & $ 0.913^{+0.045}_{-0.052} $ & $ -2.80 $ & $ -4.418 $ & $ 7.194 $ \\
J1546+2054  & $ 4633 \pm  55 $ & $ 0.652^{+0.030}_{-0.030} $ & $ -2.56 $ & $ -4.263 $ & $ 7.155 $ \\
J1556$-$0806  & $ 4876 \pm 110 $ & $ 1.054^{+0.004}_{-0.004} $ & $ -2.54 $ & $ -4.577 $ & $ 6.393 $ \\
J1559+7314  & $ 4227 \pm  39 $ & $ 0.477^{+0.008}_{-0.009} $ & $ -2.43 $ & $ -4.267 $ & $ 6.073 $ \\
J1602+0856  & $ 4797 \pm 135 $ & $ 0.895^{+0.040}_{-0.044} $ & $ -2.67 $ & $ -4.426 $ & $ 7.326 $ \\
J1610+0619  & $ 4956 \pm 273 $ & $ 1.027^{+0.042}_{-0.050} $ & $ -4.15 $ & $ -4.515 $ & $ 6.55 $ \\
J1612+5128  & $ 4670 \pm \dots $ & $ 1.061^{+0.037}_{-0.042} $ & $ -3.59 $ & $ -4.662 $ & $ 6.463 $ \\
J1632+2426  & $ 4711 \pm  35 $ & $ 0.899^{+0.007}_{-0.006} $ & $ -0.51 $ & $ -4.461 $ & $ 7.413 $ \\
J1633+3829  & $ 4569 \pm 159 $ & $ 0.768^{+0.080}_{-0.093} $ & $ -2.30 $ & $ -4.389 $ & $ 7.73 $ \\
J1639+0106  & $ 4590 \pm  79 $ & $ 0.624^{+0.052}_{-0.055} $ & $ -2.74 $ & $ -4.254 $ & $ 7.009 $ \\
J1653+6253  & $ 4993 \pm  33 $ & $ 1.049^{+0.002}_{-0.002} $ & $ -2.74 $ & $ -4.530 $ & $ 6.35 $ \\
J1722+5752  & $ 5223 \pm  50 $ & $ 0.688^{+0.010}_{-0.011} $ & $ -1.46 $ & $ -4.084 $ & $ 6.156 $ \\
J1727+0808  & $ 4682 \pm  85 $ & $ 0.551^{+0.036}_{-0.036} $ & $ -2.84 $ & $ -4.156 $ & $ 6.092 $ \\
J1753+0758  & $ 4271 \pm  23 $ & $ 0.558^{+0.009}_{-0.009} $ & $ -2.28 $ & $ -4.322 $ & $ 6.965 $ \\
J1824+1213 & $ 3414 \pm 71 $ & $ 0.338^{+0.006}_{-0.006} $ & $ -0.02 $ & $ -4.442 $ & $ 11.21 $ \\
J1830+2529  & $ 4630 \pm 139 $ & $ 0.983^{+0.035}_{-0.039} $ & $ -2.56 $ & $ -4.582 $ & $ 7.104 $ \\
J1922+0233  & $ 4436 \pm  53 $ & $ 1.065^{+0.007}_{-0.007} $ & $ -1.73 $ & $ -4.756 $ & $ 6.563 $ \\
J1922$-$0402  & $ 4441 \pm  81 $ & $ 0.731^{+0.033}_{-0.034} $ & $ -2.03 $ & $ -4.406 $ & $ 7.831 $ \\
J1944$-$0425  & $ 4499 \pm 165 $ & $ 0.483^{+0.034}_{-0.044} $ & $ -2.71 $ & $ -4.112 $ & $ 9.921 $ \\
J1951+4026  & $ 4034 \pm  38 $ & $ 0.435^{+0.004}_{-0.005} $ & $ -1.42 $ & $ -4.310 $ & $ 6.006 $ \\
J2012$-$2720  & $ 4444 \pm  43 $ & $ 0.565^{+0.024}_{-0.026} $ & $ -2.08 $ & $ -4.258 $ & $ 6.709 $ \\
J2035+4054  & $ 3021 \pm  65 $ & $ 0.240^{+0.009}_{-0.009} $ & $ -0.27 $ & $ -4.538 $ & $ 8.972 $ \\
J2056+7218  & $ 4710\pm 3639 $ & $ 1.087^{+0.024}_{-0.027} $ & $ -3.59 $ & $ -4.682 $ & $ 6.191 $ \\
J2138$-$0056  & $ 4799 \pm  79 $ & $ 0.536^{+0.015}_{-0.015} $ & $ -1.74 $ & $ -4.100 $ & $ 5.666 $ \\
J2148+2601  & $ 4639 \pm  83 $ & $ 0.565^{+0.052}_{-0.054} $ & $ -2.89 $ & $ -4.184 $ & $ 6.336 $ \\
J2148$-$2821  & $ 4600 \pm  95 $ & $ 0.954^{+0.057}_{-0.066} $ & $ -2.55 $ & $ -4.561 $ & $ 7.295 $ \\
J2150$-$0439  & $ 4549 \pm  26 $ & $ 0.600^{+0.009}_{-0.009} $ & $ -2.47 $ & $ -4.249 $ & $ 6.875 $ \\
J2217+4241  & $ 4537 \pm  84 $ & $ 0.700^{+0.028}_{-0.030} $ & $ -2.03 $ & $ -4.341 $ & $ 7.584 $ \\
J2237+2220  & $ 4614 \pm  57 $ & $ 0.525^{+0.043}_{-0.043} $ & $ -2.95 $ & $ -4.158 $ & $ 5.883 $ \\
J2239+0018  & $ 4616 \pm  61 $ & $ 0.646^{+0.038}_{-0.040} $ & $ -2.73 $ & $ -4.263 $ & $ 7.138 $ \\
J2242+0048  & $ 3862 \pm  39 $ & $ 0.362^{+0.009}_{-0.009} $ & $ -0.67 $ & $ -4.251 $ & $ 9.72 $ \\
J2305+3922  & $ 4550 \pm  33 $ & $ 0.698^{+0.004}_{-0.004} $ & $ -2.73 $ & $ -4.334 $ & $ 7.556 $ \\
J2317+1830  & $ 4557 \pm 63 $ & $ 1.076^{+0.007}_{-0.008} $ & $ -0.01 $ & $ -4.724 $ & $ 6.388 $ \\
J2332+0959  & $ 4741 \pm  42 $ & $ 0.639^{+0.030}_{-0.031} $ & $ -2.58 $ & $ -4.211 $ & $ 6.851 $ \\
J2337$-$2158  & $ 4531 \pm  60 $ & $ 0.491^{+0.024}_{-0.030} $ & $ -2.34 $ & $ -4.109 $ & $ 9.89 $ \\
J2340+6117  & $ 4569 \pm 102 $ & $ 0.736^{+0.018}_{-0.018} $ & $ -2.72 $ & $ -4.361 $ & $ 7.667 $ \\
J2340+6902  & $ 5044 \pm 468 $ & $ 0.999^{+0.020}_{-0.020} $ & $ -4.20 $ & $ -4.452 $ & $ 6.651 $ \\
J2355+4419  & $ 4273 \pm  68 $ & $ 0.543^{+0.036}_{-0.036} $ & $ -2.30 $ & $ -4.308 $ & $ 6.805 $ \\
\enddata
\end{deluxetable*}

The bottom panel of Figure \ref{correl} shows that the massive white
dwarfs identified in the upper panel also tend to have the smallest
hydrogen abundances around $\logh\sim-3.5$. In contrast, the IR-faint
white dwarfs with mild CIA absorption and normal masses (green
symbols) are characterized by hydrogen abundances around
$\logh\sim-2.5$.  This trend of $\logh$ with mass is most likely
related to the origin, and thus the progenitors, of these IR-faint
white dwarfs. If they represent the outcome of convectively mixed DA
stars, our results indicate that more massive white dwarfs have
smaller hydrogen abundances at the photosphere as a result of a larger
dilution in the stellar envelope. We come back to this point in
Section \ref{origin}.

We also observe in Figure \ref{correl} IR-faint white dwarfs with
equal amounts of hydrogen and helium in their atmospheres
($\logh\sim0$), spanning a large range in stellar mass from 0.2 to 1.0
\msun. Two of these objects are the DZ white dwarfs J1824+1213 and
J2317+1830, recently analyzed by \citet{hollands21}. Given that the
presence of metals in cool DZ stars is generally interpreted as
evidence for accretion from external sources such as the interstellar
medium, tidally disrupted asteroids, small planets, or even comets,
the large hydrogen abundance measured in these stars may also be the
result of accretion from such external sources, possibly containing
water (see, e.g., \citealt{fusillo17}). The large hydrogen abundances
in all other objects in our sample most likely have the same origin,
even if they show no metal, because heavy elements eventually diffuse
at the bottom of the mixed H/He convection zone, while hydrogen always
remain within the stellar envelope.  Interestingly, we obtain similar
stellar parameters for these two DZ white dwarfs if we use the model
atmospheres from \citet{blouin18a}, because in such hydrogen-enriched
atmospheres, the dominant source of opacity in the infrared is the
H$_2$-H$_2$ CIA opacity (instead of H$_2$-He CIA), and both sets of
model atmospheres in this case rely on the same opacity calculations.

\begin{figure}
\hspace{0.in}
\includegraphics[width=3.5in, clip=true, trim=0.5in 2.8in 0.3in 3.3in]{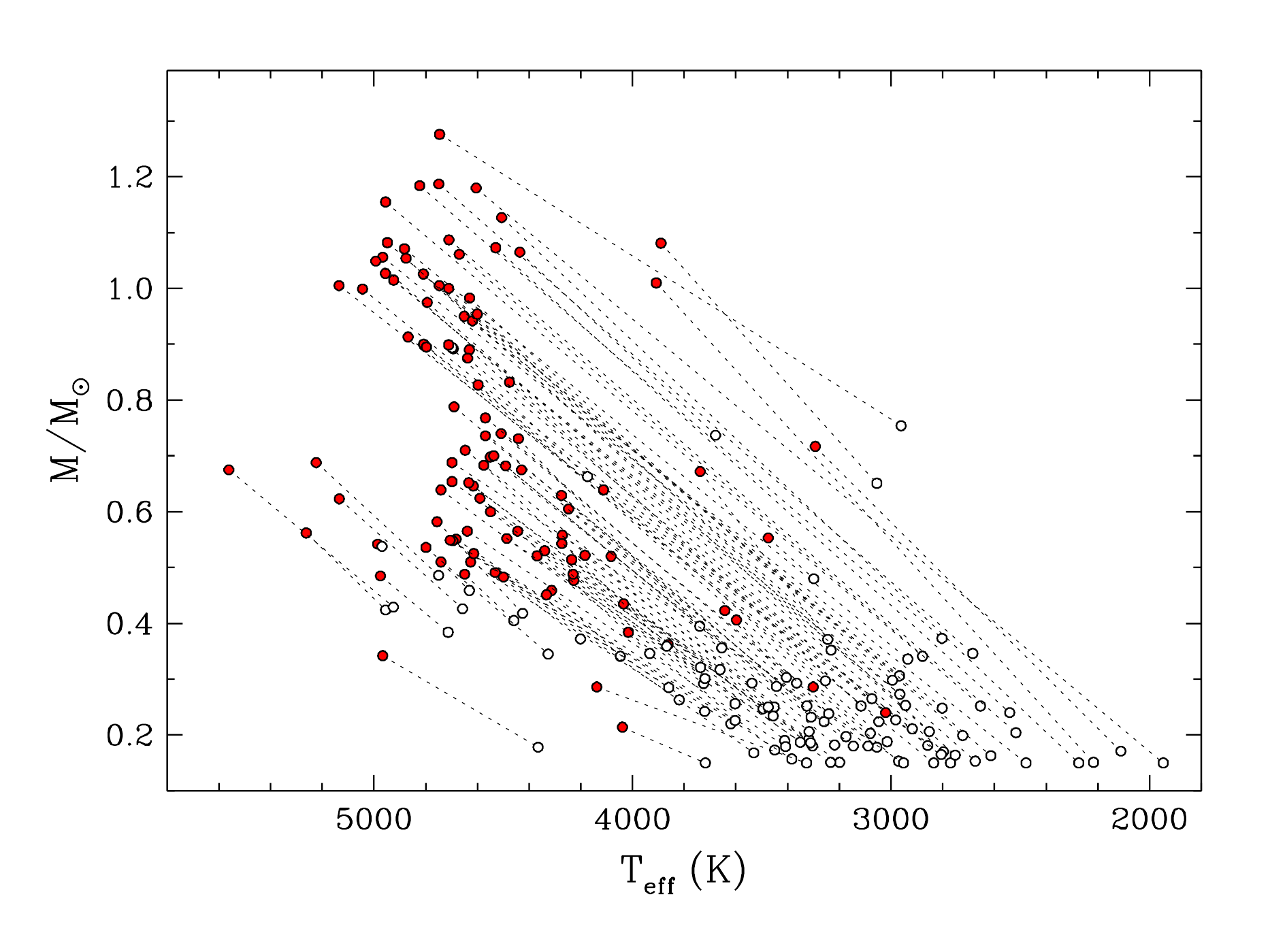}
\caption{Comparison of masses and effective temperatures obtained with
  the original (white symbols) and revised (red symbols) model atmospheres. Solutions for
  each object are connected by a dotted line.
\label{compare_M_Teff}}
\end{figure}

For completeness, we compare in Figure \ref{compare_M_Teff} the masses
and effective temperatures obtained with the original and revised
models; the changes in $\logh$ values are not as important (see, e.g.,
Figure \ref{fig_improve}) and not discussed any further. As was
described previously in the case of J2305+3922 displayed in Figure
\ref{fig_improve}, both the $M$ and $\Te$ values using our revised
models have increased significantly and systematically with respect to
the values obtained with the original model grid, although the
differences are larger in the most massive objects where the
high-density correction to the He$^-$ free-free absorption coefficient
is more important. Note also that with the original model grid, the
extremely low masses inferred at 0.15 \msun\ represent only upper
limits since this corresponds to the lowest mass value of our cooling
sequences assuming a He-core (see Figure \ref{grid}).

\subsection{Kinematics}

\begin{figure*}
\hspace{-0.2in}
\includegraphics[width=3.5in, clip=true, trim=0.3in 1.5in 0.4in 1.4in]{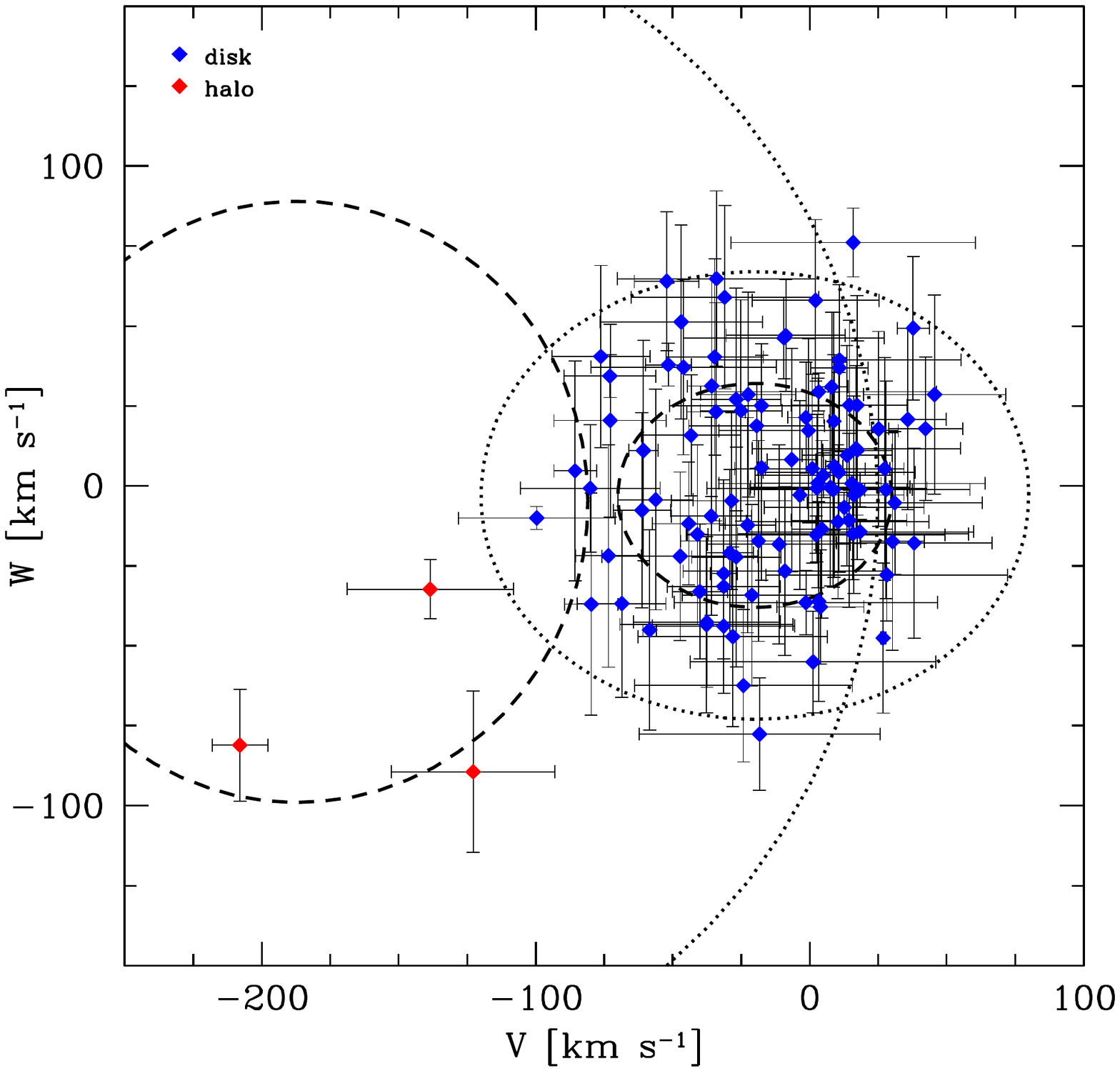}
\includegraphics[width=3.5in, clip=true, trim=0.3in 1.5in 0.4in 1.4in]{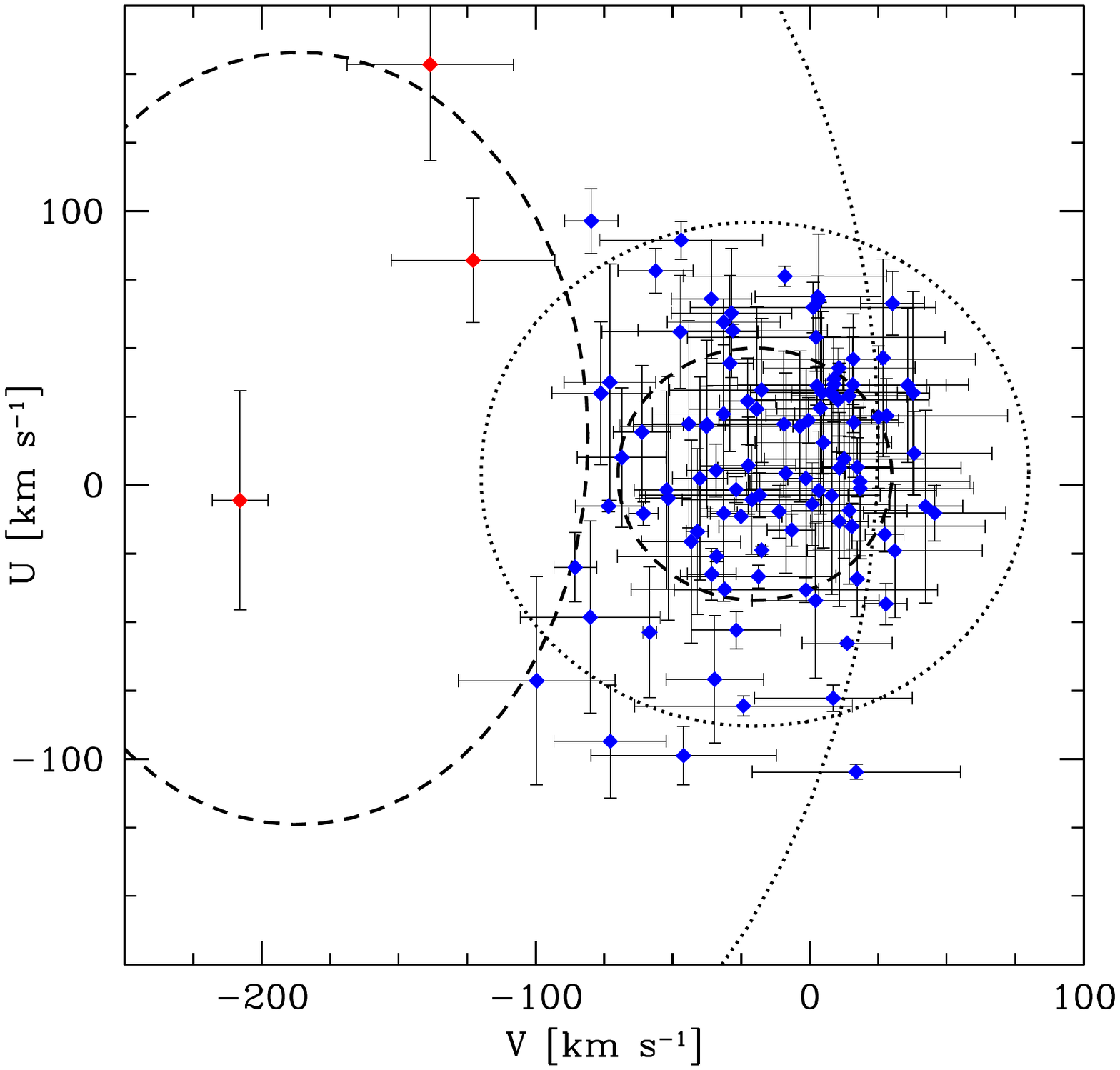}
\caption{Velocity distribution of the IR-faint white dwarf sample,
  plotted in Galactic cartesian velocity components $U$ (in the
  direction of the Galactic center), $V$ (in the direction of Galactic
  rotation), and $W$ (in the direction of the north Galactic pole).  The
  1 and $2\sigma$ velocity ellipsoid values for stellar thick disk and
  halo populations \citep{chiba00} are shown as the dashed and dotted
  lines, respectively. Error bars show the dispersion of the $UVW$ velocity
  components assuming the stars have thick disk radial velocities.
\label{figuvw}}
\end{figure*}

Kinematics can provide additional insights into the nature of IR-faint
white dwarfs. Figure \ref{figuvw} plots the distribution of Galactic
$U$, $V$, and $W$ velocity components for the IR-faint white dwarf
sample, along with the $1\sigma$ (dashed) and $2\sigma$ (dotted)
velocity ellipsoid values for the thick-disk and halo \citep{chiba00}.
The thick disk velocity ellipsoid is $(\sigma_U, \sigma_V,
\sigma_W)=(46,50,35)$ km s$^{-1}$. Since we do not have any radial
velocity constraints on the IR-faint white dwarfs, we assume zero
radial velocity. The tangential velocities clearly center on the disk,
except for three objects, WD 0343+247 (J0346+2455), LHS 2139
(J0925+0018), and J1824+1213, which are clearly halo stars.

The 2D velocity dispersion of the 105 stars, 60 km s$^{-1}$, is
practically identical to the \citet{chiba00} thick disk value. This
motivated us to draw hypothetical radial velocities from the radial
component of the thick disk velocity ellipsoid, 10,000 times for each
star. The mean radial velocity of each star remains zero in this
approach, but the dispersion (44 km s$^{-1}$, on average) provides a
quantitative measure of how plausible radial velocities may impact the
kinematics. For completeness, we also re-draw the measured proper
motions and parallaxes using the full Gaia covariance matrix.

The error bars in Figure \ref{figuvw} show the dispersion of the $UVW$
velocity components assuming the stars have thick disk radial
velocities.  Note that radial velocities project into different $UVW$
velocity components depending on a star's location on the sky (i.e. at
the pole, or anti-center, etc).  Thick disk radial velocities inflate
the velocity of any thin disk star, of course, but any thin disk star
will remain consistent with the disk. It is possible that halo stars
may remain hidden in the sample. Still, the overall sample remains
clustered around the thick disk velocity ellipsoid, and we conclude
that the IR-faint white dwarfs are a thick disk population on the
basis of their kinematics. This is consistent with our relatively
large white dwarf cooling age estimates presented in Table
\ref{tabfit}.

\subsection{On the Origin of IR-faint White Dwarfs}\label{origin}

\begin{figure*}
\hspace{0.5in}
\includegraphics[width=6.0in, clip=true, trim=0.0in 3.0in 0.0in
  3.0in]{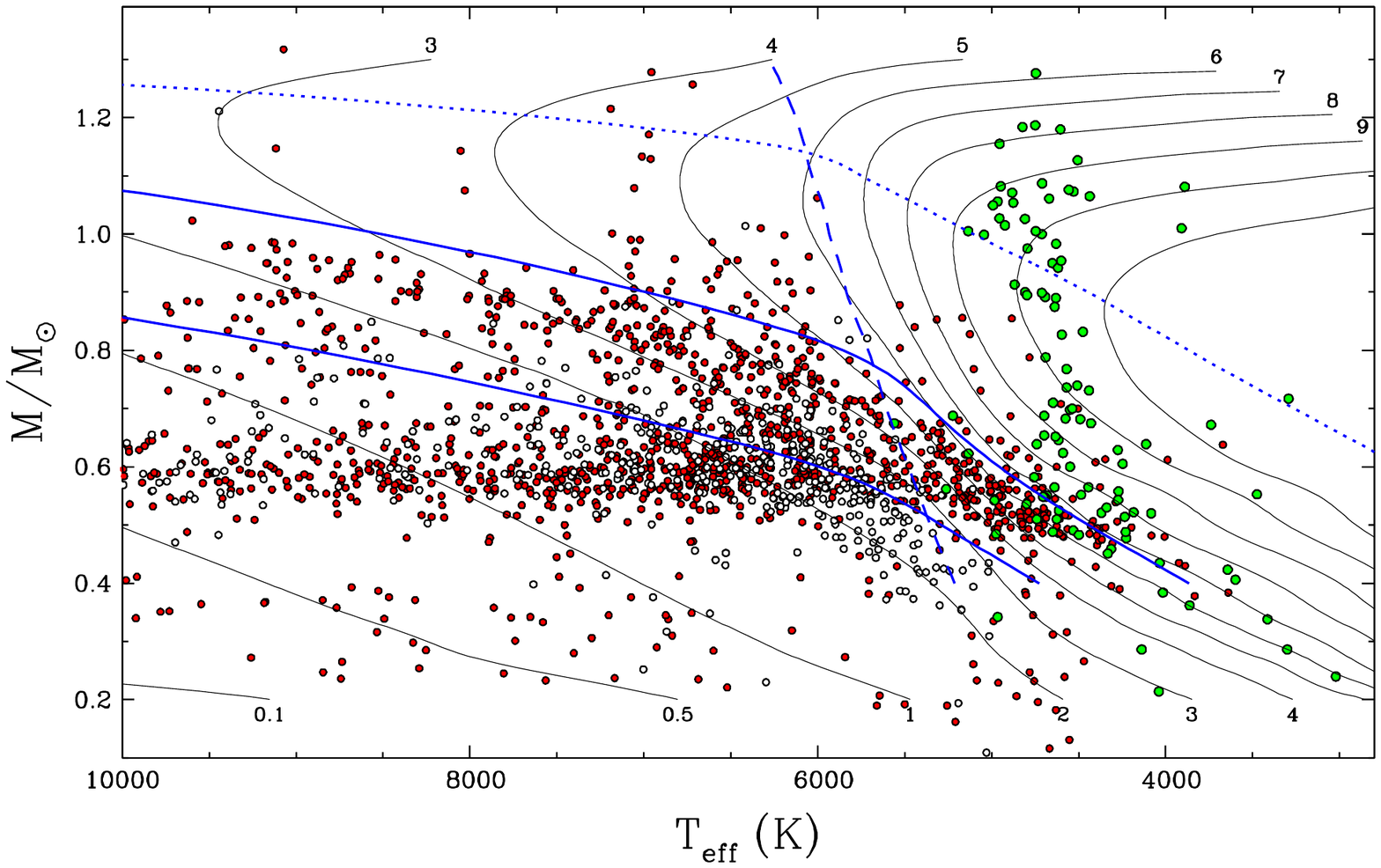}
\caption{Stellar masses as a function of effective temperature for all
  spectroscopically confirmed white dwarfs in the 100 pc sample and
  the SDSS footprint from \citet{kilic20}; DA and non-DA white dwarfs
  are shown by red and white circles, respectively. Our sample of 105
  IR-faint white dwarfs is shown by green circles.  Solid curves are
  theoretical isochrones (white dwarf cooling age only), labeled in
  units of $10^9$ yr, obtained from the cooling sequences of
  \citet{bedard20} with C/O-core compositions, $\qhe=10^{-2}$, and
  $\qh=10^{-4}$. The lower blue solid curve indicates the onset of
  crystallization at the center of evolving models, and the upper one
  indicates the locations where 80\% of the total mass has
  solidified. The dashed curve indicates the onset of convective
  coupling, while the dotted curve corresponds to the transition
  between the classical to the quantum regime in the ionic plasma (see
  text).
\label{correltm_Kilic}}
\end{figure*}

We show in Figure \ref{correltm_Kilic} the location of the 105
IR-faint white dwarfs in our sample in a $M$ vs $\Te$ diagram,
together with the spectroscopically confirmed white dwarfs in the 100
pc sample and the SDSS footprint from \citet[][see their Figure
  18]{kilic20}. Note that in Kilic et al., the DQ and DZ white dwarfs
have been analyzed using detailed model atmospheres appropriate for
these stars, but they are treated simply as non-DAs (white circles) in
Figure \ref{correltm_Kilic}. So even though the Kilic et al.~sample is
restricted to the SDSS footprint and is thus considerably smaller than
the large Gaia sample displayed in Figure 11 of \citet{bergeron19},
the analysis from Bergeron et al. relies on pure H and pure He model
atmospheres only, and their results are therefore considerably less accurate.

At the scale used in Figure \ref{correltm_Kilic}, the IR-faint white
dwarfs are found in a rather narrow range of effective temperatures.
The IR-faint objects with normal masses around 0.6 \msun\ appear as a
natural extension of the DA and non-DA white dwarfs found at higher
temperatures. More striking in this figure is the excess of massive
white dwarfs in the IR-faint population, which appears even more
extreme here. There is also an apparent gap in $\Te$ between the
coolest massive DA stars in Figure \ref{correltm_Kilic} and the
massive IR-faint white dwarfs. We have to be careful not to
overinterpret these results, however, because there are various
selection effects at play with these samples. First, the sample from
\citet{kilic20} is restricted to the SDSS footprint, while ours covers
the Pan-STARRS $3\pi$ survey. Moreover, the objects displayed in
Figure \ref{correltm_Kilic} do not include white dwarf candidates
without spectroscopic confirmation, many of which are still present in
the 100 pc sample, and even more so if they are massive, with smaller
radii, and thus lower luminosities. Also, as mentioned in Section
\ref{sec:sample}, although we restricted our analysis to a distance of
100 pc, we performed a deeper search of all white dwarf candidates
with $5\sigma$ significant parallax measurements from Gaia DR2. Hence
the most extreme IR-faint objects in our sample, which also happen to
be more massive than average, are most likely over represented in
Figure \ref{correltm_Kilic} with respect to the rest of the white
dwarfs displayed here.

\begin{figure*}
\includegraphics[width=3.4in, clip=true, trim=1.0in 1.7in 1.0in 1.9in]{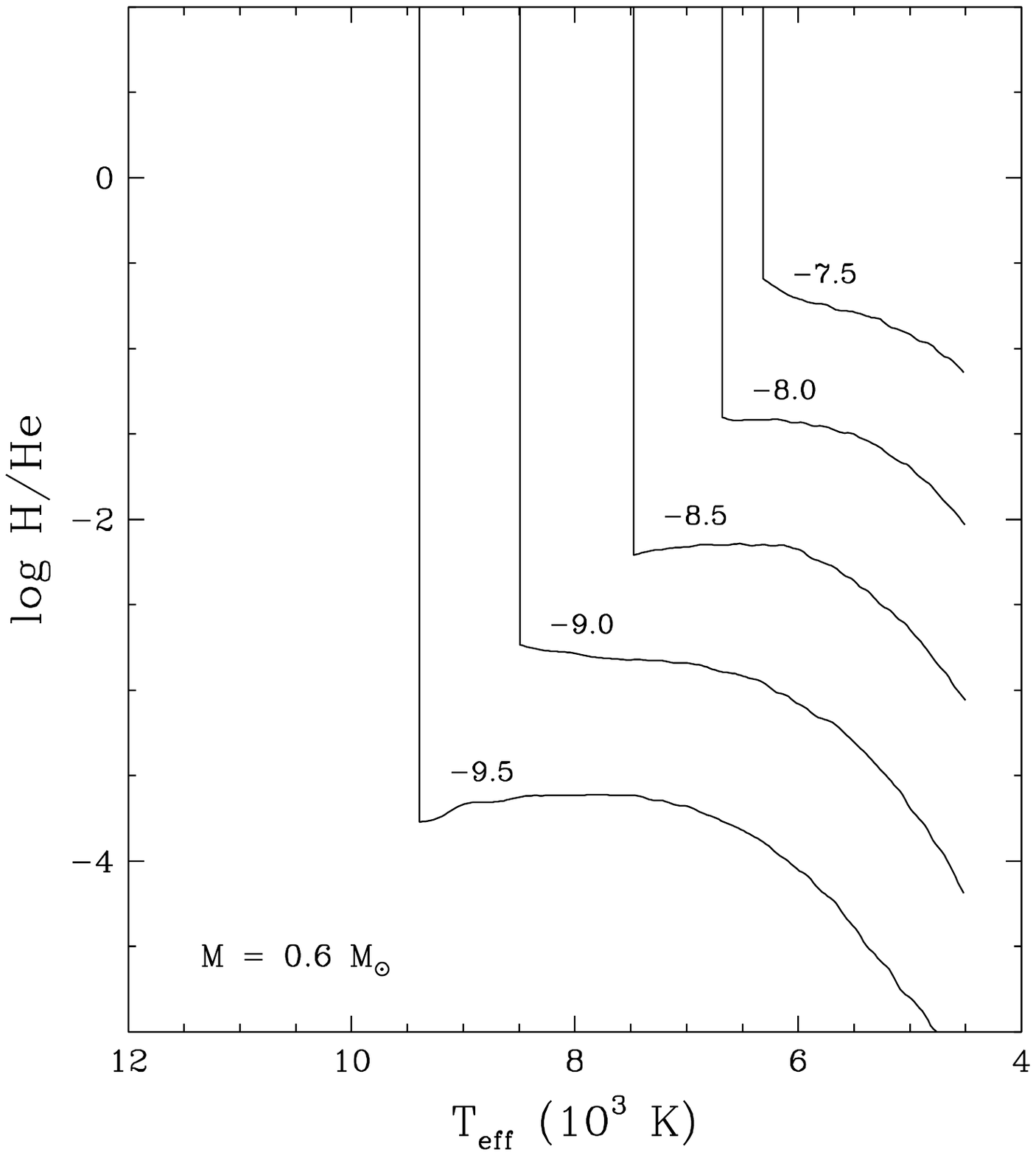}
\includegraphics[width=3.4in, clip=true, trim=1.0in 1.7in 1.0in 1.9in]{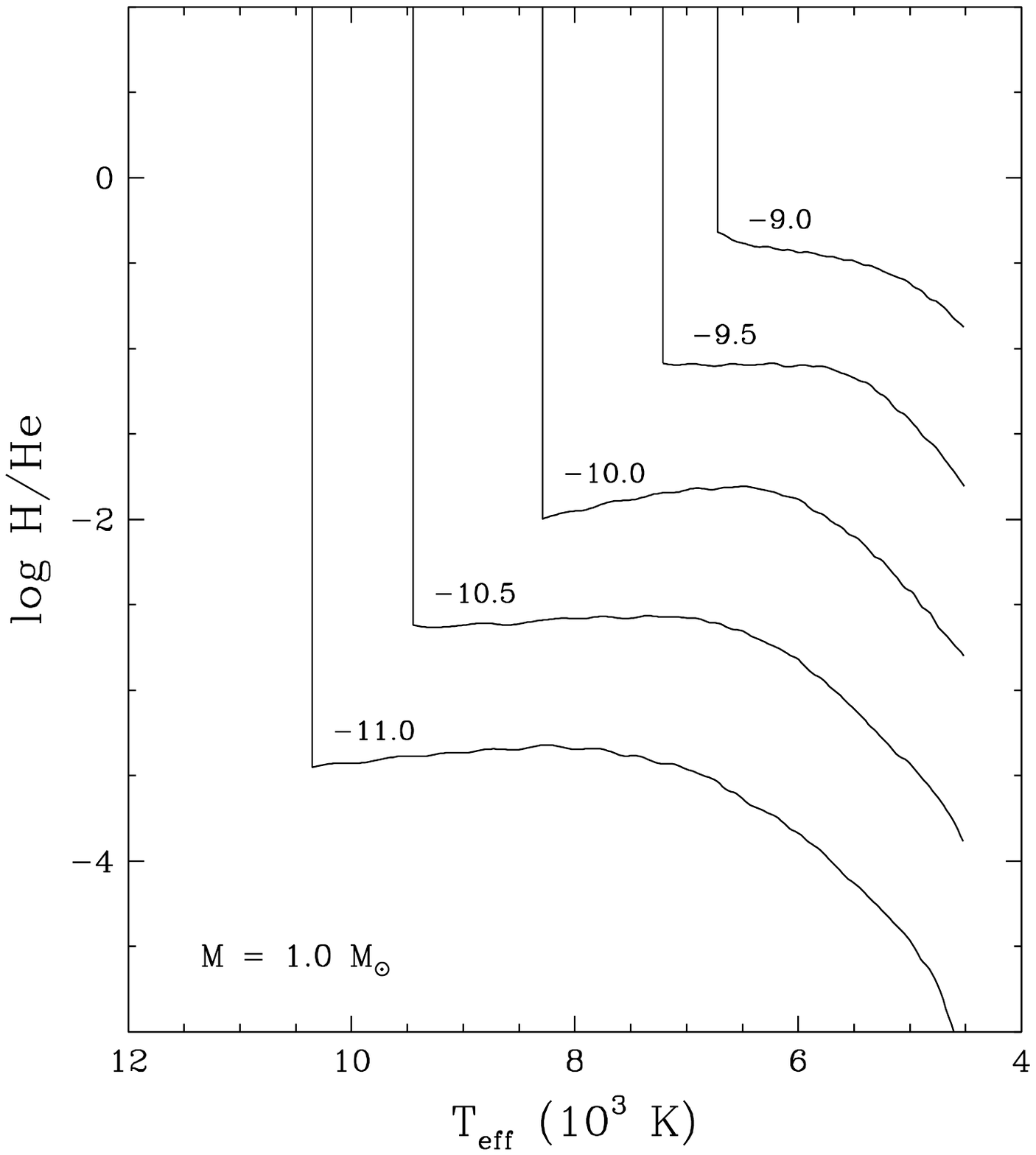}
\caption{Variation of the H/He ratio as a function of $\Te$ for DA
  models with 0.6 and 1.0 \msun\ and with various hydrogen layer masses
  given in each panel ($\log\qh=\log M_{\rm H}/M_\star$). The vertical
  lines indicate the mixing temperatures where the H/He ratio goes
  from infinity down to the value immediately after mixing. The
  subsequent evolution is governed by the change in the size of the
  mixed H/He convection zone.\label{convz}}
\end{figure*}

Nevertheless, the massive IR-faint white dwarfs in Figure
\ref{correltm_Kilic} occupy a region the $M$ - $\Te$ plane where few
or no objects have ever been reported\footnote{The few objects
  observed in Figure 11 of \citet{bergeron19} are actually IR-faint
  white dwarfs in our sample, but that have not been recognized and
  analyzed properly.}. Obviously there are none in the Kilic et
al.~sample. All cool, and massive white dwarfs in this particular
region of the $M$ vs $\Te$ diagram have mixed H/He atmospheres and
show strong IR-flux deficiencies. If we interpret these objects as the
result of convectively mixed DA stars, we encounter one obvious
problem related to the measured values of the hydrogen abundance in
these massive IR-faint white dwarfs. Figure \ref{convz} shows the
variation of the H/He ratio as a function of $\Te$ for DA models with
0.6 and 1.0 \msun\ and with various hydrogen layer masses
($\log\qh=\log M_{\rm H}/M_\star$). These envelope calculations are
similar to those of \citet{rolland18}. Upon mixing, the H/He ratio
goes from infinity down to a value set by the size of the mixed H/He
convection zone after mixing. The H/He ratio then continues to vary
with decreasing effective temperature as a result of changes in the
size of the mixed H/He convection zone. Our results indicate that in a
1.0 \msun\ white dwarf at $\Te=4800$~K with a photospheric value of
$\logh\sim-3.5$ --- typical of the massive IR-faint white dwarfs in
our sample ---, the total hydrogen mass is $\log\qh\sim-10.4$. The DA
progenitor with such a small hydrogen layer mass would mix at a
temperature of $\Te=9200$~K and cool off as a non-DA star. The problem
then of course is that most, if not all, massive white dwarfs in
Figure \ref{correltm_Kilic} are DA stars (see also Figure 11 of
\citealt{bergeron19}). Given that there are remaining uncertainties in
our model atmosphere calculations, in particular at the high densities
encountered in these massive white dwarfs, it is possible that the
hydrogen-to-helium abundance ratios are underestimated in our
analysis. For the same reasons, it would be precarious at this point
to estimate the total amount of hydrogen in individual objects until
better models become available.

In contrast, the IR-faint white dwarfs with mild CIA absorption and
normal masses (green symbols in Figure \ref{correl}), and thus lower
atmospheric pressure, are characterized by hydrogen abundances around
$\logh\sim-2.5$, indicating that their DA progenitors had thicker
hydrogen layers, and thus mixed at much lower effective temperatures,
around $\Te\sim7500$~K according to the results of Figure
\ref{convz}. And in this case, there are plenty of non-DA stars with
normal masses that could be the immediate progenitors of these
IR-faint white dwarfs.

Also shown in Figure \ref{correltm_Kilic} are the regions in the $M$ -
$\Te$ plane where crystallization occurs: the lower solid blue curve
marks the region where crystallization starts at the center of the
white dwarf, while the upper solid blue curve marks the limit of 80\%
crystallization of the core resulting from the solidification front
moving upward in the star with further cooling. The dashed curve marks
the onset of convective coupling, when the convection zone first
reaches into the degenerate interior, where all of the thermal energy
of the star resides. As discussed in more detail in
\citet{bergeron19}, the onset of convective coupling is weakly
mass-dependent, and clearly interacts with the manifestations of
crystallization, as illustrated by the behavior of the curved
isochrones at high mass and low temperatures in Figure
\ref{correltm_Kilic}. The massive IR-faint white dwarfs in our sample
located in this specific range of parameters have undoubtedly entered
the so-called Debye cooling phase, i.e., the rapidly cooling phase
resulting from the transition, in the solid phase, from the classical
regime where the specific heat of a solid is independent of
temperature, to the quantum regime where it goes down from that
constant value with decreasing temperature (as indicated by the blue
dotted curve in Figure \ref{correltm_Kilic}). Here, we defined this
transition from the classical to the quantum regime by isolating the
evolving model where the central temperature becomes equal to 1/20 the
central Debye temperature ($\theta_D/T=20$), as defined in
\citet{althaus07}. In the Debye cooling phase, the specific heat
decreases quickly with cooling, which depletes rapidly the reservoir
of thermal energy and produces the extreme increase of the cooling
rate, as observed in Figure \ref{correltm_Kilic}. It is thus perhaps
surprising to have uncovered such a large number of massive IR-faint
white dwarfs caught in this rapidly cooling phase.

\section{CONCLUSION}\label{sec:conclusion}

We presented a detailed model atmosphere analysis of 105 IR-faint
white dwarfs using the so-called photometric technique, based on Gaia
astrometry combined with Pan-STARRS and near-infrared photometry. In particular,
we identified 37 new IR-faint white dwarfs with strong flux deficits
in the optical, 30 of which have follow-up Gemini optical
spectroscopy, and 33 additional white dwarfs with milder deficits.  We
showed that the extremely low temperatures and stellar masses
previously reported in the literature (see, e.g., the case of LHS 3250
analyzed by \citealt{bergeron02}) were related to inaccuracies in the
model atmospheres. We convincingly demonstrated that the key physical
ingredient missing in the previous generations of model atmospheres
was the high-density correction to the He$^-$ free-free absorption coefficient
described in \citet{iglesias02}. Despite the significant improvements
of our new models over our previous model grid, both qualitatively and
quantitatively, we also stressed the importance that much work remains
to be done on the theoretical front in terms of the equation of state
and the H$_2$-He CIA opacity calculations.

The two major uncertainties in IR-faint white dwarf models are (1) the
atmospheric density, which depends on factors like continuum opacities
and the pressure ionization of helium, and (2) the shape of the CIA
opacities at a given density \citep{abel12,blouin17}. The only way to
untangle these uncertainties and improve the physics in our models is
to obtain spectra of IR-faint white dwarfs in the infrared. The
molecular absorption features that dominate the infrared spectra of
these objects have never been resolved. Yet they are predicted to
exhibit structures that contain critical information on the physical
conditions that characterize the atmospheres of IR-faint white dwarfs.

Recent first-principles calculations predict that the shape of the CIA
features and the overall infrared spectral energy distribution are
highly sensitive to the atmospheric density \citep[see Figure 10
  in][]{blouin17}. The James Webb Space Telescope provides an
excellent opportunity to obtain such spectra, and for the first time
constrain the shape of the CIA opacities in the high density
environments of helium-dominated atmosphere white dwarfs. Laboratory
experiments may also help constrain the opacity of helium under the
conditions prevalent at the photospheres of IR-faint white dwarfs
\citep{mcwilliams15}. In this work, we have shown how sensitive the
inferred properties of IR-faint white dwarfs are to helium continuum
opacities; experimental measurements of dense helium opacities would
help confirm the nature of those stars.

Regardless of the problems with the model atmospheres, there are a few
things that we can safely conclude:

1- The majority of the IR-faint white dwarfs form a tight sequence
around $M_g = 16$ in the color-magnitude diagram. Pure hydrogen
atmosphere composition can safely be ruled out based on the location
of these objects in the color-magnitude diagrams and the strength of
the CIA observed in the optical bands. The observed sequence can be
explained as the cooling sequence of mixed H/He atmospheres. The
majority of the warm DC white dwarfs with $\Te> 6000$~K require trace
amounts of hydrogen to explain their location in the Gaia
color-magnitude diagrams \citep{bergeron19}, hence an IR-faint white
dwarf sequence due to H$_2$-He CIA seems unavoidable.

2- Not all white dwarfs become IR-faint. In fact, IR-faint white
dwarfs represent only a small fraction of the overall cool white dwarf
population in the solar neighborhood. This indicates that most DC
white dwarfs below 5000 K have hydrogen-dominated atmospheres,
otherwise there would be significantly more objects that display
H$_2$-He CIA.

3- There is a lack of continuous IR-faint sequences in the
color-magnitude diagram. As discussed above, the majority of the
IR-faint white dwarfs are clustered on a sequence at $M_g \sim 16$,
and there are several IR-faint white dwarfs with $M_g \sim 17$ that
may or may not form another sequence. Given Gaia's limiting magnitude
of $G\sim21$, our survey volume is limited for these objects (see Figure
  \ref{colorgyz}).

Now if we take the results we obtain with our revised model
atmospheres at face value, we are forced to conclude that the vast
majority of IR-faint white dwarfs are not as cool as previously
estimated. Most of the IR-faint white dwarfs with extremely strong
flux deficits in the optical also have large masses well above 0.8
\msun, and they occupy a region in the $M$ vs $\Te$ plane where very
few objects --- if none --- have ever been identified.  These must
necessarily be in the Debye cooling phase.

Some of the progenitors of the IR-faint white dwarfs are likely DA
white dwarfs that got convectively mixed.  \citet{bedard22} computed
for the first time the spectral evolution of a helium-rich DO white
dwarf turning into a pure hydrogen DA star through the float-up
process, and then into a helium-dominated DC white dwarf with trace
amounts of hydrogen through convective mixing. This DO-DA-DC
transition model predicts a final surface hydrogen abundance of $\logh
= -4.0$ around 8000 K, and naturally explains the increase in the
number of helium atmosphere white dwarfs at low effective temperatures
\citep{blouin19b,cunningham20} and the observed bifurcation in the
Gaia color-magnitude diagrams \citep{gaia18}. \citet{bedard22} note
that the most important parameter for the spectral evolution is the
mass fraction of hydrogen in the initial models, with a larger amount
of hydrogen resulting in a larger final hydrogen abundance in the
atmosphere.  Their evolutionary calculations do not include the
IR-faint temperature regime, but strongly suggest that convectively
mixed DA white dwarfs turn into IR-faint white dwarfs when they are
sufficiently cool to show H$_2$-He CIA. A more thorough exploration of
the parameter space should shed some light on the progenitors of the
IR-faint white dwarfs analyzed in this paper. Until such calculations
become available, we can nevertheless conclude that the IR-faint white
dwarfs with the largest hydrogen abundances in our sample cannot be
explained as the result of convectively mixed DA stars alone. External
sources of hydrogen must also be invoked, most likely from tidally
disrupted asteroids, small planets, and comets.

We would like to thank the anonymous referee for a careful reading of
our manuscript and several constructive comments that helped improve
this paper. This work was supported in part by the NSF under grant
AST-1906379, the NSERC Canada, by the Fund FRQ-NT (Qu\'ebec), and by
NASA under grant 80NSSC22K0479. SB is a Banting Postdoctoral Fellow
and a CITA National Fellow, supported by NSERC.

Based on observations obtained at the international Gemini
Observatory, a program of NSF’s NOIRLab, which is managed by the
Association of Universities for Research in Astronomy (AURA) under a
cooperative agreement with the National Science Foundation. on behalf
of the Gemini Observatory partnership: the National Science Foundation
(United States), National Research Council (Canada), Agencia Nacional
de Investigaci\'{o}n y Desarrollo (Chile), Ministerio de Ciencia,
Tecnolog\'{i}a e Innovaci\'{o}n (Argentina), Minist\'{e}rio da
Ci\^{e}ncia, Tecnologia, Inova\c{c}\~{o}es e Comunica\c{c}\~{o}es
(Brazil), and Korea Astronomy and Space Science Institute (Republic of
Korea).

\facilities{Gemini North (GMOS), Gemini South (GMOS)}

\bibliography{ms}{}
\bibliographystyle{apj}

\appendix

\begin{figure*}
\includegraphics[width=2.4in]{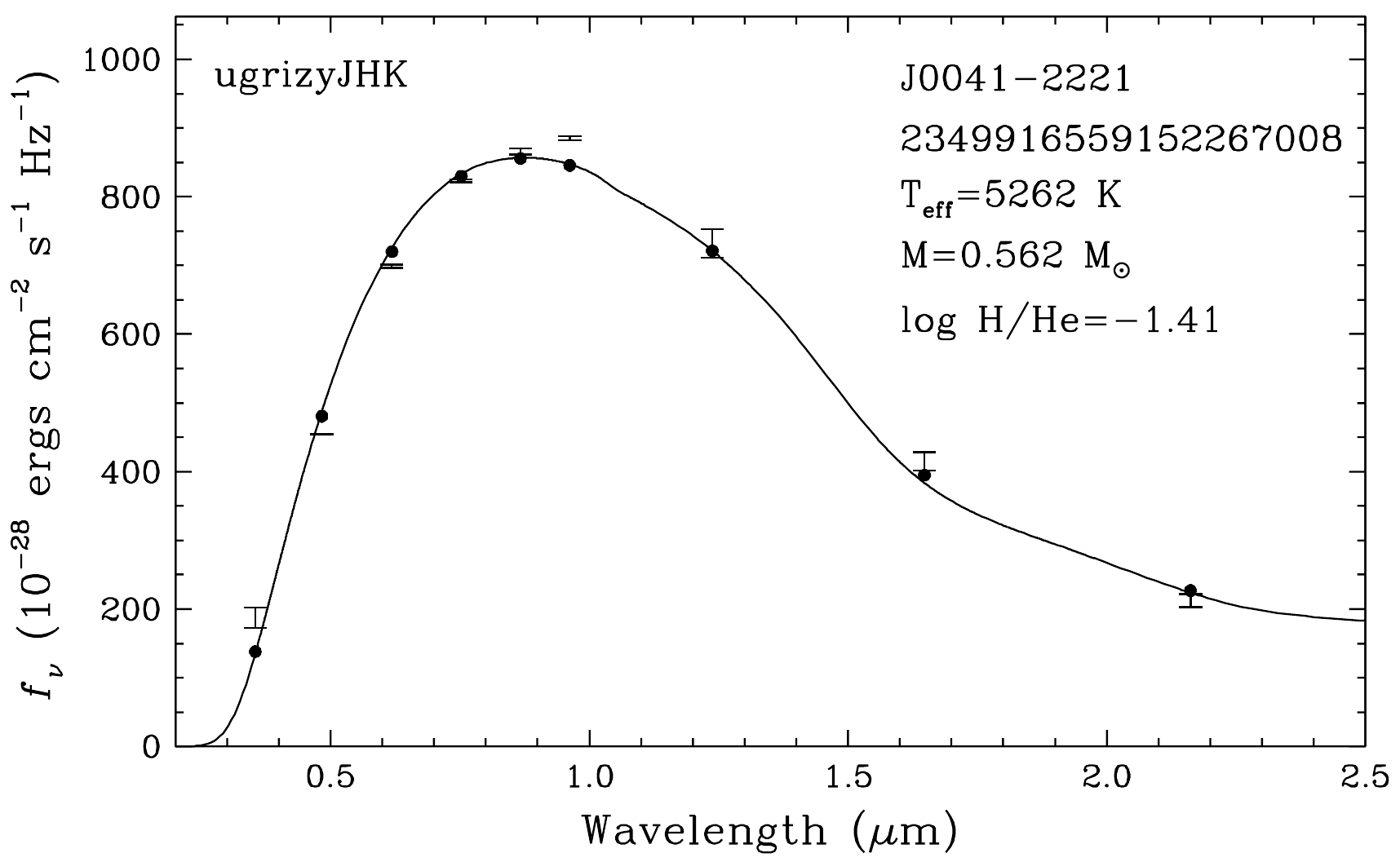}
\includegraphics[width=2.4in]{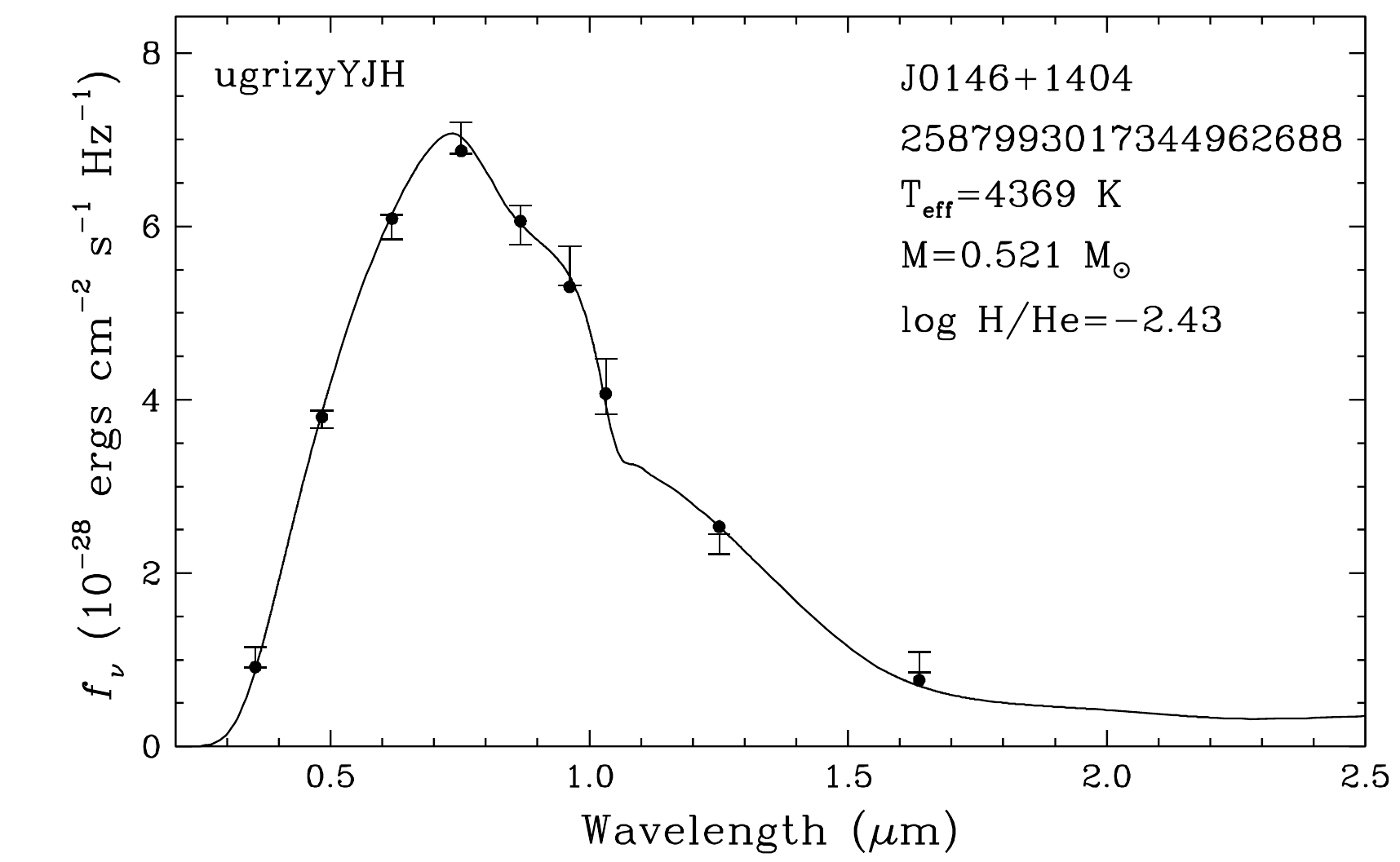}
\includegraphics[width=2.4in]{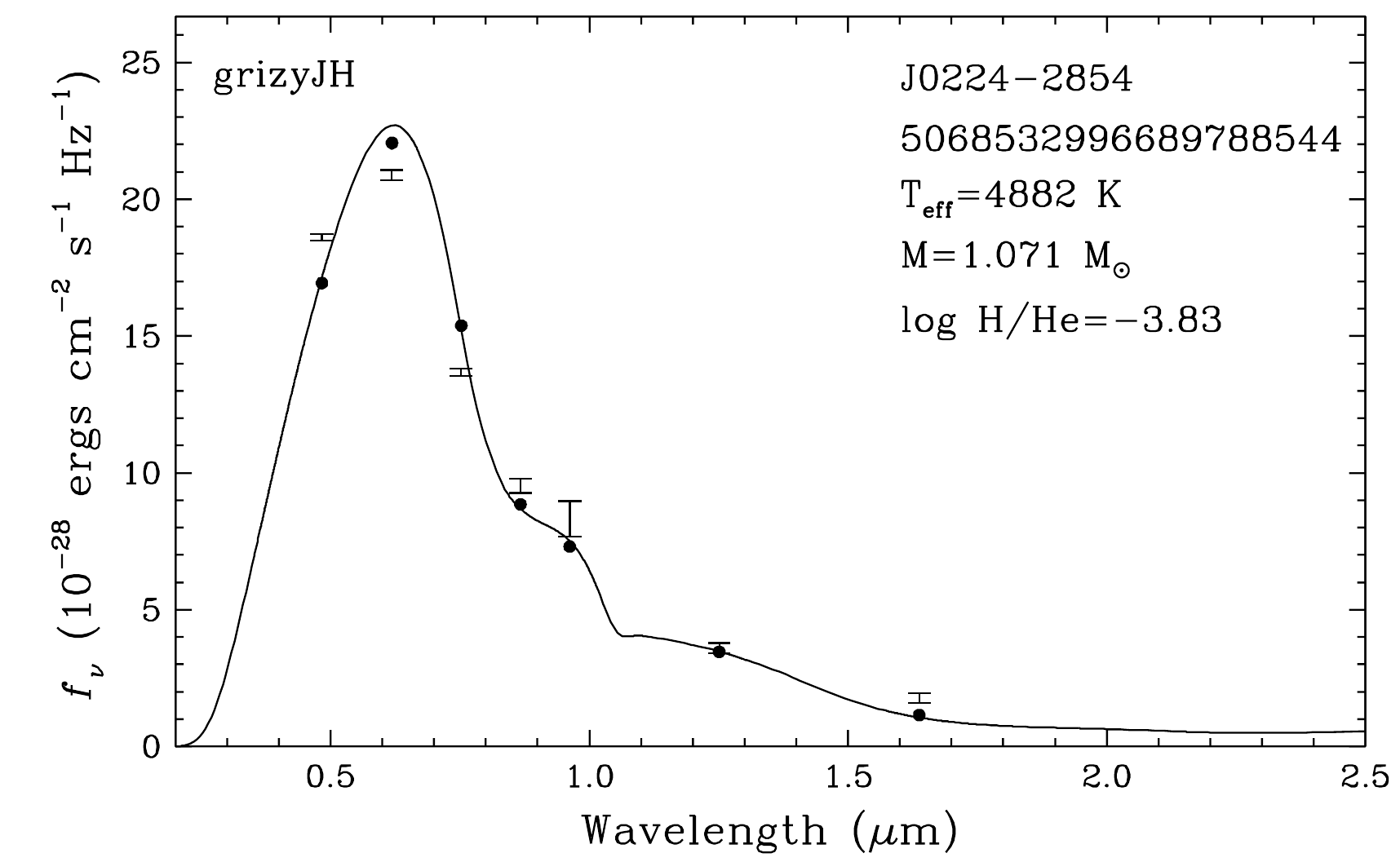}
\includegraphics[width=2.4in]{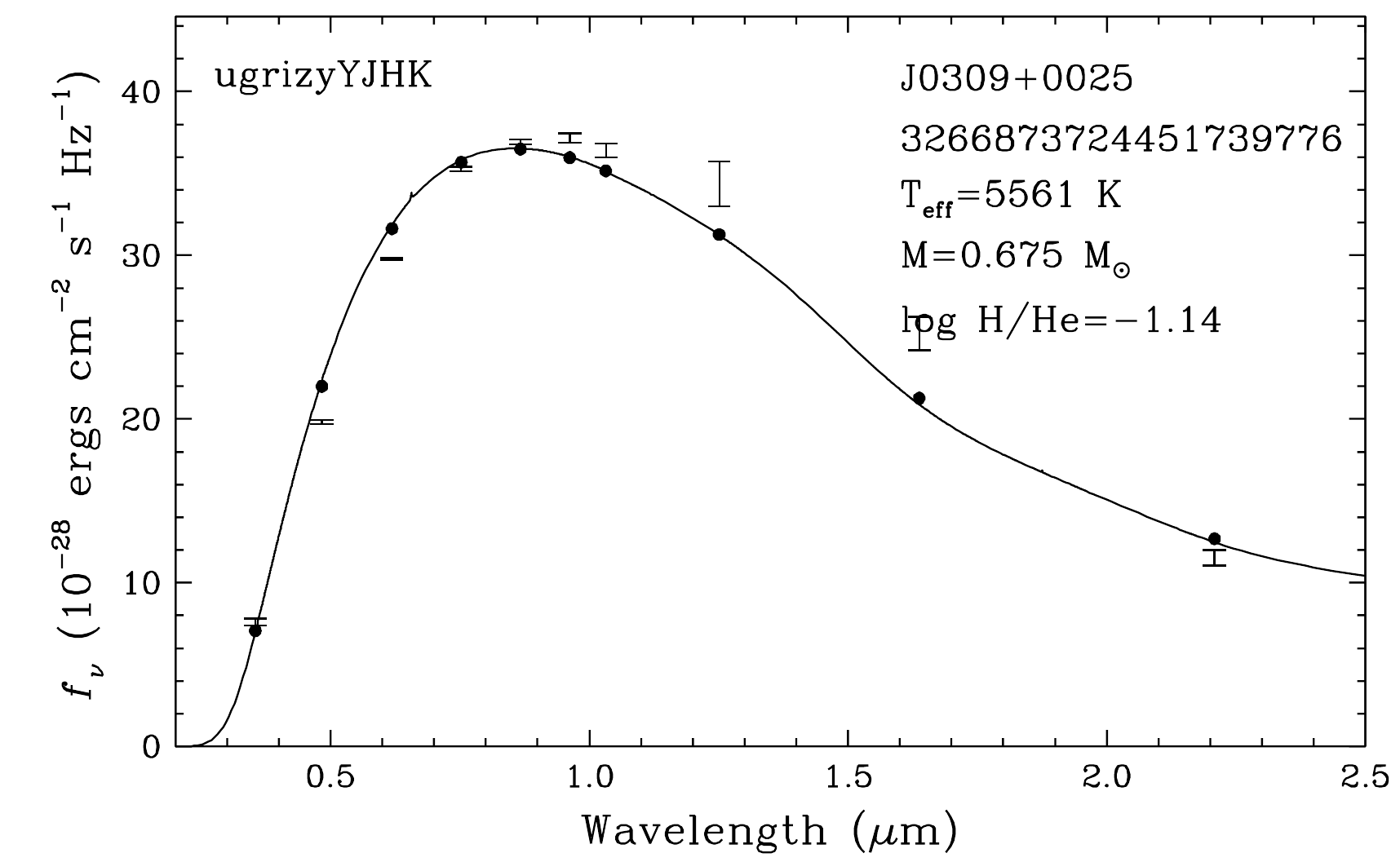}
\includegraphics[width=2.4in]{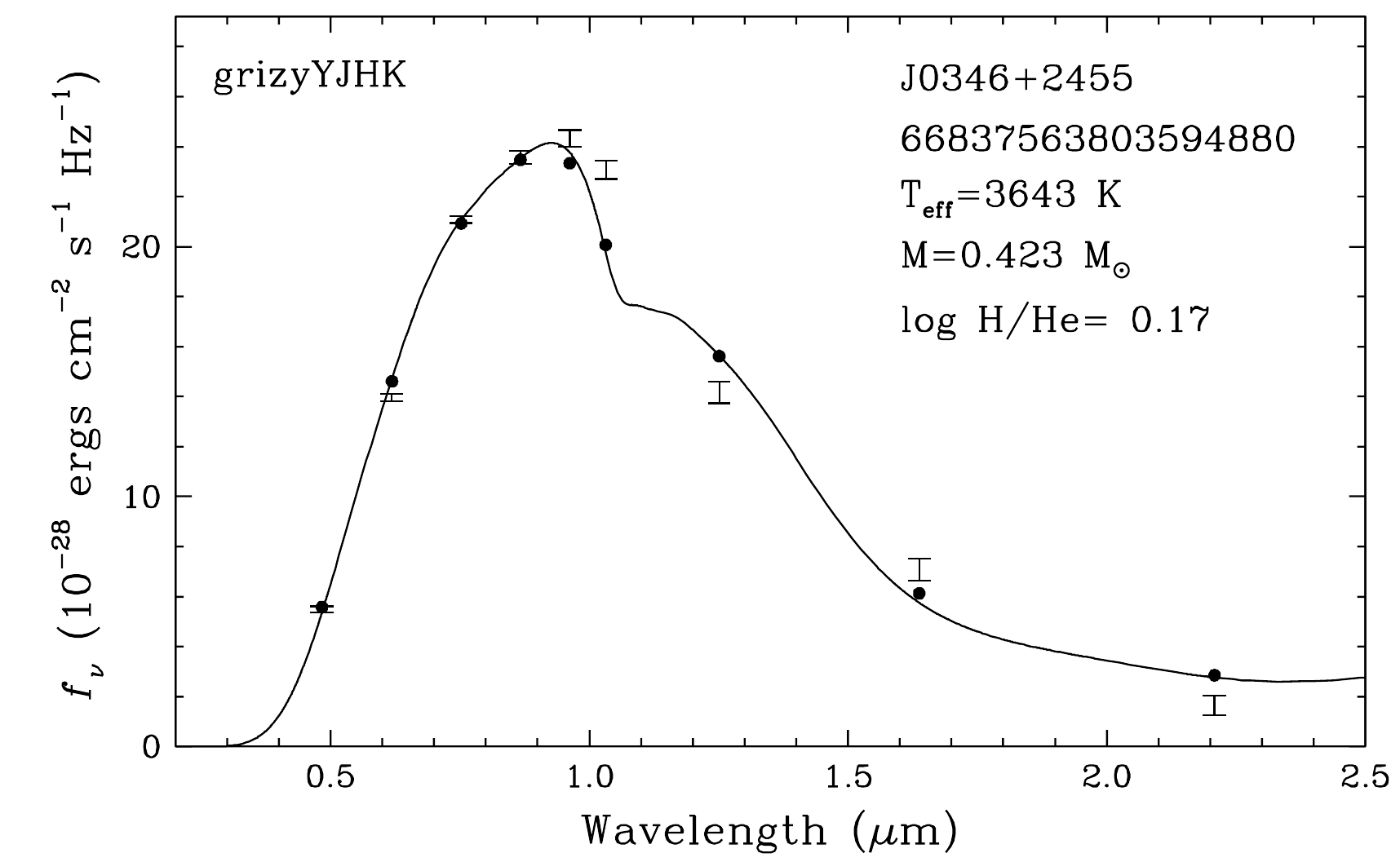}
\includegraphics[width=2.4in]{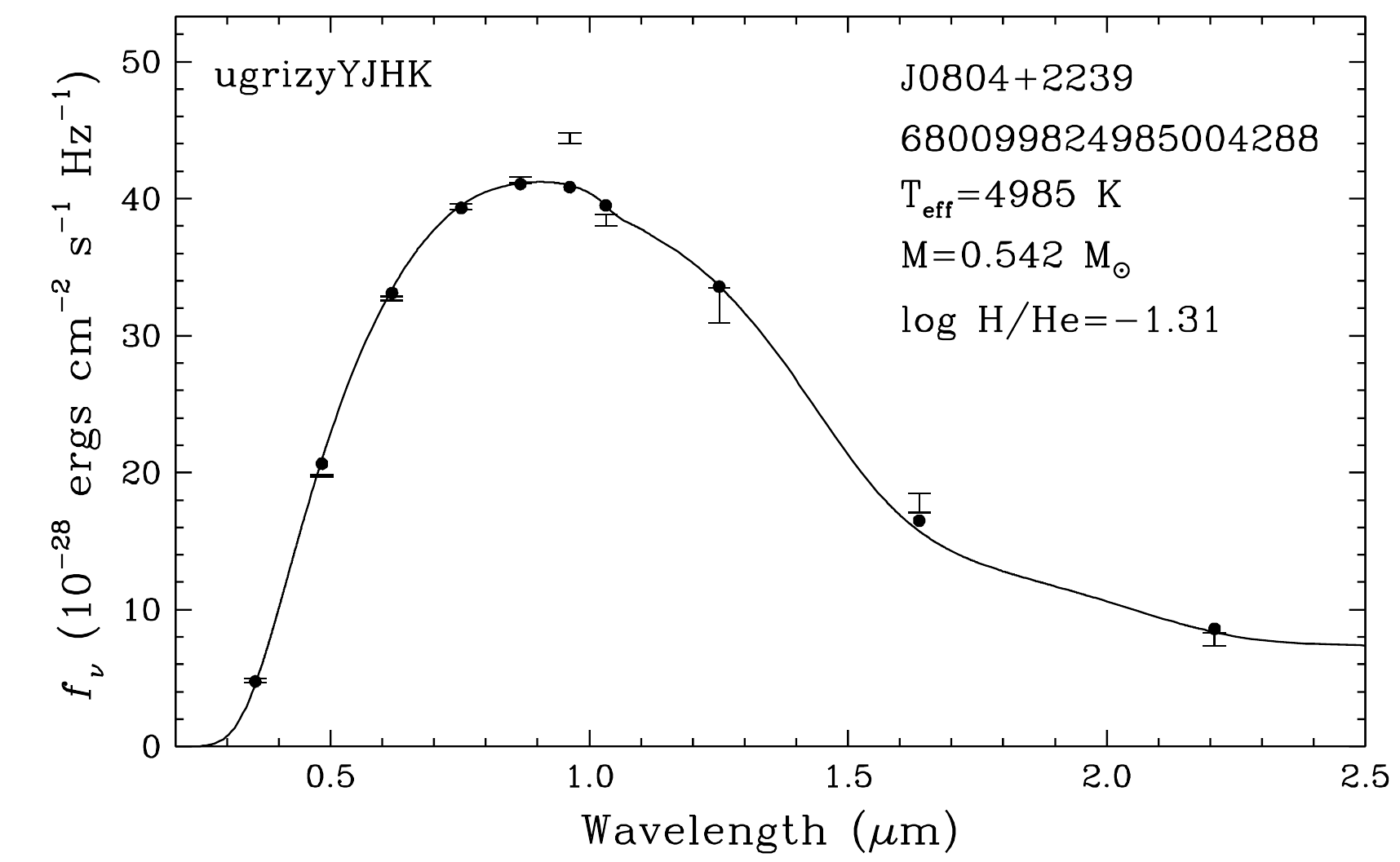}
\includegraphics[width=2.4in]{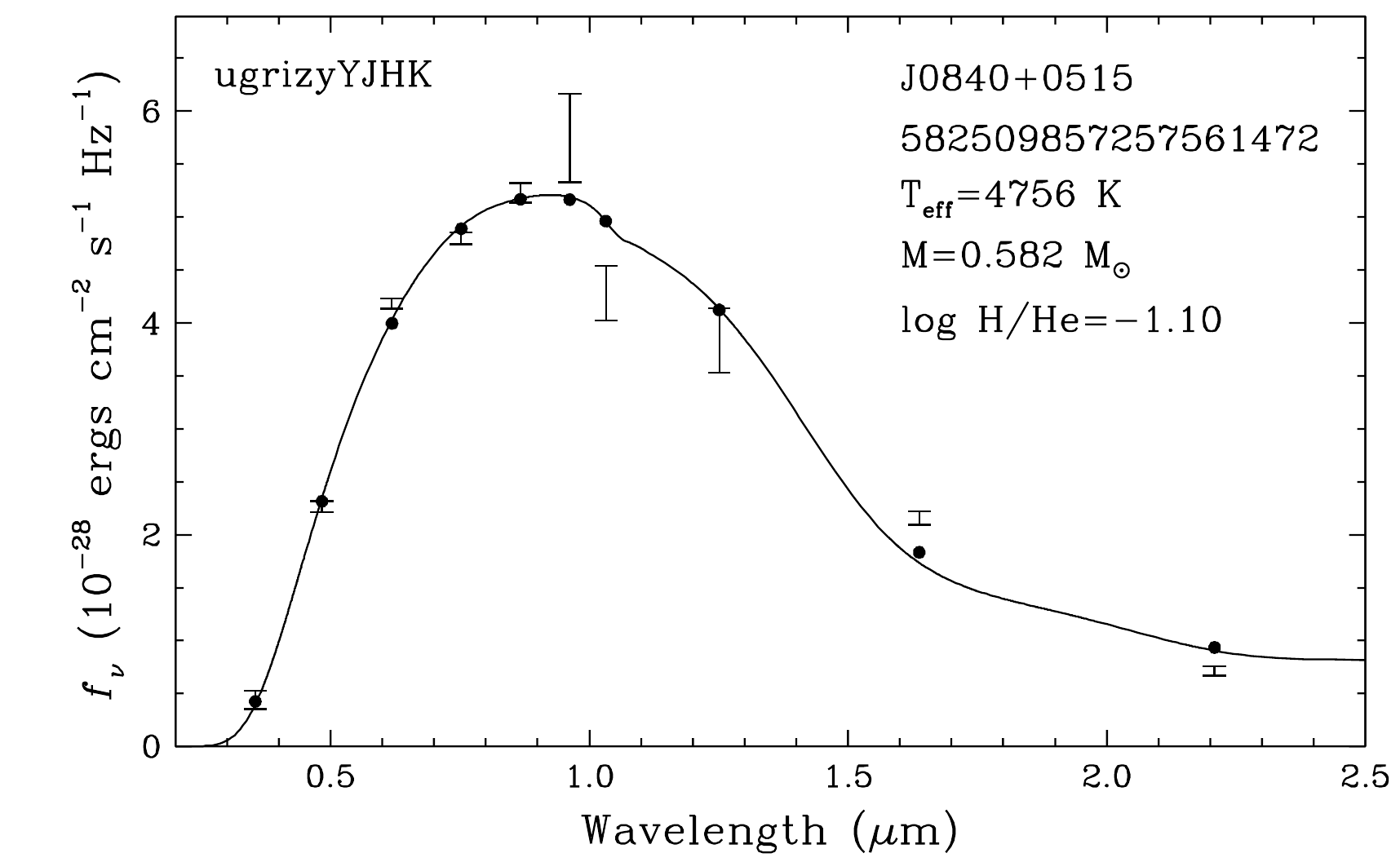}
\includegraphics[width=2.4in]{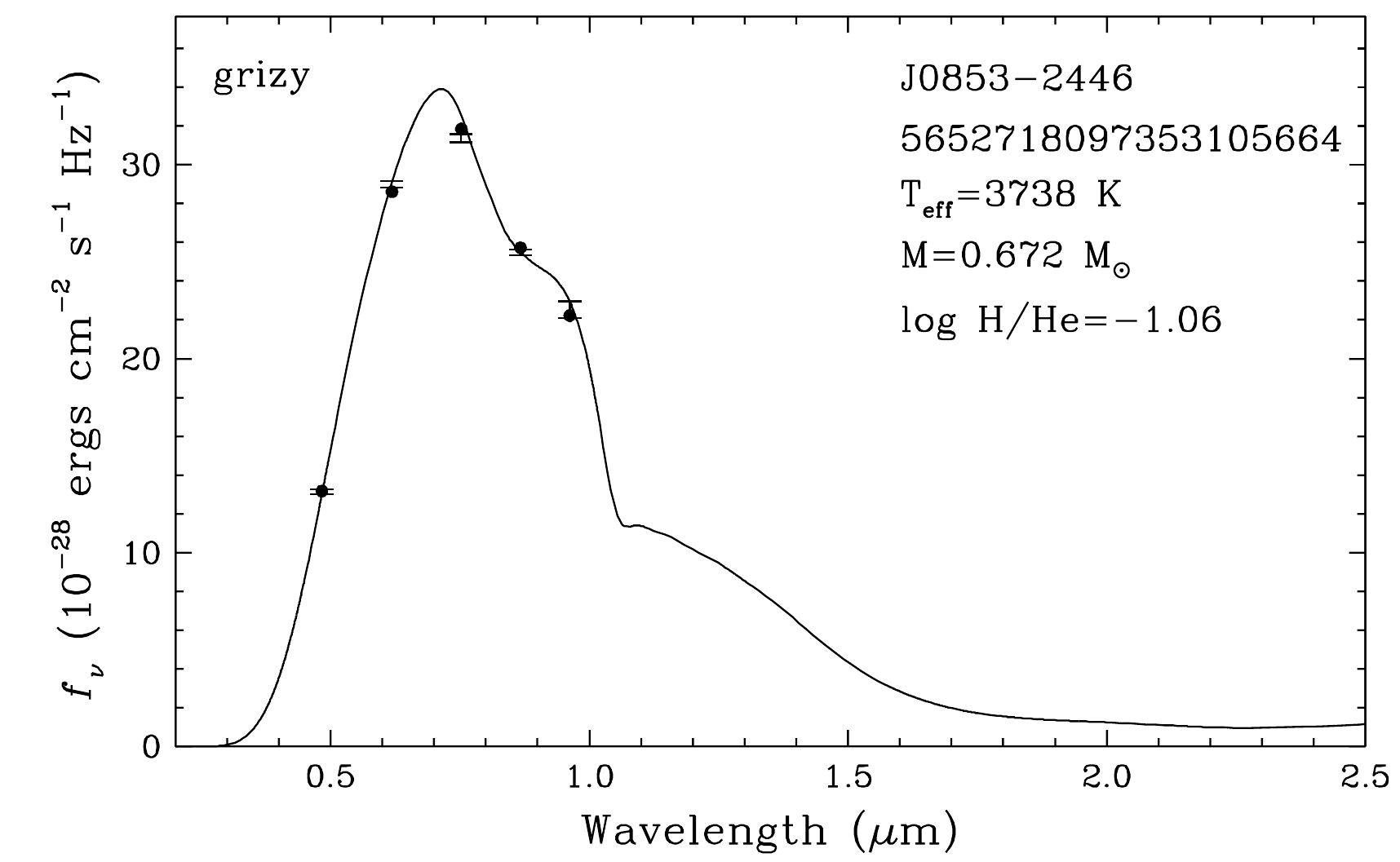}
\includegraphics[width=2.4in]{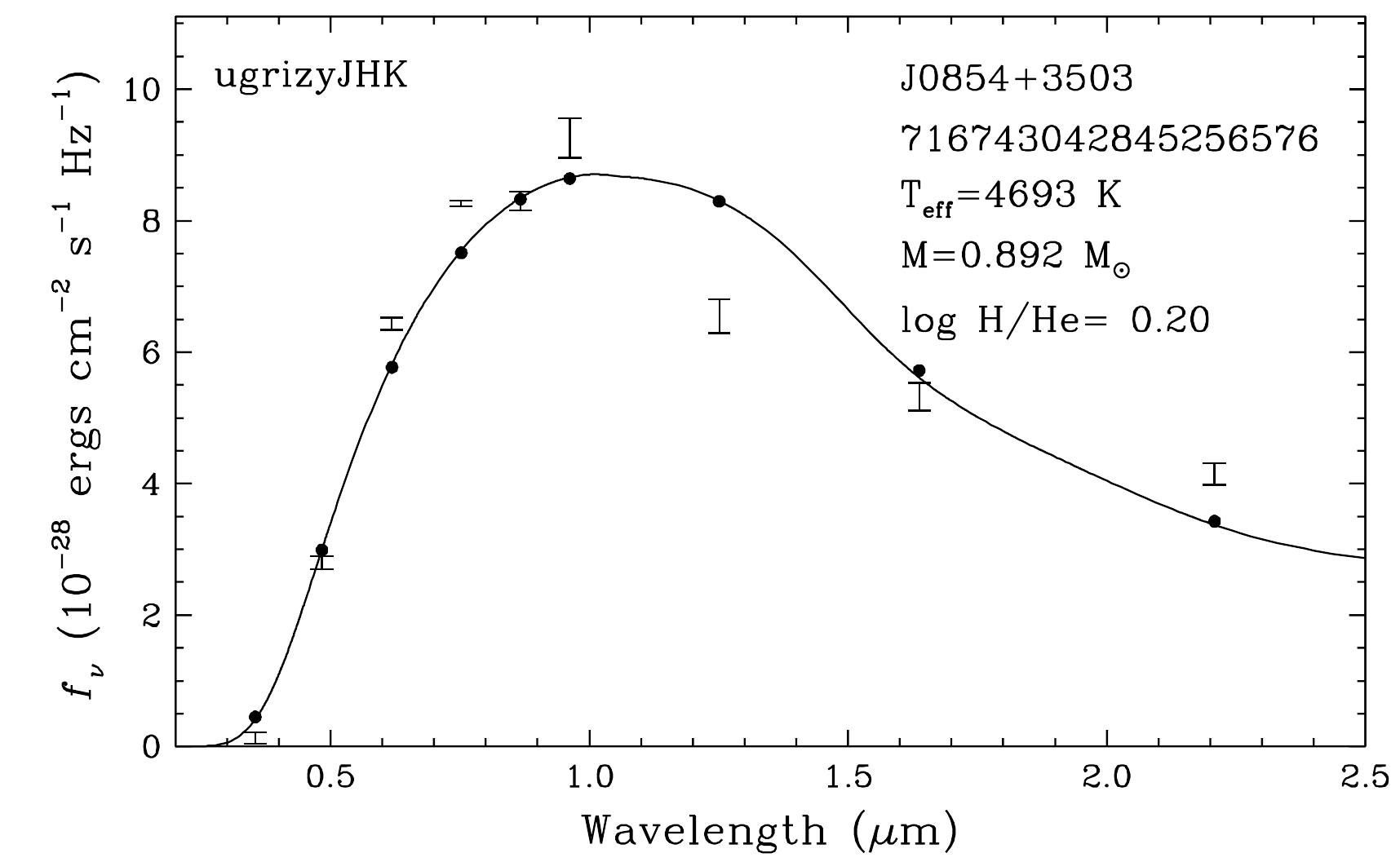}
\includegraphics[width=2.4in]{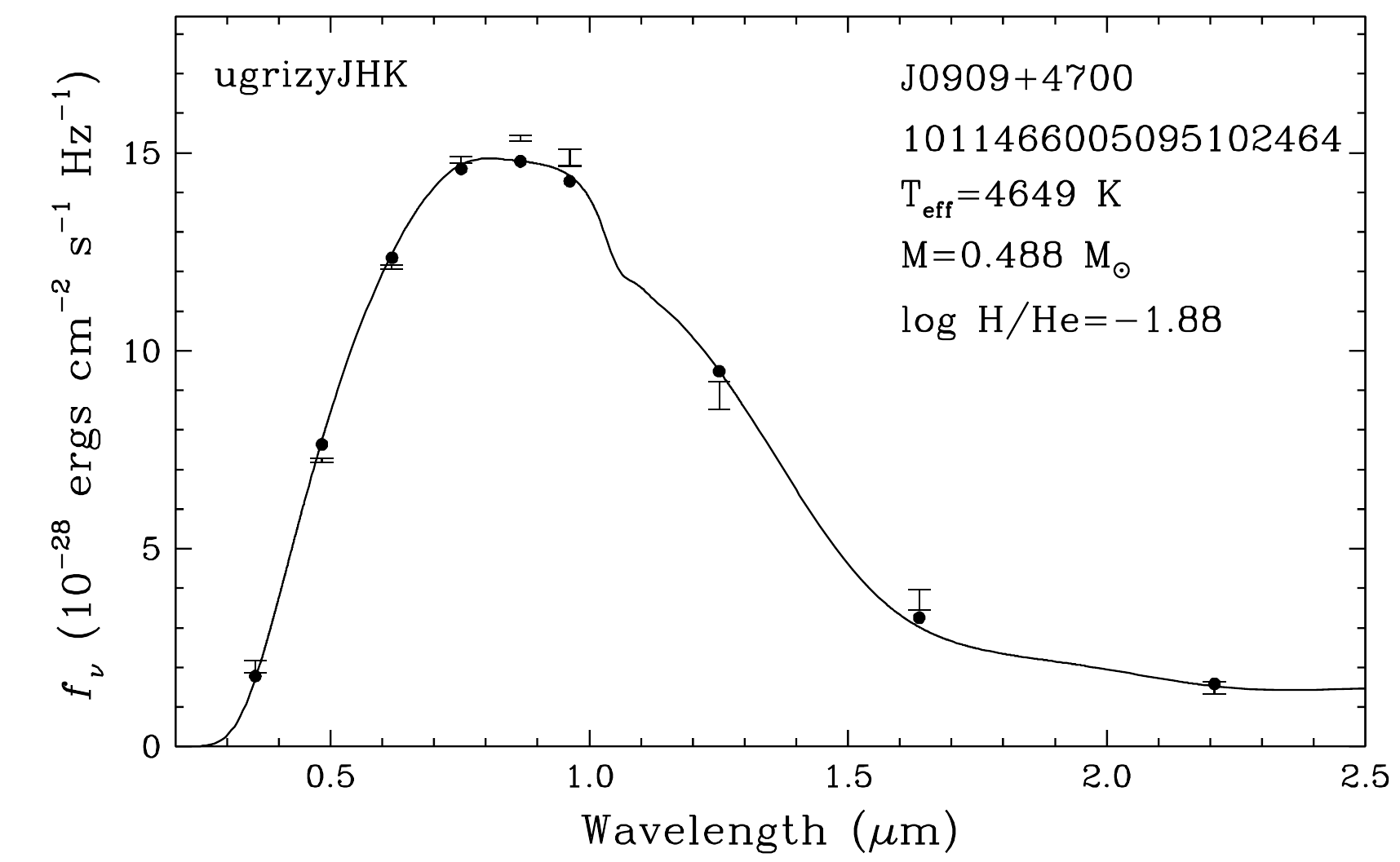}
\includegraphics[width=2.4in]{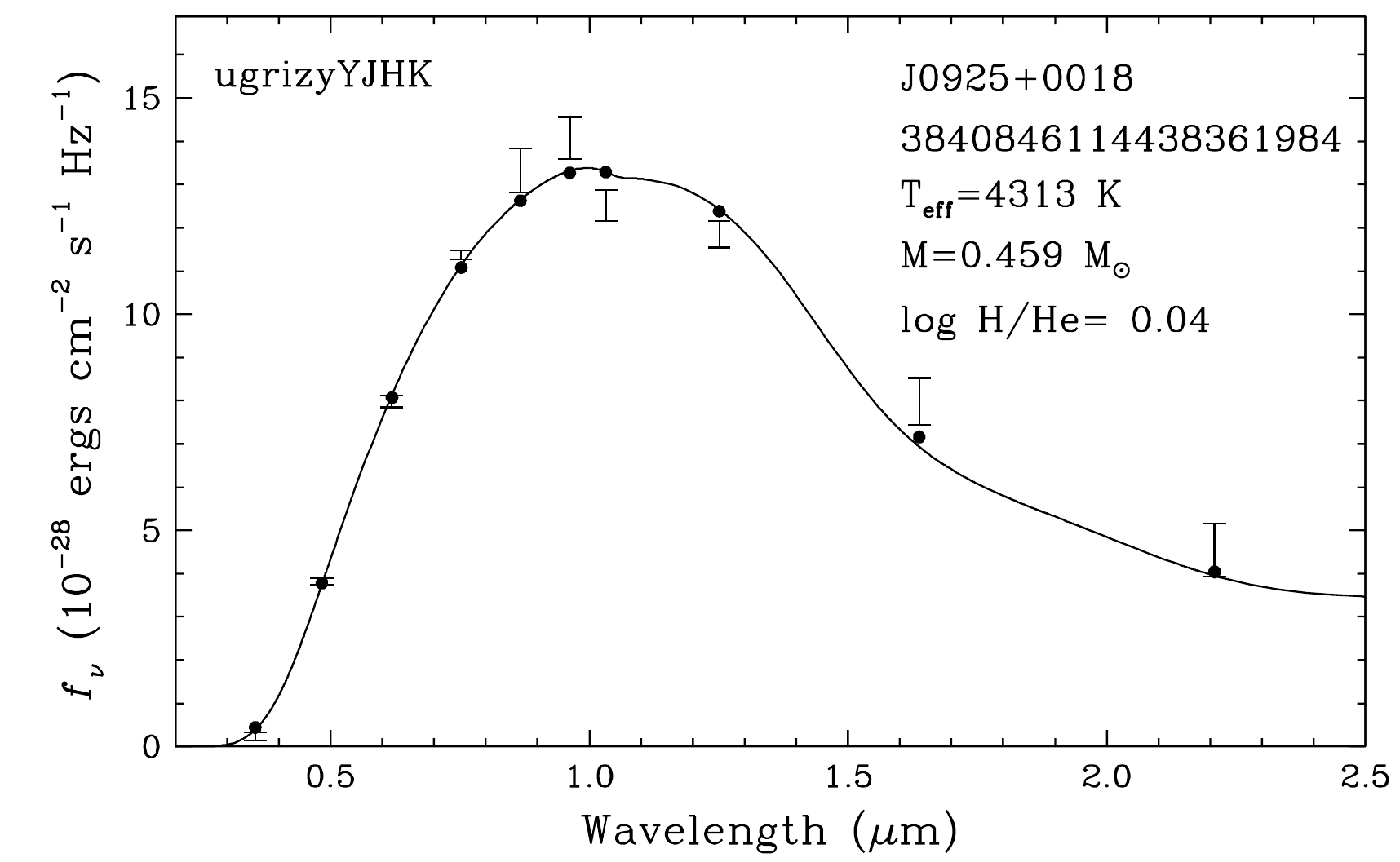}
\includegraphics[width=2.4in]{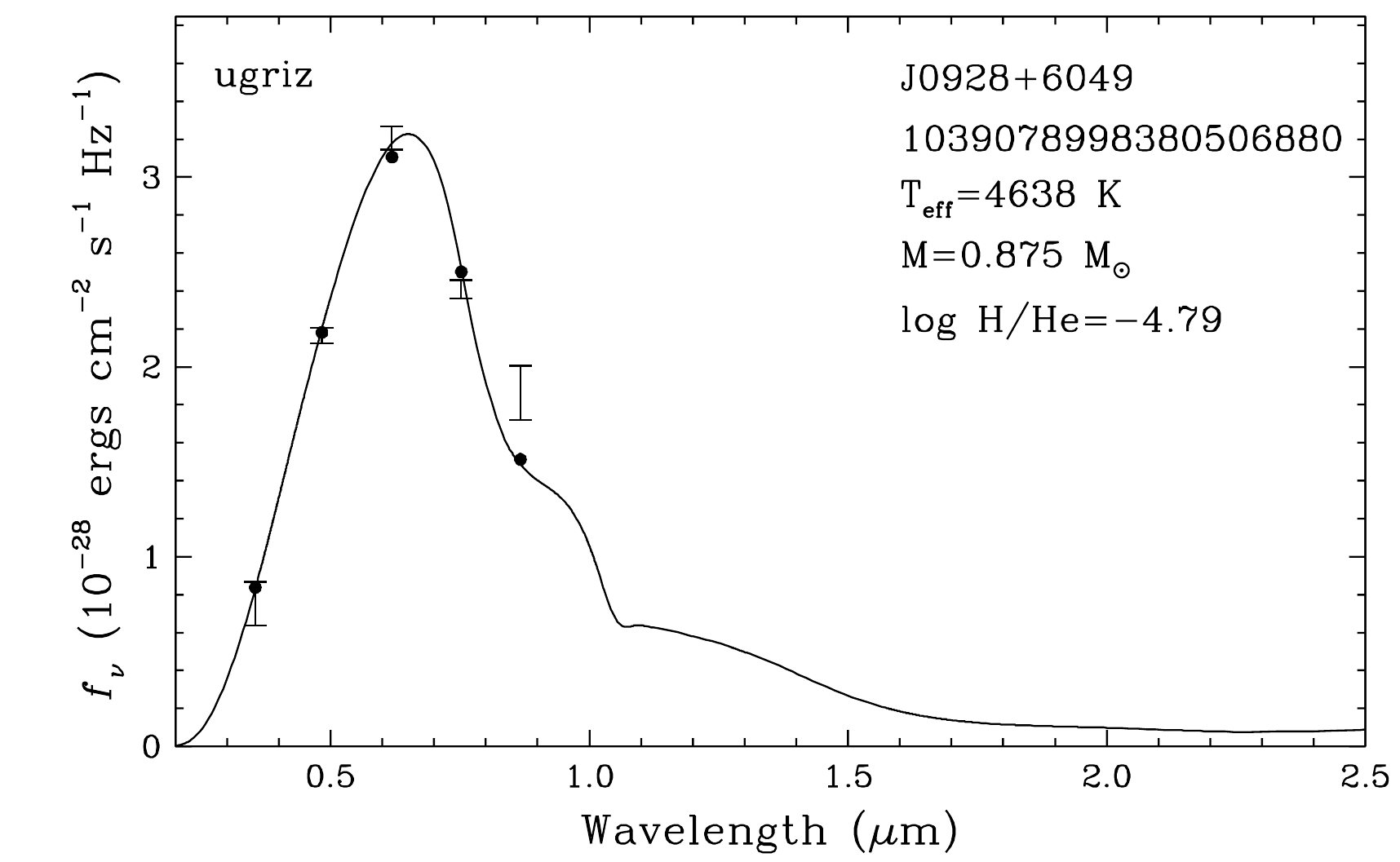}
\includegraphics[width=2.4in]{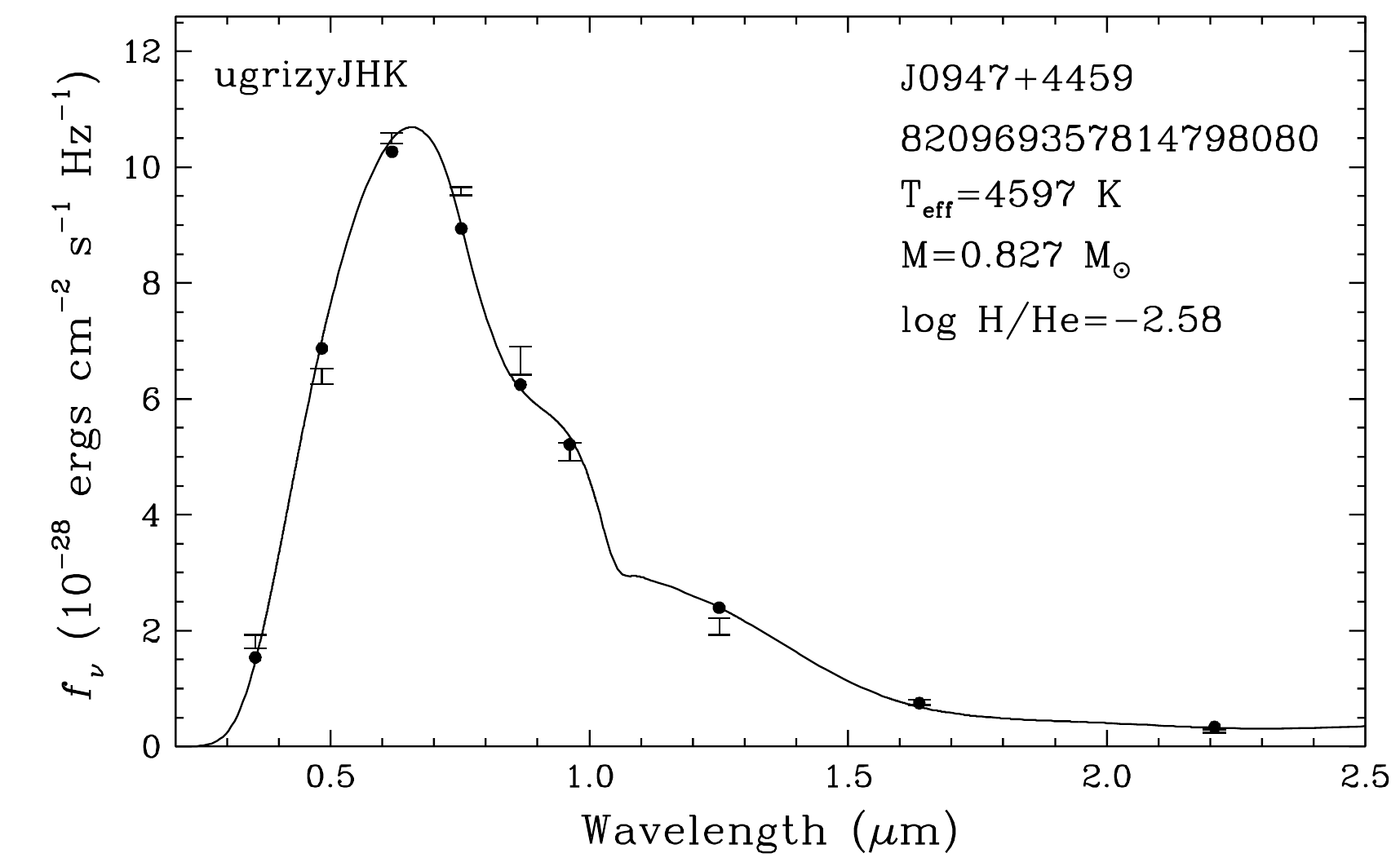}
\includegraphics[width=2.4in]{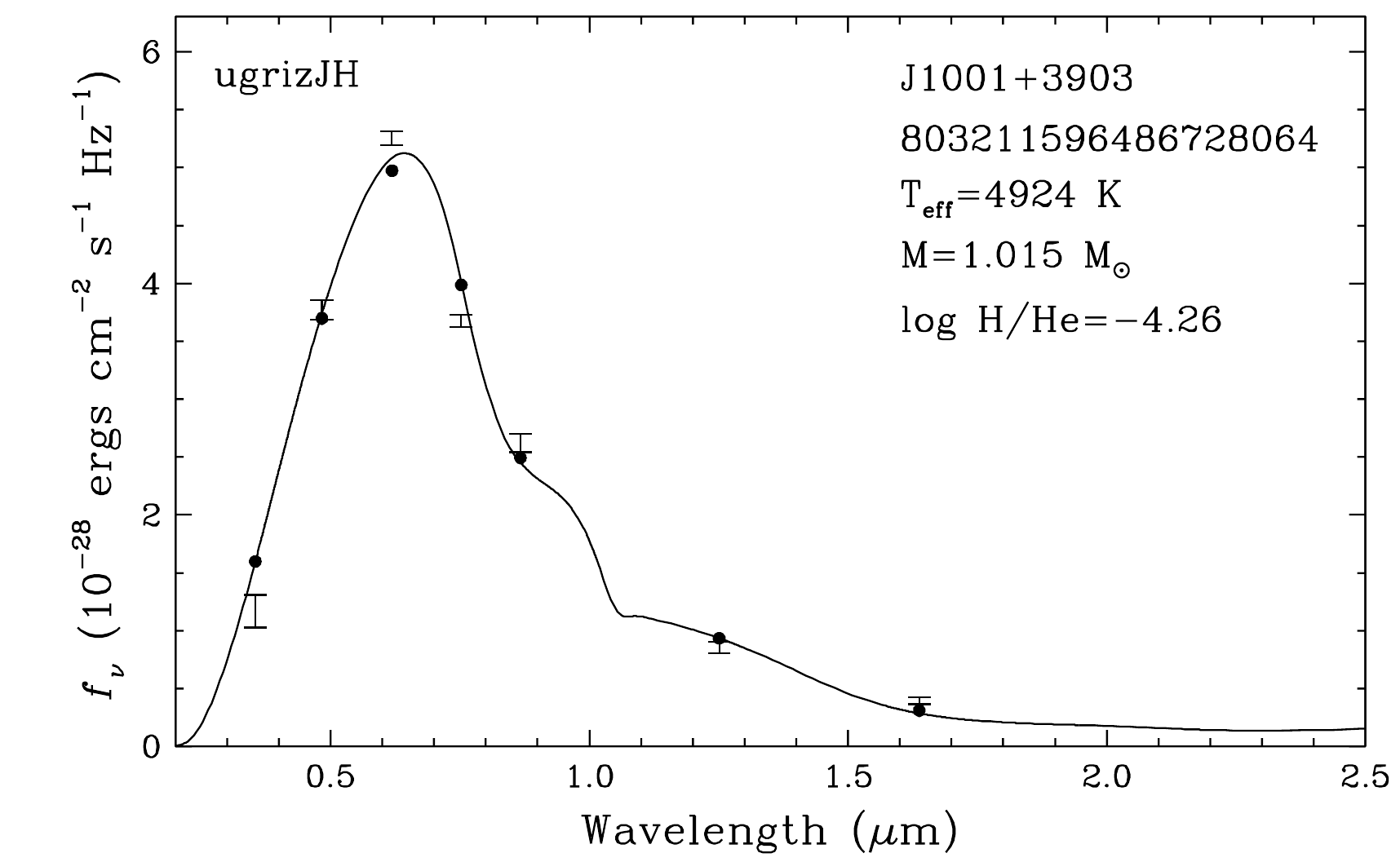}
\includegraphics[width=2.4in]{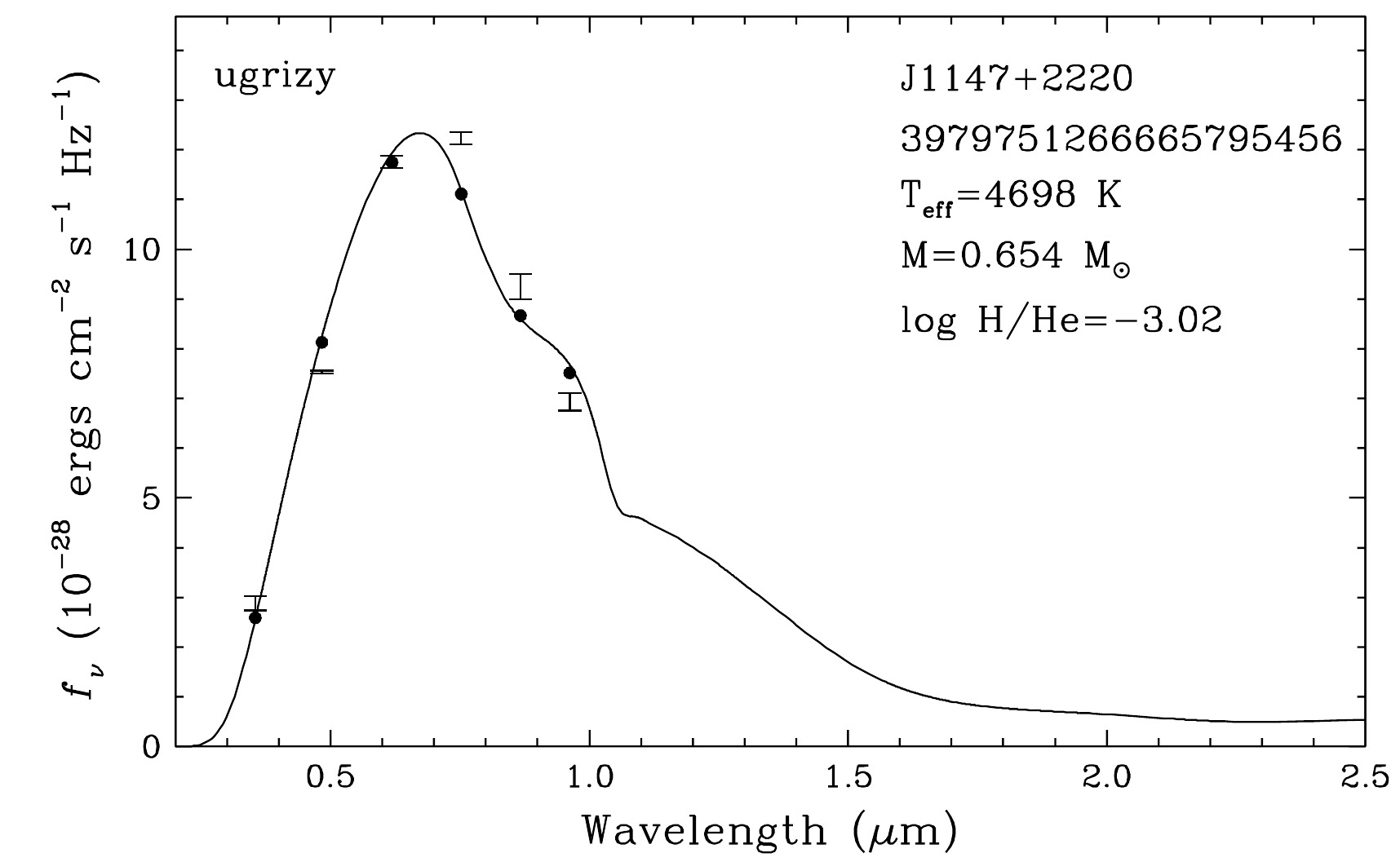}
\includegraphics[width=2.4in]{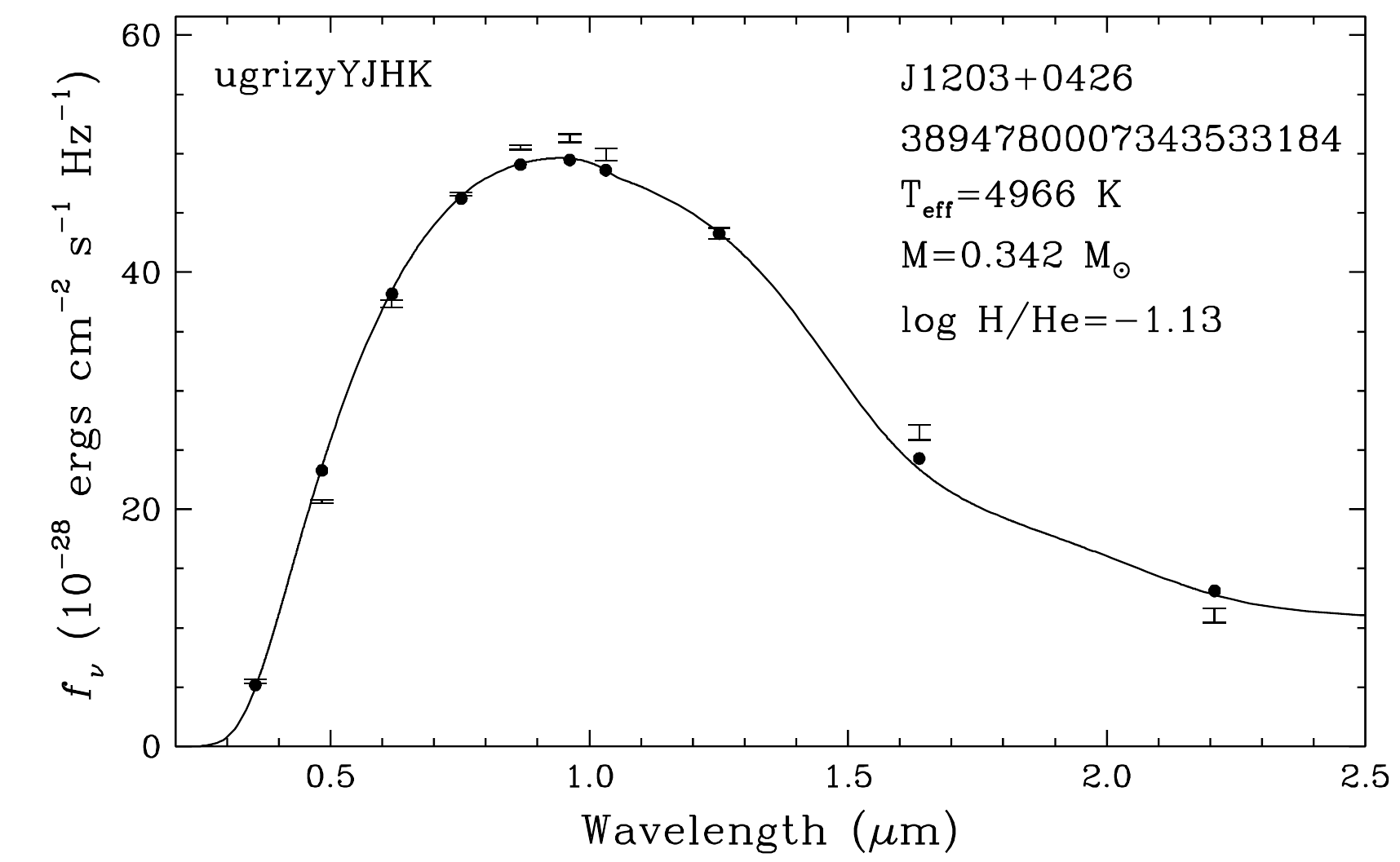}
\includegraphics[width=2.4in]{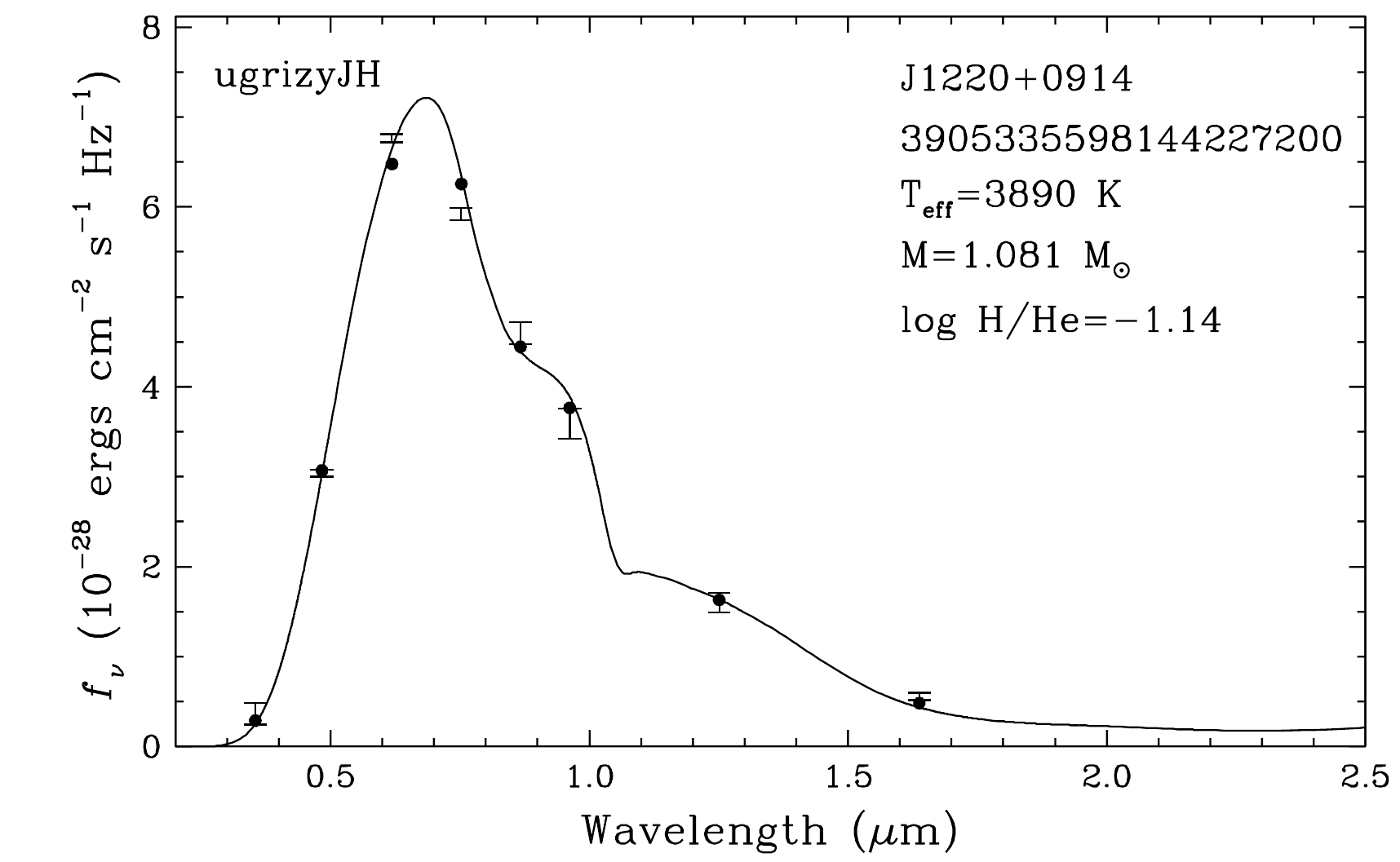}
\includegraphics[width=2.4in]{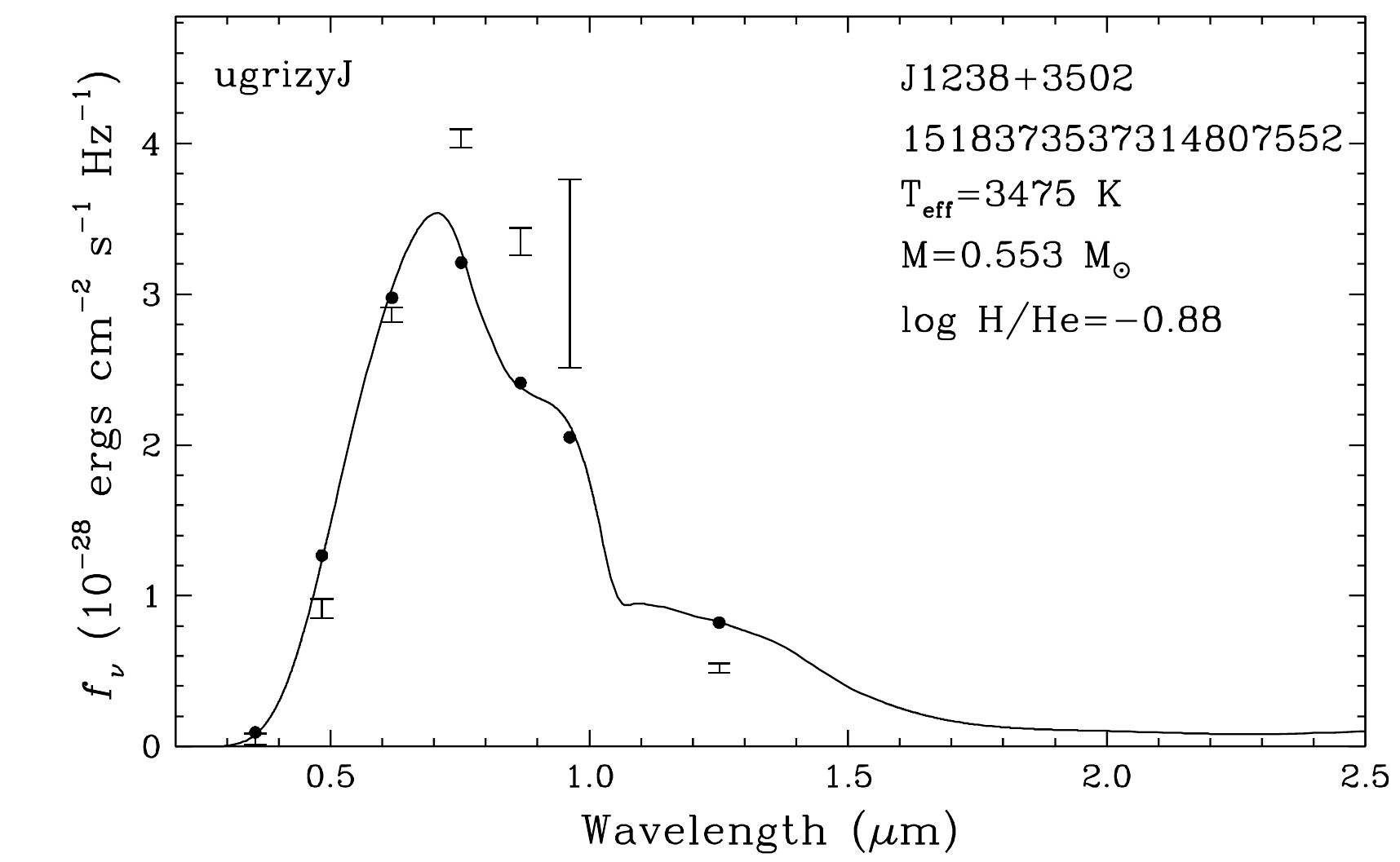}
\caption{Fits to the spectral energy distributions of previously known
  IR-faint white dwarfs listed in Table \ref{tabold}. Error bars show
  the observed photometry. Black lines show the monochromatic fluxes
  for the best-fit model for each star, and dots show the synthetic
  photometry of those models in each filter.\label{pre}}
\end{figure*}

\begin{figure*}
\addtocounter{figure}{-1}
\includegraphics[width=2.4in]{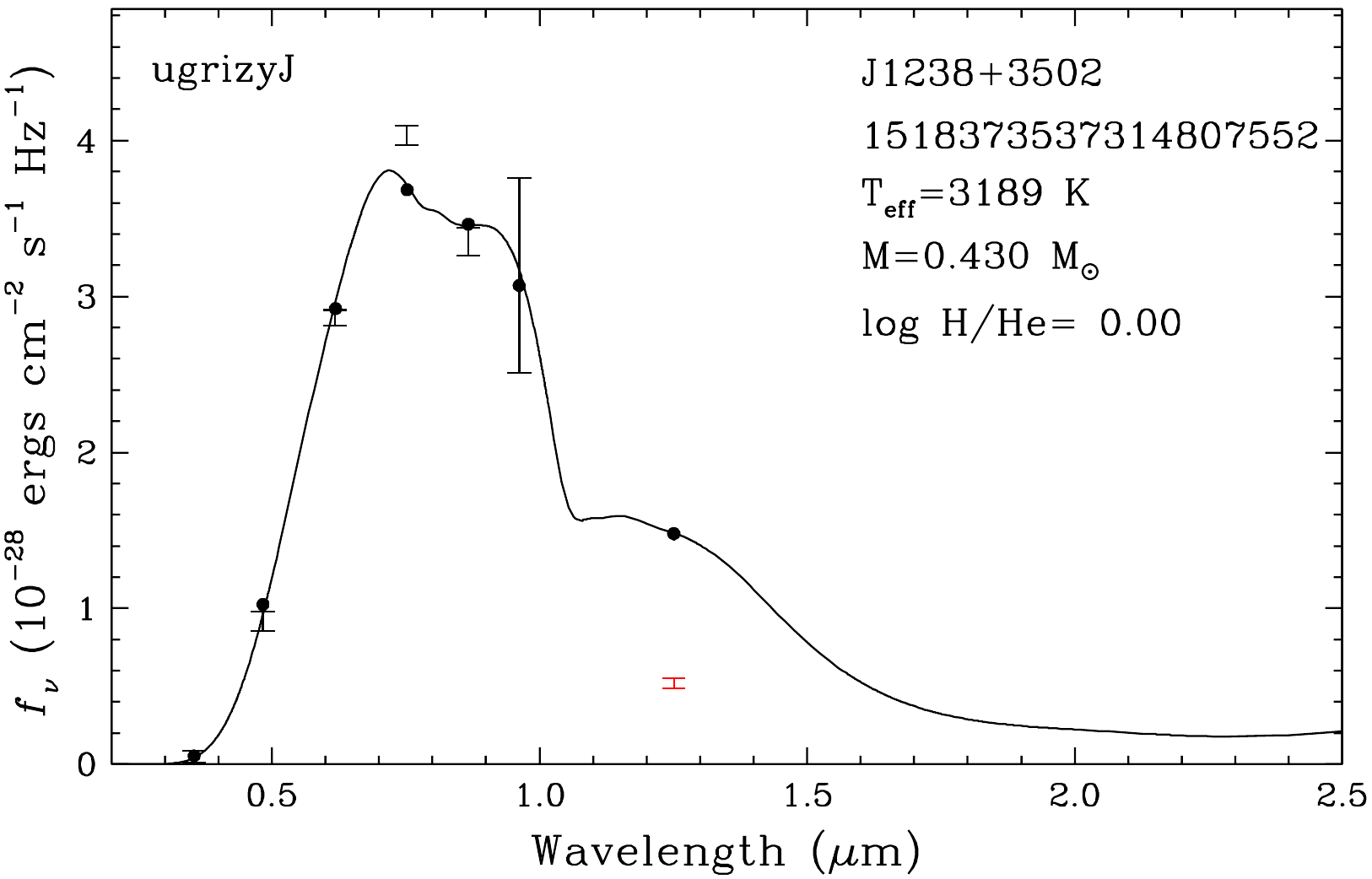}
\includegraphics[width=2.4in]{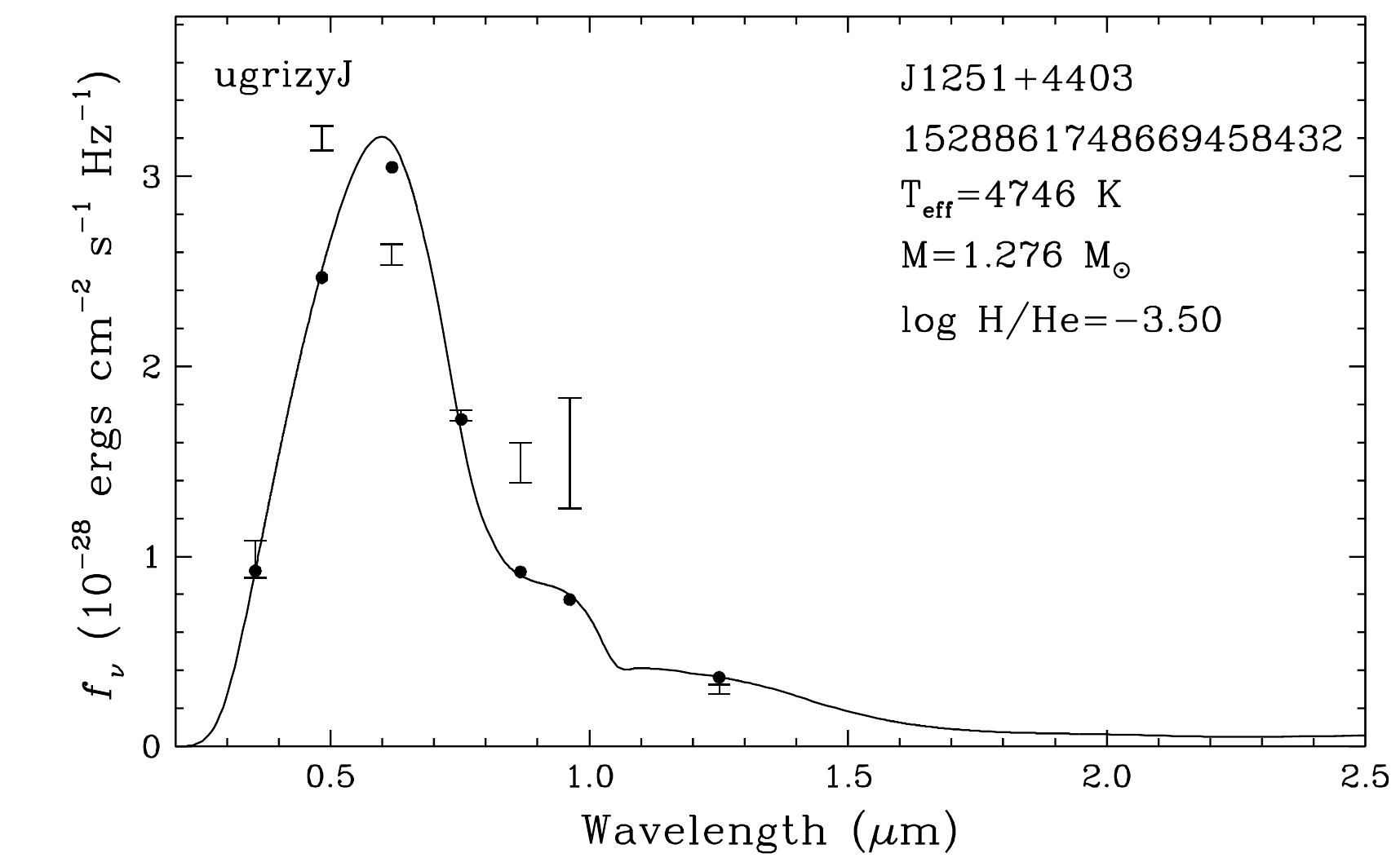}
\includegraphics[width=2.4in]{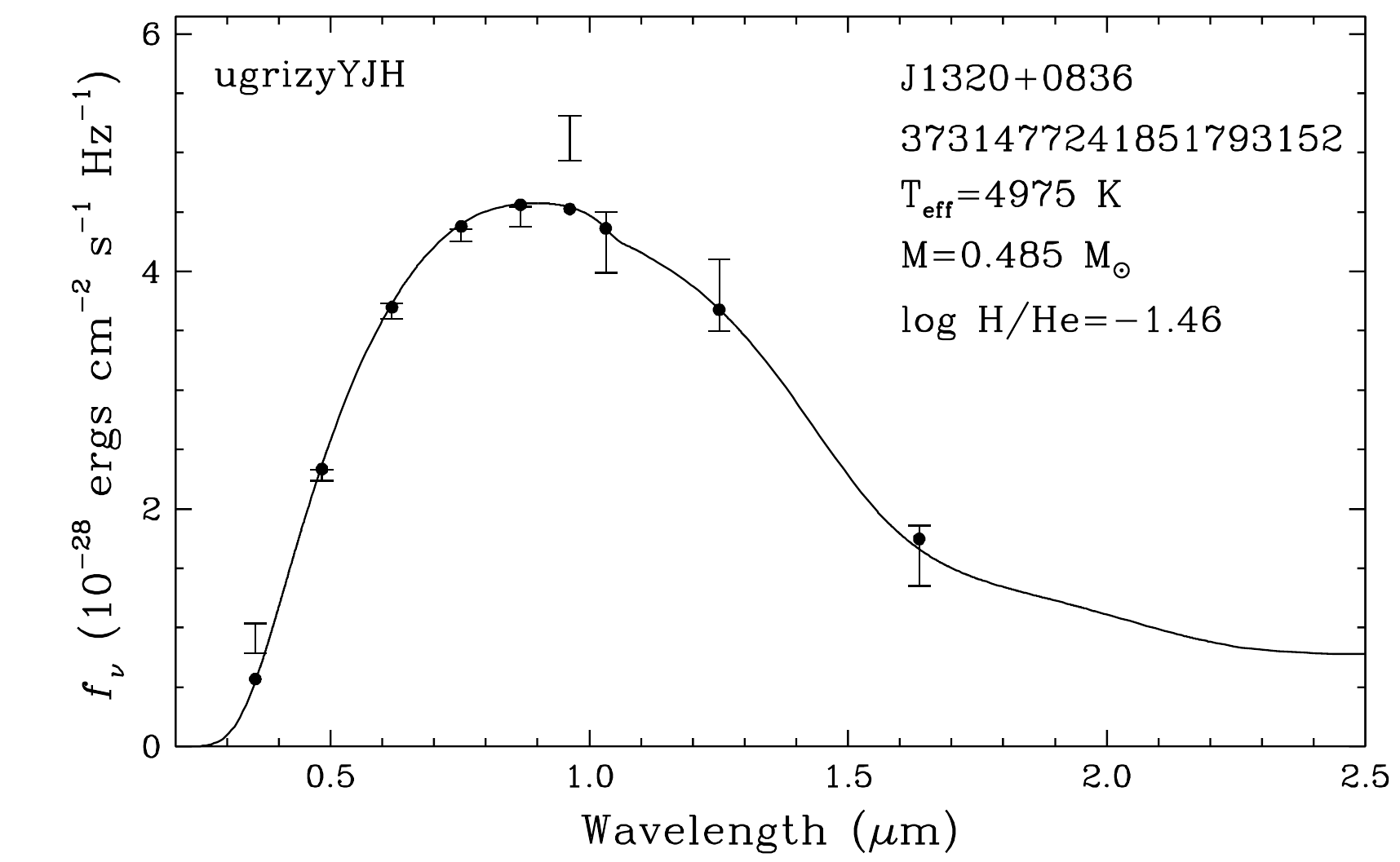}
\includegraphics[width=2.4in]{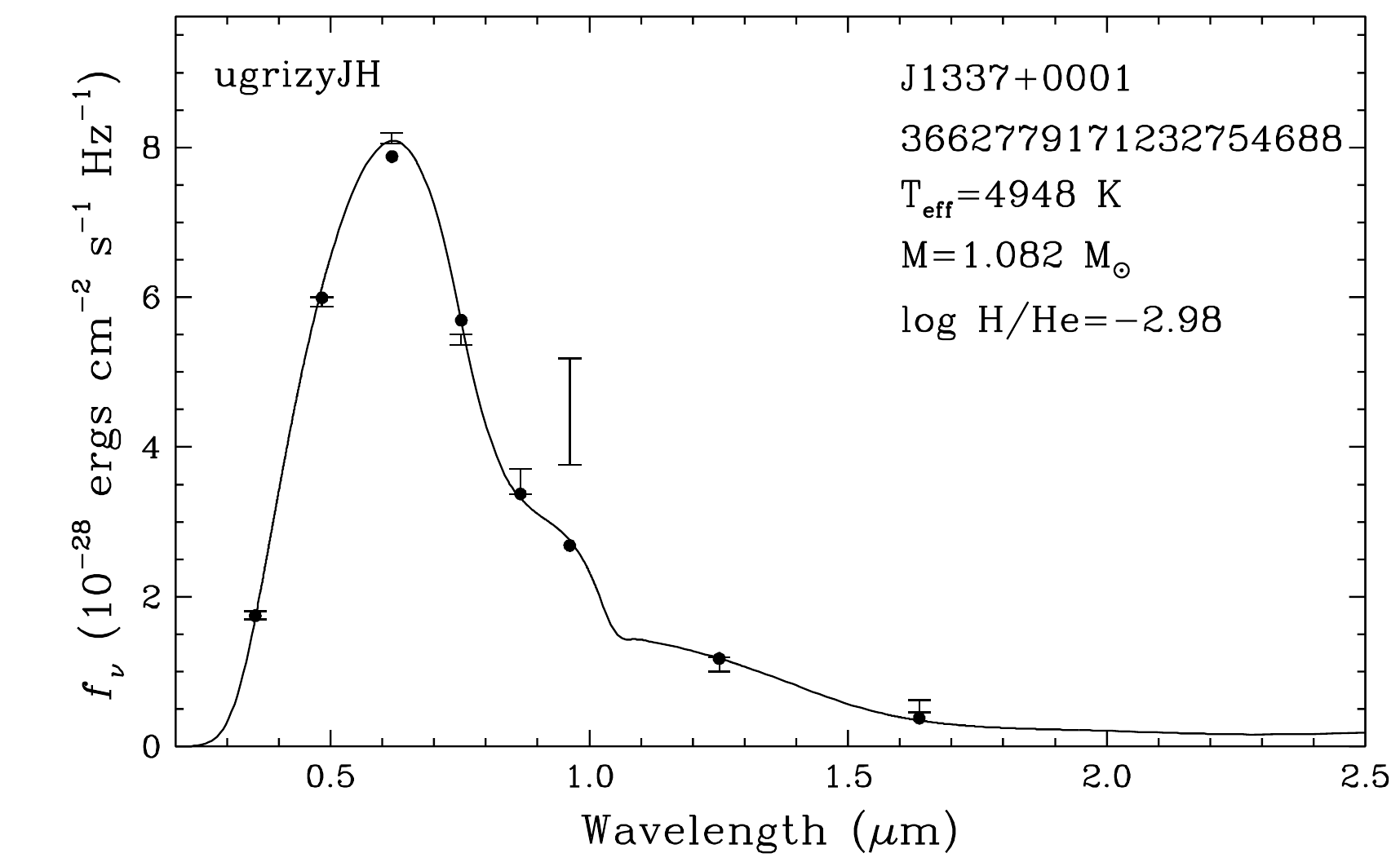}
\includegraphics[width=2.4in]{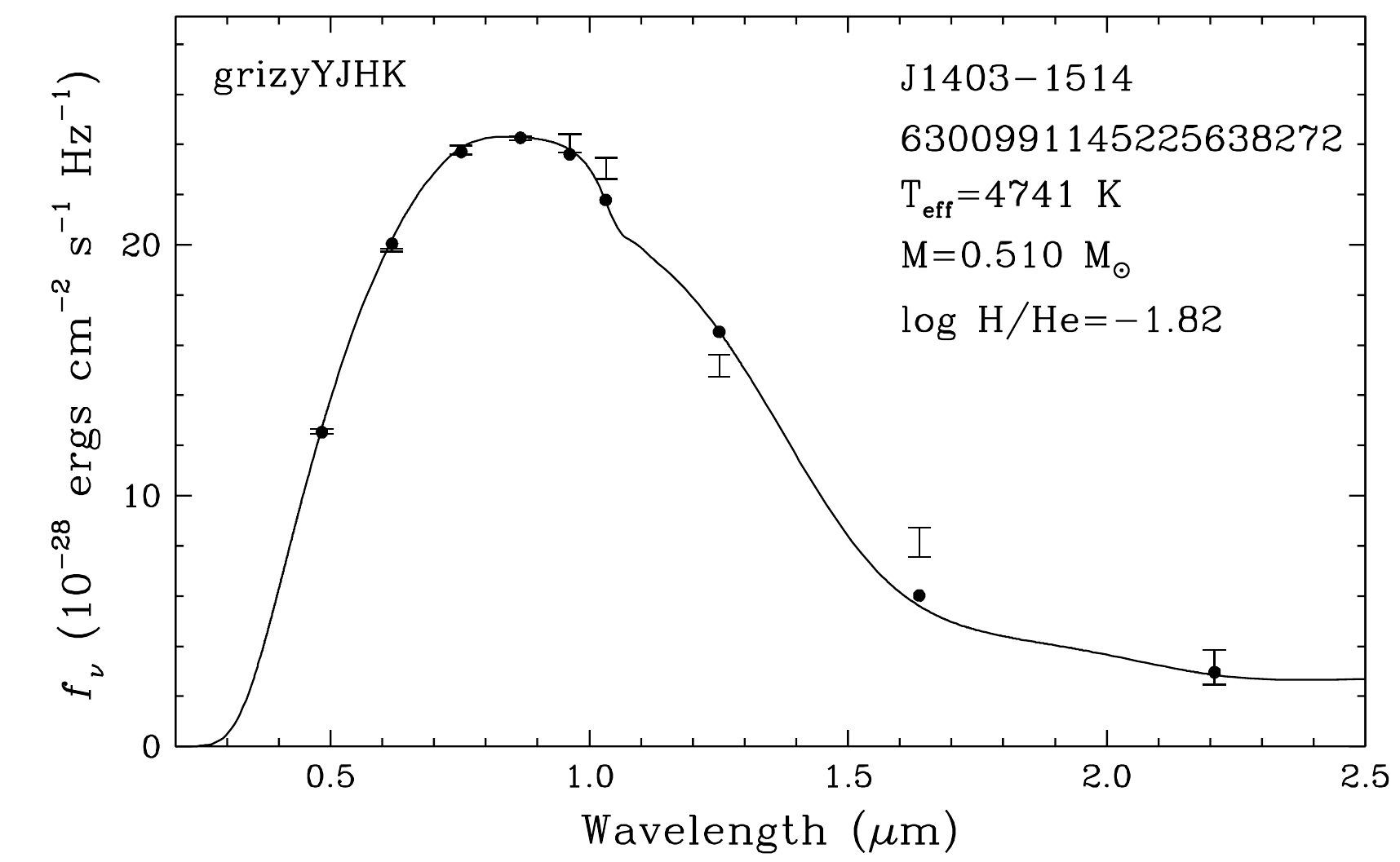}
\includegraphics[width=2.4in]{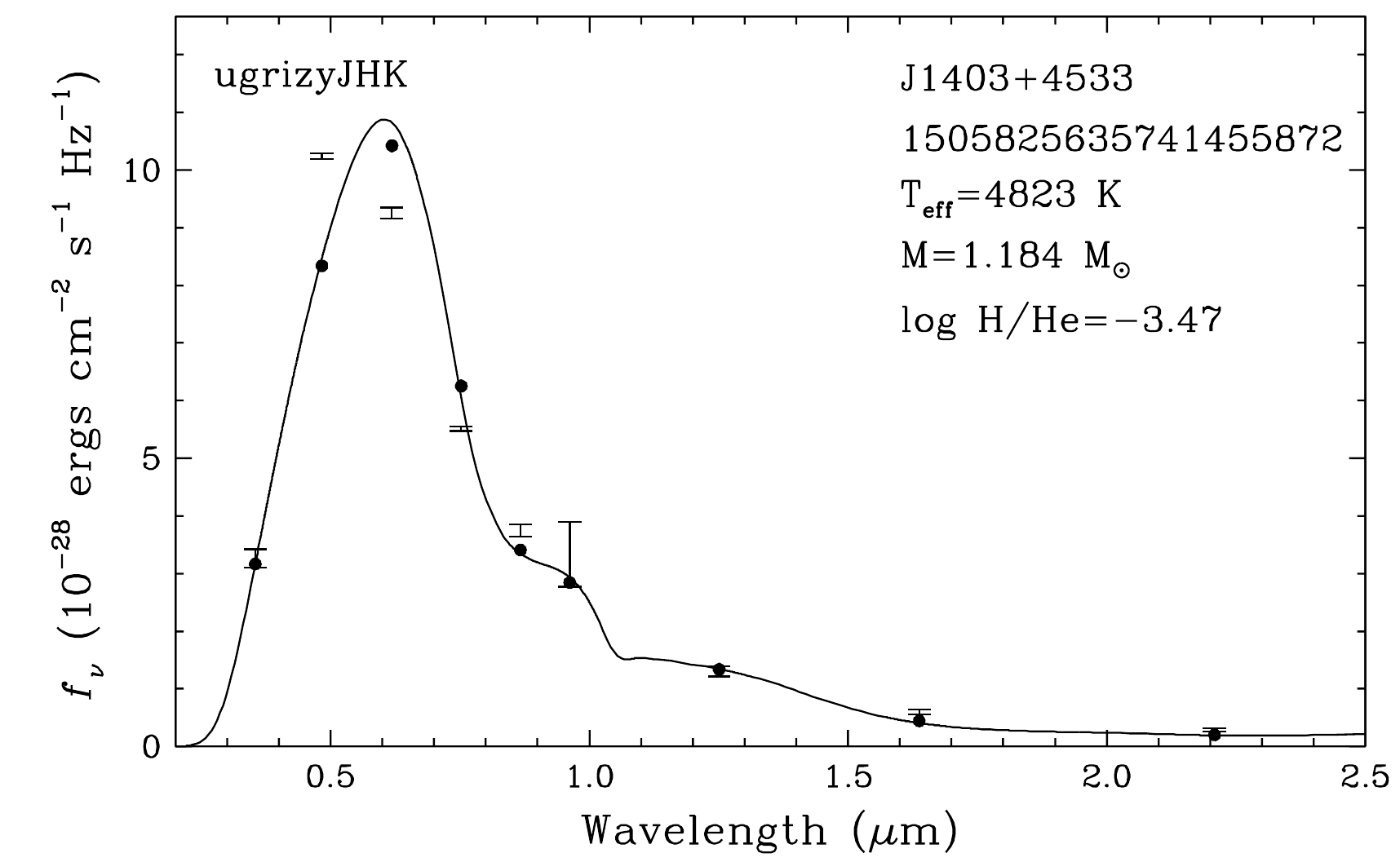}
\includegraphics[width=2.4in]{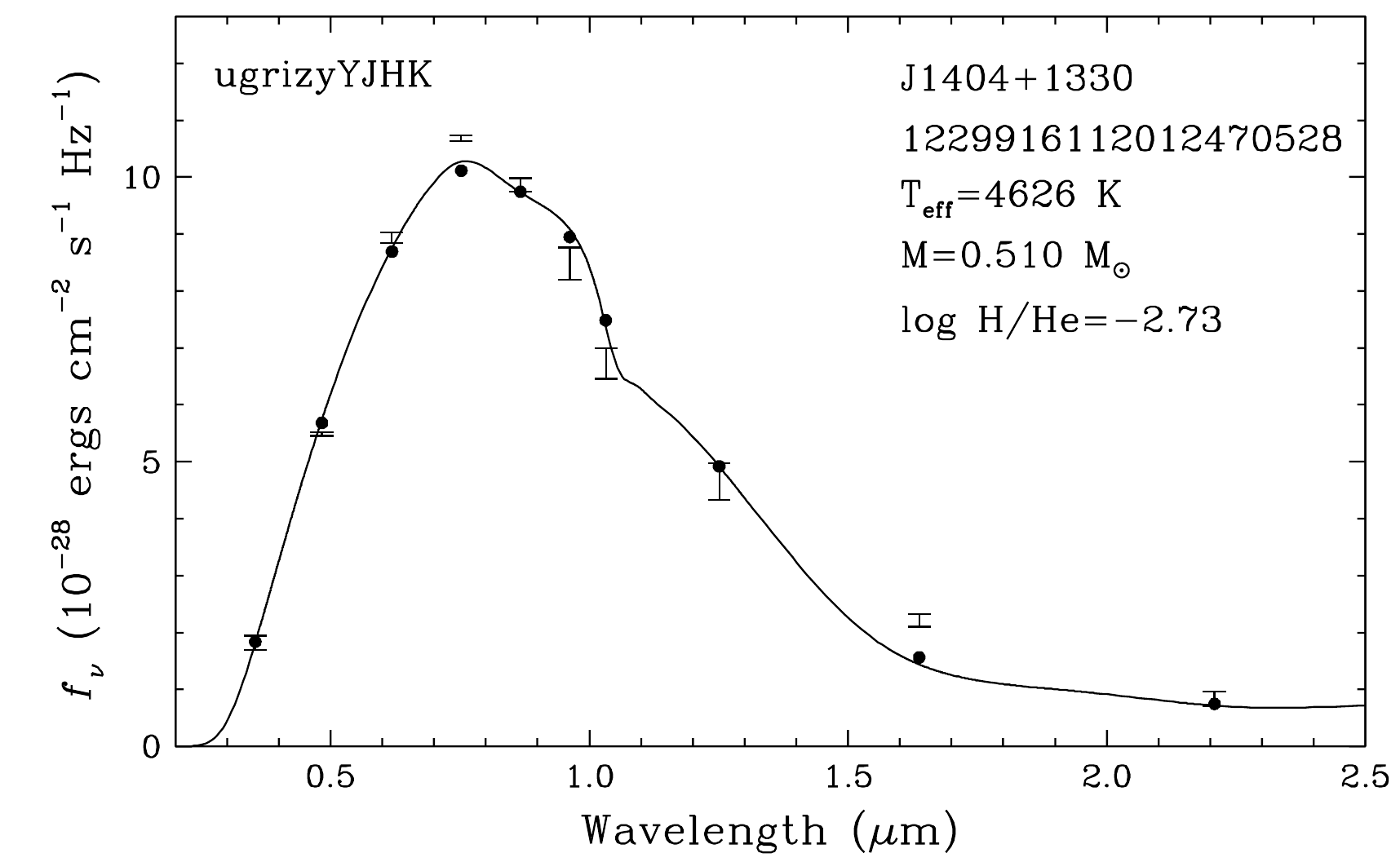}
\includegraphics[width=2.4in]{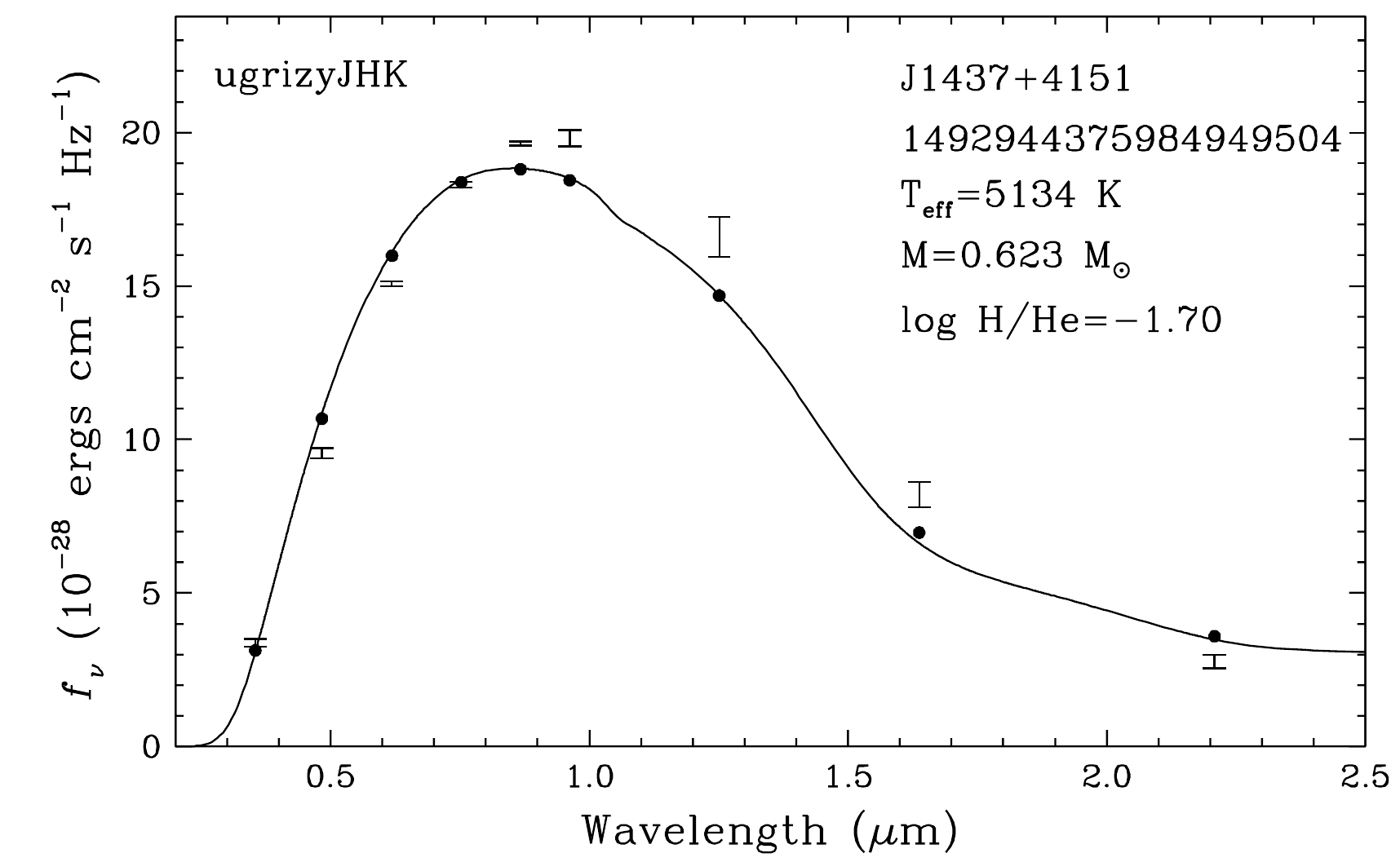}
\includegraphics[width=2.4in]{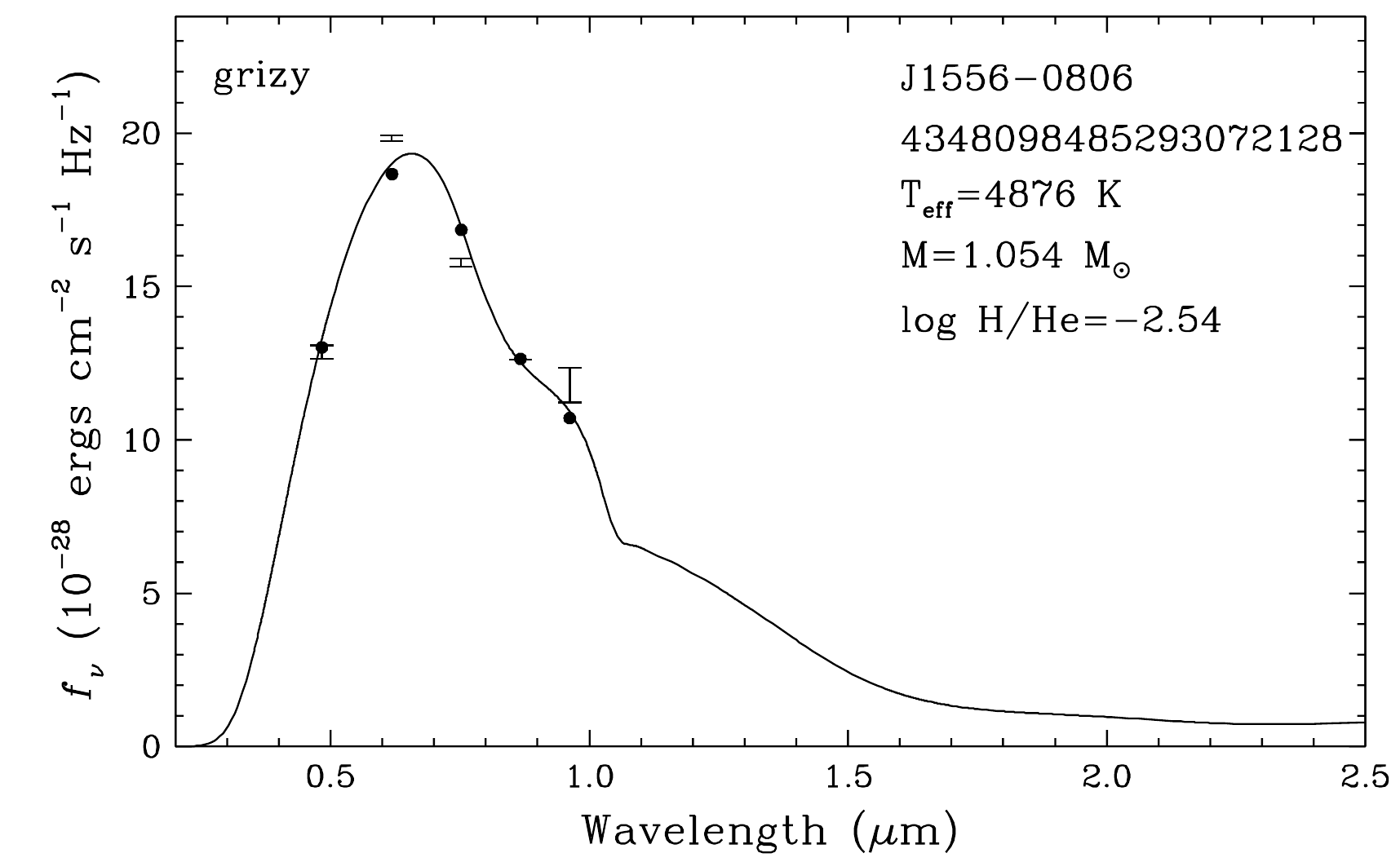}
\includegraphics[width=2.4in]{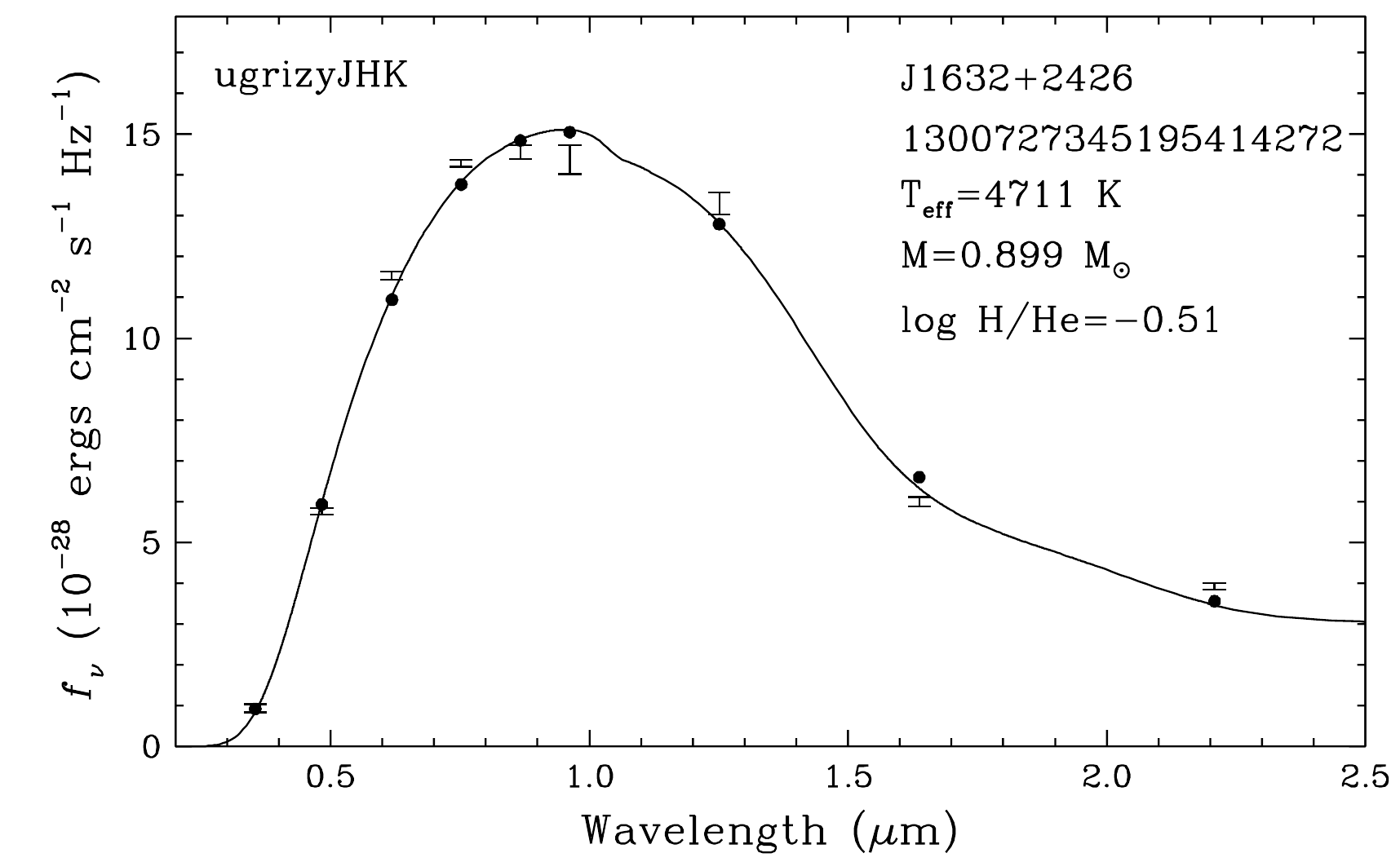}
\includegraphics[width=2.4in]{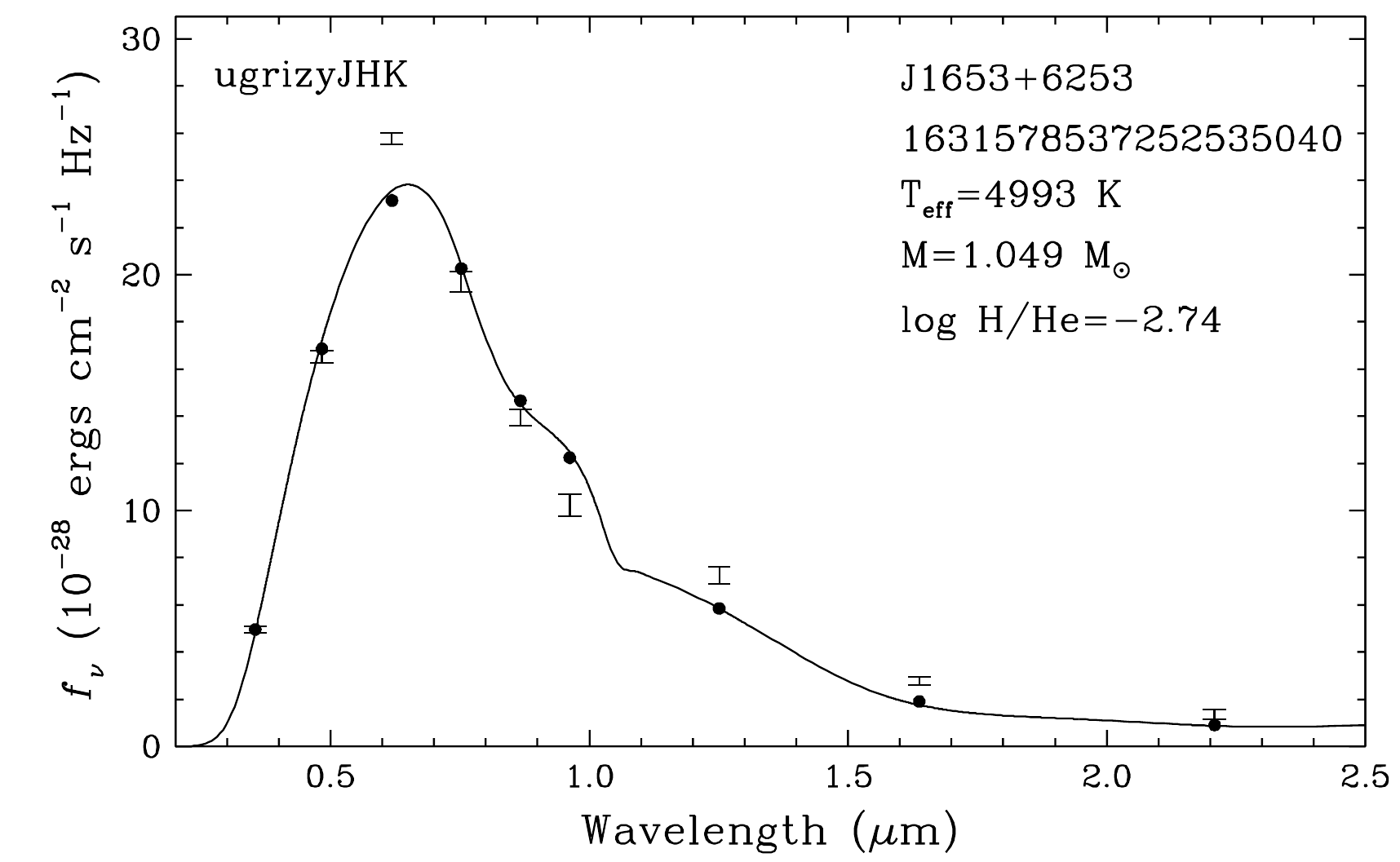}
\includegraphics[width=2.4in]{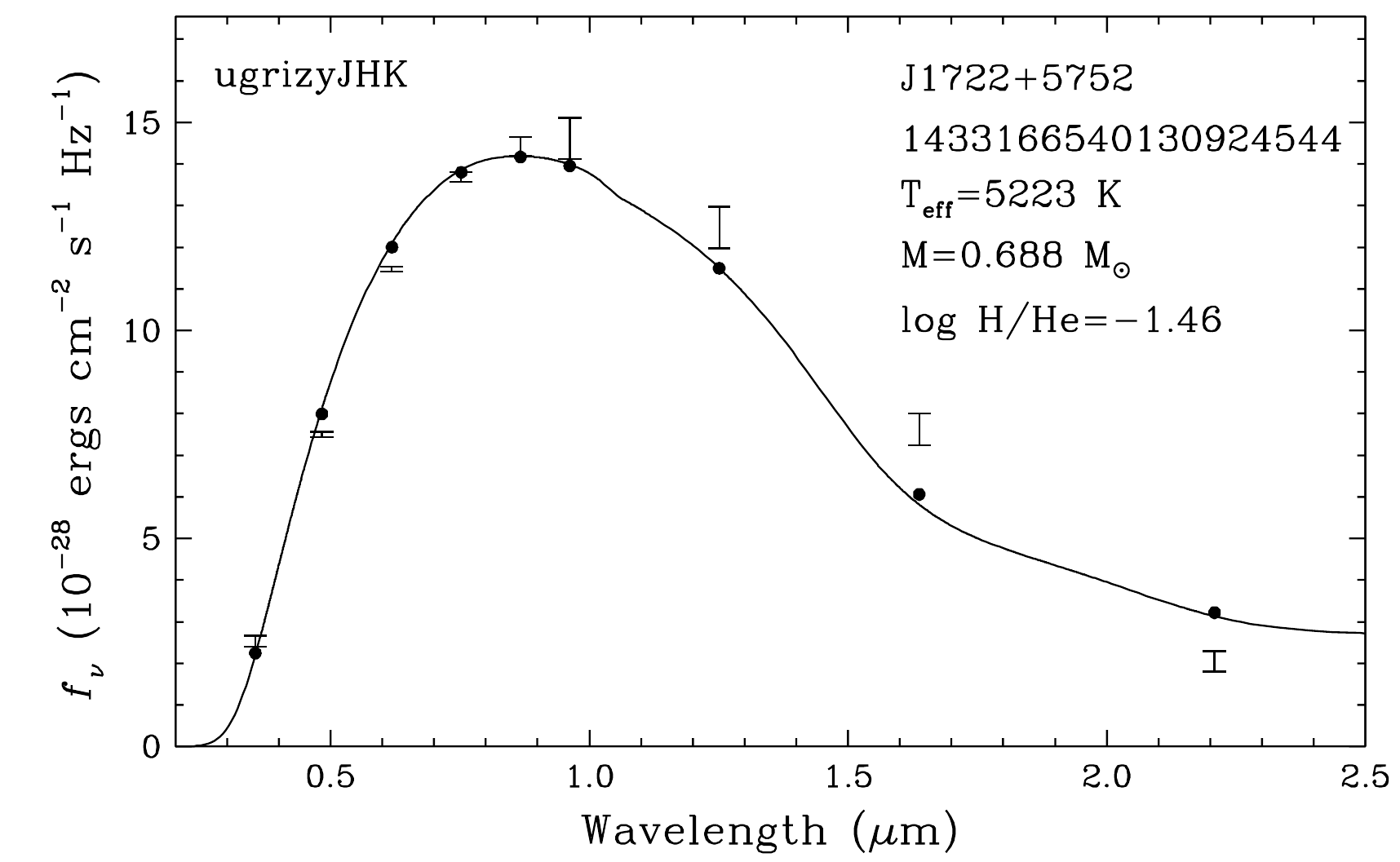}
\includegraphics[width=2.4in]{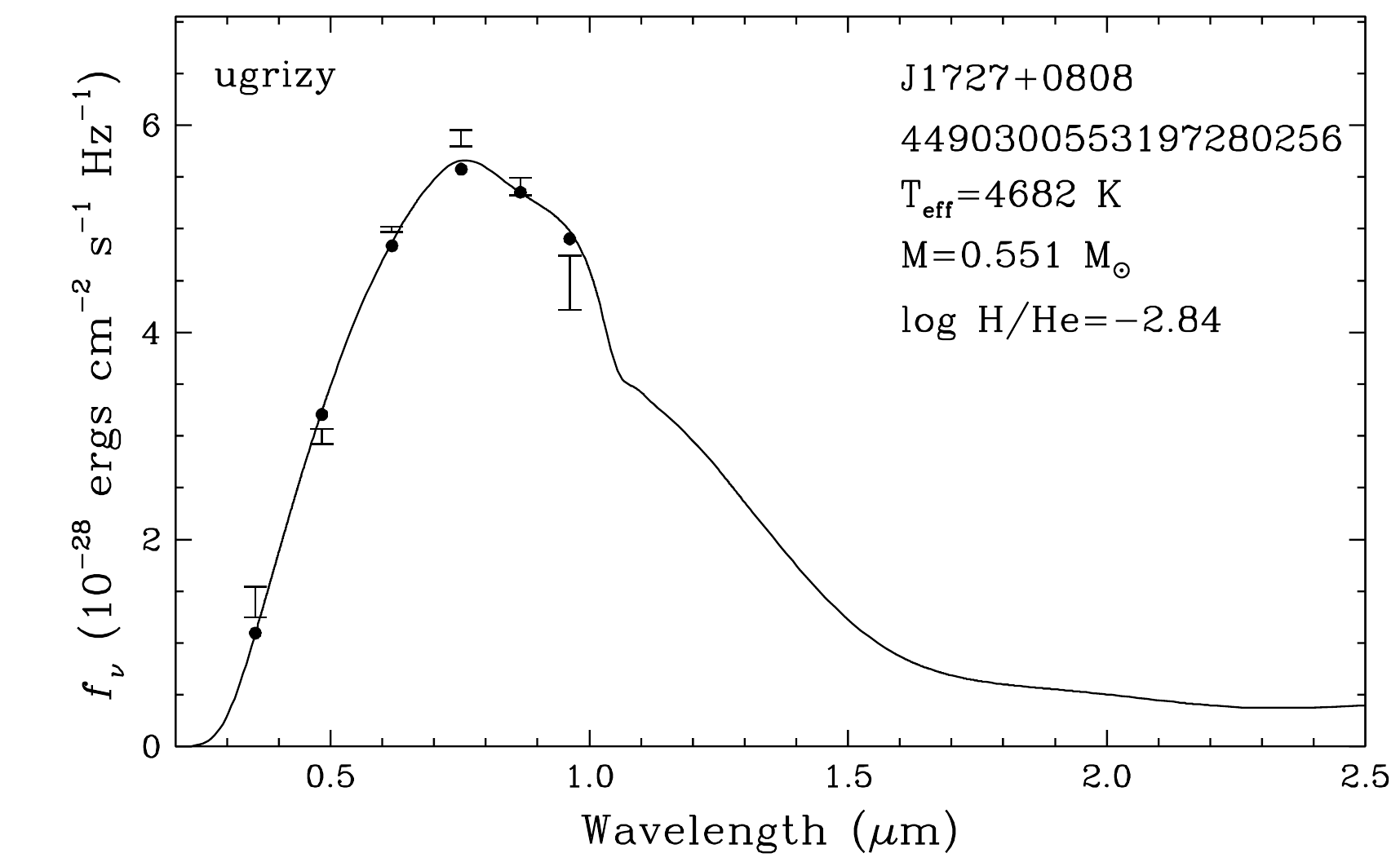}
\includegraphics[width=2.4in]{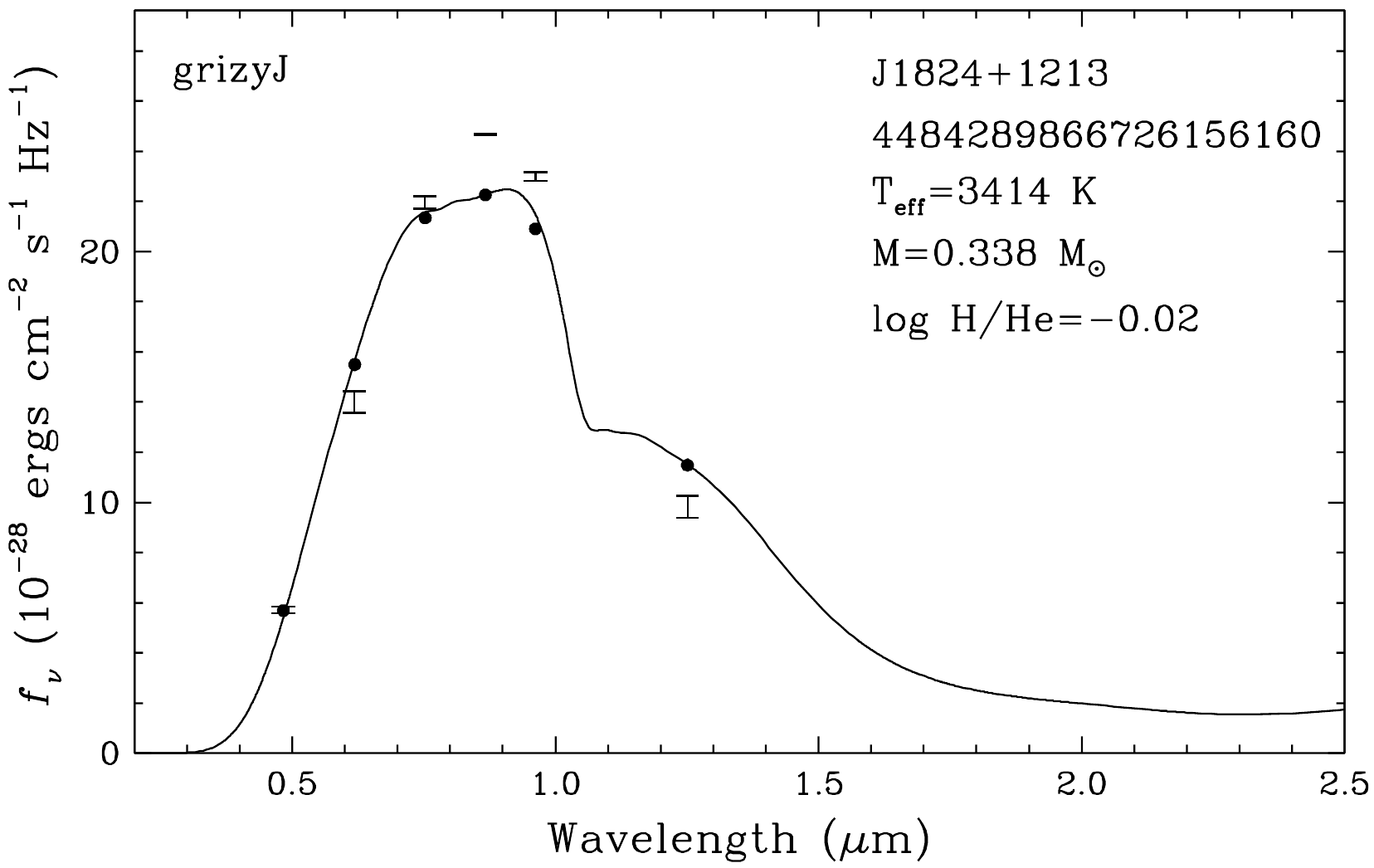}
\includegraphics[width=2.4in]{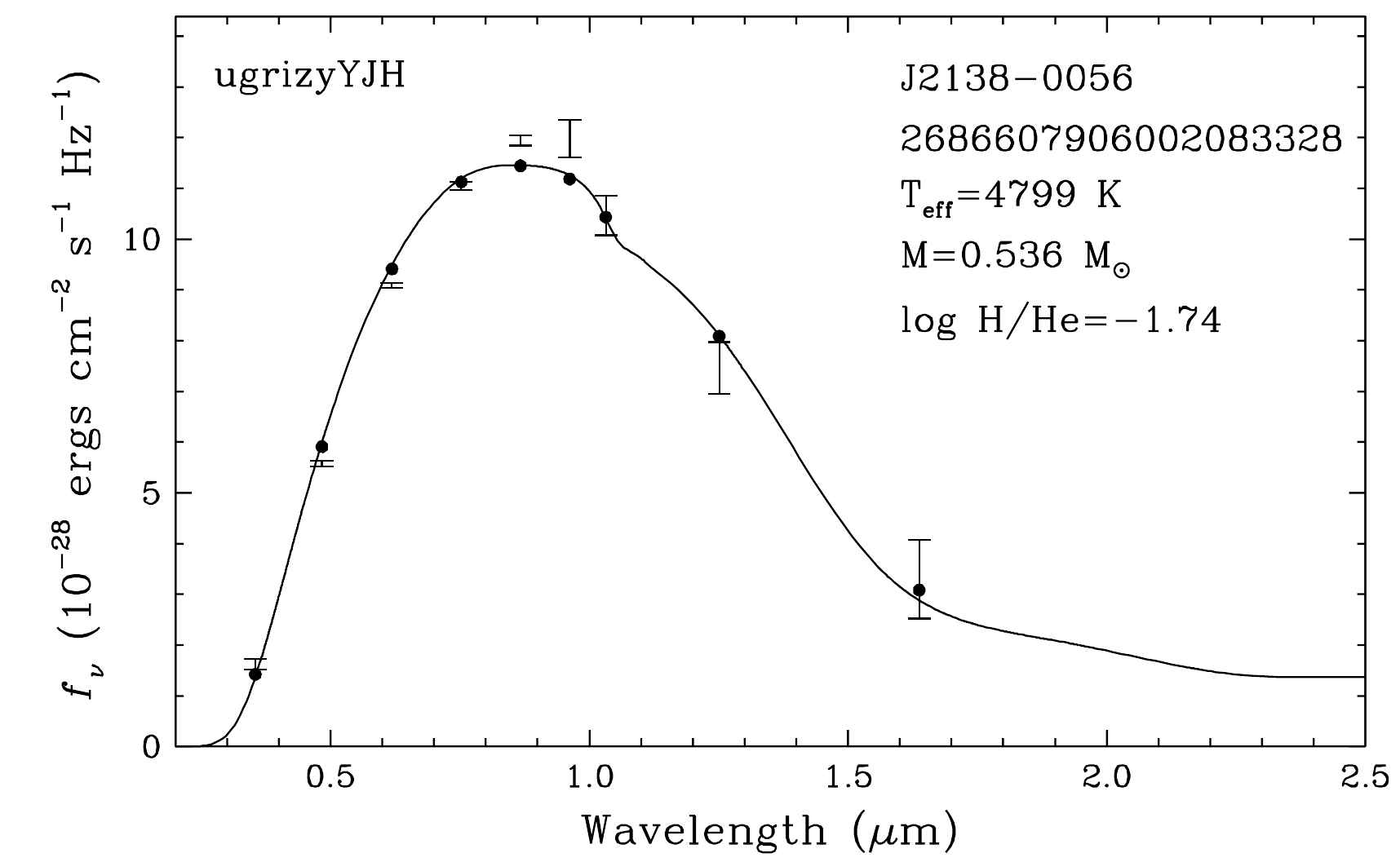}
\includegraphics[width=2.4in]{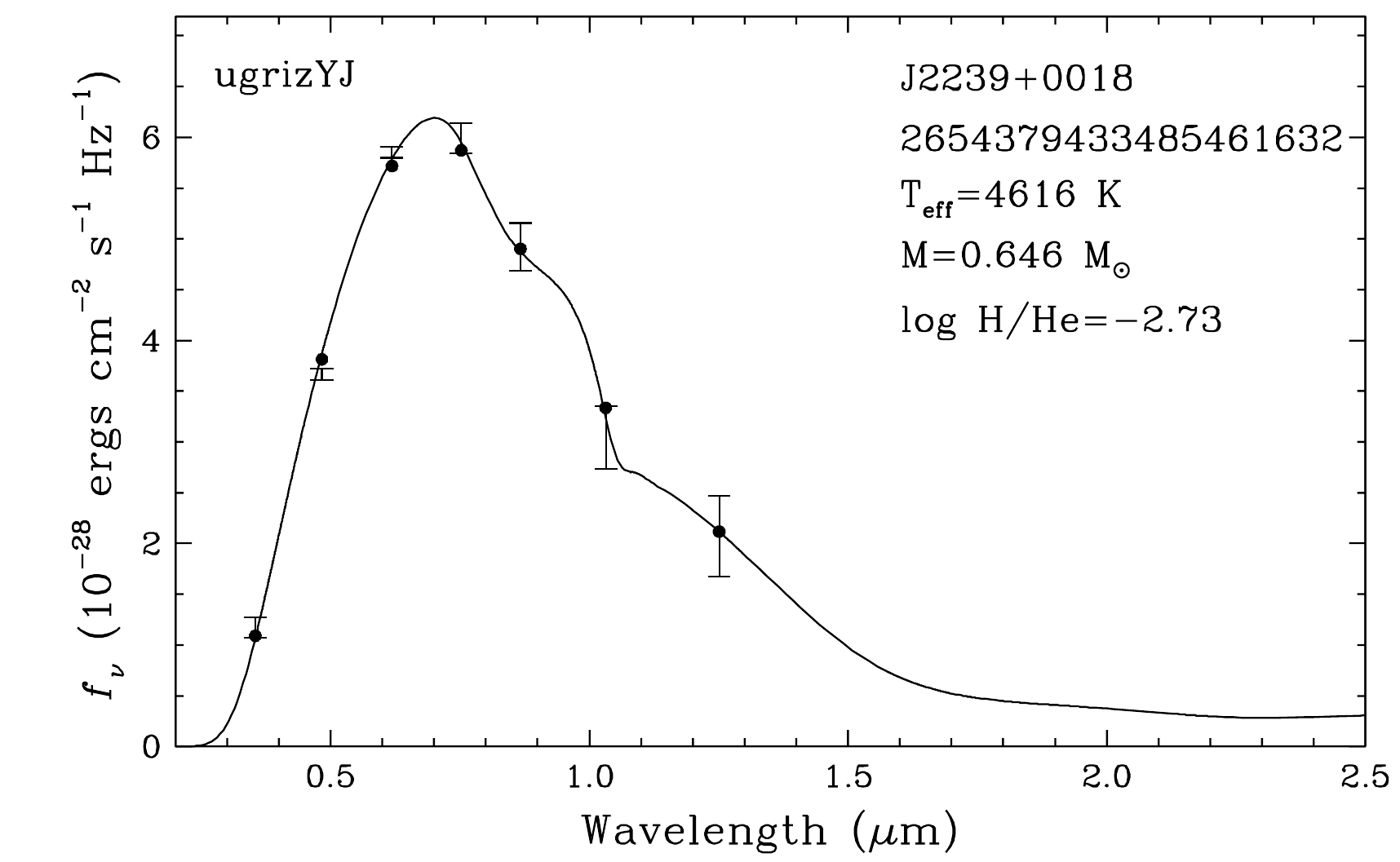}
\includegraphics[width=2.4in]{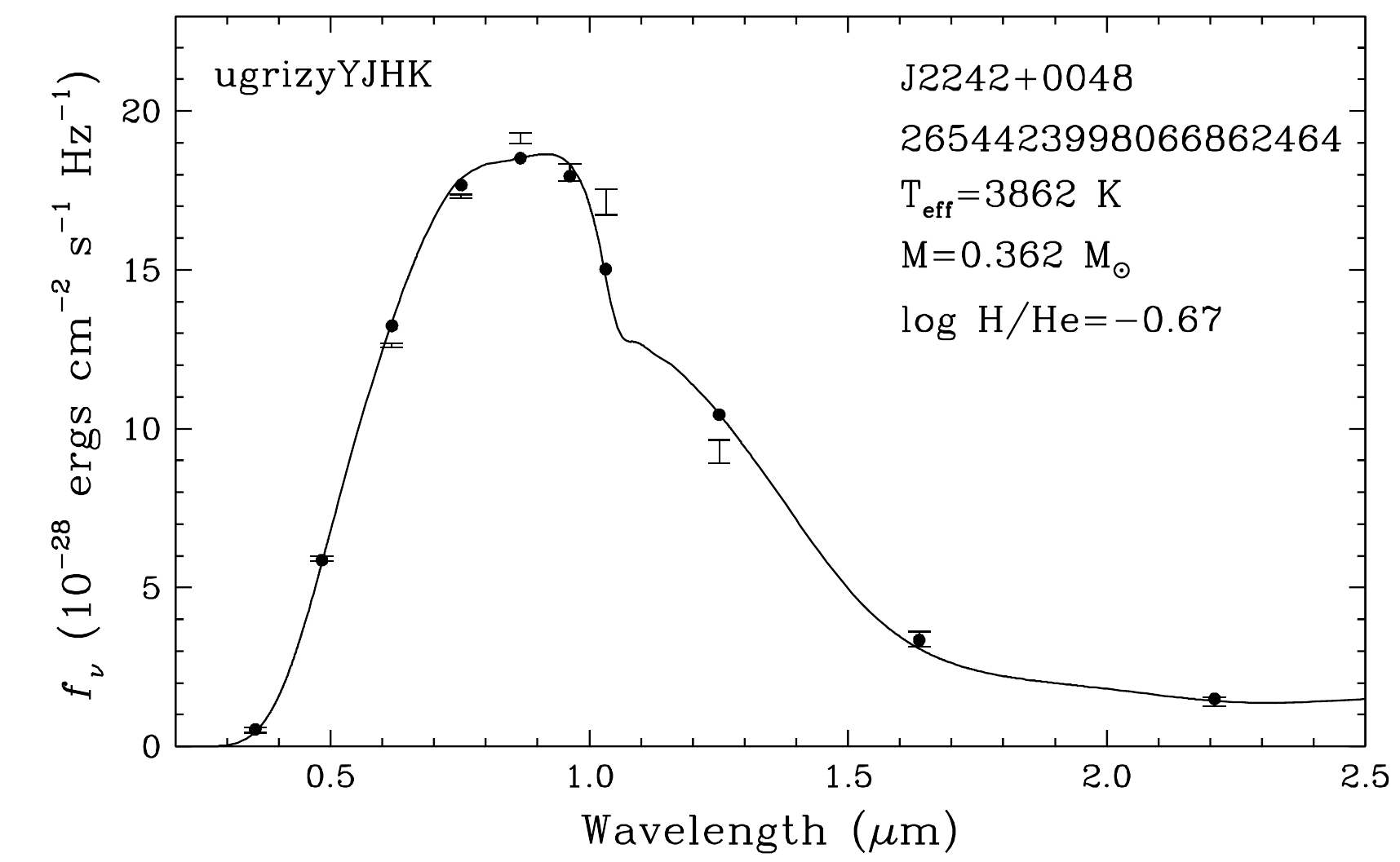}
\includegraphics[width=2.4in]{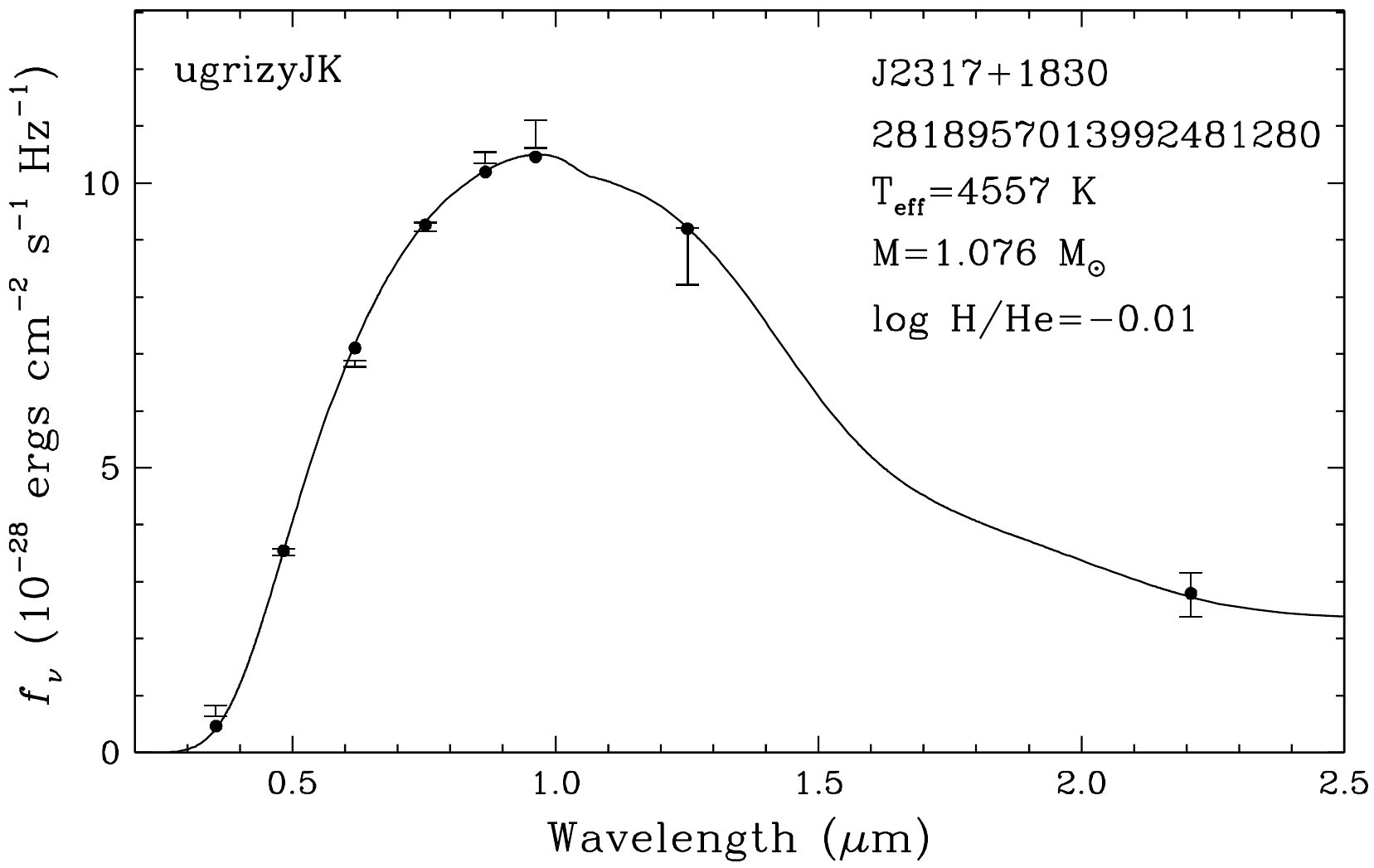}
\caption{Continued.}
\end{figure*}

\begin{figure*}
\includegraphics[width=2.4in]{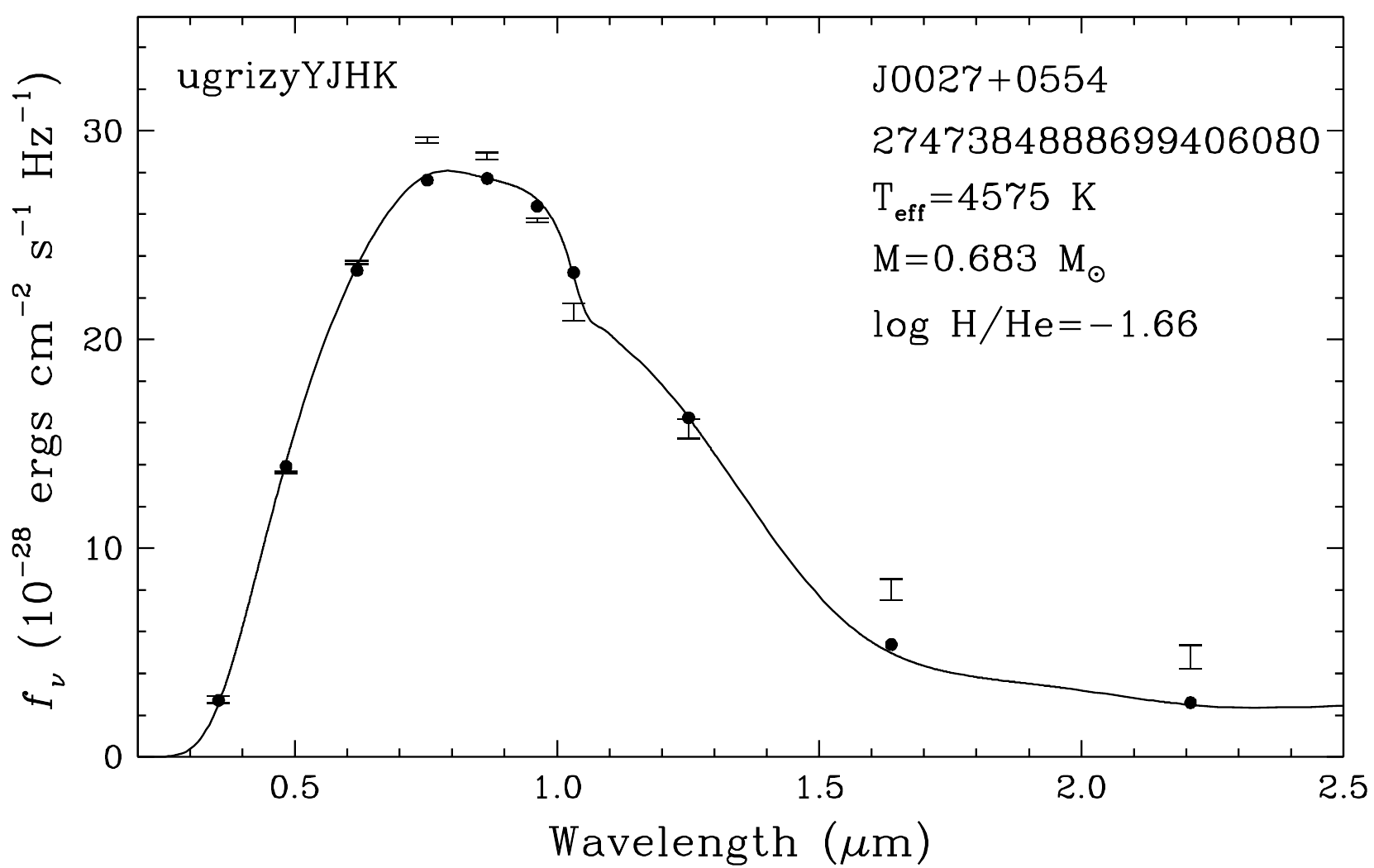}
\includegraphics[width=2.4in]{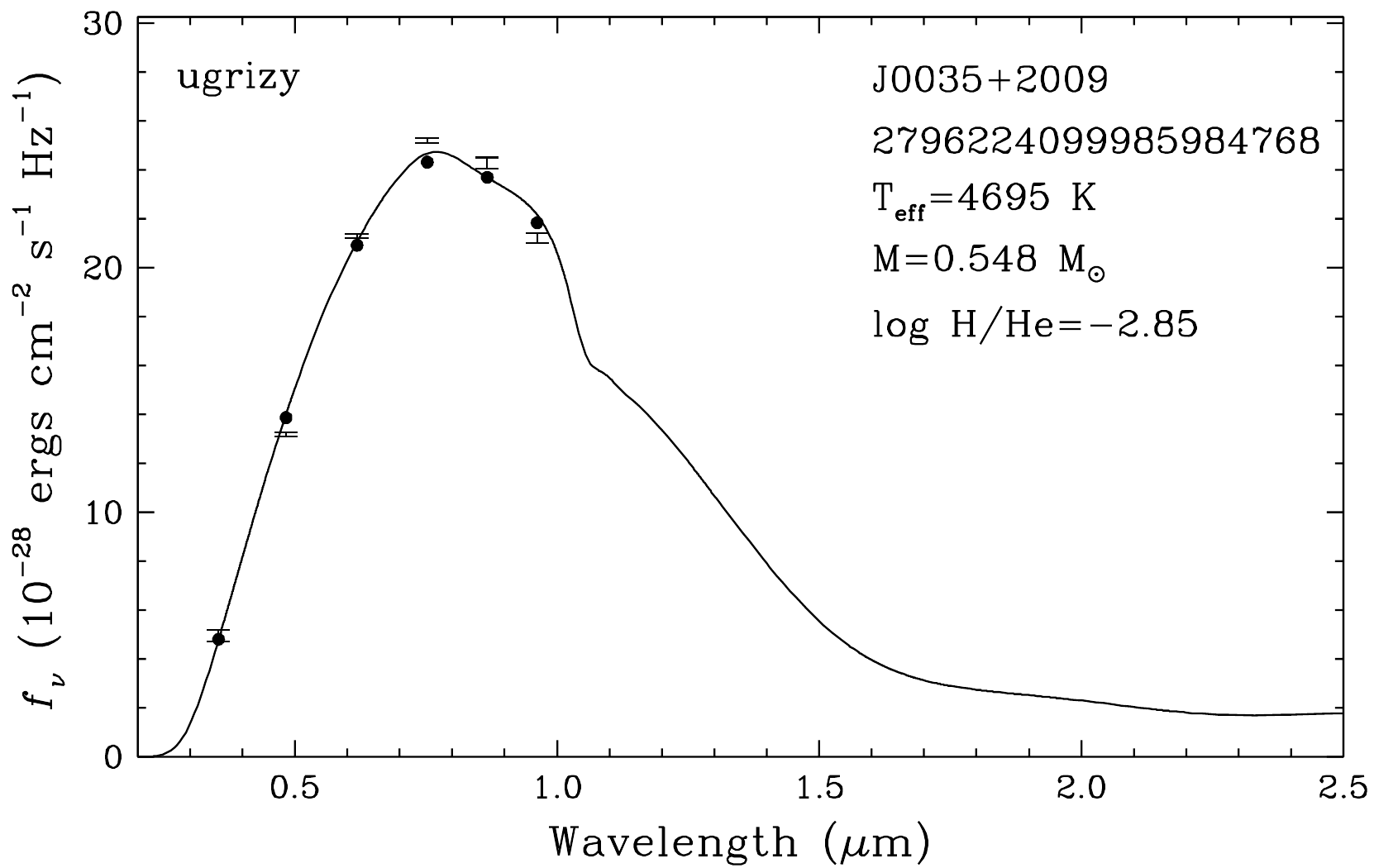}
\includegraphics[width=2.4in]{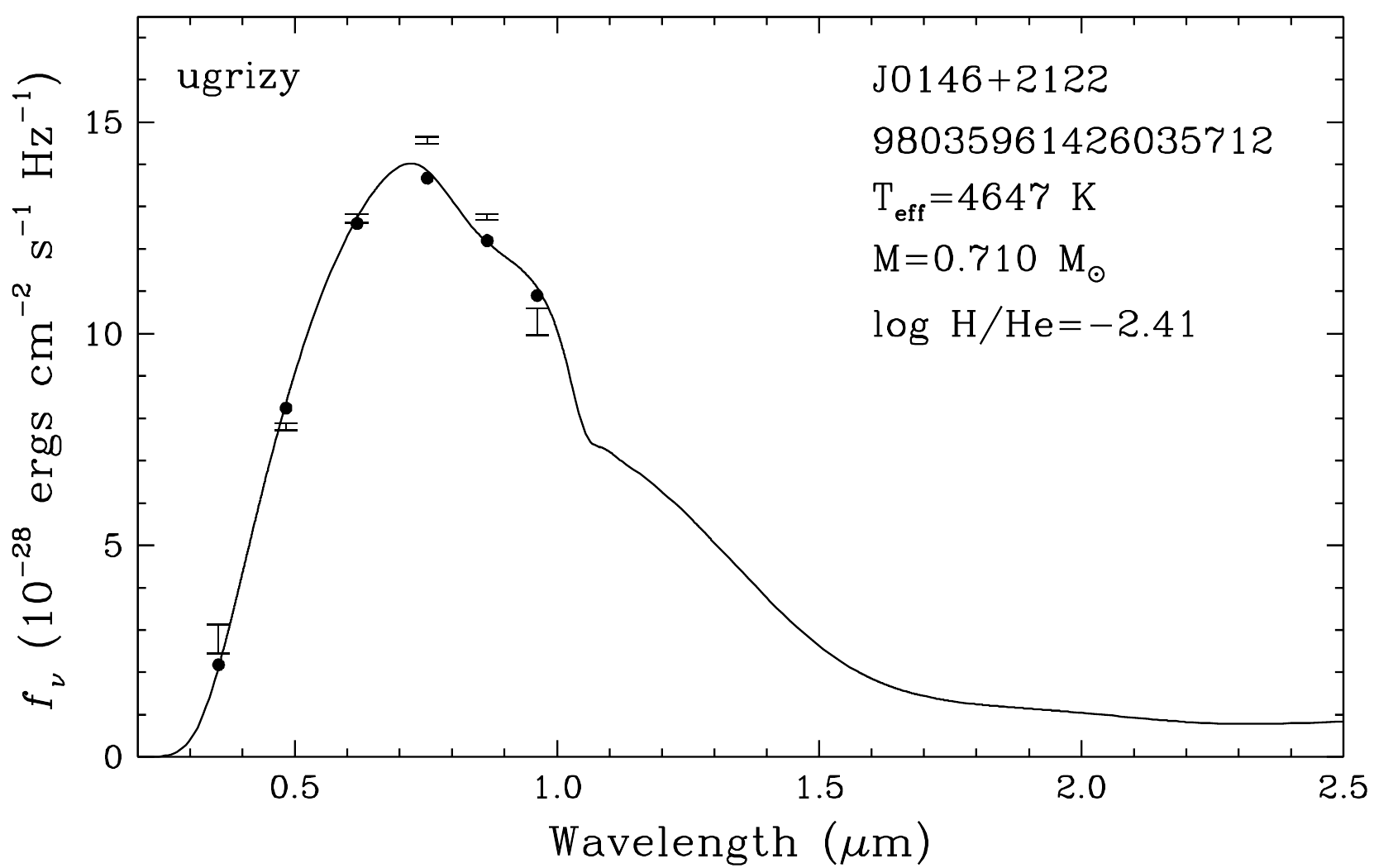}
\includegraphics[width=2.4in]{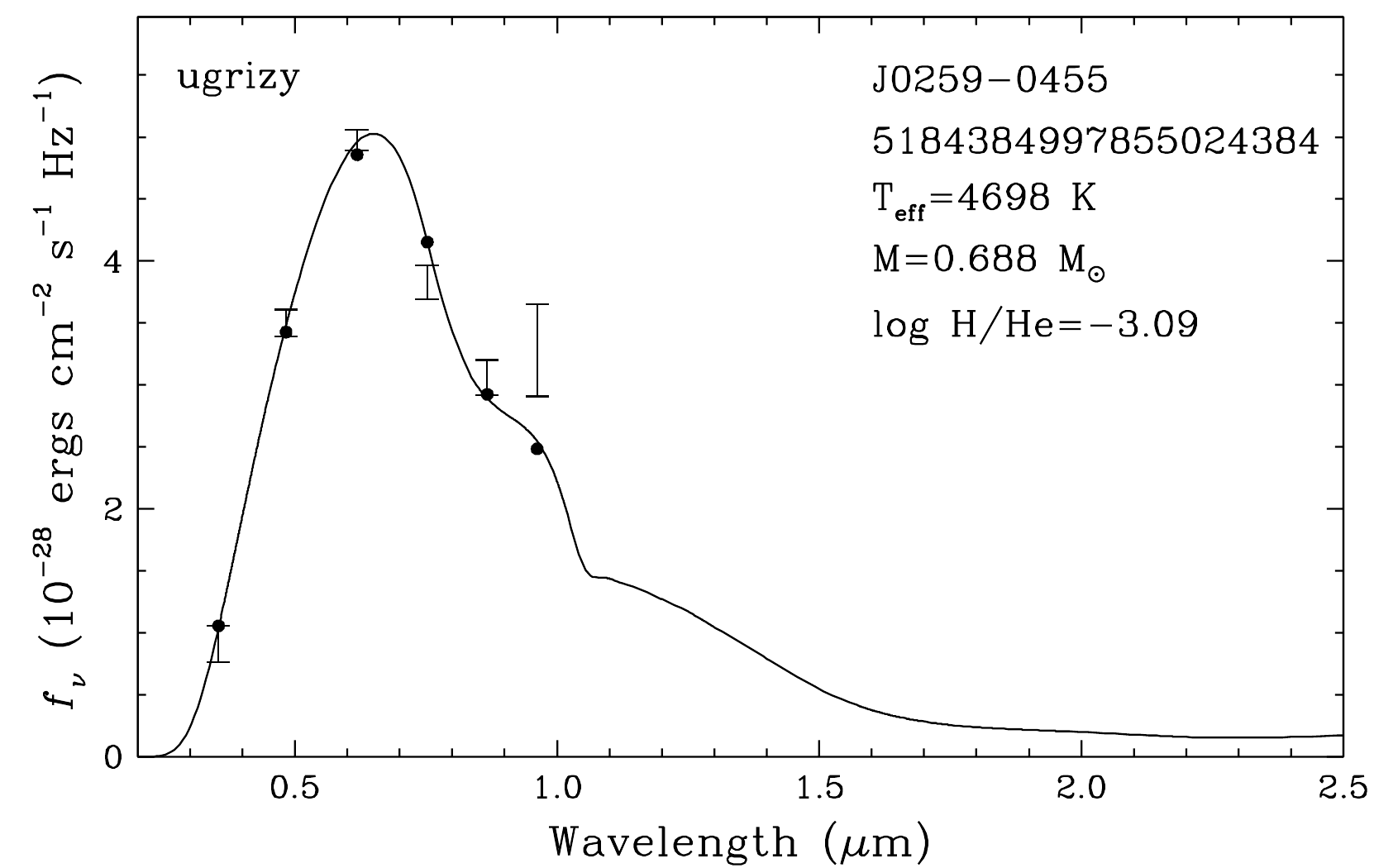}
\includegraphics[width=2.4in]{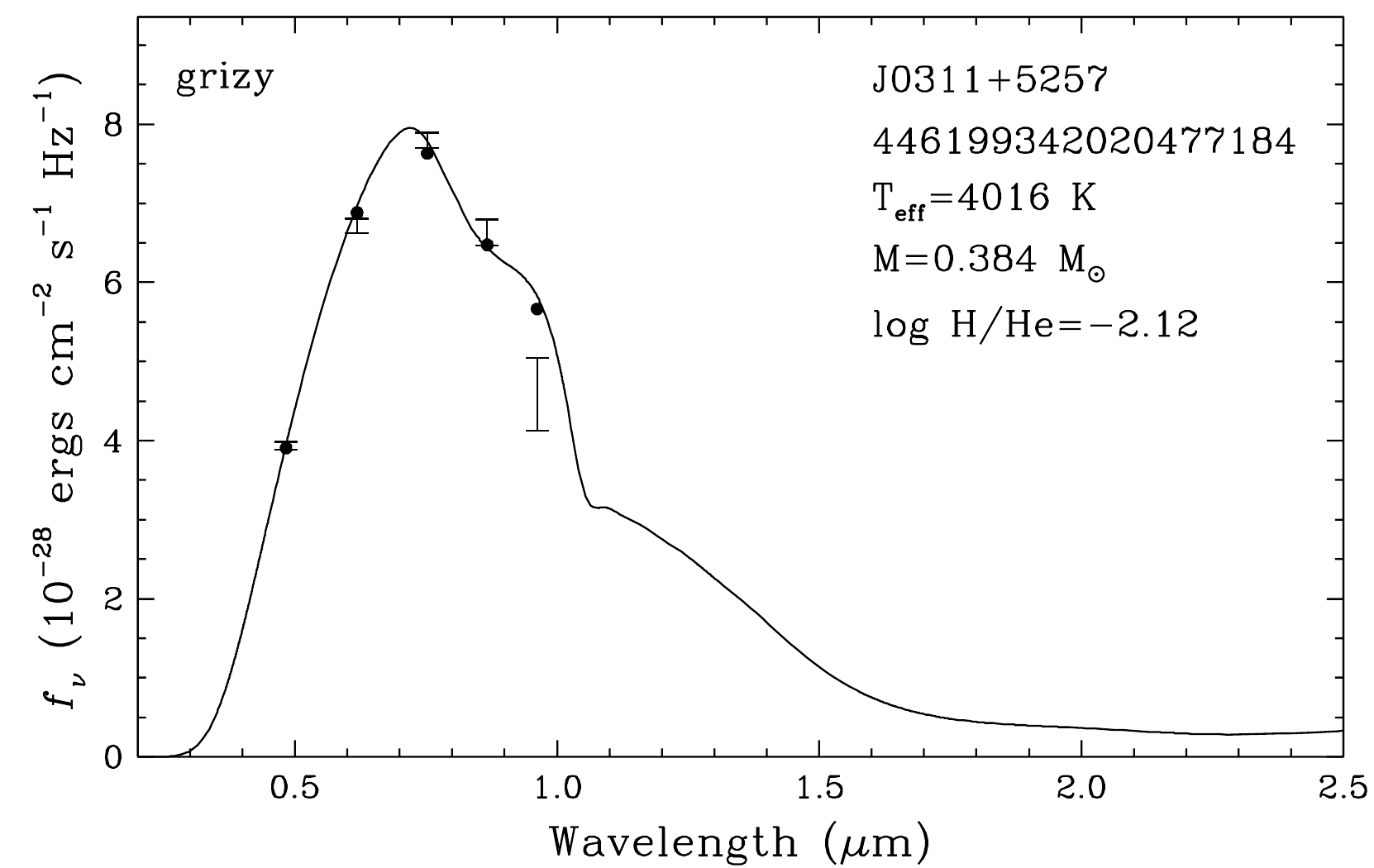}
\includegraphics[width=2.4in]{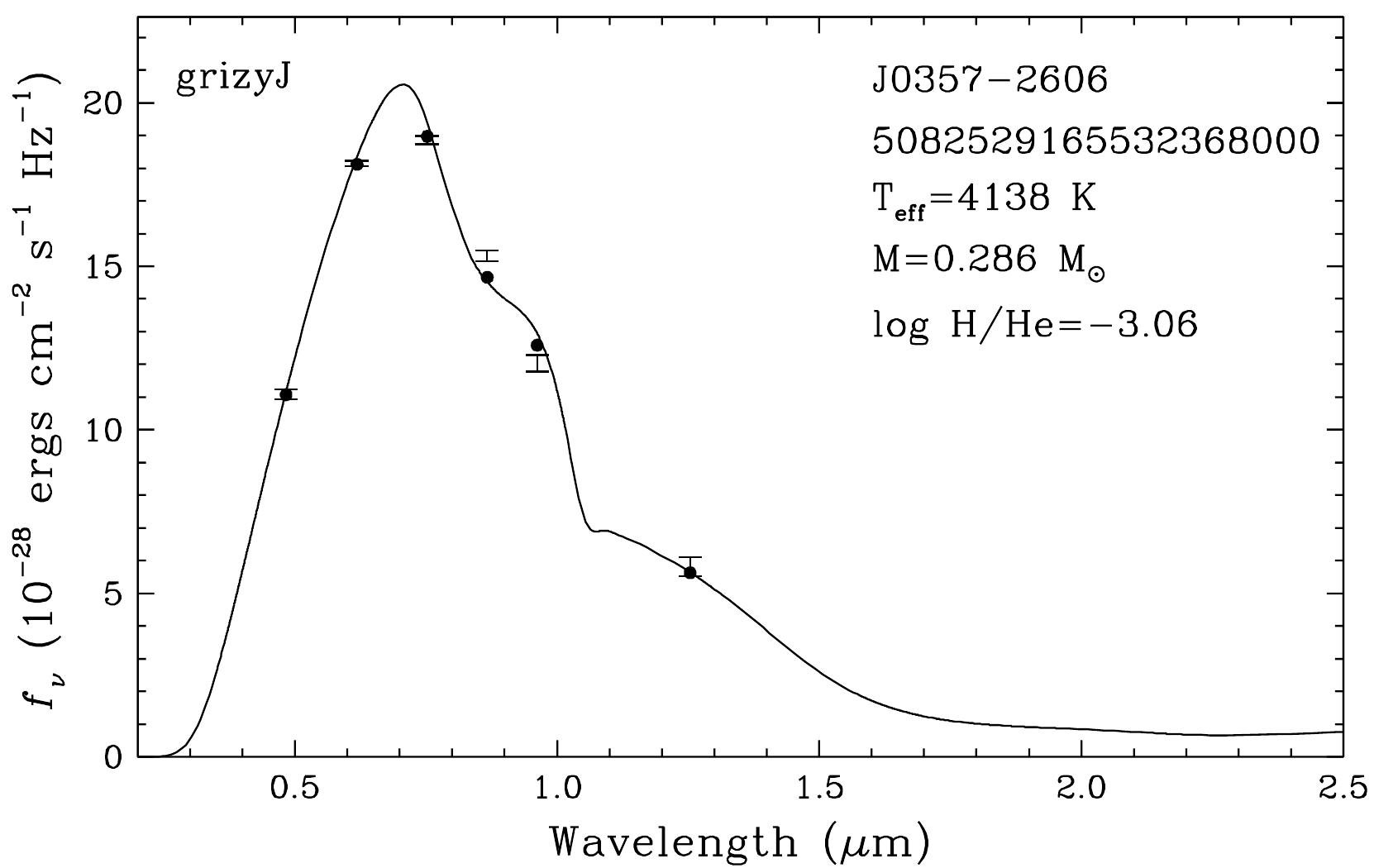}
\includegraphics[width=2.4in]{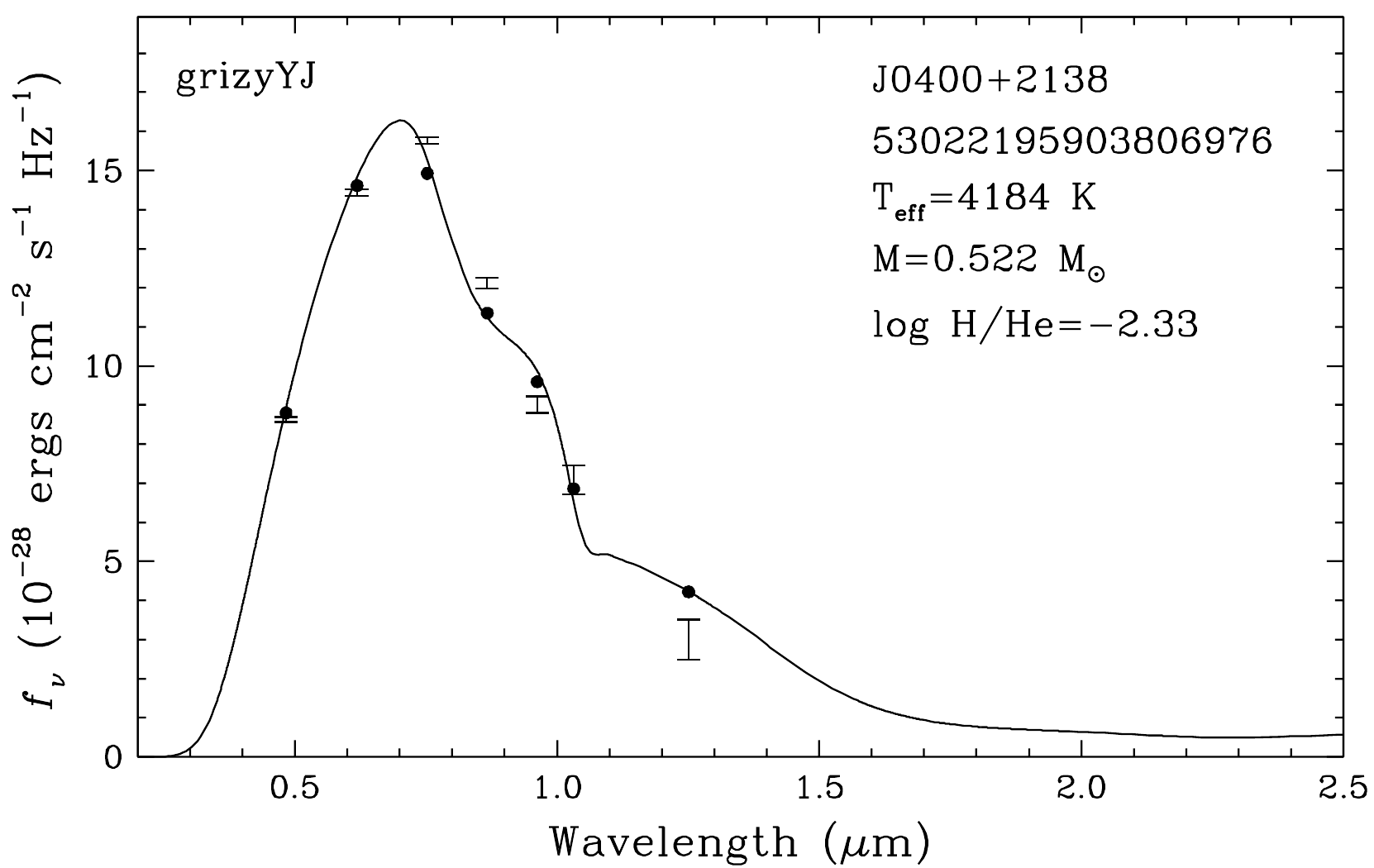}
\includegraphics[width=2.4in]{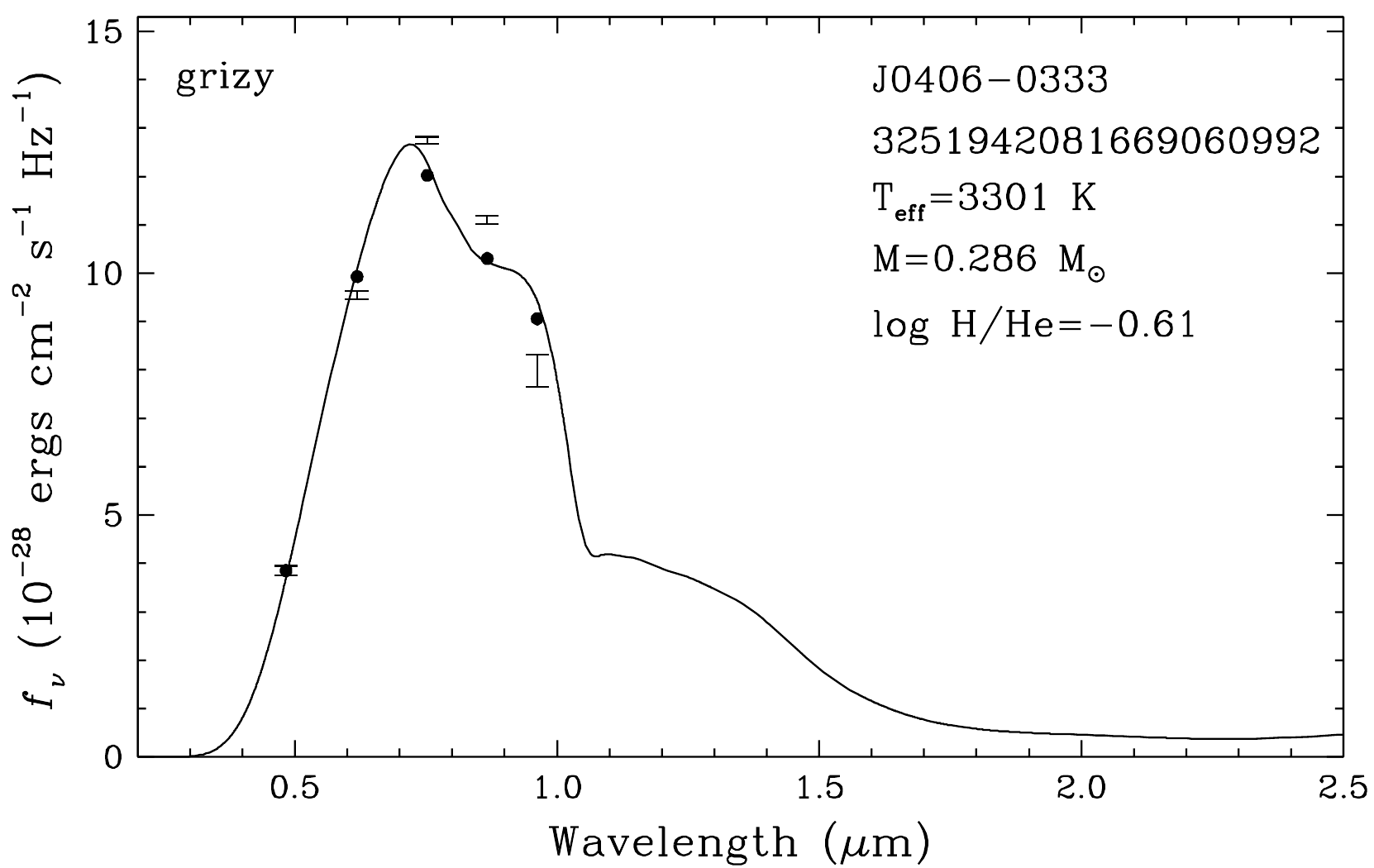}
\includegraphics[width=2.4in]{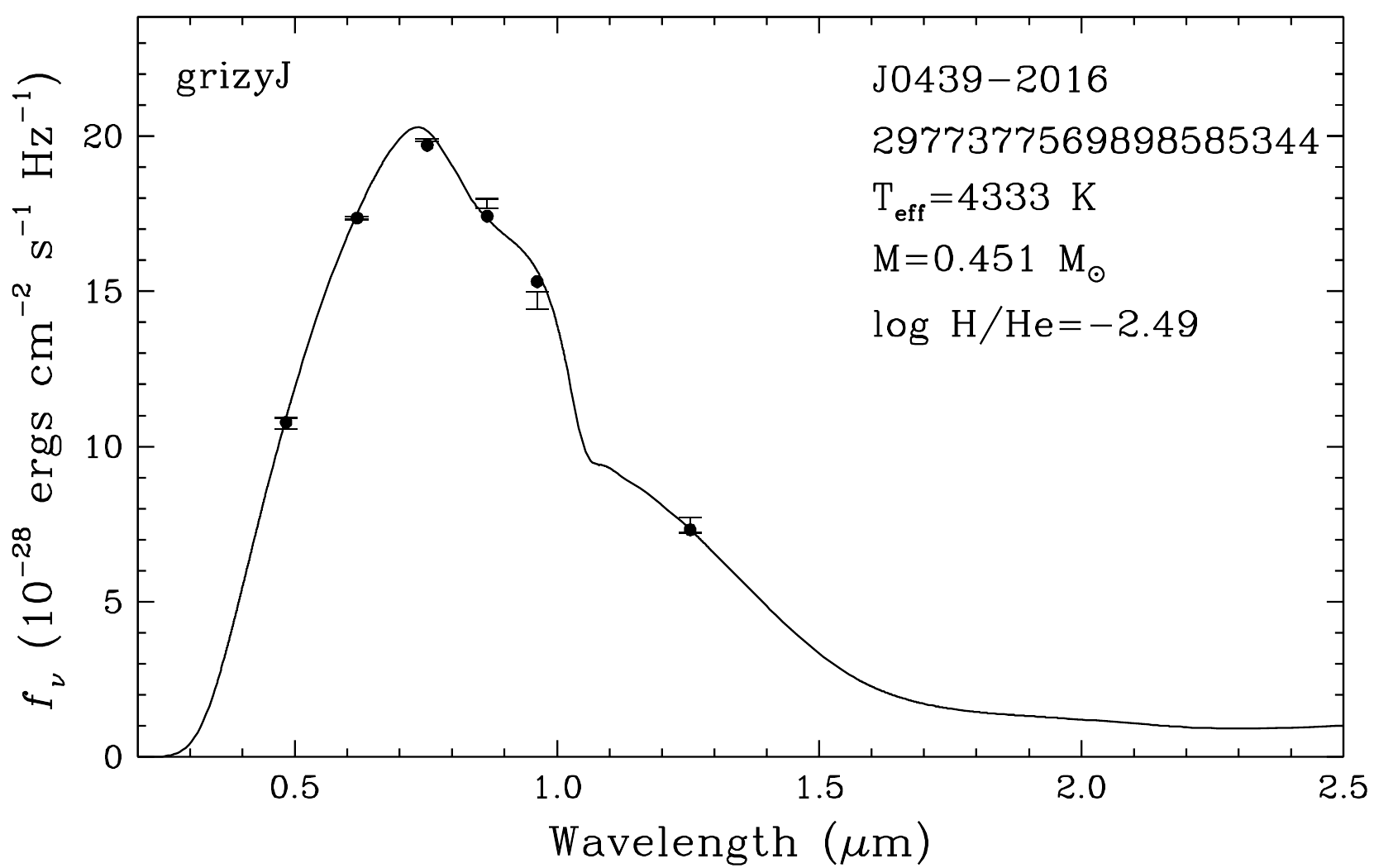}
\includegraphics[width=2.4in]{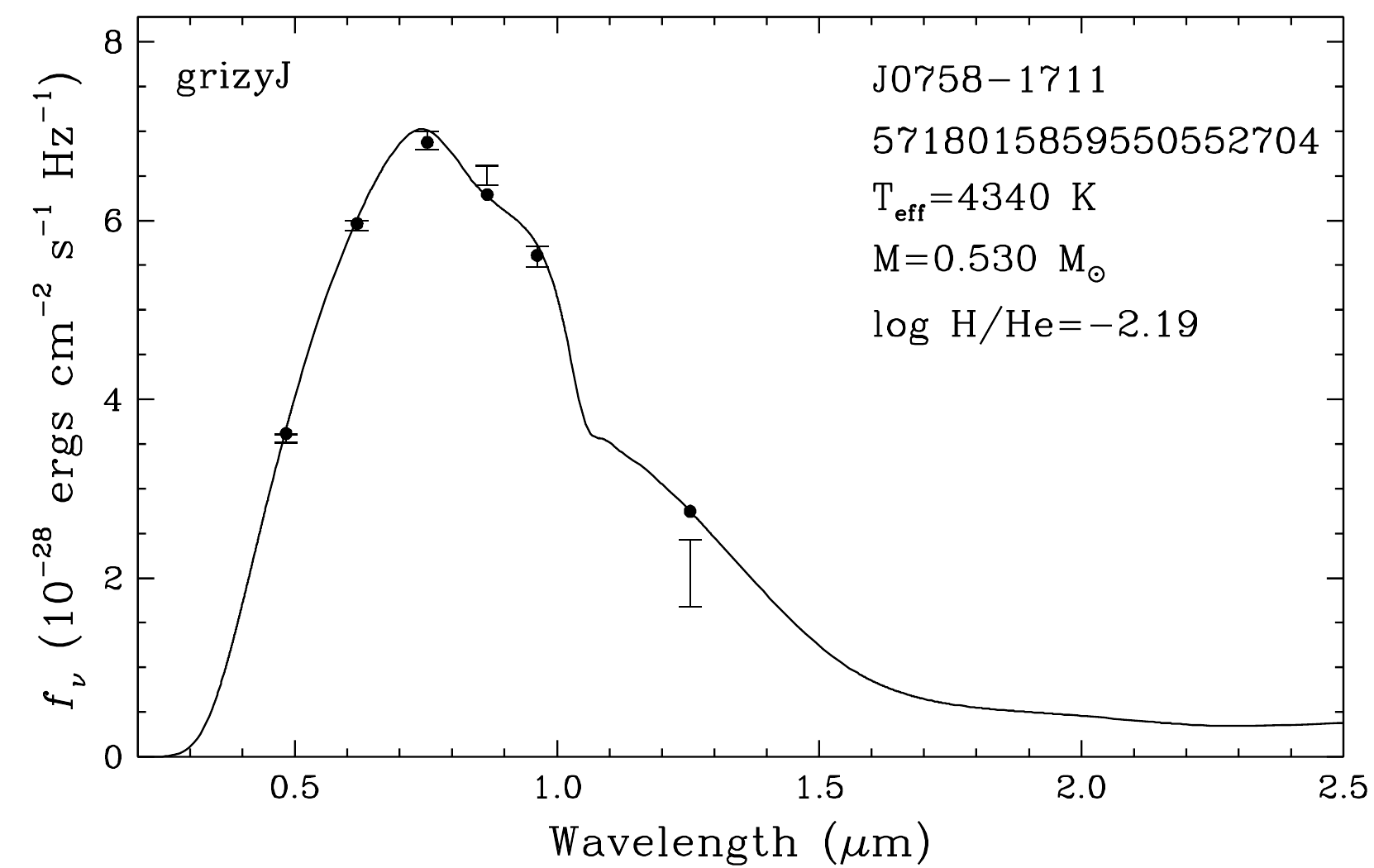}
\includegraphics[width=2.4in]{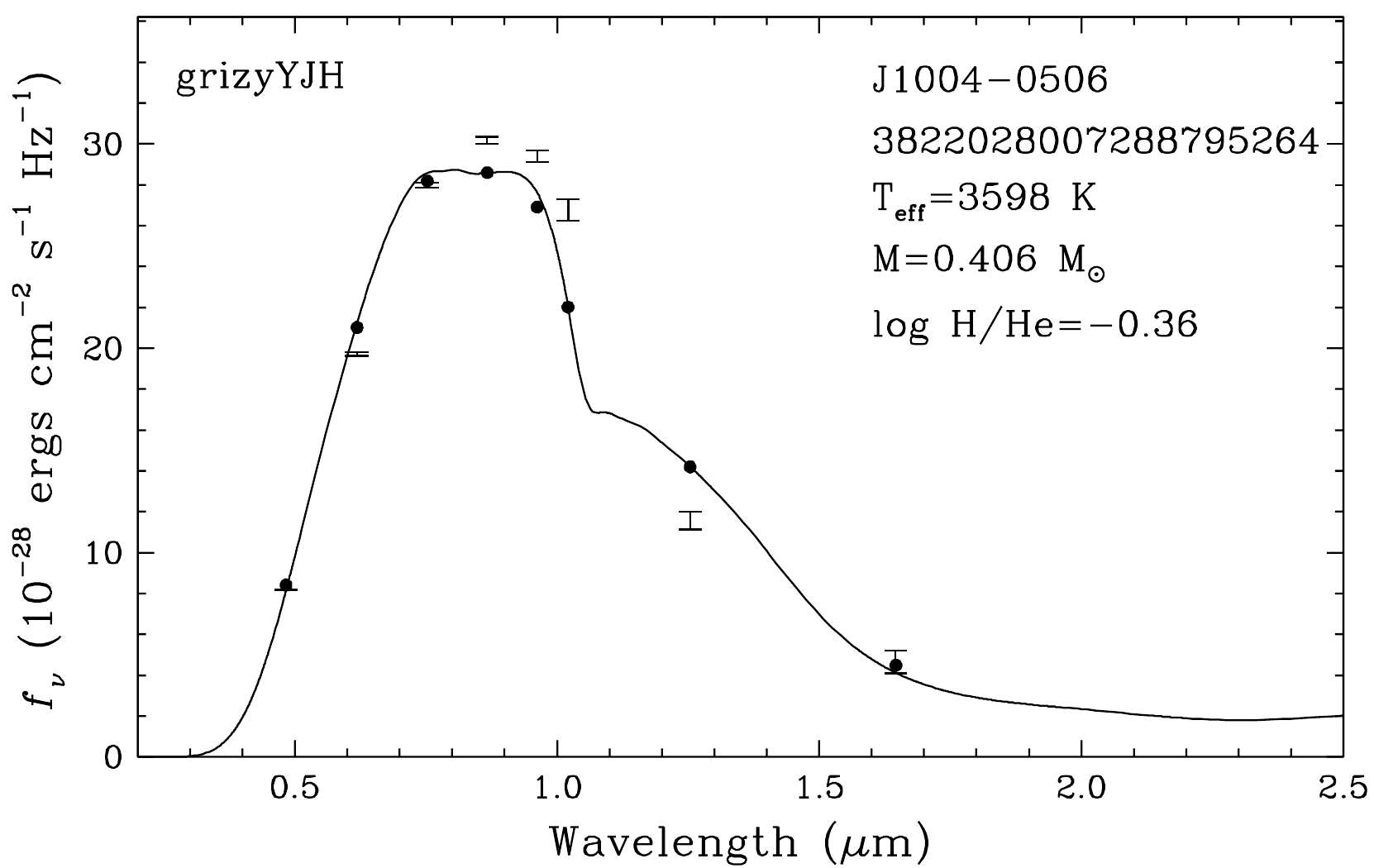}
\includegraphics[width=2.4in]{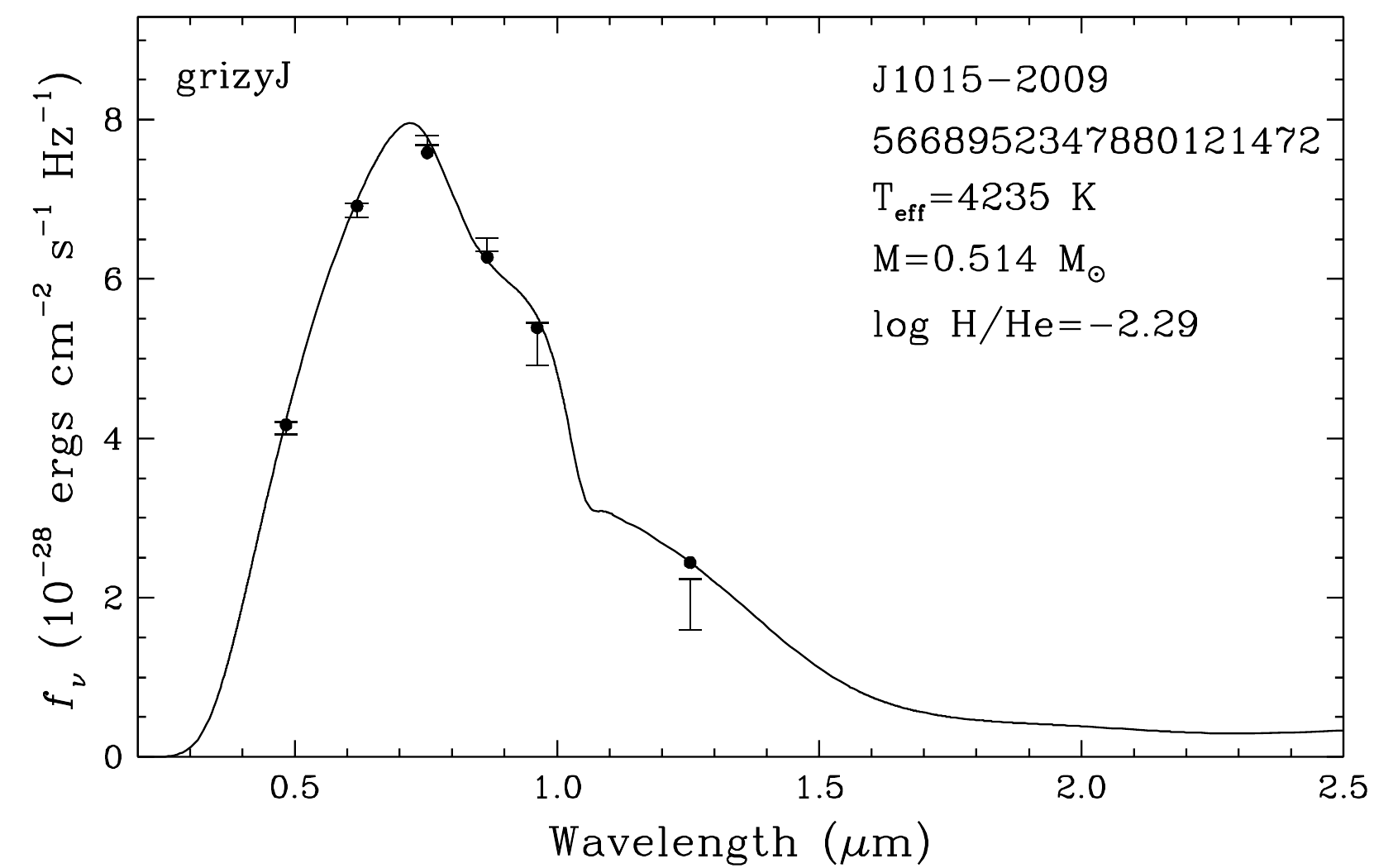}
\includegraphics[width=2.4in]{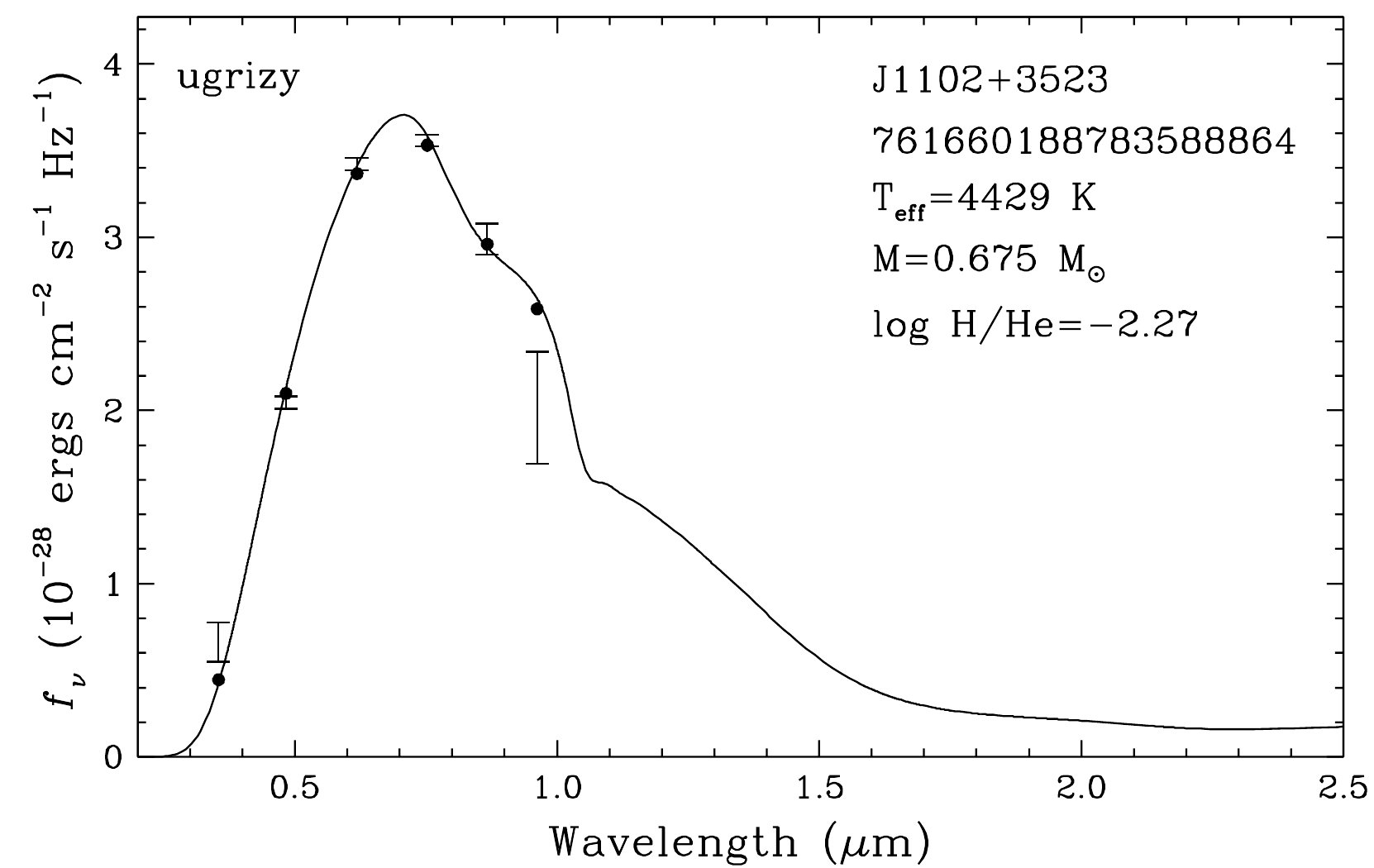}
\includegraphics[width=2.4in]{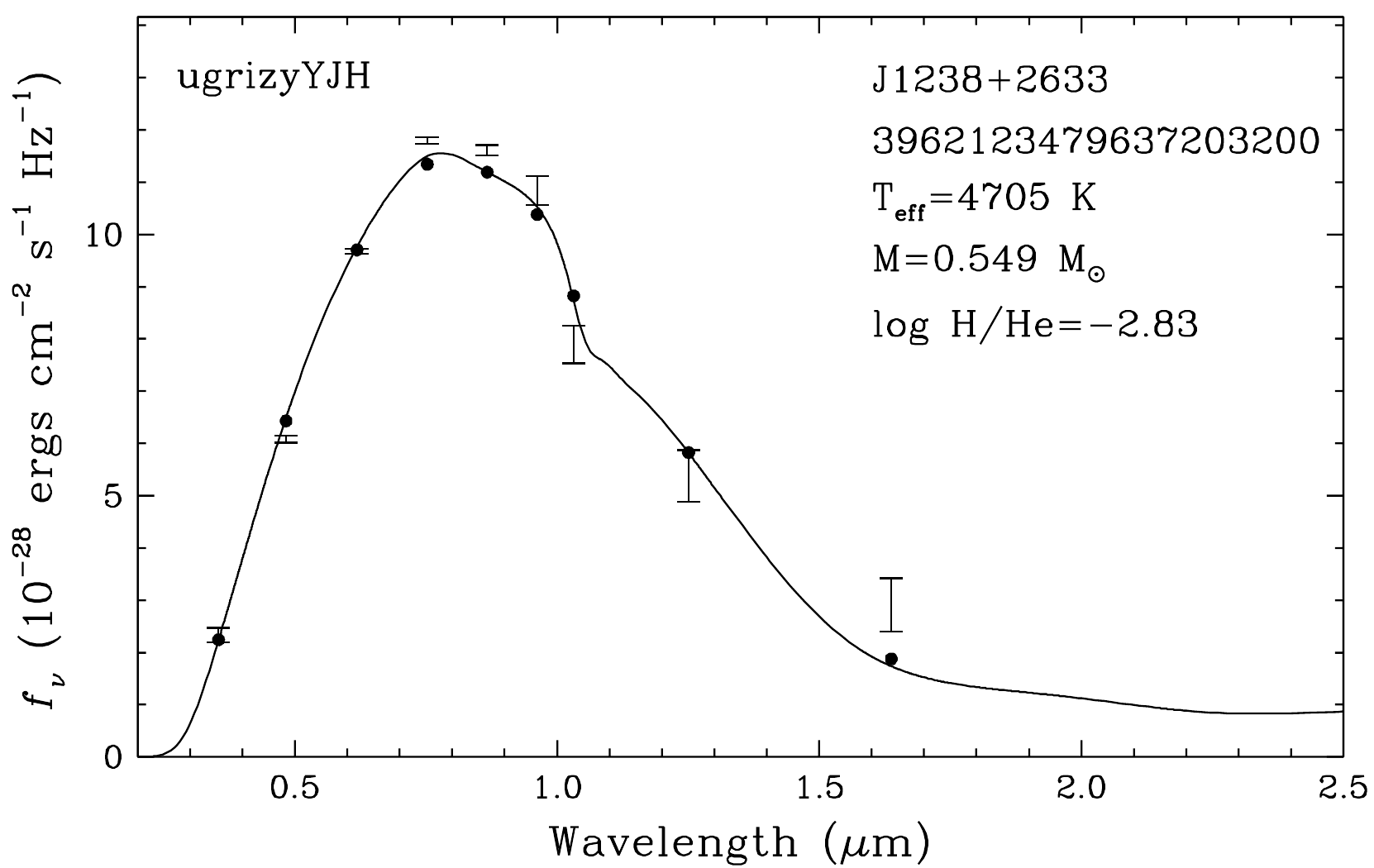}
\includegraphics[width=2.4in]{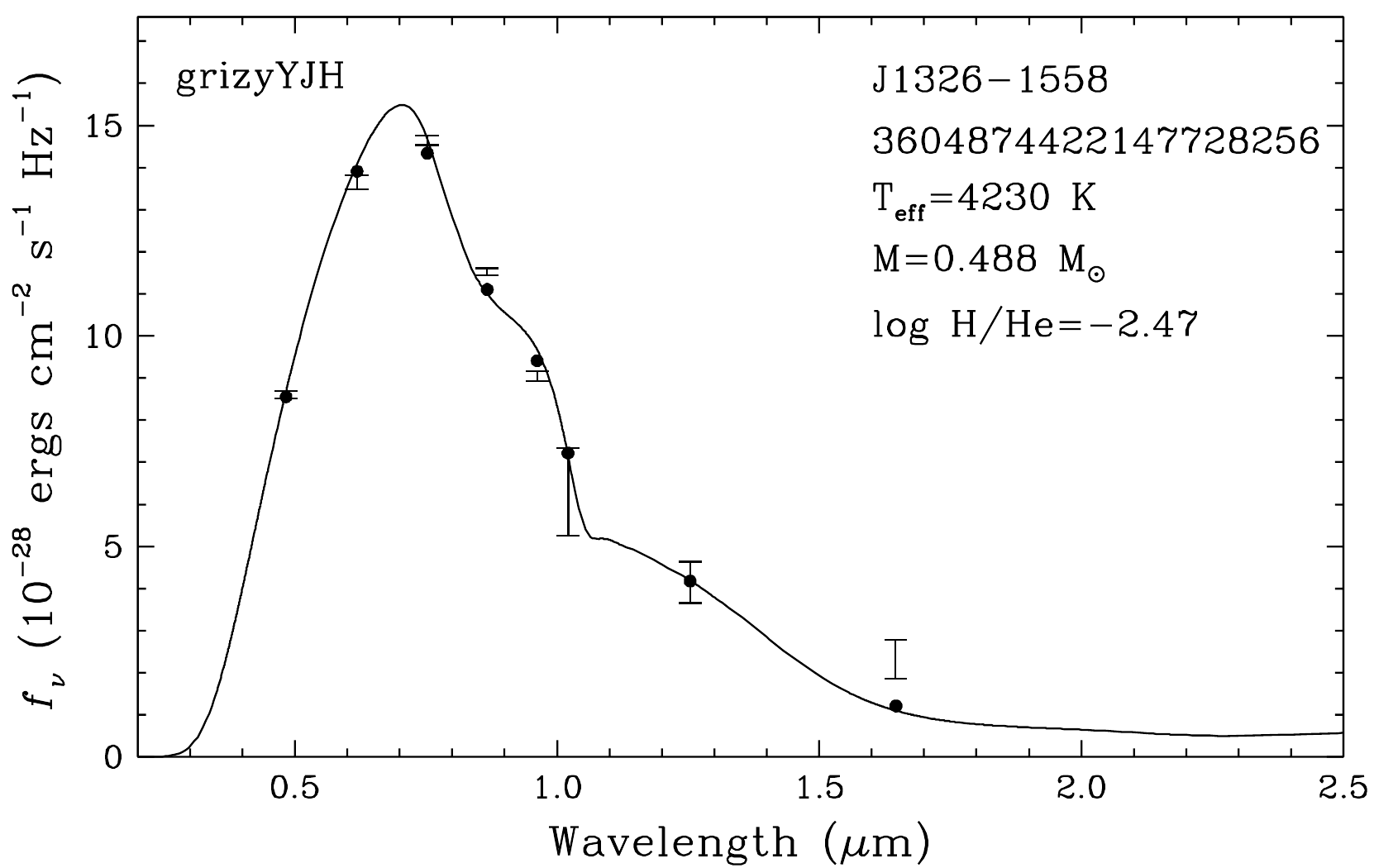}
\includegraphics[width=2.4in]{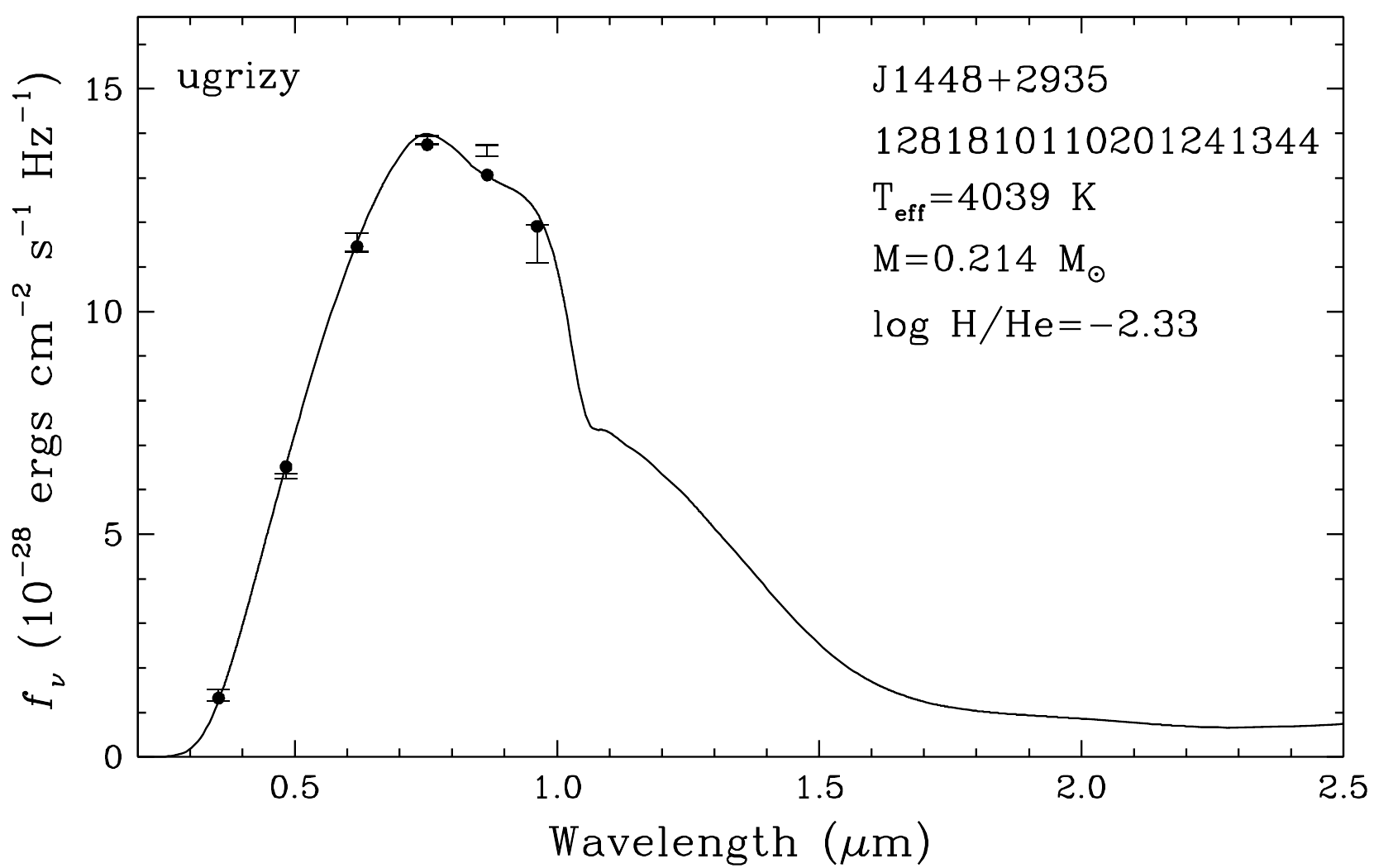}
\includegraphics[width=2.4in]{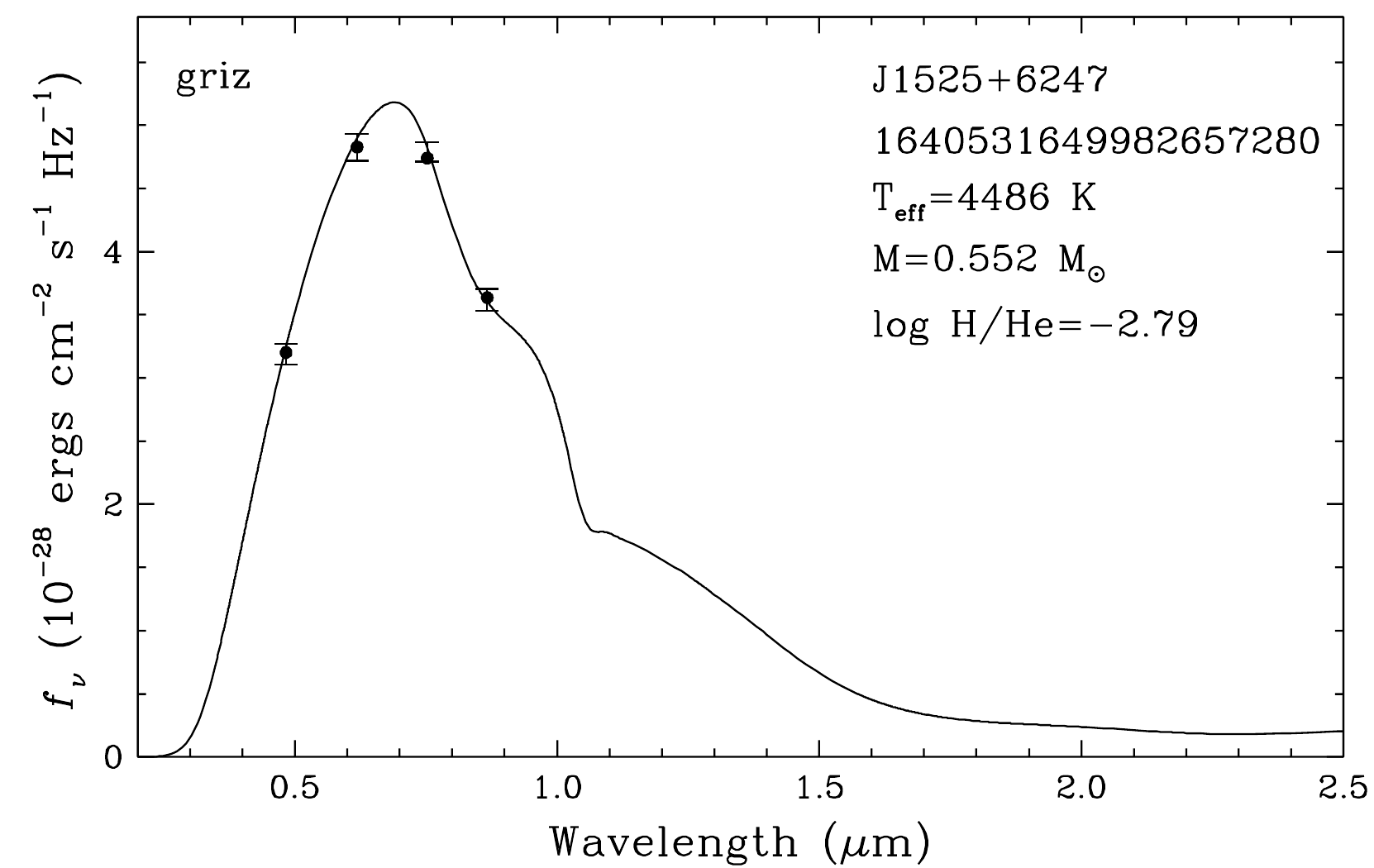}
\includegraphics[width=2.4in]{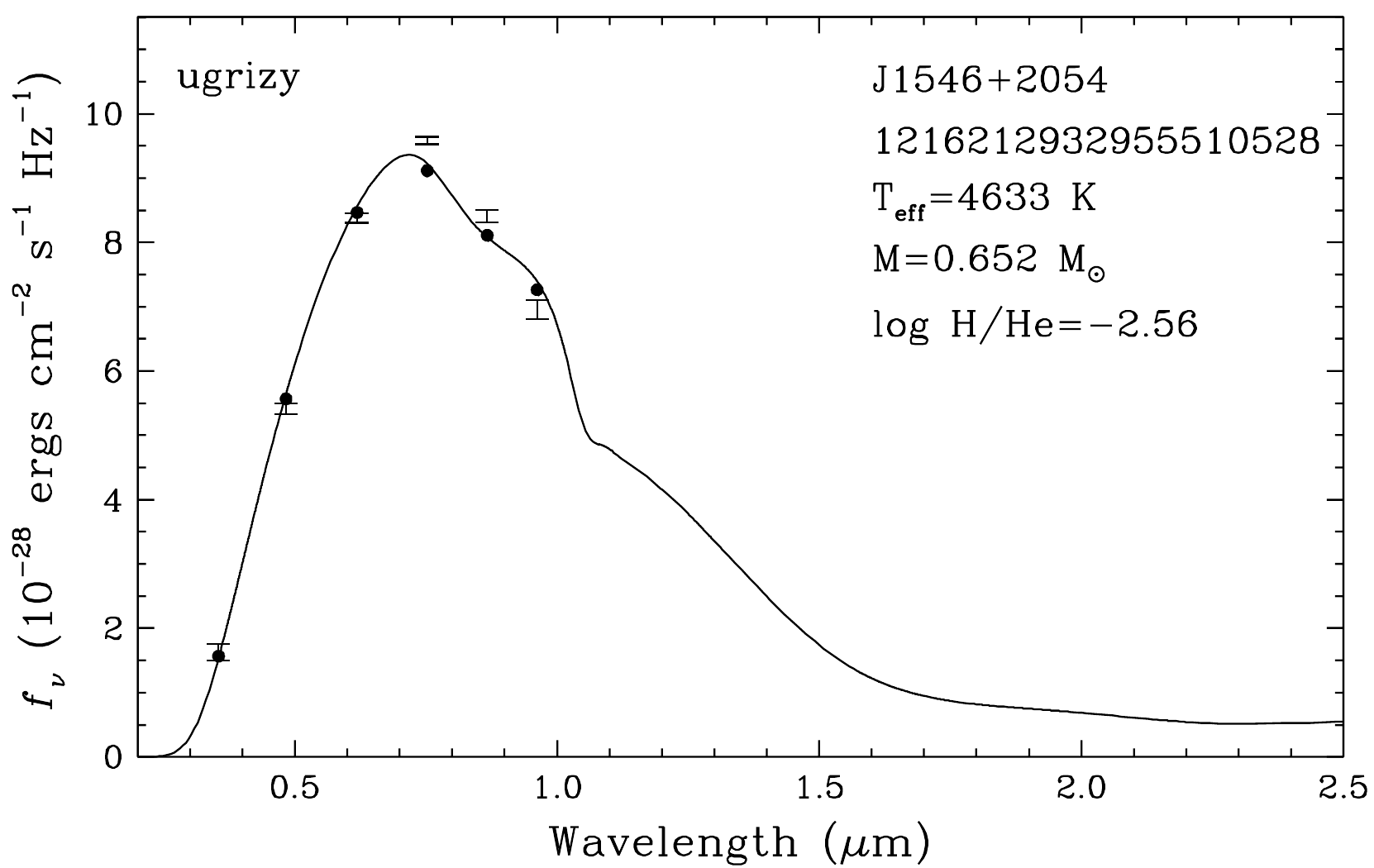}
\caption{Additional IR-faint white dwarf candidates identified in our
  analysis. Three of these objects, J0027+0554, J1004$-$0506, and
  J1951+4026, are spectroscopically confirmed DC white dwarfs.\label{fitcan}}
\end{figure*}

\begin{figure*}
\addtocounter{figure}{-1}
\includegraphics[width=2.4in]{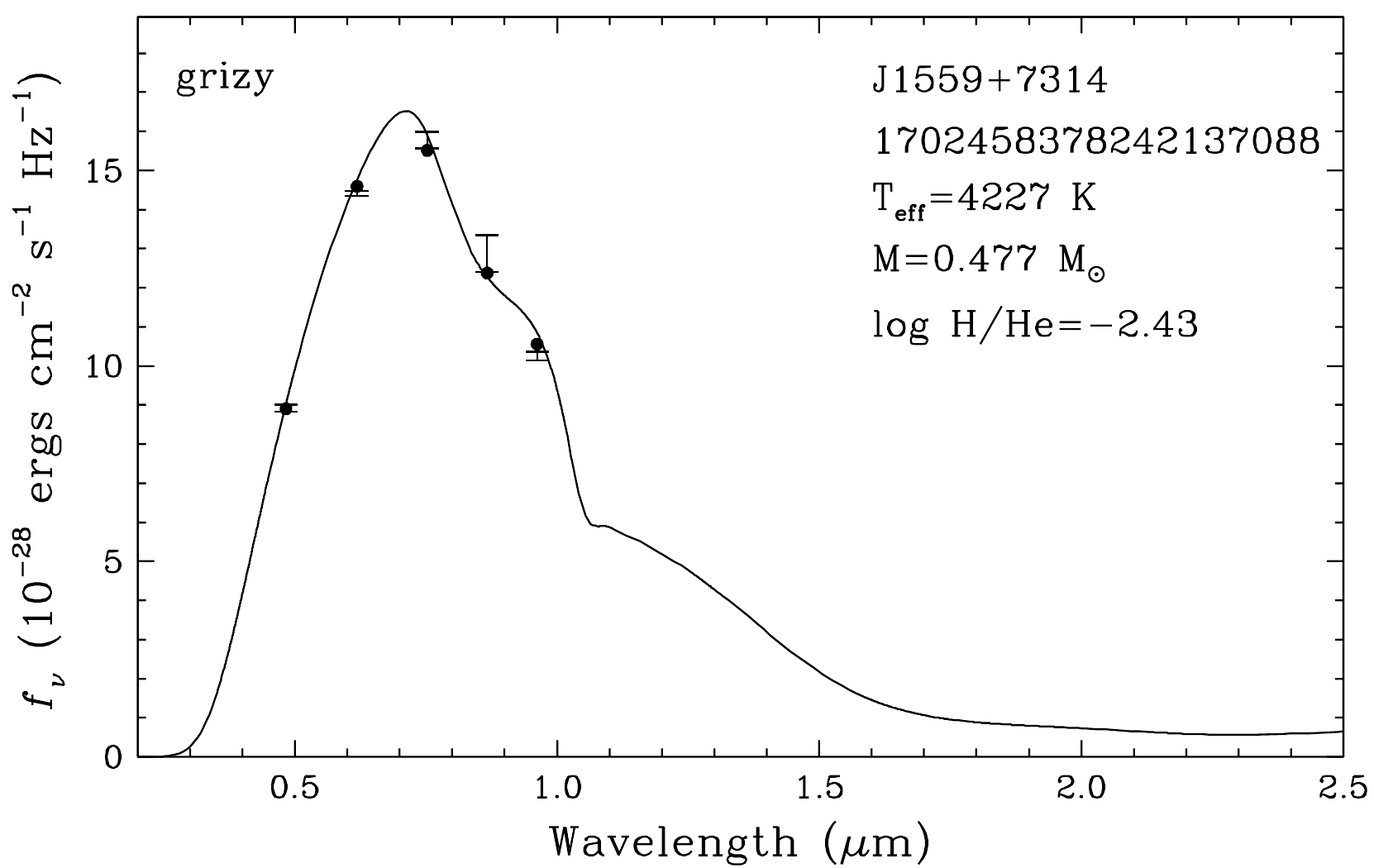}
\includegraphics[width=2.4in]{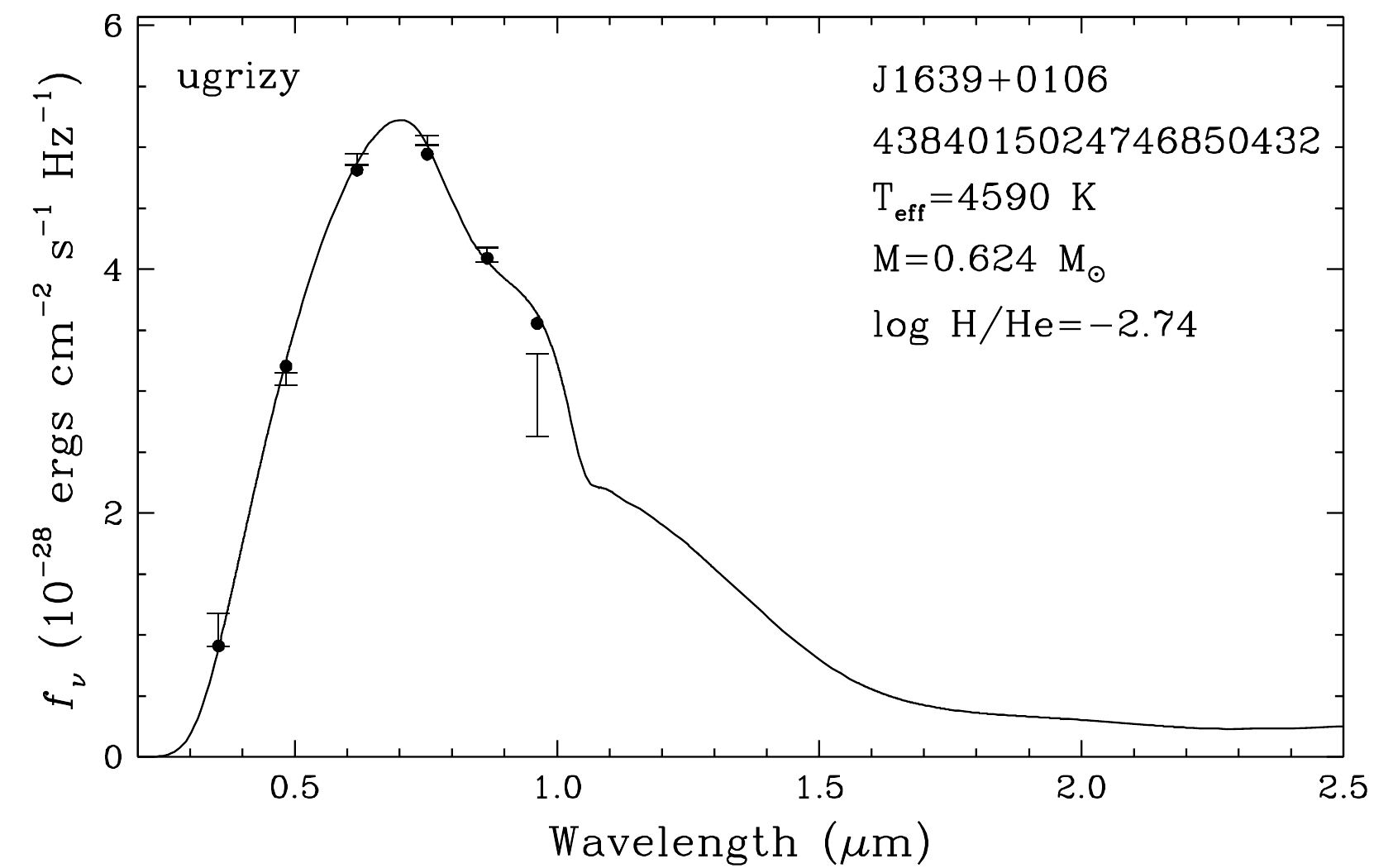}
\includegraphics[width=2.4in]{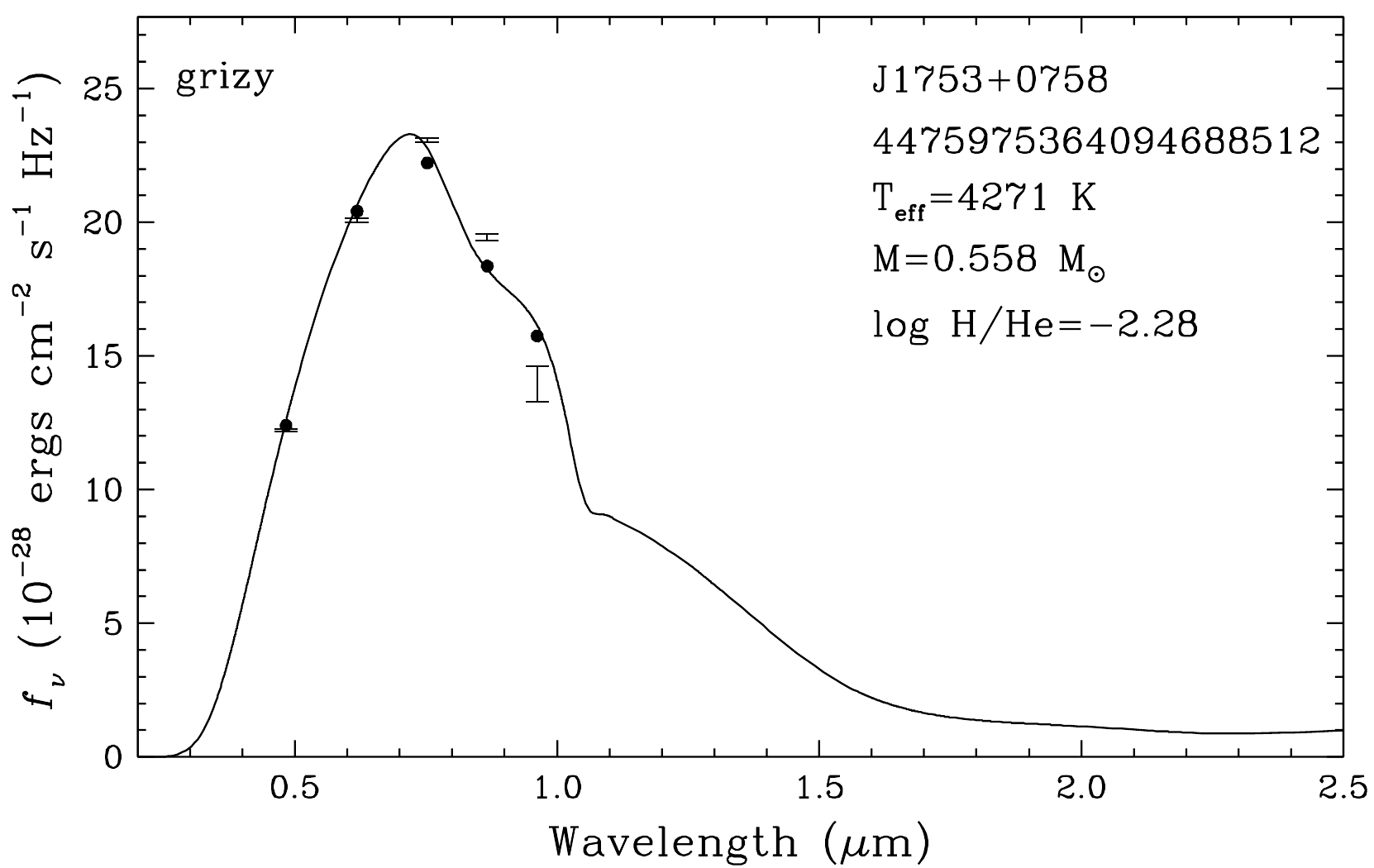}
\includegraphics[width=2.4in]{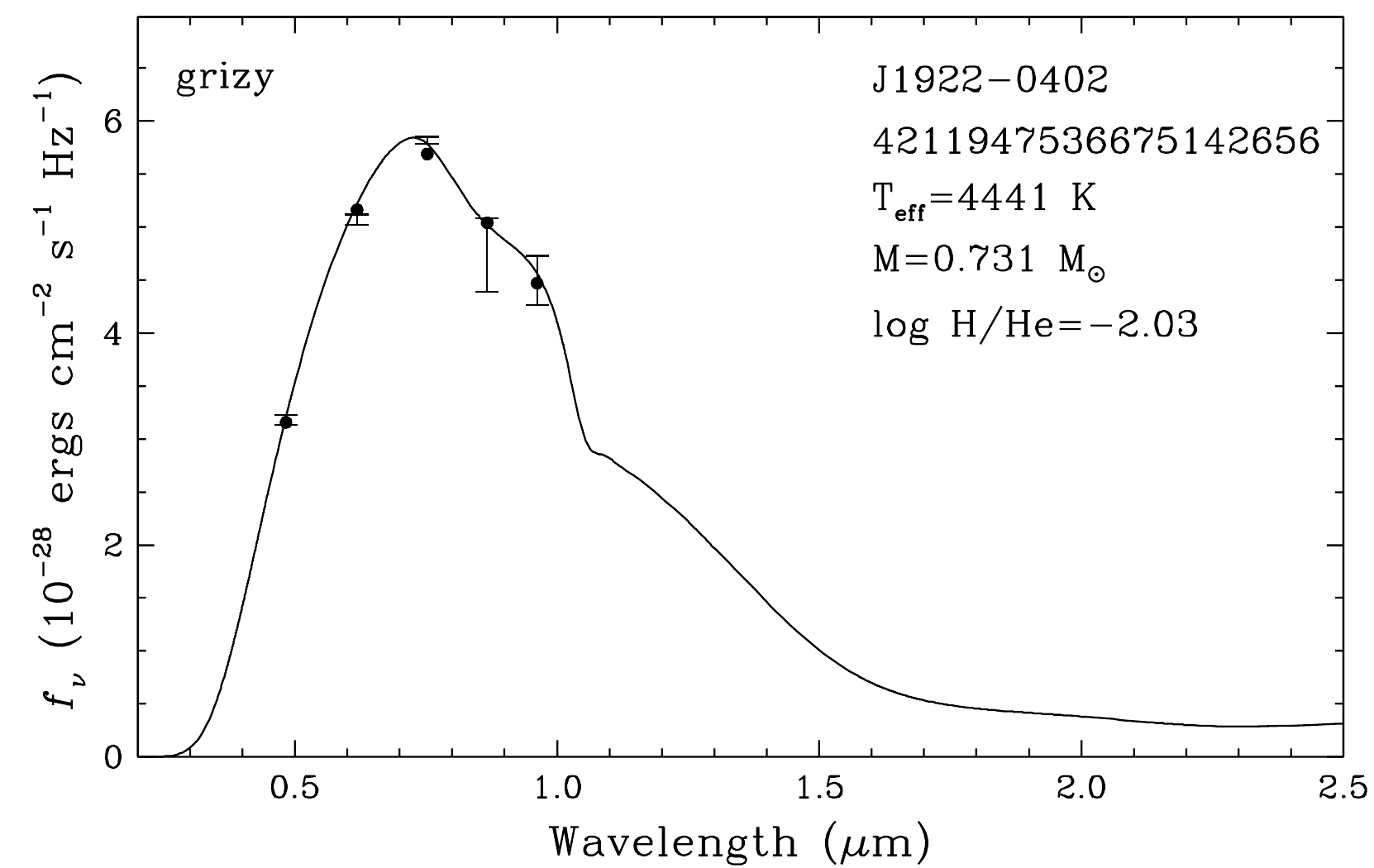}
\includegraphics[width=2.4in]{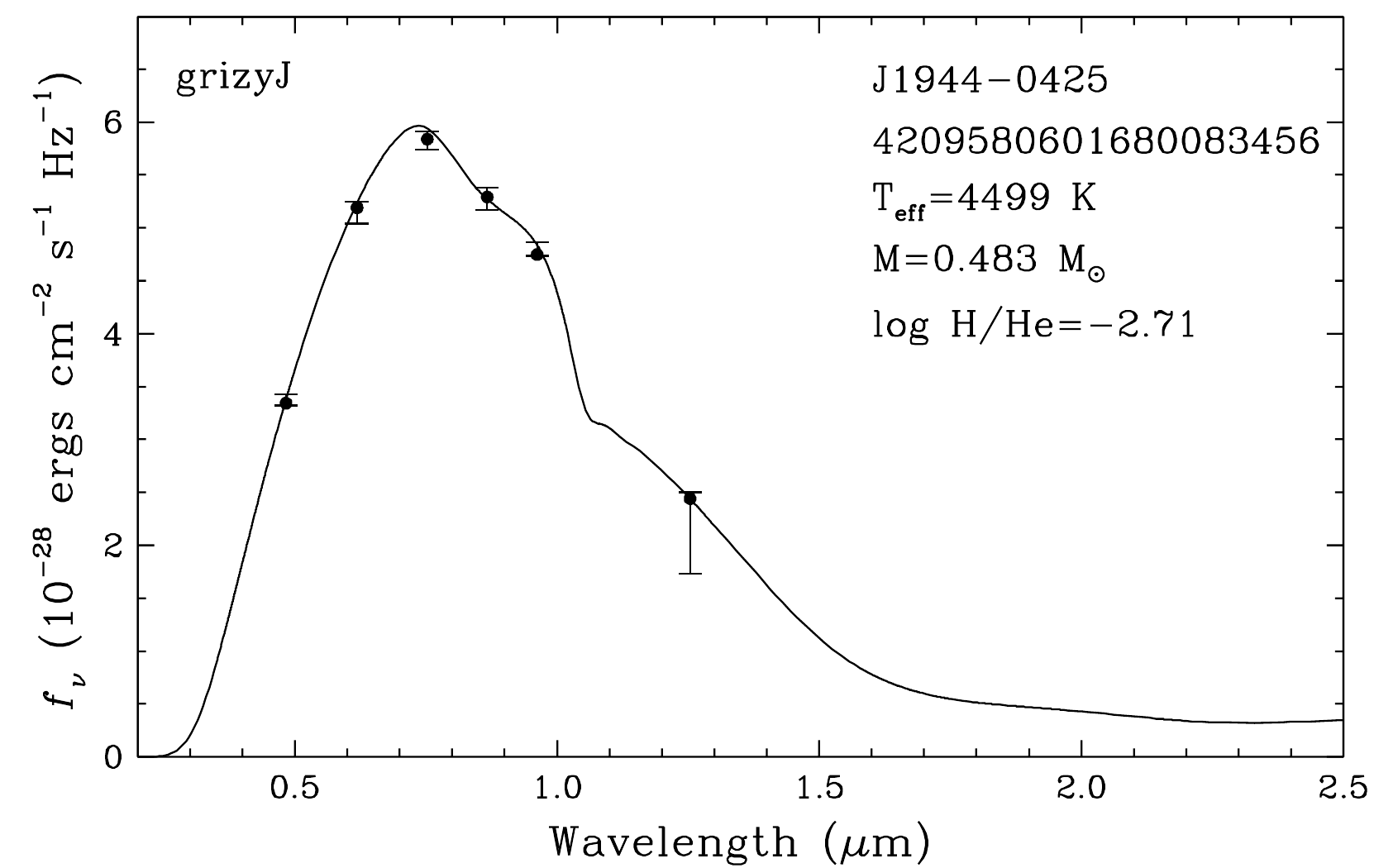}
\includegraphics[width=2.4in]{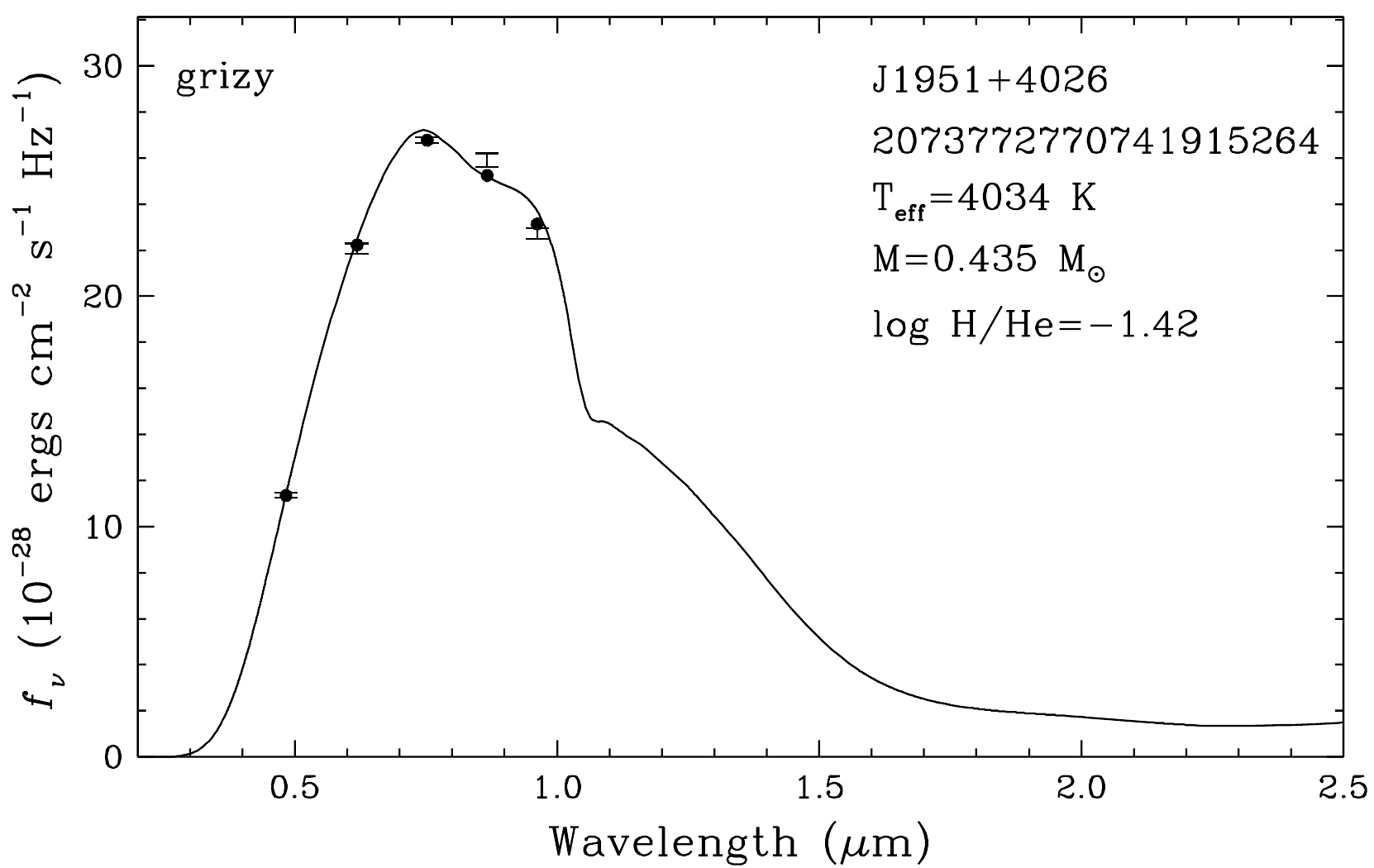}
\includegraphics[width=2.4in]{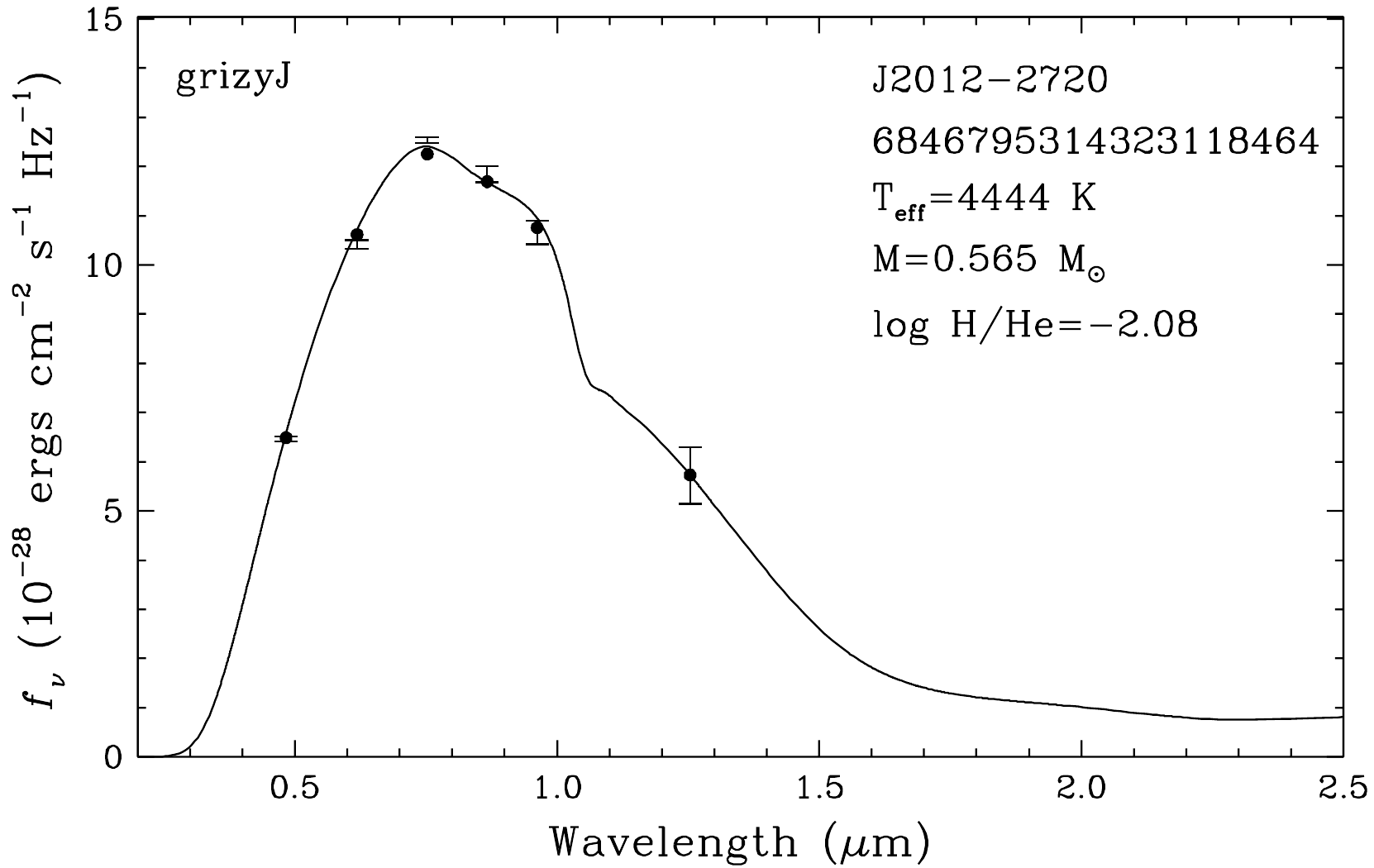}
\includegraphics[width=2.4in]{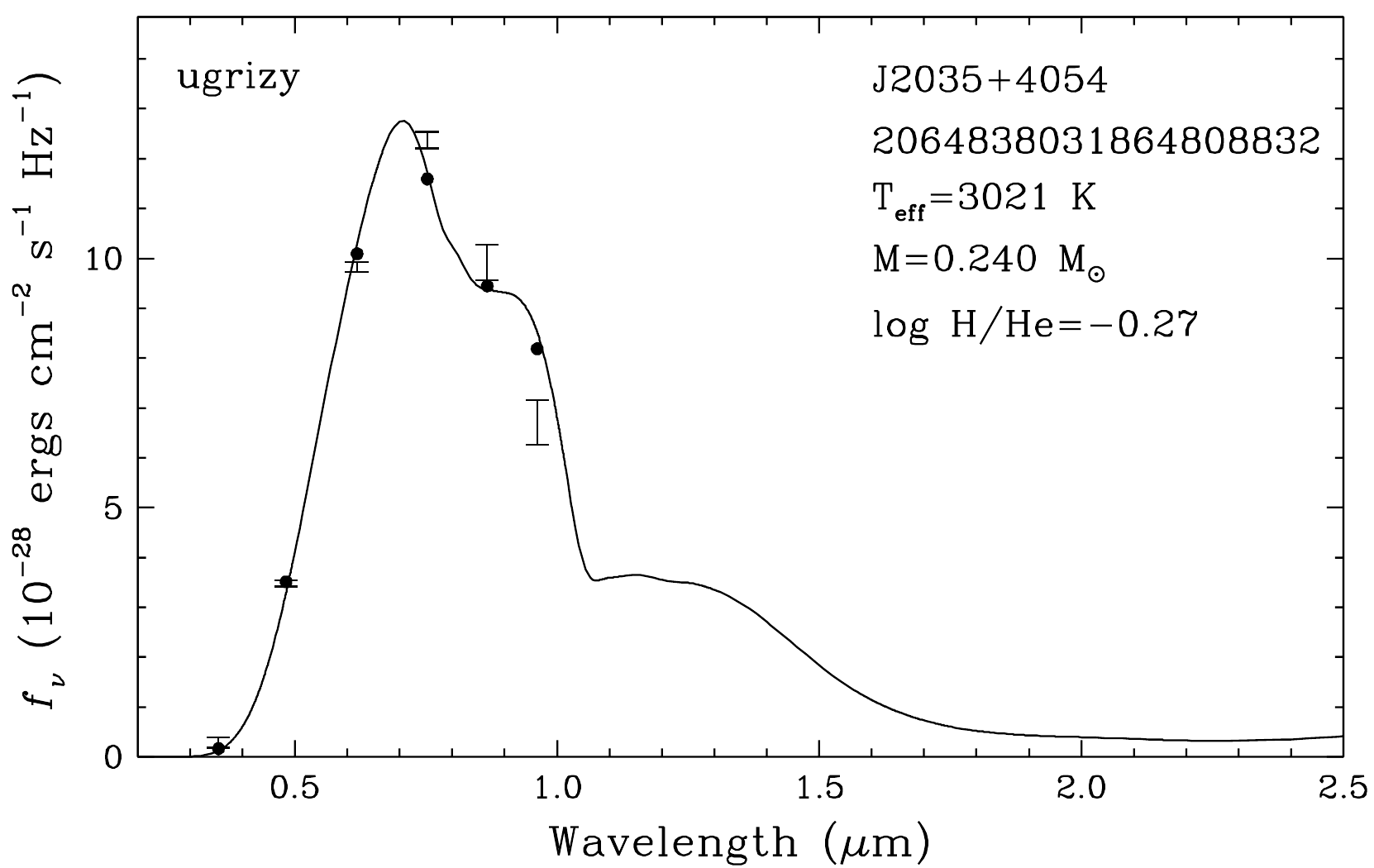}
\includegraphics[width=2.4in]{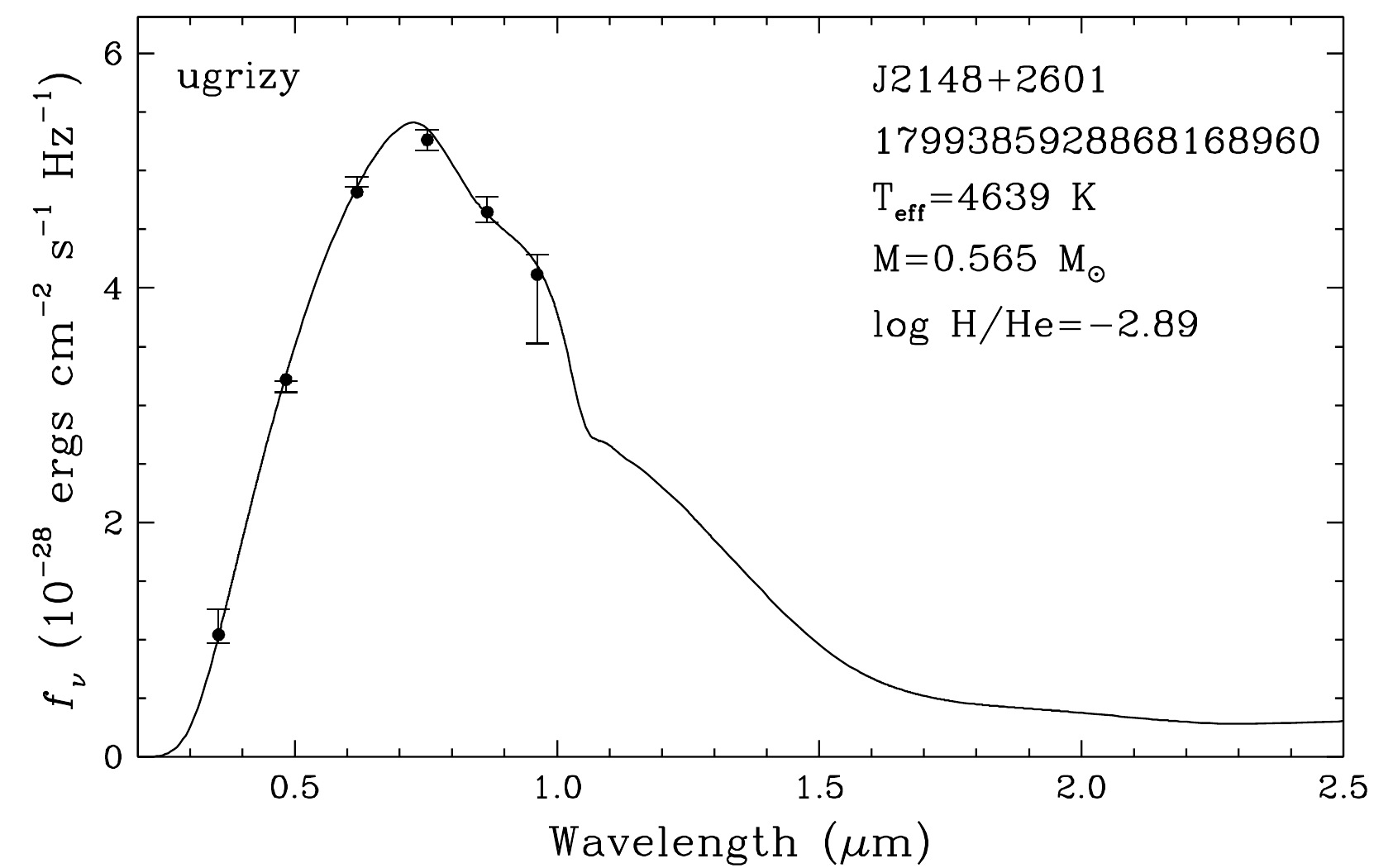}
\includegraphics[width=2.4in]{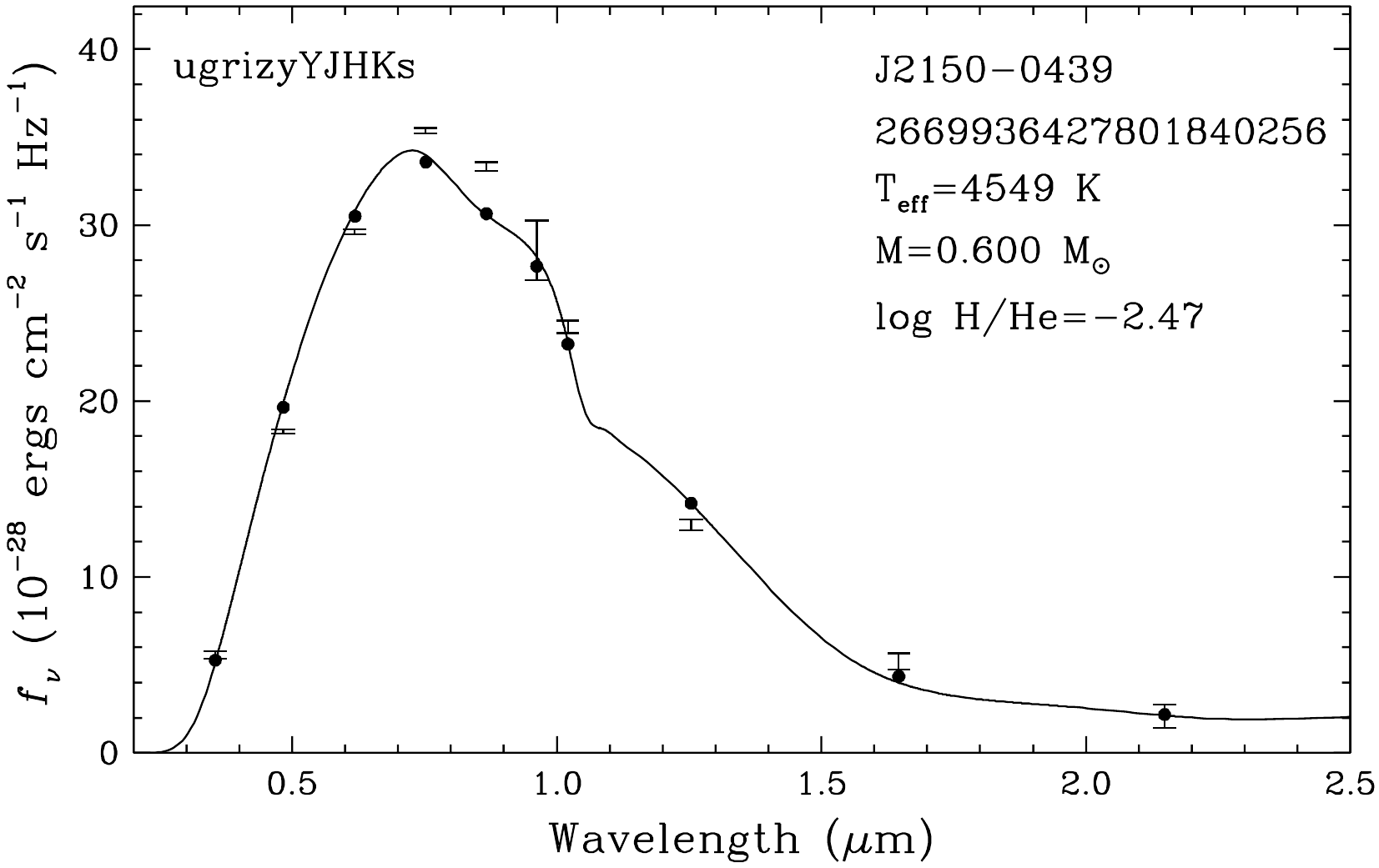}
\includegraphics[width=2.4in]{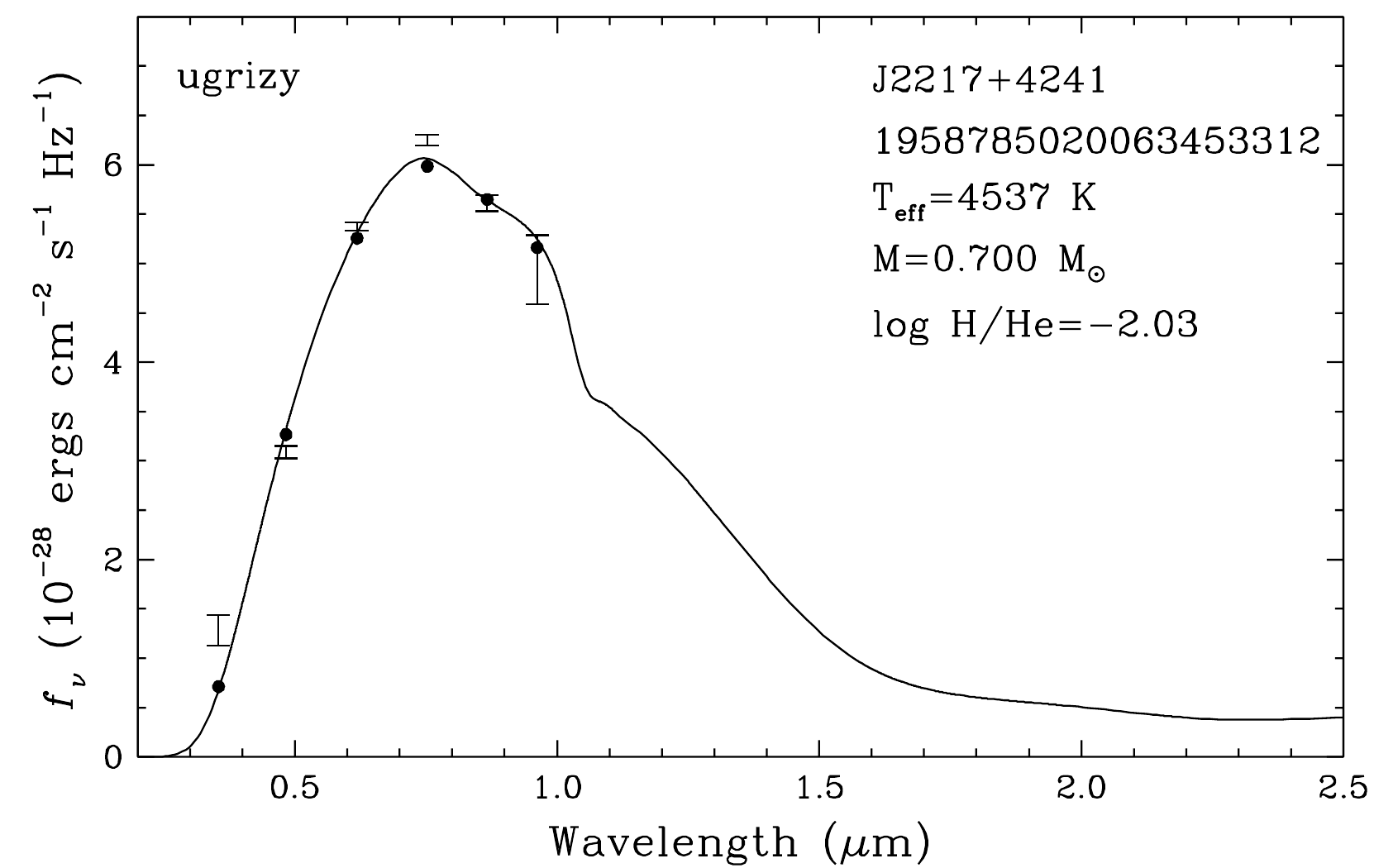}
\includegraphics[width=2.4in]{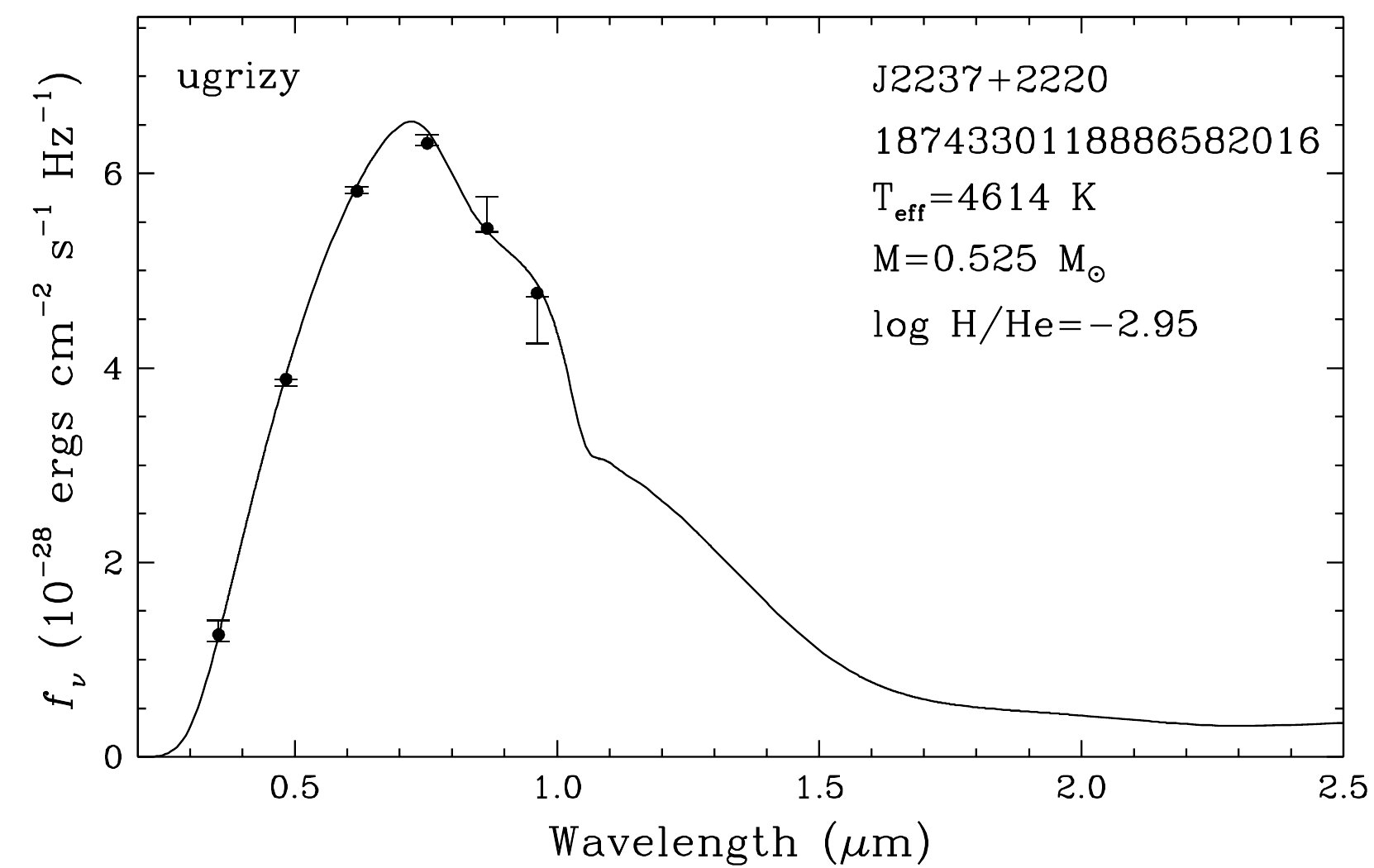}
\includegraphics[width=2.4in]{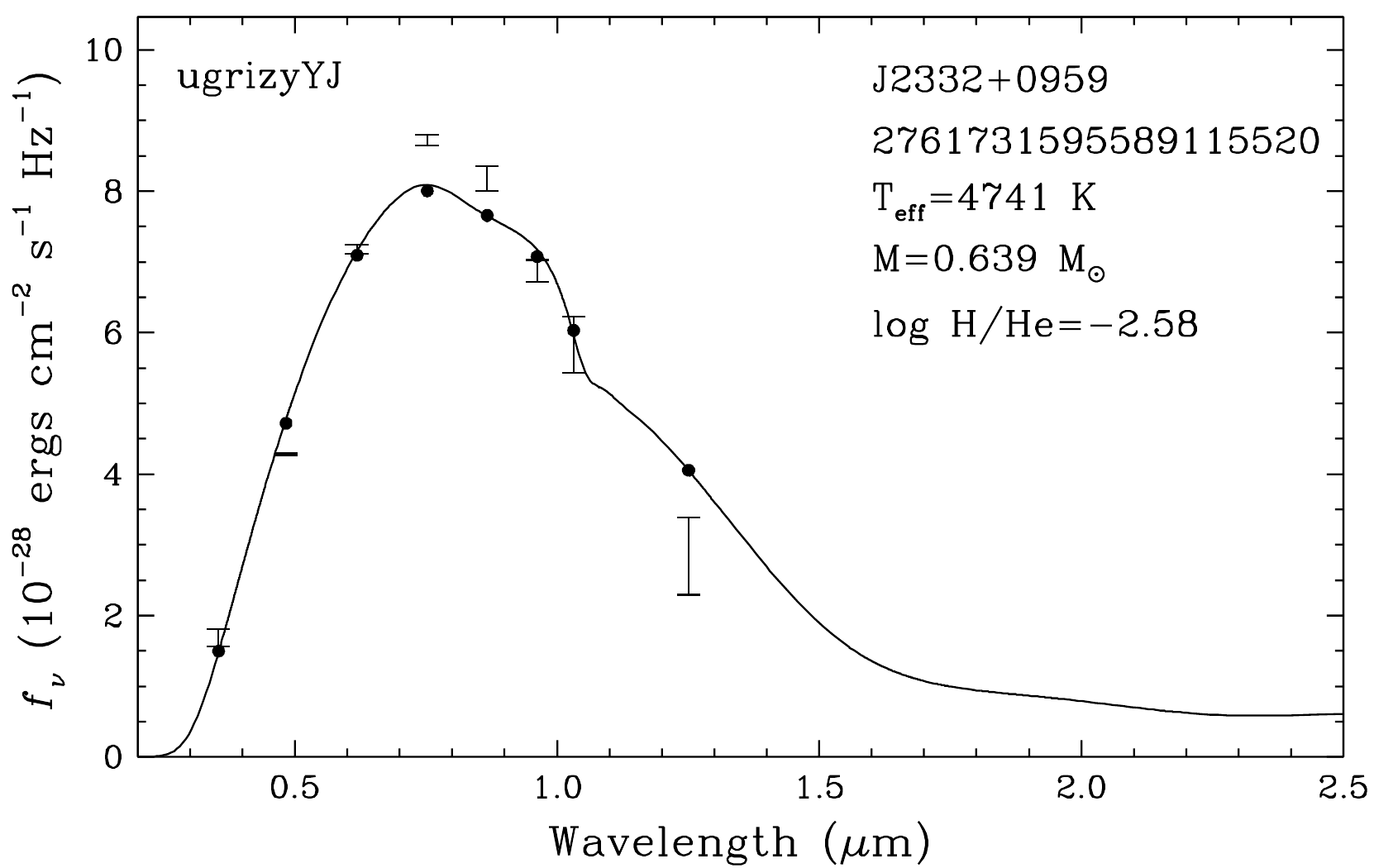}
\includegraphics[width=2.4in]{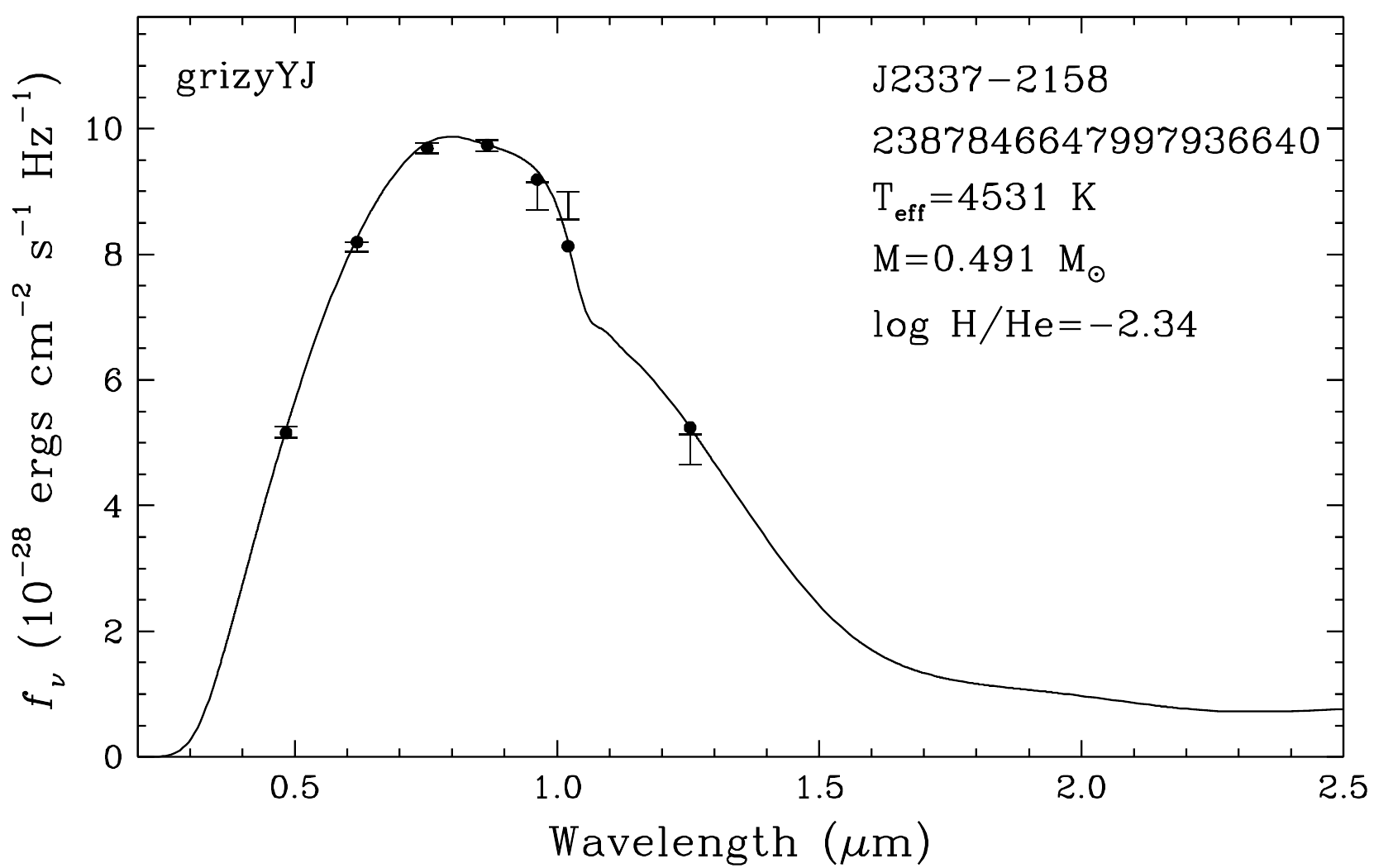}
\includegraphics[width=2.4in]{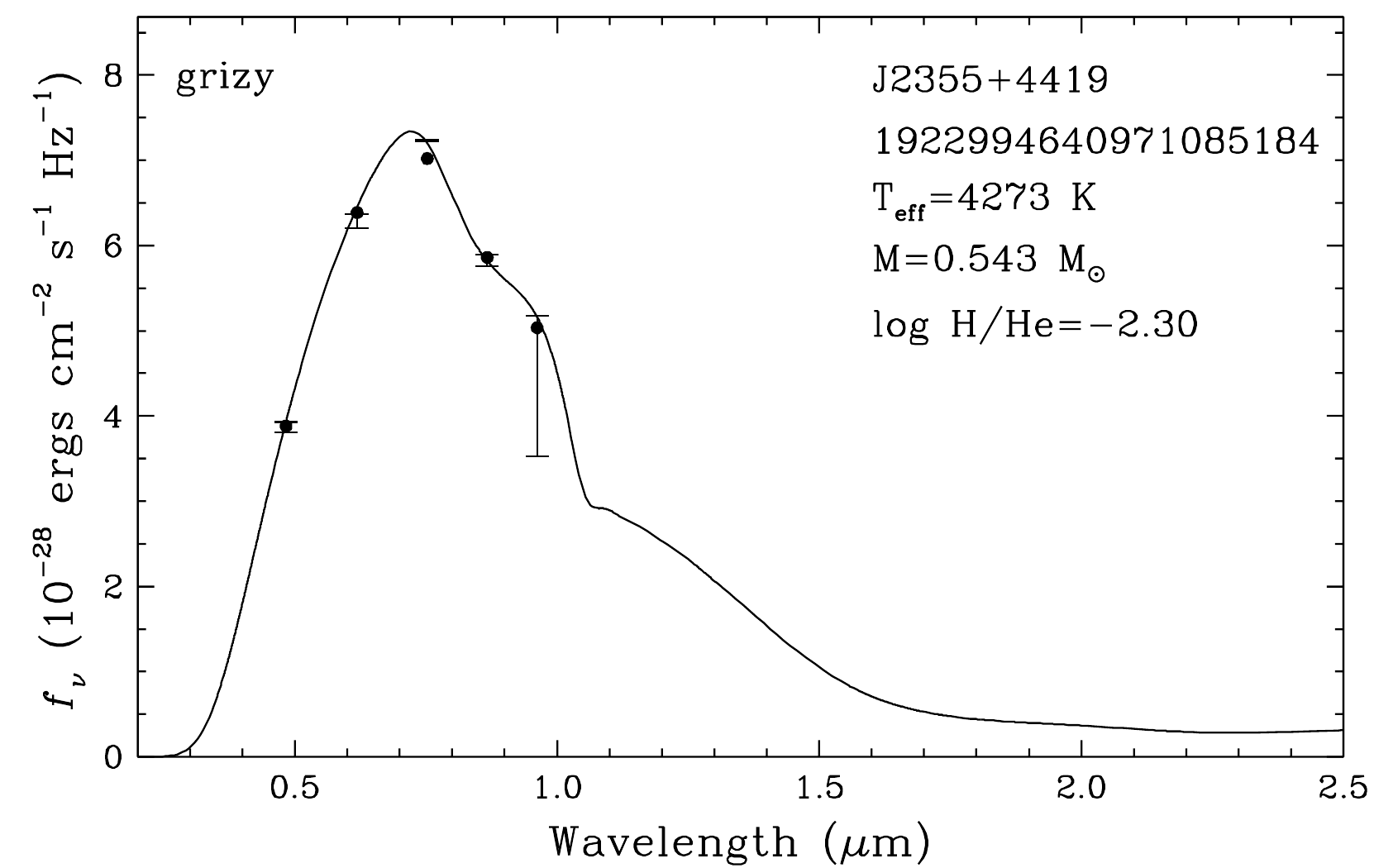}
\caption{Continued.}
\end{figure*}

\end{document}